\documentclass[ngerman,english,BCOR=6mm,cdgeometry=no,cdtitle=false,cdfont=no,DIV=13,bibliography=totoc,captions=tableheading, twoside=false]{tudscrbook}
\usepackage[T1]{fontenc}
\usepackage{selinput}\SelectInputMappings{adieresis={ä},germandbls={ß}}
\usepackage[english]{babel}
\usepackage{isodate}
\usepackage{blindtext}
\usepackage{siunitx}
\usepackage{mathtools}
\usepackage{array}
\usepackage{overpic}
\usepackage{epstopdf}
\usepackage{soul,color}
\usepackage{textgreek}
\usepackage{tikz}
\usepackage{lpic}
\usepackage{nicefrac}
\usepackage{ellipse}
\usepackage{amssymb}
\usepackage{physics}
\usepackage{ragged2e}
\usepackage[style=numeric-comp,backend=biber,natbib=true,sortlocale=de_DE,sorting=none,url=false,eprint=false]{biblatex}
\usepackage[colorlinks=true,linktoc=all,linkcolor=black,citecolor=black,urlcolor=black]{hyperref}
\usepackage[nogroupskip,toc]{glossaries}

\glsnoexpandfields
\setacronymstyle{long-short}
\newacronym{DE-LCI}{DE-LCI}{dispersion-encoded low-coherence interferometry}
\newacronym{MEMS}{MEMS}{micro-electromagnetical systems}
\newacronym{OPD}{OPD}{optical path difference}
\newacronym{DE}{DE}{dispersive element}
\newacronym{LCI}{LCI}{low-coherence interferometry}
\newacronym{SLD}{SLD}{superluminescent diode}
\newacronym{RMS}{RMS}{root-mean square error}
\newacronym{ROI}{ROI}{region-of-interest}
\newacronym{DR}{DR}{dynamic range}
\newacronym{STFT}{STFT}{short-time Fourier-transform}
\newacronym{SNR}{SNR}{signal-to-noise ratio}
\newacronym{FFT}{FFT}{Fast Fourier-transform}
\newacronym{SSE}{SSE}{error sum of squares}
\newacronym{AFM}{AFM}{atomic force mircroscopy}
\newacronym{CLSM}{CLSM}{confocal laser scanning microscopy}
\newacronym{NA}{NA}{numerical aperture}
\newacronym{DHM}{DHM}{digital holographic microscopy}
\newacronym{PSI}{PSI}{phase-shifting interferometry}
\newacronym{CSI}{CSI}{coherence scanning interferometry}
\newacronym{OCT}{OCT}{optical-coherence tomography}
\newacronym{TD-OCT}{TD-OCT}{time-domain optical-coherence tomography}
\newacronym{SD-OCT}{SD-OCT}{spectral-domain optical-coherence tomography}
\newacronym{FD-OCT}{FD-OCT}{frequency-domain optical-coherence tomography}
\newacronym{DEFR-OCT}{DEFR-OCT}{dispersion-encoded full-range OCT}
\newacronym{FF-OCT}{FF-OCT}{full-field OCT}
\newacronym{DSC}{DSC}{differential scanning caliometry}
\newacronym{DMA}{DMA}{dynamic mechanical analysis}
\newacronym{FTIR}{FTIR}{Fourier-transform infrared spectroscopy}
\newacronym{NMR}{NMR}{nuclear magnetic resonance}
\newacronym{ToF-SIMS}{ToF-SIMS}{time-of-flight secondary ion mass spectroscopy}
\newacronym{ITO}{ITO}{indium-tin oxide}
\newacronym{PTB}{PTB}{Physikalisch-Technische Bundesanstalt}
\newacronym{ACF}{ACF}{auto-correlation function}
\newacronym{PSDf}{PSDf}{power-spectral-density function}
\newacronym{Si}{Si}{silicon}
\newacronym{Ti}{Ti}{titanium}
\newacronym{GaAs}{GaAs}{gallium arsenide}
\newacronym{InGaAs}{InGaAs}{indium gallium arsenide}
\newacronym{ZnSe}{ZnSe}{zinc selenide}
\newacronym{VIS}{VIS}{visible spectral range}
\newacronym{NIR}{NIR}{near-infrared spectral range}
\newacronym{CMOS}{CMOS}{complementary metal-oxide-semiconductor}
\newacronym{SEM}{SEM}{scanning electron microscope}
\newacronym{WDM}{WDM}{wavelength division multiplexing}
\newacronym{SE}{SE}{spectroscopic ellipsometry}
\newacronym{SC}{SC}{supercontinuum white-light source}
\newacronym{LDP}{LDP}{laser-driven plasma light source}
\newacronym{SU-8}{SU-8}{epoxy-based negative photoresist}
\newacronym{EVA}{EVA}{ethylene-vinyl acetate}
\newacronym{PV}{PV}{photo voltaics}
\newacronym{WPDE}{WPDE}{wrapped phase derivative evaluation}
\newacronym{RDOT}{RDOT}{relative derived optical thickness}
\newacronym{FWP}{FWP}{forward propagator}
\newacronym{PET}{PET}{polyethylene terephthalate}
\newacronym{ASC}{ASC}{amplified supercontinuum light source}
\newacronym{CdS}{CdS}{cadmium sulfide}
\newacronym{CIGS}{CIGS}{copper indium gallium selenide}
\newacronym{Ga}{Ga}{gallium}

\glsnoexpandfields
\newglossary[symbols]{symbol}{sym}{symbl}{List of symbols} 

\glsnoexpandfields
\newglossaryentry{CoherLen}{type=symbol, name={$l_c$}, description={coherence length}, symbol={$l_c$}}
\newglossaryentry{CenWaveLen}{type=symbol, name={$\lambda_c$}, description={center wavelength of a light source}, symbol={$\lambda_c$}}
\newglossaryentry{WaveLenRange}{type=symbol, name={$\Delta \lambda$}, description={wavelength range of a light source}, symbol={$\Delta \lambda$}}
\newglossaryentry{SampThickness}{type=symbol, name={$t_{smp}$}, description={sample thickness}, symbol={$t_{smp}$}}
\newglossaryentry{RefrInd}{type=symbol, name={$n(\lambda)$}, description={real part of the refractive index with respect to the wavelength}, symbol={$n(\lambda)$}}
\newglossaryentry{RefrIndImag}{type=symbol, name={$k(\lambda)$}, description={real part of the refractive index with respect to the wavelength}, symbol={$k(\lambda)$}}
\newglossaryentry{RefrIndSample}{type=symbol, name={$n_{smp}(\lambda)$}, description={refractive index of a sample with respect to the wavelength}, symbol={$n_{smp}(\lambda)$}}
\newglossaryentry{MolRef}{type=symbol, name={$R_m$}, description={molar refractivity}, symbol={$R_m$}}
\newglossaryentry{RealRefInd}{type=symbol, name={$n =$\,$n(\lambda)$}, description={real part of the refractive index}, symbol={$n$}}
\newglossaryentry{DeltaRealRefInd}{type=symbol, name={$\Delta n$}, description={change in the refractive index}, symbol={$\Delta n$}}
\newglossaryentry{MolMass}{type=symbol, name={$M_w$}, description={mass-averaged molar mass}, symbol={$M_w$}}
\newglossaryentry{Density}{type=symbol, name={$\rho$}, description={density}, symbol={$\rho$}}
\newglossaryentry{Stress}{type=symbol, name={$\sigma$}, description={mechanical stress}, symbol={$\sigma$}}
\newglossaryentry{Strain}{type=symbol, name={$\varepsilon$}, description={mechanical strain}, symbol={$\varepsilon$}}
\newglossaryentry{PhiDMA}{type=symbol, name={$\Phi$}, description={phase difference in dynamic mechanical analysis}, symbol={$\Phi$}}
\newglossaryentry{CompModulus}{type=symbol, name={$E^*$}, description={complex modulus of a sample in dynamic mechanical analysis}, symbol={$E^*$}}
\newglossaryentry{StorModulus}{type=symbol, name={$E^\prime$}, description={storage modulus of a sample in dynamic mechanical analysis}, symbol={$E^\prime$}}
\newglossaryentry{LossModulus}{type=symbol, name={$E^{\prime\prime}$}, description={loss modulus of a sample in dynamic mechanical analysis}, symbol={$E^{\prime\prime}$}}
\newglossaryentry{GlassTranTemp}{type=symbol, name={$T_G$}, description={glass transition temperature of a material}, symbol={$T_G$}}
\newglossaryentry{IPeakRaman}{type=symbol, name={$I_{peak}$}, description={intensity of a line peak in Raman spectroscopy}, symbol={$I_{peak}$}}
\newglossaryentry{SEpsi}{type=symbol, name={$\Psi$}, description={amplitude quotient in spectroscopic ellipsometry}, symbol={$\Psi$}}
\newglossaryentry{SEdelta}{type=symbol, name={$\Delta$}, description={phase difference in spectroscopic ellipsometry}, symbol={$\Delta$}}
\newglossaryentry{RefIndAbs}{type=symbol, name={$n_{abs}$}, description={absolute refractive index of a material}, symbol={$n_{abs}$}}
\newglossaryentry{RefIndRel}{type=symbol, name={$n_{rel}$}, description={relative refractive index}, symbol={$n_{rel}$}}
\newglossaryentry{RefIndAir}{type=symbol, name={$n_{air}$}, description={refractive index of air}, symbol={$n_{air}$}}
\newglossaryentry{VelLightVac}{type=symbol, name={$c_0$}, description={velocity of light in vacuum}, symbol={$c_0$}}
\newglossaryentry{VelLight}{type=symbol, name={$c$}, description={velocity of light in a material}, symbol={$c$}}
\newglossaryentry{ImagRefInd}{type=symbol, name={$k^\prime$}, description={imaginary part of the refractive index}, symbol={$k^\prime$}}
\newglossaryentry{NumMol}{type=symbol, name={$N$}, description={number of molecules per unit volume}, symbol={$N$}}
\newglossaryentry{CharElec}{type=symbol, name={$e$}, description={charge of an electron}, symbol={$e$}}
\newglossaryentry{MassElec}{type=symbol, name={$m$}, description={mass of an electron}, symbol={$m$}}
\newglossaryentry{VacPerm}{type=symbol, name={$\varepsilon_0$}, description={vacuum permittivity}, symbol={$\varepsilon_0$}}
\newglossaryentry{AbsStren}{type=symbol, name={$F_j$}, description={strength of the absorption at the resonance frequency}, symbol={$F_j$}}
\newglossaryentry{FricForce}{type=symbol, name={$\gamma_j$}, description={measure of the frictional force at the resonance frequency}, symbol={$\gamma_j$}}
\newglossaryentry{Omega}{type=symbol, name={$\omega$}, description={angular optical frequency}, symbol={$\omega$}}
\newglossaryentry{OmegaJ}{type=symbol, name={$\omega_j$}, description={resonance frequency}, symbol={$\omega_j$}}
\newglossaryentry{Lambda}{type=symbol, name={$\lambda$}, description={wavelength}, symbol={$\lambda$}}
\newglossaryentry{LambdaJ}{type=symbol, name={$\lambda_j$}, description={resonance wavelength}, symbol={$\lambda_j$}}
\newglossaryentry{MatIndAbrrev}{type=symbol, name={$A_j$}, description={material specific factor of the refractive index}, symbol={$A_j$}}
\newglossaryentry{LambdaD}{type=symbol, name={$\lambda_D$}, description={wavelength of the sodium D-line = 589.592\,nm}, symbol={$\lambda_D$}}
\newglossaryentry{Pressure}{type=symbol, name={$p$}, description={air pressure}, symbol={$p$}}
\newglossaryentry{DeltaPressure}{type=symbol, name={$\text{d}p$}, description={differential air pressure}, symbol={d$p$}}
\newglossaryentry{BasePressure}{type=symbol, name={$p_0$}, description={reference air pressure for spectroscopy}, symbol={$p_0$}}
\newglossaryentry{Temp}{type=symbol, name={$T$}, description={air temperature}, symbol={$T$}}
\newglossaryentry{DeltaTemp}{type=symbol, name={$\text{d}T$}, description={differential air temperature}, symbol={d$T$}}
\newglossaryentry{AlphaTemp}{type=symbol, name={$\alpha$}, description={thermal expansion coefficient}, symbol={$\alpha$}}
\newglossaryentry{ParVapor}{type=symbol, name={$w$}, description={partial pressure of vapor}, symbol={$w$}}
\newglossaryentry{LambdaK}{type=symbol, name={$\lambda_k$}, description={wavelength of different Fraunhofer lines}, symbol={$\lambda_k$}}
\newglossaryentry{TempI}{type=symbol, name={$T_i$}, description={air temperature for different measurement points to determine the thermo-optic coefficient}, symbol={$T_i$}}
\newglossaryentry{Temp0}{type=symbol, name={$T_0$}, description={reference air temperature to determine the thermo-optic coefficient}, symbol={$T_0$}}
\newglossaryentry{ThermOpCoeffs}{type=symbol, name={$D_0,D_1,D_2,E_0,E_1$}, description={thermo-optic coefficients}, symbol={$D_0,D_1,D_2,E_0$ and $E_1$}}
\newglossaryentry{Lambda0}{type=symbol, name={$\lambda_0$}, description={reference wavelength}, symbol={$\lambda_0$}}
\newglossaryentry{RefIndPar}{type=symbol, name={$n_\parallel$}, description={refractive index parallel to the stress plane}, symbol={$n_\parallel$}}
\newglossaryentry{RefIndPerp}{type=symbol, name={$n_\perp$}, description={refractive index perpendicular to the stress plane}, symbol={$n_\perp$}}
\newglossaryentry{K}{type=symbol, name={$K$}, description={stress optical coefficient}, symbol={$K$}}
\newglossaryentry{KPar}{type=symbol, name={$K_\parallel$}, description={stress optical coefficient parallel to the stress plane}, symbol={$K_\parallel$}}
\newglossaryentry{KPerp}{type=symbol, name={$K_\perp$}, description={stress optical coefficient perpendicular to the stress plane}, symbol={$K_\perp$}}
\newglossaryentry{SOPathDiff}{type=symbol, name={$\Delta s$}, description={optical path difference in stress optics}, symbol={$\Delta s$}}
\newglossaryentry{SOPhaseDiff}{type=symbol, name={$\Delta \Phi$}, description={phase difference in stress optics}, symbol={$\Delta \Phi$}}
\newglossaryentry{SOSampLen}{type=symbol, name={$l$}, description={sample length in stress optics}, symbol={$l$}}
\newglossaryentry{StressThermOpCoeffs}{type=symbol, name={$A_0,A_1,A_2,B$}, description={stress-thermo-optical coefficients}, symbol={$A_0,A_1,A_2$ and $B$}}

\newglossaryentry{Eout}{type=symbol, name={$E_{out}$}, description={resulting electric field of an interferometer}, symbol={$E_{out}$}}
\newglossaryentry{E0}{type=symbol, name={$E_{0}$}, description={incoming electric field}, symbol={$E_{0}$}}
\newglossaryentry{E1}{type=symbol, name={$E_{1}$}, description={electric field in the first arm of an interferometer}, symbol={$E_{1}$}}
\newglossaryentry{E2}{type=symbol, name={$E_{2}$}, description={electric field in the second arm of an interferometer}, symbol={$E_{2}$}}
\newglossaryentry{DeltaInter}{type=symbol, name={$\delta$}, description={optical path difference}, symbol={$\delta$}}
\newglossaryentry{PhaseDispFree}{type=symbol, name={$\varphi_0$}, description={phase of a dispersion-free interferometric signal}, symbol={$\varphi_0$}}
\newglossaryentry{PhaseDispInter}{type=symbol, name={$\varphi$}, description={phase of a dispersive interferometric signal}, symbol={$\varphi$}}
\newglossaryentry{Time}{type=symbol, name={$t$}, description={time}, symbol={$t$}}
\newglossaryentry{Frequency}{type=symbol, name={$f$}, description={optical frequency}, symbol={$f$}}
\newglossaryentry{I}{type=symbol, name={$I$}, description={resulting intensity of an interferometer}, symbol={$I$}}
\newglossaryentry{ILambda}{type=symbol, name={$I(\lambda)$}, description={spectral intensity of an interferometer}, symbol={$I(\lambda)$}}
\newglossaryentry{I0}{type=symbol, name={$I_0$}, description={initial intensity of a light source}, symbol={$I_0$}}
\newglossaryentry{InterContrast}{type=symbol, name={$\gamma(\lambda)$}, description={spectral contrast of interference fringes}, symbol={$\gamma(\lambda)$}}
\newglossaryentry{RefIndDE}{type=symbol, name={$n^{DE}(\lambda)$}, description={refractive index of a dispersive element}, symbol={$n^{DE}(\lambda)$}}
\newglossaryentry{GroupRefIndDE}{type=symbol, name={$n^{DE}_{g}(\lambda)$}, description={refractive index of a dispersive element}, symbol={$n^{DE}_{g}(\lambda)$}}
\newglossaryentry{ThickDE}{type=symbol, name={$t_{DE}$}, description={thickness of a dispersive element}, symbol={$t_{DE}$}}
\newglossaryentry{LambdaEQ}{type=symbol, name={$\lambda_{eq}$}, description={equalization wavelength}, symbol={$\lambda_{eq}$}}
\newglossaryentry{MeasRange}{type=symbol, name={$\Delta z(\lambda)$ = $\Delta z$}, description={axial measurement range of the dispersion-encoded low-coherence interferometer}, symbol={$\Delta z(\lambda)$}}
\newglossaryentry{NumPix}{type=symbol, name={$n_p$}, description={number of spectrometer pixels}, symbol={$n_p$}}
\newglossaryentry{GratConst}{type=symbol, name={$g$}, description={grating constant}, symbol={$g$}}
\newglossaryentry{ResSpectro}{type=symbol, name={$\Delta r_{spec}$}, description={spectral resolution of a spectrometer}, symbol={$\Delta r_{spec}$}}
\newglossaryentry{ResDE}{type=symbol, name={$r_{DE}$}, description={resolution of a dispersion-enhanced low-coherence interferometry due to the dispersive element}, symbol={$r_{DE}$}}
\newglossaryentry{ResSys}{type=symbol, name={$r_{sys}$}, description={system resolution of dispersion-enhanced low-coherence interferometry}, symbol={$r_{sys}$}}
\newglossaryentry{ResFit}{type=symbol, name={$r_{fit}$}, description={resolution of dispersion-enhanced low-coherence interferometry due to fitting of spectral intensity data}, symbol={$r_{fit}$}}
\newglossaryentry{DeltaDelta}{type=symbol, name={$\Delta \delta$}, description={path length uncertainty / introduced or simulated steps of the optical path length difference}, symbol={$\Delta \delta$}}
\newglossaryentry{MeasHeight}{type=symbol, name={$z_{meas}$}, description={measured height with a dispersion-encoded low-coherence interferometer}, symbol={$z_{meas}$}}
\newglossaryentry{DynRange}{type=symbol, name={DR}, description={dynamic range}, symbol={DR}}
\newglossaryentry{NoiseLightSource}{type=symbol, name={$P_{LS}$}, description={power of noise of the light source}, symbol={$P_{LS}$}}
\newglossaryentry{NoiseCam}{type=symbol, name={$P_{cam}$}, description={power of noise of the camera chip}, symbol={$P_{cam}$}}
\newglossaryentry{NoiseAmp}{type=symbol, name={$P_{amp}$}, description={power of noise of the cameras amplifier circuit}, symbol={$P_{amp}$}}
\newglossaryentry{NoiseAD}{type=symbol, name={$P_{A/D}$}, description={power of noise of the A/D conversion of the camera}, symbol={$P_{A/D}$}}
\newglossaryentry{NoiseSpecSigIn}{type=symbol, name={$S_{i}$}, description={spectral signal entering the imaging spectrometer}, symbol={$S_{i}$}}
\newglossaryentry{NoiseSpecSigOut}{type=symbol, name={$S_{o}$}, description={spectral signal recorded after processing in the imaging spectrometer}, symbol={$S_{o}$}}
\newglossaryentry{DeltaI}{type=symbol, name={$\Delta I$}, description={relative spectral intensity noise}, symbol={$\Delta I$}}
\newglossaryentry{RSquareS}{type=symbol, name={$R_{s}^2$}, description={coefficient of determination for a Gaussian fit of the spatially-resolved distribution of the relative intensity noise}, symbol={$R_{s}^2$}}
\newglossaryentry{NumFitPoints}{type=symbol, name={$n$}, description={number of data points used for fitting of spectral intensity data}, symbol={$n$}}
\newglossaryentry{DeltGuess}{type=symbol, name={$\delta_{guess}$}, description={guessed start value for $\delta$}, symbol={$\delta_{guess}$}}
\newglossaryentry{DeltSim}{type=symbol, name={$\delta_{sim}$}, description={simulated value for $\delta$}, symbol={$\delta_{sim}$}}
\newglossaryentry{DeltFit}{type=symbol, name={$\delta_{fit}$}, description={fitted value for $\delta$}, symbol={$\delta_{fit}$}}
\newglossaryentry{DeltRange}{type=symbol, name={$\delta_{rng}$}, description={range of values for $\delta$ in a brute force calculation}, symbol={$\delta_{rng}$}}
\newglossaryentry{SigmaSSE}{type=symbol, name={$\sigma_{sse}$}, description={squared sum of errors for the brute force calculation of $\delta$}, symbol={$\sigma_{sse}$}}
\newglossaryentry{DeltaW}{type=symbol, name={$\Delta w$}, description={width of the window in STFT}, symbol={$\Delta w$}}
\newglossaryentry{DeltaO}{type=symbol, name={$\Delta o$}, description={overlap of successive windows in STFT}, symbol={$\Delta o$}}
\newglossaryentry{DirSTFT}{type=symbol, name={$d_s$}, description={direction of window movement in in STFT}, symbol={$d_s$}}
\newglossaryentry{SpecSliceSTFT}{type=symbol, name={$S^\prime_n$}, description={spectral slice in STFT}, symbol={$S^\prime_n$}}
\newglossaryentry{FourierSpecSliceSTFT}{type=symbol, name={$\mathfrak{F}(S^\prime_n)$}, description={Fourier-transformed spectral slice in STFT}, symbol={$\mathfrak{F}(S^\prime_n)$}}
\newglossaryentry{Variance}{type=symbol, name={$\sigma^2$}, description={variance for the fitted spectral intensity data}, symbol={$\sigma^2$}}
\newglossaryentry{SSEmin}{type=symbol, name={$SSE_{min}$}, description={minimal squared sum of error for the fitted spectral intensity data}, symbol={$SSE_{min}$}}
\newglossaryentry{FitParams}{type=symbol, name={$m$}, description={number of fit parameters}, symbol={$m$}}
\newglossaryentry{ResFitExp}{type=symbol, name={$r_{fit}^{exp}$}, description={experimentally evaluated resolution of dispersion-enhanced low-coherence interferometry due to fitting of spectral intensity data}, symbol={$r_{fit}^{exp}$}}
\newglossaryentry{HeightNom}{type=symbol, name={$z_{nom}$}, description={nominal height}, symbol={$z_{nom}$}}
\newglossaryentry{HeightMean}{type=symbol, name={$z_{mean}$}, description={mean height}, symbol={$z_{mean}$}}
\newglossaryentry{HeightMin}{type=symbol, name={$\Delta z_{min}$}, description={measured resolution minimum}, symbol={$\Delta z_{min}$}}
\newglossaryentry{NoisePower}{type=symbol, name={$P$}, description={power of a measured signal}, symbol={$P$}}
\newglossaryentry{NumSpectra}{type=symbol, name={$S$}, description={number of measured spectra}, symbol={$S$}}
\newglossaryentry{SNRmeas}{type=symbol, name={$SNR_{meas}$}, description={measured signal-to-noise ratio}, symbol={$SNR_{meas}$}}
\newglossaryentry{SNRsimu}{type=symbol, name={$SNR_{simu}$}, description={simulated signal-to-noise ratio}, symbol={$SNR_{simu}$}}
\newglossaryentry{SurfProfile}{type=symbol, name={$z$ = $z(x)$}, description={surface profile}, symbol={$z$}}
\newglossaryentry{RoughnessRa}{type=symbol, name={Ra}, description={arithmetical mean height (roughness)}, symbol={Ra}}
\newglossaryentry{RoughnessRt}{type=symbol, name={Rt}, description={total height of a profile (roughness)}, symbol={Rt}}
\newglossaryentry{RoughnessRq}{type=symbol, name={Ra}, description={root-mean square height (roughness)}, symbol={Rq}}
\newglossaryentry{Repeatability}{type=symbol, name={$\sigma_z(x)$}, description={repeatability of surface profile measurements along a spatial axis $x$}, symbol={$\sigma_z(x)$}}
\newglossaryentry{MeanRepeatability}{type=symbol, name={$\overline{\sigma_z(x)}$}, description={mean repeatability of surface profile measurements along a spatial axis $x$}, symbol={$\overline{\sigma_z(x)}$}}
\newglossaryentry{NumSurfProfiles}{type=symbol, name={$N$}, description={number of surface profiles used to calculate the repeatability}, symbol={$N$}}
\newglossaryentry{MeanSurfProfiles}{type=symbol, name={$\overline{z(x)}$}, description={mean profile height to calculate the repeatability}, symbol={$\overline{z(x)}$}}
\newglossaryentry{FeatureHeight}{type=symbol, name={$h_i$}, description={height of a feature on a surface profile}, symbol={$h_i$}}
\newglossaryentry{MeanHeight}{type=symbol, name={$\overline{h}$}, description={mean height of a feature on a surface profile}, symbol={$\overline{h}$}}
\newglossaryentry{TtestHat}{type=symbol, name={$\hat{t}$}, description={calculated Student t-value}, symbol={$\hat{t}$}}
\newglossaryentry{Ttest}{type=symbol, name={$t$}, description={tabulated Student t-value}, symbol={$t$}}
\newglossaryentry{FeatureHeightNom}{type=symbol, name={$h_{nom}$}, description={nominal height of a feature on a surface profile}, symbol={$h_{nom}$}}
\newglossaryentry{FeatureHeightStep}{type=symbol, name={$h_{stp}$}, description={step height on a surface profile}, symbol={$h_{stp}$}}
\newglossaryentry{LambCFilter}{type=symbol, name={$\Lambda_c$}, description={filter wavelength for waviness and roughness}, symbol={$\Lambda_c$}}
\newglossaryentry{LambSFilter}{type=symbol, name={$\Lambda_s$}, description={filter wavelength for form correction}, symbol={$\Lambda_s$}}
\newglossaryentry{EvaLenRough}{type=symbol, name={$L_c$}, description={evaluation length for roughness measurements}, symbol={$L_c$}}
\newglossaryentry{TauX}{type=symbol, name={$\tau_x$}, description={correlation length of the auto-correlation function}, symbol={$\tau_x$}}
\newglossaryentry{SigmaRMS}{type=symbol, name={$\sigma_{rms}$}, description={square-root of the amplitude of the auto-correlation function / RMS roughness}, symbol={$\sigma_{rms}$}}
\newglossaryentry{Mag}{type=symbol, name={$M$}, description={magnification of the imaging system}, symbol={$M$}}
\newglossaryentry{RefIndSMP}{type=symbol, name={$n^{smp}(\lambda)$}, description={refractive index of a sample}, symbol={$n^{smp}(\lambda)$}}
\newglossaryentry{GroupRefIndSMP}{type=symbol, name={$n^{smp}_{g}(\lambda)$}, description={refractive index of a sample}, symbol={$n^{smp}_{g}(\lambda)$}}
\newglossaryentry{ThickSMP}{type=symbol, name={$t_{smp}$}, description={thickness of a sample}, symbol={$t_{smp}$}}
\newglossaryentry{WavLenPMP}{type=symbol, name={$\lambda_{pmp}$}, description={pump wavelength for the ASC light source}, symbol={$\lambda_{pmp}$}}
\newglossaryentry{WavLenSeed}{type=symbol, name={$\lambda_{seed}$}, description={seed wavelength for the ASC light source}, symbol={$\lambda_{seed}$}}
\newglossaryentry{FocalLen}{type=symbol, name={$f_i$}, description={focal length of an optical element}, symbol={$f_i$}}
\newglossaryentry{SensorSize}{type=symbol, name={$p$}, description={physical sensor size of the imaging spectrometer}, symbol={$p$}}
\newglossaryentry{LatMeasRange}{type=symbol, name={$\Delta l$}, description={lateral measurement range in dispersion-encoded low-coherence interferometry}, symbol={$\Delta l$}}
\newglossaryentry{LambCalc}{type=symbol, name={$\lambda_{calc}$}, description={calculated wavelength for the calibration of the imaging spectrometer}, symbol={$\lambda_{calc}$}}
\newglossaryentry{PixPos}{type=symbol, name={$p_{px}$}, description={pixel positions of spectral peaks for the calibration of the imaging spectrometer}, symbol={$p_{px}$}}
\newglossaryentry{CalCoeff}{type=symbol, name={$a,b,c$}, description={calibration coefficients for the calibration of the imaging spectrometer}, symbol={$a,b\text{ and }c$}}
\newglossaryentry{DeltaFocus}{type=symbol, name={$\Delta f$}, description={focus variation}, symbol={$\Delta f$}}
\newglossaryentry{WienerU}{type=symbol, name={$U$}, description={measured profile in Fourier-space for \textsc{Wiener}-deconvolution}, symbol={$U$}}
\newglossaryentry{WienerZ}{type=symbol, name={$Z$}, description={ideal profile in Fourier-space for \textsc{Wiener}-deconvolution}, symbol={$Z$}}
\newglossaryentry{WienerH}{type=symbol, name={$H$}, description={impulse response function in Fourier-space for \textsc{Wiener}-deconvolution}, symbol={$H$}}
\newglossaryentry{WienerZPrime}{type=symbol, name={$Z^\prime$}, description={deconvoluted profile in Fourier-space for \textsc{Wiener}-deconvolution}, symbol={$Z^\prime$}}
\newglossaryentry{WienerG}{type=symbol, name={$G$}, description={wild-card function in Fourier-space for \textsc{Wiener}-deconvolution}, symbol={$G$}}
\newglossaryentry{WienerQ}{type=symbol, name={$Q$}, description={wild-card function in Fourier-space to estimated the response function for \textsc{Wiener}-deconvolution}, symbol={$Q$}}

\newglossaryentry{DelayInterfero}{type=symbol, name={$\tau(\lambda)$}, description={temporal delay of an interferometer}, symbol={$\tau(\lambda)$}}
\newglossaryentry{NomThickness}{type=symbol, name={$t_{nom}$}, description={nominal sample thickness}, symbol={$t_{nom}$}}
\newglossaryentry{SellmeierCoeffs}{type=symbol, name={$B_1,B_2,B_3,C_1,C_2,C_3$}, description={Sellmeier coefficients}, symbol={$B_1,B_2,B_3,C_1,C_2 \text{ and } C_3$}}
\newglossaryentry{CauchyCoeffs}{type=symbol, name={$A_1,A_2,A_3$}, description={Cauchy coefficients}, symbol={$A_1,A_2 \text{ and } A_3$}}
\newglossaryentry{DispParameter}{type=symbol, name={$D$}, description={Dispersion parameter}, symbol={$D \text{ = } D(\lambda)$}}
\newglossaryentry{PhaseMeas}{type=symbol, name={$\varphi_{meas}$}, description={measured spectral phase of interferometer}, symbol={$\varphi_{meas}$}}
\newglossaryentry{PhaseLocal}{type=symbol, name={$\varphi_{loc}$}, description={local measured spectral phase of interferometer within region-of-interest}, symbol={$\varphi_{loc}$}}
\newglossaryentry{DerivPhaseLocal}{type=symbol, name={$\varphi_{loc}^\prime$}, description={derivative of the local measured spectral phase of interferometer with respect to the wavelength}, symbol={$\varphi_{loc}^\prime$}}
\newglossaryentry{PhaseOffset}{type=symbol, name={$\varphi_{off}$}, description={spectral phase offset due phase wrapping}, symbol={$\varphi_{off}$}}
\newglossaryentry{LambdaLocal}{type=symbol, name={$\lambda_{loc}$}, description={local wavelength range within region-of-interest}, symbol={$\lambda_{loc}$}}
\newglossaryentry{DerivOptThickness}{type=symbol, name={$t_{OPT}^\prime$}, description={relative derived optical thickness}, symbol={$t_{OPT}^\prime$}}

\newglossaryentry{ThickSubstrate}{type=symbol, name={$t_{s}$}, description={thickness of substrate in thin-film characterization}, symbol={$t_{s}$}}
\newglossaryentry{RefIndSubstrate}{type=symbol, name={$n_{s}$}, description={refractive index of the substrate material in thin-film characterization}, symbol={$n_{s}$}}
\newglossaryentry{ThickFilm}{type=symbol, name={$t_{fn}$}, description={thickness of a thin-film layer in thin-film characterization}, symbol={$t_{fn}$}}
\newglossaryentry{RefIndFilm}{type=symbol, name={$n_{fn}$}, description={refractive index of a thin-film layer in thin-film characterization}, symbol={$n_{fn}$}}
\newglossaryentry{Ein}{type=symbol, name={$E_{in}$}, description={incoming electric field}, symbol={$E_{in}$}}
\newglossaryentry{Wavenumber}{type=symbol, name={$k^\prime$}, description={angular wavenumber}, symbol={$k^\prime$}}
\newglossaryentry{FieldCoeff}{type=symbol, name={$\Xi, \Pi, \Upsilon,\Omega$}, description={propagation coefficients of the electric field in a transfer-matrix formalism}, symbol={$\Xi, \Pi, \Upsilon \text{ and } \Omega$}}
\newglossaryentry{Min}{type=symbol, name={$M_{i}$}, description={matrix describing the electric field entering a material system when using a transfer-matrix formalism}, symbol={$M_{i}$}}
\newglossaryentry{Mout}{type=symbol, name={$M_{0}$}, description={matrix describing the electric field exiting a material system when using a transfer-matrix formalism}, symbol={$M_{o}$}}
\newglossaryentry{MTrans}{type=symbol, name={$M_{T}$}, description={matrix describing the complete electric field of a material system with a number of material boundaries when using a transfer-matrix formalism}, symbol={$M_{T}$}}
\newglossaryentry{BoundaryTFM}{type=symbol, name={$B_{ij}$}, description={matrix describing the boundary of a material when using the transfer-matrix formalism}, symbol={$B_{ij}$}}
\newglossaryentry{PropagatorTFM}{type=symbol, name={$P_{j}$}, description={matrix describing the propagation through a material when using the transfer-matrix formalism}, symbol={$P_{j}$}}
\newglossaryentry{ReflectionTFM}{type=symbol, name={$r$}, description={reflecting electric field from a material system when using the transfer-matrix formalism}, symbol={$r$}}
\newglossaryentry{BackReflectionTFM}{type=symbol, name={$r_{back}$}, description={back reflecting electric field at the end of a material system when using the transfer-matrix formalism}, symbol={$r_{back}$}}

\makenoidxglossaries
\begin{document}

\subject{\textnormal{Technische Universität Dresden}}
\title{Development and characterization of a dispersion-encoded method for low-coherence interferometry}

\author{Dipl.-Ing.\,(FH) B.Eng. \\ \textbf{Christopher Taudt}}
\date{}
\publishers{der Fakultät Elektrotechnik und Informationstechnik der Technischen Universität Dresden\\ \vfill zur Erlangung des akademischen Grades\\ \vfill \textbf{Doktoringenieur}\\(Dr.-Ing.)\\ \vfill vorgelegte Dissertation}

\lowertitleback{\textit{"The test of all knowledge is experiment."}, Richard P. Feynman\\
				\\ \textit{"Wenn man nicht scheitern kann, reproduziert man nur was schon bekannt war."}, Moritz Klenk}

\dedication{\textit{"The test of all knowledge is experiment."}, Richard P. Feynman\\ \vfill \textit{"Wenn man nicht scheitern kann, reproduziert man nur was schon bekannt war."}, Moritz Klenk}

\maketitle
 \frontmatter
\TUDoption{abstract}{section}
\begin{abstract}[pagestyle=empty, language=english]
This work discusses an extension to conventional low-coherence interferometry by the introduction of dispersion-encoding. The extension facilitates the measurement of surface height profiles with sub-nm resolution. The selection of a dispersive element for encoding allows for tuning of the axial measurement range and resolution of the setup. The approach is theoretically designed and implemented for applications such as surface profilometry, the characterization of polymeric cross-linking and as a tool for the determination of layer thicknesses in thin-film processing. During the characterization of the implemented setup, it was shown that an axial measurement range of 79.91\,\textmugreek m with a resolution of 0.1\,nm was achievable in the evaluation of surface profiles. Simultaneously, profiles of up to 1.5\,mm length could be obtained without the need for mechanical scanning. This marked a significant improvement in relation to state-of-the-art technologies in terms of dynamic range. It was also shown that axial and lateral measurement range can be decoupled partially. Additionally, functional parameters such as surface roughness were characterized with the same tool. The characterization of the degree of polymeric cross-linking  was performed as a function of the refractive index. Here, the refractive index could be acquired in a spatially-resolved manner with an index resolution of down to \num{3.36e-5}. This was achieved by the development of a novel mathematical analysis approach. For the acquisition of layer thicknesses of thin-films, an advanced setup was developed which could be used to characterize the thickness thin-films and its (flexible) substrate simultaneously.
\end{abstract}

\listoffigures
\listoftables

\clearpage
\printnoidxglossary[type=main,style = list,nonumberlist]
\printnoidxglossary[type=symbol,style = list,nonumberlist]

\tableofcontents

\mainmatter
\chapter{Introduction and motivation}
The electronics industry with all its branches such as semiconductors, organic-electronics and the photovoltaics industry, is continuously growing in terms of its economic as well as its technological influence, \cite{Stat_SemioCon_2020}. This trend is fostered by the ongoing integration of various electronic functionalities in fields such as energy generation \& distribution, transport \& mobility as well as in consumer goods. Over three decades, electronic and semiconductor products as well as processes were driven by \textsc{Moore}s law, \cite{Daintith2009}. This paradigm focused nearly exclusively on the miniaturization of structures and therefore on the increase of the number of transistors per chip area. While this paradigm was directly attributed to an increase in computing power, the  industry's scope broadened in recent years. With the introduction of the so called \textit{More-than-Moore} paradigm, the enhancement of functionality and cross-modality integration has become more important, \cite{WhitePaper_MtM}. The implementation of this approach has enabled the development of novel products where functionalities of the analog, digital as well as of the power domain were combined. Furthermore, sensors and other \gls{MEMS} have been used to enhance these products, \cite{Topaloglu2015}. Opposed to traditional geometrical scaling of chips this is also known as \textit{performance scaling}, \cite{Graef2011}. The implementation of these ideas at the product level has enabled the transformation of traditional analog components from board level to package (SiP - system in package) or chip level (SoC - system on chip). This includes a significant increase of complexity in semiconductor manufacturing, \cite{MartinBragado2018,Seebauer2010}. New and advanced processes have been developed to enable three-dimensional production and packaging. Additionally, novel concepts for the management of disturbances such as thermal and electro-magnetical influences have been developed. All of these product and process developments drove research and applications of novel materials and material combinations, \cite{Bronovets2020,Karthikeyan2020,Liu2019,Saraswat2018}. On the one hand, this enabled bold innovations in power semiconductors regarding performance and energy efficiency which are main drivers behind electro-mobility and sustainable energy, \cite{Deboy2013,Wang2013}. On the other hand, the novel and advanced manufacturing technologies imposed great challenges towards quality assurance, reliability and therefore metrology, \cite{LeachFundamentals}. According to Leach et al., \cite{OpenQuestions}, the currently available metrology still lags behind advanced production regarding precision, measurement speed and cost scaling. Where processes such as lithography could significantly scale on a cost per unit basis, metrology hardly did. This was partly due to technological reasons but also partly due to high specialization of metrology approaches with a strong focus on single applications. On the technological side, the main challenges lie in the need for systems that deliver large measurement ranges with high precision (high-dynamic range) while maintaining fast speeds to ensure process-integrated operation. Appropriate metrology has to enable the determination of geometrical parameters like critical dimensions or surface profiles alongside functional parameters such as surface roughness or defect detection. A typical case for the requirements of a process-integrated metrology tool is given by the stack of a modern solar cell module, \cite{OpenQuestions}. The cell itself consists of a number of active and barrier layers, deposited by thin-film processes, which require the layer thicknesses to be monitored in order to ensure performance. Interconnects or divides such as bus-bars, laser scribes and vias need to be observed regarding positions and dimensions. Optics with high aspect ratios on top of the cell increase the efficiency and must be measured with respect to their surface profile. Additionally, polymeric adhesives as well as a polymeric top laminate protect the compound against environmental hazards. Monitoring of the degree of cross-linking of these materials enables optimal performance over 20 and more years of operation, \cite{Badiee2014}.\\
The main scope of this work is the development of a metrology system which is capable of meeting the main requirements posted by this exemplary application in terms of resolution, dynamic range, measurement speed and flexibility. More precisely, the metrology approach aims to measure surface profiles, thin-film layer thickness as well as the degree of cross-linking in polymers. A review of relevant research literature paved the way for the development of a novel approach based on the principles of low-coherence interferometry. The enhancement of the principle by the introduction of dispersion encoding is performed in order to facilitate higher resolutions while maintaining large measurement ranges, known as high-dynamic range metrology. This approach will be called \gls{DE-LCI}. With regard to surface profilometry, the aim is to cover an axial measurement range of nearly 100\,\textmugreek m while having sub-nm resolution. Additionally, the capabilities of the novel approach regarding the characterization of thin-film thickness as well as polymeric cross-linking will be assessed. Necessary developments of the setup and the analysis algorithms facilitating all three possible measurement modes are going to be evaluated in detail in this work.\\
The design and implementation of the novel approach will furthermore focus on the robustness for thermal and mechanical influences to maintain the resolution during long measurements. Another design criterion is the preparation for process integration which can be translated to acquisition and data processing speed. This work will demonstrate efforts to capture large sections of a sample comparatively fast to conventional methods. Theoretical limits of measurement range and resolution will be investigated on a more detailed level. Also, the occurrence of these limits under practical conditions will be assessed. In particular, which transients influence the axial resolution and measurement range are investigated and which technological measures can be taken to partially decouple axial and lateral measurement ranges are looked into. Furthermore, this work will evaluate techniques to gather two- and three-dimensional information of surfaces and bulk materials without the need for mechanical scanning to reduce influences from movement and increase measurement speed. 
In terms of the characterization of polymeric cross-linking, it is investigated how the wavelength-dependent refractive index can be used as a measure for cross-linking. Consequently, the resolution of the \gls{DE-LCI} approach with regard to the refractive index will be examined. An exploration of the capabilities to resolve nm-sized layers of thin-film materials will conclude the work.

\chapter{Related works and basic considerations}

\section{Profilometry}\label{StateOfTheArtProfilometry}
The characterization and measurement of surface profiles is one of the most basic metrology tasks in industrial manufacturing. What started with mechanical stylus profilometers has developed with ultrasonic transducers towards optical instruments. These are capable of appropriate resolutions to enable nanometrology, \cite{LeachFundamentals}. As surface metrology is well established, basic terms and parameters are defined in corresponding norms such as ISO 4288 (\textit{assessment of surface texture}) and ISO 4287 (\textit{terms, definitions and surface texture parameters}), \cite{ISO4288,ISO4287}. Established optical instruments are classified as areal integrating, line profiling and areal topography instruments in ISO 25187, \cite{ISO25178}. According to Leach et al. \cite{LeachFundamentals}, sensors with the capability to record areal information about surface shape, waviness and roughness simultaneously are the most desirable in nanometrology. These sensors are usually defined as 3D-sensors where the topography information is gathered as the local distance $z$ at a specific coordinate $(x,y)$. Apart from the classification mentioned in \cite{ISO25178}, numerous variants exists, \cite{Knauer}. On the most basic level, the common denominator for optical approaches is that all approaches follow a simple interaction model where light from a light source interacts with an object of interest. The result of this interaction can be observed separately on an observation plane. The properties of the light source such as intensity, polarization state, temporal behavior, coherence and spectral behavior can be utilized to encode information about the object surface topography. Depending on the surface structure of the object of interest and the design of the observation method, different limitations or artifacts occur. One example is the occurrence of so called speckles on optical rough surfaces under coherent illumination. While most metrology approaches suffer from this effect in terms of a decreased \gls{SNR}, some techniques such as coherence scanning interferometry or speckle interferometry make use of this effect.
Production integrated metrology becomes increasingly important in areas where the critical dimensions decrease from the \textmugreek m- to the nm-range. Industrial sectors such as \gls{MEMS}, power semiconductors or photovoltaics require a continuous, precise monitoring of production to maintain quality, \cite{LeachFundamentals}. A variety of measurement technologies to characterize, e.g. surface roughness, topography or film thickness have been reported, \cite{Leach}. Apart from specific measurement machines with pm resolution for research purposes which required high efforts in building, calibration and maintenance, a broad range of technologies is known which is more applicable as production-accompanying tool, \cite{Manske2019}. The following section is intended to give an overview of these approaches.\\
Häussler and Ettl classified optical 3D sensors in \cite{Leach} mainly according to their limitations and their dominant noise sources. Following that, four main classes of sensors can be associated:
\begin{itemize}
	\item Type I sensors (e.g. laser triangulation)
	\item Type II sensors (e.g. coherence scanning interferometry)
	\item Type III sensors (e.g. phase-shifting interferometry)
	\item Type IV sensors (e.g. deflectometry)
\end{itemize}
As Type I sensors mostly incorporate technologies which measure a lateral perspective shift on local details, the uncertainties are determined by the uncertainty of this shift measurement. This means that the uncertainty scales with the inverse square of the distance to the detail, \cite{Leach}. Techniques include laser triangulation, fringe projection, confocal microscopy and others. For Type II sensors such as coherence scanning interferometry, the measurement relies not on the detection of of the phase of the signal but on the correlation of single speckles reflected from a rough surface. Therefore the statistical noise, mainly composed of the standard deviation of the object surface, is the primary source of noise, \cite{Haeusler1999, Ettl2001}. In contrast, Type III sensors involve classical interferometric technologies such as phase-shifting interferometry (\cite{SchreiberBruning}) and digital holographic microscopy \cite{KimDHM2010}. These are phase measuring technologies which enable sub-nm axial resolution and are only limited to photon noise. Furthermore, \cite{Leach} separates deflectometry as a Type IV sensor. With limitations of photon noise and developments towards nanoscale accuracy this technology is promising for industrial applications, \cite{LeachFundamentals}.\\
This section analyses important technologies from every sensor class as well as \gls{AFM} as one important, high-resolution, stylus-based technology in order to evaluate open questions in nano-profilometry. All of them have been widely qualified for the measurement of $nm$-scale surface parameters in a lab environment, \cite{ZABIHIcomparison,Bowe}.\\

\subsection{Atomic force microscopy}
\gls{AFM}, although not an optical metrology approach, has become an important and highly precise tool in nano-engineering and science, \cite{Chen2019,Giessibl,AFMoriginal}. It is a technological development which goes beyond the capabilities of classical stylus instruments in terms of its resolution. It provides an axial resolution down to 0.01\,nm while the lateral resolution can be in the range of 0.1\,-\,10\,nm, \cite{Gan2009}. These features made it exceptionally important in areas such as cell biology and molecular sciences where it enables the precise study of molecular processes and interactions, \cite{Chtcheglova2018, Krieg2019}. Industrial and technical applications make use of it for the characterization of nano-structured materials and features, \cite{AFMapplication1,AFMapplication2,AFMapplication3}. \Gls{AFM} can deliver interesting additional features like force sensing and surface modification, \cite{AFMforce}. Several developments such as multi-cantilever arrangements and resolution-enhancing methods also have enabled the technology to investigate cm\textsuperscript{2}-large structures with high aspect ratios.  In this context, aspect ratios of a axial measurement range of 15\,\textmugreek m with a deviation of $\pm$\,60 nm could be achieved while the lateral measurement range was 15\,x\,100\,\textmugreek m\textsuperscript{2}, \cite{AFMHighAspectRatio}. Using an array of cantilevers in a 32x1 chip design, Minne et al. \cite{AFMlargeRange} were able to scan large areas of 2000\,x\,2000\,\textmugreek m\textsuperscript{2} within 30 minutes while the axial measurement range was 2.5\,\textmugreek m at a resolution of 3.5\,nm. Most of these developments have the need for complex setups and measurements last several minutes. Additionally, AFM is sensitive to systematic errors and contamination on samples. These circumstances prevent using \gls{AFM}-based approaches for production accompanying tasks.

\subsection{Confocal laser scanning microscopy}
Another established method for the characterization of surface profiles is \gls{CLSM}, \cite{Lu2019,Buajarern}. By aligning the focal plane of the illumination of a single point on a sample with the imaging plane of that same point on a pinhole, information is solely captured from this point, \cite{CLSMReview}. By scanning a depth range, the height profile of a sample can be acquired with a high \gls{SNR} even if the surface is strongly scattering. As scanning has to be performed in the lateral dimension as well, the acquisition of a full areal height map requires a long time in the range of minutes. In order to achieve a high axial resolution, the so called \textit{auto-focus method} is applied. In this method, the focal plane is adjusted for the highest intensity for a given point. The method is repeated in a given axial range of the sample while the position of the focal points is recorded correspondingly, \cite{Udupa}. A faster, yet less precise, option is the so called \textit{intensity method}. This method correlates a measured intensity to a calibrated height-intensity curve. According to  \cite{Udupa}, this results in measurement times in the order of $\nicefrac{1}{50}$ compared to the \textit{auto-focus method}. Also, due to the limited linearity of the calibration curve, the measurement range and precision is also significantly reduced. Furthermore, the calibration curve has to be recaptured for every new sample material in combination with the microscope objective used for the investigation.\\
Depending on the translation stages and magnification, a lateral resolution of (100\,-\,500)\,nm is typically achieved to capture surface parameters such as roughness, waviness and form error in an areal fashion, \cite{UdupaII}. Larger lateral measurement ranges such as 1.3\,x\,1.3\,mm are usually captured with a lower resolution of e.g. 10\,\textmugreek m in order to speed up the measurement process. Buajarern et al., \cite{Buajarern}, have shown, that the axial resolution is not only limited by the resolution of the translation stages used to perform the auto-focus determination but even more by the depth of focus / resolution of the microscope objective used for imaging. Using high \gls{NA}, large magnification objectives, the axial resolution is optimal but restricted to about 150\,nm.\\
Different works have shown the applicability of \gls{CLSM} to characterizations in biology, for dental as well as engineering materials \cite{Halbhuber,Paepegaey,TingSun}. Besides the need for mechanical scanning and the interdependence of measurement range and resolution, a major disadvantage of \gls{CLSM} is the high effort necessary for adjustment of parameters in order to achieve optimal results. Tomovich et al., \cite{Tomovich}, have demonstrated that the adjustment of parameters is highly dependent on the reflective and structural properties of the sample. From this perspective, \gls{CLSM} is a versatile tool in research labs but is limited in the usage for production accompanying tasks.\\
In addition to classical confocal microscopy, several developments have been made in order to circumvent some of the drawbacks such as the need for scanning. Chromatic confocal microscopy is one approach. It particularly targets scanning along the $z$-dimension. While dispersion in the optical system leads to different foci for each wavelength, height information can be decoded in spectral variations, \cite{Lu2019,Seppa2018}. In terms of measurement range and resolution, both, the spectral range and spectral intensity stability, are important. Typical light sources include halogen and xenon lamps as well as LED sources. Other works have demonstrated the usability of supercontinuum sources as well, \cite{Shi2004}. Using the linearized wavelength calibration, the calculation of height can be performed in different ways. The fastest and computational most efficient yet the technique with the lowest resolution is to determine the maximum intensity of every spectral line. The highest computational effort is necessary when fitting the intensity distribution with e.g. a Gaussian approach but these approaches enable the highest possible resolutions. A possible compromise between speed and accuracy is the so called bary-center calculation to estimate the height, \cite{Leach}. The limitations regarding the lateral resolution and the resolution of slopes are very similar to other microscopy techniques. Depending on the light sources, the \gls{NA} of the objective and the size of the pinhole, typical spot sizes are 5\,-\,10\,\textmugreek m for vertical ranges of $<$\,1\,mm and 10\,-\,30\,\textmugreek m for vertical ranges $>$\,1\,mm, \cite{Leach} can be achieved. The resolution of slopes is usually defined by the half aperture angle where the range is between $\pm$\,\ang{18} to $\pm$\,\ang{44} for \glspl{NA} of 0.3 to 0.7 respectively.\\
The extension of this method in order to capture areal information normally relies on scanning approaches. Some occurring problems are due to dynamic stitching errors and spherical aberrations from scanning, \cite{Cha2000,Ruprecht2004}. \\
The method can also be used to measure the profile and thickness of multiple, semi-transparent layers. In this case, the analysis has to take the refractive index of the transparent material at the different wavelengths into account. Faster approaches utilize multi-probe setups in order to parallelize data acquisition. Nonetheless, they are strongly limited in the lateral resolution, \cite{Kim2019,Kim2019b}.

\subsection{Digital holographic microscopy}
\Gls{DHM} is a fast, robust and full-frame-capturing technology to evaluate the surface topography of samples. While surface data from an interferometric approach is captured by a camera as full-field information (spatially, intensity, phase) within the integration time of the device, all aberration correction, exact focusing and surface reconstruction can be done digitally within post-production. Stroboscopic illumination can be used additionally to speed up acquisition from conventional frame rates such as 20\,-\,60\,fps to 25\,MHz, \cite{Leach}. The axial accuracy  is dependent on the calibration of the light sources wavelength/frequency  which leads to achievable axial resolutions of 0.1 nm with stabilized laser sources, \cite{Leach}. One of the limiting factors is the axial measurement range  which can be extended by the use of multiple wavelengths to several \textmugreek ms while keeping a sub-nm resolution, \cite{Kuehn2007}. Furthermore, the combination with other techniques, such as reflectometry, enables the analysis of parameters such as layer thickness, refractive index profiles and topography of multilayered structures, \cite{Colomb2010}.
The signal formation and analysis is generally described in two steps where one is the data acquisition and the other is the so called reconstruction, \cite{KimDHM2010}.
A typical holographic setup contains e.g. a Mach-Zehnder interferometer as its main component. In this setup, the object wave, which is either reflected or transmitted from the object, is imaged close to the camera plane. The reference wave interferes with the object wave and its intensity can be detected and analyzed. The subsequent reconstruction is performed in two steps. First, the complex wave is reconstructed in order to separate the real and virtual hologram from the zeroth-order signal. This can be performed in hardware by introducing defined phase jumps (in inline-holography) or by introducing an angle between the object and the reference wave (in off-axis holography). Furthermore, the separation of the zeroth-order term (\cite{Pavillon2009}) or the twin-image term (\cite{Weng2010}) can be done by FFT approaches in software. Afterwards, the filtered hologram is digitally illuminated by a reference wave. During a second process step, numerical propagation is performed by approaches like the single Fourier approach (\cite{Ferraro2003}), the angular spectrum approach (\cite{deNicola2005}) and the convolution approach (\cite{Colomb2006b}). In typical implementations, two approaches to define numerical lenses for the correction of higher order aberrations are used
\begin{itemize}
	\item recording of a physical reference hologram on a plane sample and by transmission through air, \cite{Ferraro2003, Colomb2006b} 
	\item fitting of Zernike polynomials in an assumed flat area of a measured sample, \cite{Colomb2006a,Colomb2006c, Miccio2007}.
\end{itemize}
It is quite common to use combinations of both solutions in order to achieve optimal results. The reference wave is used to correct all phase and tilt aberrations as well as for curvature of waves and de-focus correction. In particular, this technique has the advantage that only a single hologram needs to be acquired while the correct focal point can be tuned by numerical propagation. This can also be used to increase the depth of field, \cite{Ferraro2005}. Furthermore, it indicates that \gls{DHM} is a full-field metrology where all relevant information is captured in one image acquisition.\\
A common problem in \gls{DHM} is $2\pi$/phase ambiguity as it limits the usable measurement range. An efficient method to overcome this problem is the so called multiple-wavelength holography, \cite{Cheremkhin2015,Yang2020}. In this approach, data is captured with slightly different wavelengths which allows the numerical construction of a new, artificial wavelength which can be used to 'construct' unambiguous measurement ranges. \gls{DHM} is also capable of recording dynamic events which occur in a repetitive fashion. For this purpose,  the events of interest are observed using stroboscopic techniques. It could be shown that an acquisition of events with a repetition frequency of up to 25 MHz is possible, \cite{Cuche2009}. Extensions to \gls{DHM} in order to evaluate functional parameters such as reflectometry have been reported, \cite{Leach}. However, complex wavefront reconstructions or fitting procedures are necessary to separate the information of different layers. Currently, this prevents the use of this approach in industry.

\subsection{Phase-shifting interferometry}
\Gls{PSI}, is an more elaborate variant of traditional (high coherence) interferometry for the determination of surface information, \cite{SchreiberBruning}. Depending on the light source, the detector and the optical setup, lateral resolution of 2 \textmugreek m and axial resolutions  up to 1\,nm are reachable, \cite{Leach} S. 167. \gls{PSI} enhances classical interferometry by introducing multiple controlled phase shifts in order to determine the phase-dependent surface height of an object, \cite{Schmit2007} . Furthermore, it overcomes the directional ambiguity of interferometric signals by generating quadrature signals, since back-reflected light from a surface in combination with light from the reference arm generates an areal interferometric fringe pattern, \cite{Leach}. Due to some simplifications, data analysis of the generated quadrature signals can be accelerated by fitting sine and cosine signal components to determine the phase with high accuracy. A common approach for the data analysis of these signals is the so called auto-correlation or synchronous detection, \cite{Leach}. Both, sine and cosine components are described as integrals of a full cycle of the phase shift. Depending on the number of introduced phase shifts, various algorithms can be used, \cite{SchreiberBruning}. Furthermore, the use of non-linear, e.g. sinusoidal manipulation of the phase shifts has been reported to be accurate and computational more efficient than linear manipulation, \cite{deGroot2009}. In addition to temporal \gls{PSI}, other works have described alternative methods to encode and detect phase-shifts such as polarization encoding, \cite{Massig1992, SafraniPhaseShift}.\\
A common problem in interferometry occurs in \gls{PSI} is the determination of phase using an $arctan$ function which is only possible in the range of $0-2\pi$. Surface heights corresponding to multiples of $2\pi$ will therefore be calculated within the given range resulting in phase jumps, known as a wrapped phase. Several algorithms are currently being researched, \cite{RodriguezZurita2020,Phuc2018,Estrada2020,Servin2016}. More recent approaches on phase-shift interferometry extend the standard setup with multiple sensors, \cite{SafraniPhaseShift, Liu2020,RodriguezZurita2020}.  By capturing interference signals at different polarizations, phase angles and wavelengths, an axial resolution of 2\,nm on a 3\,\textmugreek m height step in an area of 0.5\,x\,0.5\,\textmugreek m could be achieved. Due to the need for multiple channels, high efforts have to be made to synchronize and handle the data of up to 6 cameras. The evaluation of the phase shift of the total interference contrast (TIC) is an alternative phase-shifting technique which is capable of providing sub-nm axial resolution in a full-field manner, \cite{VaupelTIC}. As it is not based on coherence properties of the light source it is reliant on the knowledge of an material model of the tested sample. Therefore, the method is limited to mainly laboratory usage, \cite{VaupelConference}.

\subsection{Coherence scanning interferometry}\label{Lit:SubSec:CSI}
Belonging to so called type II sensors, \gls{CSI}  distinctively deviates from classical interferometry. This approach utilizes light of low temporal coherence not to measure the phase of the signal, but to estimate the correlogram of individual speckles on a surface, \cite{Leach}. The technique traditionally scans one interferometer arm in order to evaluate the signal contrast which is equivalent to the envelope of the correlogram, \cite{Schmit2007}. The coherence length of the light source is the main limitation as it restricts the possible measurement range. Thus, only height variations within the coherence length can be detected continuously. This can be an issue on very rough surfaces or steep slopes, \cite{HaeusslerPatent}. Usually setups are designed to capture the full lateral information using a camera which detects the correlogram envelope at every point $P(x,y)$, while a scanner ensures the axial movement in the height range, \cite{Leach2008, Petzing2010}. While the typical scanning range is (10\,-\,200)\,\textmugreek m with piezo-based scanners and $>$\,1\,mm for mechanical scanners, the interferometer part of the system is most commonly designed as an microscope objective of the Michelson, Mirau or Linnik type. The approach is also known by the names of coherence radar, coherence scanning, white-light interferometry, vertical scanning interferometry and others, \cite{Schmit2007}.\\
According to \cite{Abdulhalim2001}, the simplest model for the description of the signal in \gls{CSI} is an incoherent superposition. The signal shape in this case is determined by the spectral and spatial distribution of light in the pupil plane. Using this model, several boundary cases can be found. For example, the properties of a system with low \gls{NA} (e.g. 0.2), narrow bandwidth (e.g. $\Delta \lambda$\,=\,100\,nm) will basically be dominated by the spectral distribution of the light source with high fringe visibility at the zero scan position. In contrast, if the system has a high \gls{NA} (e.g. 0.6) and narrow bandwidth (e.g. $\Delta \lambda$\,=\,20\,nm), the coherence properties are dominated by the spatial distribution in the pupil plane, hence the focusing. In real systems, both effects will be present, whereas it can be stated that in low \gls{NA} systems the spectral contribution and in high \gls{NA} systems the spatial contribution of the light source primarily influences the signal.\\
One of the distinguishing features of \gls{CSI} is the lack of a $2\pi$ ambiguity. This makes it well suited for measurements of rough surfaces and steep slopes. In general, it can be said that the slowly modulated envelope of the interference fringes is an indicator of the signal strength as well as a measure for the surface profile. But it has to be clear that the peak of the envelope is not equal to the maximum height of the sample. It represents only the point where the \gls{OPD} equals zero. More precisely, it is the point where the dispersion affected path length difference in the setup equals zero. Therefore, the approximation of the envelope signal is only an idealization, where in reality dispersion effects and scan dependent distortions have to be taken into account. 
Signal processing in \gls{CSI} is usually performed in multiple steps, where the start of the scan position is determined by the maximum signal strength or fringe visibility. The scan range is determined by sectioning the images according to the expected height of the sample. Several methods to perform the envelope detection including peak finding of the maximum have evolved over the years. While \cite{Caber1993} have shown the application of digital filters and an demodulation technique, other groups such as \cite{Haneishi} and \cite{Kino1990} have shown how to remove the carrier signal to separate the envelope. According to approaches of \cite{Larkin1996} and \cite{AiNovak} who performed a centroid determination of the square of the signal derivative, a higher resistance to noise on the signal is achievable.\\
Noise, e.g. from optical aberrations, the scanning process, diffraction or vibration can significantly influence the sensitivity of the technique. Therefore, more advanced signal processing algorithms use the result of the envelope detection only as a rough estimate of the fringe order and call the result a first topography map. Based on this, a fine estimation using a phase analysis of the fringes results in a phase map . This method increases the accuracy significantly although other problems such as $2\pi$ ambiguities on thin film structures, surface roughness or sharp edges occur during phase analysis. Some algorithms take this into account, \cite{Harasaki2000, deGroot2002}.\\
More sophisticated approaches developed methods for the simultaneous envelope detection and phase analysis by the correlation of the intensity data with a complex kernel. These kernels have been derived from \gls{PSI} algorithms and use different approaches such as wavelet techniques or least-squares fitting, \cite{Larkin1996,Sandoz1997,Lee-Bennett2004,deGroot2008}. Furthermore, recent works have also demonstrated that sub-nm axial resolution is possible through advanced signal modeling techniques, although limitations due to scanning and the relatively small axial measurement range remain, \cite{Freischlad2019, Thomas2019}. Coherence Scanning Interferometry is well suited for the determination of the surface topography on rough surfaces, \cite{Thomas2020}. One major disadvantage is the necessity for scanning along one physical dimension of a sample as this introduces error and slows down the measurement procedure.

\subsection{Low-coherence interferometry}
One of the most advanced technologies for high resolution surface profilometry and tomography is \gls{LCI}. \Gls{OCT} is a commonly known implementation, \cite{Moon2018,HuangOCT}. Although OCT was initially used in medicine, developing it as as a  tool for industrial purposes have become increasingly important, \cite{Kirsten2018,Gambichler2011, Hitzenberger2018,KirstenHiRes}\\
The range of different OCT approaches reaches from \gls{TD-OCT} and \gls{SD-OCT} to \gls{FD-OCT}, \cite{OCTreview}. While most approaches are implemented as point sensors with the possibility to mechanically scan a surface, they vary in speed, measurement range and accuracy. More fundamentally, the accuracy and measurement range in the axial dimension are defined by the coherence length \glssymbol{CoherLen} of the light source used, \cite{octBook} pp. 74.\\
\begin{equation}
l_c= \frac{2 \cdot \ln{2}}{\pi} \frac{\lambda_c^2}{\Delta \lambda}
\end{equation}
where \glssymbol{CenWaveLen} denotes the center wavelength of the light source and \glssymbol{WaveLenRange} its spectral range, assuming that the light source has a Gaussian-shaped spectrum. The typical axial measurement range, using broadband light sources, is a few hundred \textmugreek m with a resolution of about 1\,\textmugreek m, \cite{Froehly2012}. These values are usually fixed within the design of the specific setup and can not be adjusted during operation. In order to achieve higher axial measurement ranges as well as a higher resolution some hardware problems have to be overcome. The most notable influences on the resolution are are environmental disturbances on the interferometer arms as well as the repeatability of the scanning system used to gather areal information, \cite{KochSubNM}. An extension of the measurement range in TD-OCT is achieved by the introduction of multiple reflecting surfaces in the reference arm, \cite{McNamaraMRoct}. This modification enables the measurement of a large axial range while keeping actual mechanical scanning to a minimum. Another approach merges both reference and sample arm in a \gls{FD-OCT} configuration, \cite{KochSubNM}. In this approach, a reflective surface in the vicinity of the sample is used as a reference. Environmental changes have the same influence on both optical paths. This hardware adaption in combination with algorithmic frequency and phase evaluation leads to an accuracy  of 0.1\,nm in a measurement range of about 3\,mm in the axial dimension.\\
Other approaches aiming to extend the measurement range and to increase the resolution rely on dispersion compensation and are known by the term \gls{DEFR-OCT}, \cite{HoferFRoct,HermannFRocts}. The authors applied a numerical dispersion compensation to remove complex conjugates and therefore extend the measurement range. Another advantage is, that the point of highest sensitivity of the setup is shifted to the center of the measurement range. As the initial algorithm relied on multiple iterations of Fourier transforms, new approaches have been developed to speed up processing times. Some of them introduce artificial dispersion into the setup which can be compensated by the algorithm with only one Fourier transform and a convolution, \cite{KoettigDEFR}. Current works applying these ideas show high accuracy in the tomographic analysis of nano-structured domain walls in ferro-electric media, \cite{KirstenHiRes}.\\
In contrast to medical applications, industrial measurement tasks often demand the characterization of relatively large areas (mm\textsuperscript{2} instead of \textmugreek m\textsuperscript{2}). Scanning of samples is therefore necessary, but introduces issues regarding the accuracy and repeatability of results, \cite{OCT_surface}. Approaches to overcome these limitations are based on a setup incorporating a camera and a high resolution translation stage, are summarized by the term \textit{\gls{FF-OCT}}, \cite{DuboisFullField,Auksorius2020,Berthelon2017,Recher2020}. In this particular, approach scanning is still required in the reference arm in order to observe changes in the interference data due to different path lengths.\\
More recently, full-field approaches that avoid scanning altogether have been developed. Based on a hyperspectral imager and an etalon, \textit{Zhu et al.}, \cite{ZhuSingleShot} have shown high precision surface measurements. While one dimension of the sample is encoded in the bandwidth of the etalon-generated spectral slices , the second dimension is imaged on the camera directly. The surface profile is calculated from the phase information of the interference fringes. The resolution and measurement range of this setup is predominantly defined by the wavelength spacing of the etalon as well as by the resolution of the spectrometer. Both tuning options are opposing each other.\\
A more advanced method substitutes the etalon for a microlens array in order to decode one areal dimension on the hyperspectral imager which increases the measurement range and light efficiency, \cite{RuizHSI}. The setup was capable of measuring an axial range of up to 880\,\textmugreek m with an resolution of 0.49\,\textmugreek m while having a lateral measurement area of 3.5\,x\,3.5\,mm. Further developments by the same group have demonstrated the possibility to acquire 2500 independent probing points which increased the measurement range, light efficiency as well as the tilt angle acceptance, \cite{Reichold2019}. One drawback is that the method is only able to make use of about 50\,\% of the detector size to image  \textmugreek m\textsuperscript{2}-sized samples which decreases its lateral resolution. The axial measurement range was about 825\,\textmugreek m where a resolution of 6\,nm was achieved. According to these works, the dynamic range (DR) is defined as the inverse ratio of the resolution to the measurement range. The analyzed works of recent, high-dynamic range approaches to LCI and OCT demonstrate a progression in technology and dynamic range, Tab.\,\ref{RelatedWorks:Table_comparison_DR}.
\begingroup
\renewcommand{\arraystretch}{1.5}
\begin {table}[ht]
\caption[Comparison of current LCI approaches]{Comparison of current LCI approaches regarding measurement range, resolution and dynamic range} \label{RelatedWorks:Table_comparison_DR}
\begin{tabular}{p{2.8cm}p{2.cm}p{2.cm}p{2.cm}p{3cm}}
	
	Authors & Axial measurement range & Axial resolution & DR & Remarks \\
	\hline \hline
	Pavlíček / Häußler (\cite{PavlicekHaeussler2005}, 2005) & 900 \textmugreek m & 0.05 \textmugreek m & 18000 & \RaggedRight point sensor, scanning necessary\\
	\hline
	E. Koch et al. (\cite{KochSubNM}, 2005) & 3 mm & 0.1 nm & \num{3e7} & \RaggedRight point sensor, scanning necessary\\
	\hline
	Zhu et al. (\cite{ZhuSingleShot}, 2012) & 250 \textmugreek m & 0.1 \textmugreek m &  2500 & \RaggedRight etalon-based hyper-spectral imaging\\
	\hline
	Ruiz et al. (\cite{RuizHSI}, 2017) & 880 \textmugreek m & 0.49 \textmugreek m &  1871 & \RaggedRight microlens-based hyper-spectral imaging\\
	\hline
	Reichold et al. (\cite{Reichold2019}, 2019) & 825 \textmugreek m & 6 nm &  137 500 & \RaggedRight microlens-based hyper-spectral imaging\\
	\hline
\end{tabular}
\end{table}
\endgroup
In direct comparison, it can be deduced that one aim is to develop a full-field, areal approach to surface profilometry which is capable of measuring large distances with sub-nm resolution as Koch et al. have shown in a point sensor. Recent works such as of Reichold et al. have already demonstrated the potential of spectrally-encoded interferometry but still lack sub-nm resolution.\\
The current work aims to develop a system which is able to detect surface profiles with sub-nm resolution and a high-dynamic range in the axial as well as lateral dimension incorporating sub-nm resolution.

\section{Polymer cross-linking characterization}\label{StateOfTheArtPolymer}
The control of mechanical, electrical and optical properties of polymers during fabrication is necessary to ensure their performance, \cite{Ehrenstein}. As cross-linking is a crucial process step in order to optimize properties and fabrication parameters, it is necessary to monitor its degree, \cite{Oreski}. Long-term mechanical resistance, temperature stability as well as functional parameters such as refractive index are adjusted with the cross-linking process, \cite{Diss_Crosslinking,Hirschl, Feng-Ferris}.\\
Interesting industrial applications of polymers are centered around lithographic processing such as resist coatings and optical waveguides, \cite{Woods2014,Cumpston1999}. Polymer-based optical waveguides are usually processed by patterning (photoresist-based or direct lithography), soft lithography or printing techniques \cite{Ma2002, Wolfer2014} in order to achieve defined cross-linking and refractive index differences.
From the analysis of reaction kinetics of polymers, it is known that the density of a material as well as its refractive index changes during cross-linking, \cite{Diss_Crosslinking}. This relationship can be described by the Lorentz-Lorenz equation, \cite{Kudo, Tyng2013},
\begin{equation}
R_m = \frac{(n^2 - 1)M_w}{(n^2 + 2)\rho}
\end{equation}
where \glssymbol{MolRef} represents the molar refractivity, \glssymbol{RealRefInd} the refractive index, \glssymbol{MolMass} the mass-averaged molar mass and \glssymbol{Density} the density.
Although it is known that the Lorentz-Lorenz relation is only an approximation, it has been proven applicable to a variety of polymers and was correlated with other methods such as hardness measurements, \cite{Tyng2013, Muller2010}.
For their fabrication, optically-cured polymers have to fulfill several requirements such as optical transparency and chemical as well as thermal stability \cite{Rashed2016, Kang2001}. Advancing from conventional thermoplastics such as polymethyl methacrylate, polystyrene, polycarbonate and polyurethane, research has been geared towards the development of new polymers, which exhibit lower absorption losses and higher stability, \cite{Ma2002}. Promising classes of polymers are halogenated polyacrylates \cite{Joehnck2000}, fluorinated polyimides \cite{Kobayashi1998} or polysiloxanes \cite{Cai2008}. In particular, applications such as polymeric waveguides or direct laser writing on wafers make use of this effect to generate functional properties with refractive index changes of about $10^{-2}$, \cite{2Photon_waveguides}. {\v Z}ukauskas et al. applied this effect to generate gradient-index lens elements with a size of 50\,x\,50\,x\,10 \textmugreek m\textsuperscript{3}, \cite{3D_laser_writing}.\\
In order to characterize these functional properties alongside with the degree of cross-linking and their spatial distribution, different metrology approaches are known from literature.

\subsection{Soxhlet-type extraction}
 A very common method to determine the degree of cross-linking is the Soxhlet-type extraction,  \cite{ethylene-standard}. For that purpose, a sample is exposed to a solvent, typically a xylene isomer, in which non-cross-linked material will dissolve after a few hours. After drying, the degree of cross-linking can be calculated from the respective weights of cross-linked and non-cross-linked material. The result is very precise whereas the process is time consuming, destructive and not spatially resolving. Oreski et al. found that the time for the extraction is at least 18 hrs while drying takes another 24 hrs, \cite{Oreski2017}. Furthermore, Hirschl et al. found that the extraction time and other process parameters can have a huge influence on the repeatability of the measured degree of cross-linking, especially in weakly cross-linked samples. They determined that the repeatability ranges from 2-4 $\%$, \cite{Hirschl2015}. 

\subsection{Differential scanning caliometry} 
A method to determine cross-linking in polymers as a measure of thermal features such as  heat-flow, melting point and reaction enthalpy is \gls{DSC}, \cite{Stark,Ehrenstein,Xia}. Similar to chemical-based methods, this method does not allow spatially resolved measurements, works destructively and is time consuming. A typical measurement cycle in the so called \textit{dual-run mode} takes 2\,x\,45\,min during which a defined heating profile is applied, \cite{Oreski2017}. A comparative study has shown that different approaches for referencing the measurements to other methods might apply and also, that errors in the repeatability can be $10 \%$ and larger, \cite{Hirschl2013}. Especially weakly cross-linked samples require slower heating profiles, and hence longer measurement times, and produce inherently larger errors, \cite{Stark, Ogier2017}.

\subsection{Dynamic Mechanical Analysis}
\Gls{DMA} is a well established laboratory approach to characterize thermal and mechanical properties of materials. It is especially well suited for visco-elastic materials such as polymers, \cite{Ehrenstein}. During analysis, a small dynamic force (e.g. in tension, compression, bending or shear mode) is applied to a well defined sample, typically in a sinusoidal fashion, \cite{Menard2008}. The measurement time, the temperature as well as the oscillation frequency are typically variable parameters dependent on the sample types to be measured, \cite{DINENISO6721}. In response to the introduced stress \glssymbol{Stress}, the samples strain \glssymbol{Strain} is measured dynamically. Here, these values are measured as the amplitudes of the sinusoidal signals. Furthermore, the phase difference between both signals is measured as \glssymbol{PhiDMA}. From these, the complex modulus of the sample \glssymbol{CompModulus} $= \frac{\sigma}{\varepsilon}$ can be calculated. It holds information on the real and imaginary components which are used to characterize the elastic properties of a material. The real component \glssymbol{StorModulus} is known as the storage modulus which is proportional to the elastic deformation work stored by the sample during deformation. The imaginary component, known as the loss modulus \glssymbol{LossModulus}, is a measure for the thermo-mechanical loss of deformation work due to internal friction among other effects. Both parameters can be referred to the phase difference $\Phi$ by
\begin{eqnarray}
 E^\prime = |E^*| \cdot \cos\Phi\\
 E^{\prime\prime} = |E^*| \cdot \sin\Phi
\end{eqnarray}
From this relation, it becomes clear that the phase difference between excitation and response is crucial for the understanding of certain material properties. While fully elastic materials, such as steel, typically show a phase difference of $\Phi$\,=\,0, fully viscous materials show a phase difference close to $\Phi$\,=\,\ang{90}. In the case of visco-elastic polymers any value in between may occur. Another important measure in \gls{DMA} is the determination of the glass-transition temperature \glssymbol{GlassTranTemp}. It describes the transition from a state where molecular networks are stiff and allow only elastic deformations to a state where non-elastic deformations are also possible. The $T_G$ can be calculated with a variety of approaches, some of which are comparable to \gls{DSC} measurements, \cite{Ehrenstein}.\\
In order to calculate the applied stress to a specific sample, the knowledge of its geometry is important. Usually standard sample sizes are fabricated for the purpose of \gls{DMA} measurements, \cite{DINENISO6721}. The investigation of fabricated products or components is not possible. As the dimensional parameters depend on the kind of stress applied, their input can be significant. Any error in the determination of the samples dimension will have a strong influence on the resulting stress on the sample as well as on comparability between samples.\\
The time for a measurement can very significantly dependent on the chosen experimental parameters. Usually, the temperature range as well as the heating rate are subject to changes while the excitation frequency is kept fixed. Typical investigations of polymeric materials work in ranges from -75\,-\,\ang{150}C with a heating rate of 2\,\nicefrac{K}{min}, \cite{Bradler2018} to ranges of -150\,-\,\ang{200}C at a heating rate of 1\,\nicefrac{K}{min}, \cite{Stark}.\\
One of the main advantages of \gls{DMA} is the ability to directly measure the degree of cross-linking, \cite{Hirschl2013}. Typically, the degree of cross-linking is determined as a weighted ratio of the storage modulus, the sample density and the monomer molecular weight, \cite{Molero2019}. Other works propose the calculation as a quotient of the logarithmic storage modulus of cross-linked and non-cross-linked material, \cite{BarszczewskaRybarek2017}.\\
\gls{DMA} is, amongst others, a standard testing procedure in material development for novel polymers, fabrication technologies or the understanding of cross-linking mechanisms, \cite{Chen2018, Wu2019, Xiong2018}. With its high accuracy, the capability to determine the degree of cross-linking directly and the long measurement times, \gls{DMA} is a predominantly laboratory-based method. Applications in production accompanying tasks are not established. 

\subsection{Spectroscopy-based methods}
Non-destructive measurements of cross-linking can be obtained by using optical metrology such as Raman spectroscopy, \cite{Peike} or luminescence spectroscopy, \cite{Schlothauer}. The analysis of spectral features of reactional groups and bonds or the luminescence intensity of characteristic peaks can be utilized to calculate the degree of cross-linking. Recent works have shown that these technologies are able to characterize cross-linking of coatings on solar cells. In a comparative study, Hirschl and co-workers, \cite{Hirschl_inline}, have demonstrated that Raman spectroscopy gives comparable results to classical methods like Soxhlet-extraction. Although, it has to be noted that the measured errors of the degree of cross-linking were up to 15\,\%, especially for samples with weak cross-linking. Furthermore, acquisition times for Raman spectra depend very much on the \gls{SNR} of relevant spectral intensity peaks and hence require a large amount of averaging. Recent studies report acquisition times for single-point measurements between 50\,-\,100 s, \cite{Hirschl_inline, 3D_laser_writing}. Peike et al. \cite{Peike} point out, that Raman analysis is very material specific and can be complex with different peaks overlaying each other. Additionally, they found that the \gls{SNR} decreases with peaks at higher wavelengths as \glssymbol{IPeakRaman}\,$\sim \nicefrac{1}{\lambda^4}$ . This can be critical for weakly cross-linked material or materials with a low number of reactional groups.\\
A recent work by Schlothauer et al. has qualified  luminescence spectroscopy as a tool for cross-linking characterization with an accuracy of 4\,-\,6\,\%, \cite{Schlothauer}. However, the method requires a large amount of averaged spectra in a point-by-point scanning fashion. Acquisition times for a 16\,x\,16\,cm\textsuperscript{2} were about 80 minutes.

\subsection{Low-coherence interferometry and other optical methods}
As refractometry is a well established method to measure refractive indices, it is also suited to evaluate cross-linking in polymers, \cite{Harris2009}. It has been used to determine cross-linking progress directly during curing within thermoset polymers used as a matrix material for composites, \cite{Fernando2006}. For the purpose of these examinations, optical fibers are declad over some probing area and integrated in a to-be cured polymeric matrix in order measure the change in refractive index in terms of a change in transmission intensity, \cite{Harris2009}. Another common approach is, the integration of cleaved optical fibers in a polymeric matrix where the Fresnel reflection is measured in relation to the change in refractive index, \cite{Vacher2004,Sampath2015}. Both of these approaches are well suited for the characterization of thermoset polymers as the change in refractive index is rather high during cross-linking with \glssymbol{DeltaRealRefInd}\,=\,2\,-\,\num{6e-2}.\\
Classical optical-coherence tomography has been used to examine structural defects such as bubbles or phase separation during cross-linking  by scanning a sample with in a few seconds, \cite{OCT_curing}. Other interferometric techniques such as spectrally-resolved white-light interferometry, frequency domain interferometry or digital holographic interferometry have been utilized to measure refractive indices with accuracies in the range of \num{e-5}\,-\,\num{e-6}, \cite{Sainz1994, Kumar1995} as well as mechanical deformations on the nm-scale in material and biomedical engineering, \cite{AN2003,TaveraRuiz2018,Kumar2016}. Methods based on phase-sensitive OCT have been used to characterize photo-elasticity on polymeric composite materials and might also be suitable for cross-linking characterization, \cite{Oh2003}. More recent works made use of the combination of LCI and confocal approaches in order to measure refractive index of transparent media from a distance, \cite{Francis2019}.

\subsection{Spatially-resolved approaches}
None of the technologies presented so far for cross-linking characterization are inherently spatially resolving. Spatial resolution for chemical methods such as \gls{DSC} or \gls{DMA} is usually realized by cutting samples into defined sub-samples, \cite{Schlothauer2017, Schlothauer2016}. Especially spectroscopic technologies such as Raman spectroscopy gain spatial resolution by scanning over a sample. Due to the need for integration over multiple spectra at each point, the measurement times of a sample having 330 x 150 probing points can be as long as 82 minutes, \cite{Schlothauer2017}. Although this technology can be used to cover an area of a few cm\textsuperscript{2}, the typical lateral resolution limit is about 2 \textmugreek m, \cite{Bokobza2019}. Consequently, other groups have shown a significant increase in lateral resolution down to 35 nm by combining \gls{FTIR} spectroscopy with AFM. This approach increases the measurement time significantly so that an area of 1 x 1.5 \textmugreek m\textsuperscript{2} was measured in 7.4 hrs, \cite{Amenabar2017, Dazzi2017}. None of these technologies have been used for cross-linking characterization so far. Among the LCI techniques, Guerrero et al., \cite{SRWLI} published an approach which was able to reach a spatial resolution of about 17 \textmugreek m. It was based on a phase estimation of intensity extremes and shows a theoretical refractive index resolution of \num{e-4} which did not take any thermal or noise influences into account. Shortcomings of the method are the restriction to measurements of the differential refractive index as well as its dependence on intensity measurements which are influenced by noise.\\
Beside these technologies, some other imaging approaches might be interesting for cross-linking analysis. Singh et al. have shown that corneal cross-linking can be indirectly imaged by using an OCT-based approach, \cite{Singh2017}. An evaluation of the mechanical stiffness was performed by analyzing  the damping vibrational response of the cornea by an OCT system. Other works have shown the applicability of \gls{NMR} imaging for polymer characterization, \cite{Koenig2000}. The resolution of \gls{NMR} is strongly dependent on the natural line width of the molecules that are measured and the magnetic field gradient. The method is primarily suited to characterize samples in a laboratory environment. A different approach would be \gls{ToF-SIMS} where spatial resolution in the axial domain is achieved. By structuring the sample under test with an ion beam, the cross-linking depth profile is acquired in depths up to a few nanometers, \cite{NaderiGohar2017}.\\
None of these technologies is suited for the in-line characterization of polymers in production or for products.\\
The main ambition in the characterization of polymeric cross-linking is to establish a measurement technique which is capable to measure non-destructive on product level, of a high refractive index resolution (better than \num{e-3}), to measure fast enough to be process-integrated and which has spatial resolution at the same time. None of the technologies known from literature combines these characteristics.

\section{Film thickness measurement}
Appropriate metrological tools for in-line characterization have to fulfill different requirements in distinction to classical lab-based technologies. In particular, tools are often required to measure a certain area ($\approx$ some \textmugreek m\textsuperscript{2}) with high axial resolution in accordance with the speed of the respective processing step and perform measurements autonomously for a large variety of materials. As the amount of data gathered is usually large, appropriate algorithms have to monitor key values constantly and only report on deviations. This applies especially to the production of high volume, complex multi-material systems in roll-to-roll processes as in the production of thin-film systems,  \cite{Osten}.\\
State-of-the-art technologies for thin-film characterization  most often incorporate variants of ellipsometry, \cite{Vedam} and reflectometry,\cite{Schenk}. While both methods are capable of being production accompanying tools \cite{Fried}, they usually cannot provide high lateral resolution, \cite{Juhasz, Major}. As experimental works showed, higher resolutions and the ability to measure multi-layer systems, one of the most challenging problems in both technologies is the knowledge of material constants and models in order to find or fit correct start parameters as well as converging criteria for fits, \cite{Hilfiker2019,Leng,Aspnes}.\\
Alternatively, optical and mechanical properties of thin-films have been characterized by interferometric approaches, \cite{Duparre}. While reaching high lateral resolutions, these technologies tend to be slow, as they rely on \textit{z}-scanning of the \gls{OPD} between the sample and reference arm, \cite{de1993three}. With the development of  \textit{k}-scanning approaches this disadvantage was overcome as mechanical scanning in the axial dimension was substituted by a tuneable light source to scan through \textit{k}-space, \cite{kinoshita1999optical,kuwamura1997wavelength}. However, these approaches are limited at both ends of the thickness measurement range. On the one end, the typical spectral scanning range of 20\,nm limits the maximum measurable thickness to about 15\,\textmugreek m. On the other end, the minimal resolvable thickness is limited by the distinguishability of Fourier peaks which are unique for every reflection of a material interface. The thickness as well as the amount of dispersion of every material contributes to the width and position of each Fourier peak. Ghim et al. demonstrated that 500\,nm is a typical minimal film thickness which can be resolved with this technique, \cite{ghim2006thin}.
\subsection{Spectral reflectometry}
A common method for the determination of film thickness and material parameters of thin-films in the semiconductor industry is spectrally-resolved reflectometry, \cite{Schenk}. This technique captures the spectrum of a broadband light source back-reflected from a sample's surface and tries to fit it to a model spectrum that was calculated beforehand. During measurement, a reference spectrum, from e.g. a pure silicon surface, is captured first in order to reference absolute intensity values later. Afterwards, the reflection spectrum of a substrate with a thin-film is captured by illuminating the sample with a broad spectrum under a defined angle $\theta$ (typical $\theta$\,=\,\ang{0}). The back-reflected spectrum is typically collected with a fiber and spectrally decomposed using a grating spectrometer. For the determination of the film thickness, the real and imaginary part of the substrates refractive index as well as of the film material has to be known. By applying e.g. a brute-force fitting routine, the error between measured and calculated spectrum is minimized to obtain the film thickness. In order to calculate all possible reflection spectra for different thicknesses of a material system, the transfer-matrix approach is typically applied, \cite{Born}. Using this approach, the $E$-field is described as a matrix for every layer of material as well as for every transition between different materials. Each matrix $M$ accounts for the reflectivity and transmission characteristics of every material and transition, \cite{Schenk}.\\
In practice, a variety of theoretical spectra can be calculated with an assumed a set of different film thicknesses in order to evaluate the measured film thickness. Usually, the \gls{RMS} is determined between measured and calculated values. Additionally, it is possible to perform calculations for other parameters like the refractive index (real and imaginary parts) in relation to the wavelength on the basis of knowledge of the film thickness. The technique is advantageous because of the low instrumental effort, the lack of sample preparation and the fast data acquisition times. However, the method is tied to the knowledge of the material model of a sample and caution is necessary when handling measurements from samples with changing material parameters which can occur in a production environment. Some research has shown, that the parallel acquisition of data under a range of polarization angles in combination with a camera might enable the usage of spectral reflectometry in process-accompanying for complex multi-layer samples, \cite{Ghim2019}. Currently, research has proven the possibility to combine technologies such as spectral reflectometry with other techniques such as low-coherence interferometry, spectroscopic ellipsometry and Raman spectroscopy, \cite{vanDuren2019,Vodak2017,Luria2020,Hlubina2009,Ohlidal2011}

\subsection{Spectroscopic ellipsometry}
A widely used method to determine the dispersion properties of thin-films is \gls{SE}. This technique evaluates the change of polarization of light reflected from one or more layers of a thin-film of material. In particular, the technique is based on the determination of the complex quotient of the reflection coefficient of both polarization components. Typically, the amplitude quotient \glssymbol{SEpsi} and the phase difference \glssymbol{SEdelta} of s- and p-polarized light are captured as a measure, \cite{FujiwaraEllipso}. The data of these measures can be gathered either at a discrete wavelength or at a broad spectral range using a spectrometer as detector. In order to determine the dispersion of a thin-film, the geometrical parameters such as film thickness and angle of incidence as well as the refractive index of the substrate material have to be known. By using an appropriate mathematical model of the material system, the real and complex part of the films' refractive index can be calculated. Under the assumption of a measurement in a broad spectral range, this approach can be used to derive the dispersion of the film material. As \cite{Vedam} proved, appropriate models can be used to determine the dispersion of various materials in multi-layer material systems as well. Park, \cite{Park} has shown that the determination of the real and imaginary part of the dielectric function, hence the refractive index, the extinction coefficient and the dispersion in the range of (190\,-\,1000)\,nm is possible. The measurements were performed in an angular range from \ang{45}\,-\,\ang{75}  in \SI{5}{\degree} intervals. In the model, the authors took into account the material properties of the surroundings, a rough surface, a thin-film of \gls{CdS} as well as for a substrate material. The film thickness of the rough surface and the \gls{CdS} layer were determined by minimizing the error between calculated and measured values for the amplitude quotient $\Psi$ and the phase difference $\Delta$. Additionally, the gathered data in relation to the measured spectral range could be used to determine other material characteristics such as characteristic peaks which are dependent on the crystal structure of the material. Furthermore, the determination of the dispersion (according to the Wemple-DiDomenico model \cite{Wemple}) could be used to evaluate the bandgap of the material.
Other works, such as \cite{Dai}, demonstrated the determination of optical properties on a variety of thin-film materials under the usage of different dispersion models. Obviously, the knowledge of material properties and the application of the correct model are important in order to obtain precise results using \gls{SE}.
Some developments proved that an application of the technology as a process monitoring tool could be possible. Fried et al. \cite{Fried} demonstrated that by a modification of the setup and the simplification of the fit function, significantly faster data acquisition and analysis are possible. For example, in order to evaluate \gls{CIGS} solar cells, a correlation of the fit values with the amount of \gls{Ga} in the material was performed. This reduces the number of adjustable fit parameters to one and speeds up the calculation. The authors show, that a modification of the fit model can be done for other materials which are relevant for solar cells as well. It was also shown, that a modification of the mechanical setup from using a collimated beam to the usage of a so called expanded beam can decrease the time needed for data acquisition. This setup allows for the illumination and data acquisition on a large area and over various angles of incidence at the same time. Also, the simultaneous acquisition of data under several wavelengths and angles was demonstrated. When the sample is translated under the measurement spot, the fast measurement of areas with several cm\textsuperscript{2} becomes possible. The authors of \cite{Juhasz,Major} showed maps of film thickness for various material systems on a wafer. Additionally, it was shown, that a line projection of the expanded beam can be utilized to perform film thickness measurement as process monitoring in a roll-to-roll production environment. Although the results are promising, it has to be clarified that this approach only works in settings were the material is well known and the production focuses on a small number of material systems (in \cite{Fried} the manufacturing of thin-film solar cells).\\
According to \cite{Aspnes}, the typical usable spectral range of \gls{SE} is (130\,-\,2000)\,nm whereas most practical setups use a reduced range. While the lateral resolution of production accompanying is rather low, experimental works have shown that a lateral resolution of 4\,\textmugreek m can be reached using a microscope objective, \cite{Leng}. However, this makes scanning of the sample a necessity in order to measure reasonably large samples.\\
Modifications of spectroscopic ellipsometry have been combined with scatterometry in order to perform the control of optically critical dimensions in the semiconductor industry, \cite{Chouaib}. The measurement of the smallest lithographic structures like the 7-nm-node, introduces new challenges to metrology as e.g. dielectric functions for layers thinner $<$\,10\,nm. According to \cite{Aspnes}, future challenges for \gls{SE} lie primarily in the increase of accuracy of the optical models as well as in the generation of new models. Furthermore, approaches for the combination with other techniques,  \cite{Hilfiker}, as well as for parallel acquisition of different parameters over a large spectral and angular range exist,  \cite{Opsal}. In this area, it will be necessary to develop new and adapted components such as light sources with shorter emitting wavelengths,\cite{Huang}.\\
Some current research has tried to minimize the disadvantages in terms of the need for scanning by developing one-shot techniques using broadband light sources, a modulated carrier-frequency from an interferometer and a common spectrometer, \cite{Dembele2020}. Other works have tried to extend known analysis models by new mathematical approaches or by the combination of data from \gls{SE} and other technologies such as reflectometry, \cite{Ohlidal2020,Necas2014,Ogieglo2020}. Some works have shown imaging \gls{SE} techniques which suffer from poor lateral resolution ($\approx$ 60\,\textmugreek m) and the need for temporal effort to perform the measurement and analyze the data ($\approx$ 8\,s per wavelength in $\Delta \lambda$\,=\,(400\,-\,700)\,nm), \cite{Chegal2004,YonghongMeng2010}. Additionally, some work has been done to combine spectral ellipsometry with low-coherence  or phase-shifting interferometry in order to get information on the surface profile of the sample alongside with the thin-film thickness, \cite{Kim2019c,Yun2018}. While the resolution for film thickness and surface profile were in the nm-range, the combination of multiple technologies imposed new obstacles in terms of data fusion.
Spectral ellipsometry is especially powerful in laboratory situations where the material model of a sample is well understood. In this case, thickness resolutions in the sub-nm can be achieved. In situations like an production environment, \gls{SE} suffers from its low-lateral resolution and time-consuming data acquisition/analysis.

\section{Material dispersion}
\begin{figure}[h]
	\centering
	\begin{tabular}{c}
		\begin{overpic}[scale=0.4,grid = false]{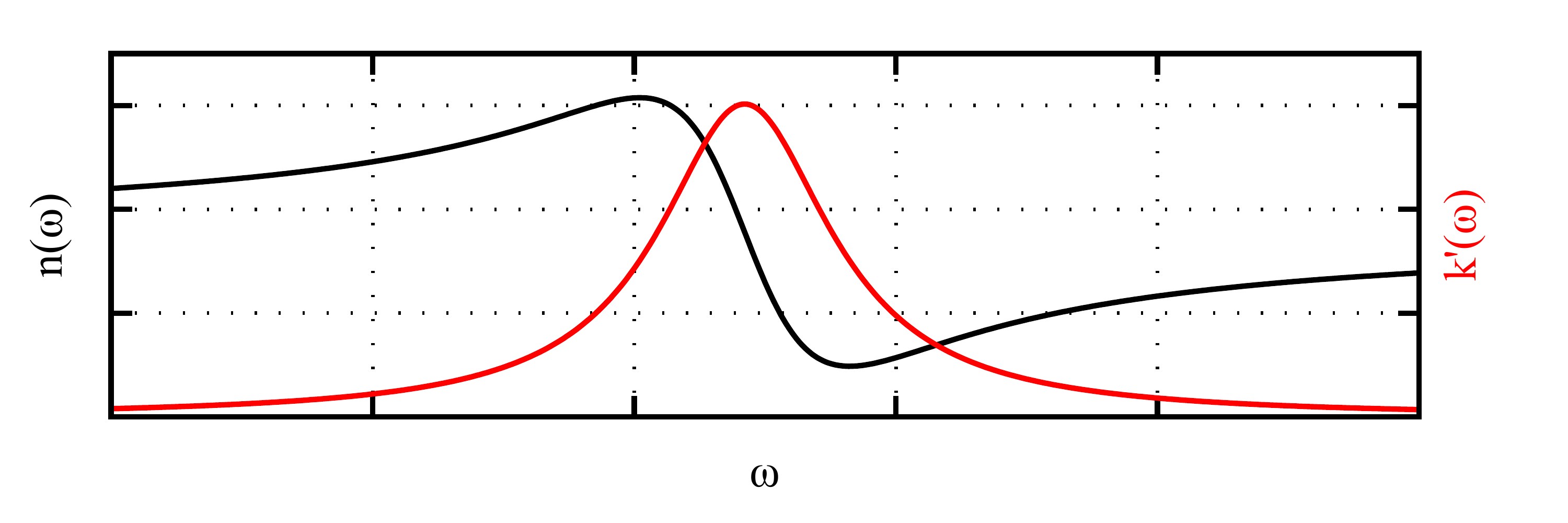}
			\put(50,9){\makebox(0,0){\textcolor{blue}{$\scriptstyle \omega_j$}}}
			\put(47.25,6.7){\tikz \draw [line width=0.5mm,color = blue,-] (0,0) -- (0,2.8);}
		\end{overpic}
	\end{tabular}
	\caption{Depiction of the slope of the real part $n(\omega)$ and the imaginary part $k^\prime(\omega)$ of the refractive index for a simplified, resonator model with a resonant frequency $\omega_j$.}\label{Pic:Kramers_Kronig}
\end{figure}

The refractive index, \glssymbol{RefIndAbs}, of a material is a measure of the refraction of light traveling through a material. Since it is defined as the relation of the vacuum light velocity \glssymbol{VelLightVac} to the light velocity in the material \glssymbol{VelLight} and can be described as the optical resistance of the material, \cite{Kohlrausch} pp. 94,
\begin{equation}\label{refractive_index}
n_{abs} = \frac{c_0}{c}.
\end{equation}
As most practical applications are operated in air rather than in vacuum, the term \textit{relative refractive index}, \glssymbol{RefIndRel}, is more common. This relative index is defined as the refractive index in relation to the refractive index of air, \glssymbol{RefIndAir},
\begin{equation}\label{relative_refractive_index}
n_{rel} = \frac{n_{abs}}{n_{air}}.
\end{equation}
The dispersion of a material describes the relation of the refractive index, and therefore the phase velocity of the light, with respect to the wavelength. This implies that the refractive index is wavelength dependent $n = n(\lambda)$. Normal and anomalous dispersion can occur. In the case of glass or most polymers, normal dispersion is present in the visible wavelength range. In this case the material shows comparatively high refractive indices ranging from (300\,-\,500)\,nm and significantly lower refractive index values in higher spectral regions (e.g. (500\,-\,900)\,nm). Anomalous dispersion consequently shows an opposing behavior. In materials with relatively low optical density such as glasses or polymers, the dispersion characteristic is closely related to the absorption behavior. In general, the material behavior can be described using a quantum mechanical model. The application of a simplified description utilizing an electro-magnetic model is feasible for materials with low optical density, \cite{Born}.
This model describes the influence of an electro-magnetic wave on the moleculary structure of a material. Each bond charge of a material's molecular structure has a specific resonant frequency \glssymbol{OmegaJ} which defines its absorption behavior. Dependent on the material properties, an incoming electro-magnetic wave excites the molecular structure which than leads to a specific refraction behavior. The resulting behavior of the real and imaginary part of the refractive index, \glssymbol{RealRefInd} and \glssymbol{ImagRefInd}, can be descried mathematically using a resonator model of the following form dependent on the angular frequency \glssymbol{Omega}, \cite{BachNeuroth},
\begin{equation}\label{material_dispersion_general}
(n- \text{i}k^\prime)^2 = \frac{Ne^2}{\varepsilon_0m}\sum_{j}^{}\frac{F_j}{\left(\omega_j^2-\omega^2 \right)+\text{i}\gamma_j\omega},
\end{equation}
where \glssymbol{NumMol} is the number of molecules per unit volume, \glssymbol{CharElec} and \glssymbol{MassElec} are the charge and mass of the electron, \glssymbol{VacPerm} is the vacuum permittivity, \glssymbol{AbsStren} is the strength of the absorption and \glssymbol{FricForce} is a measure for the frictional force at the resonance frequency.
The \textsc{Kramers-Kronig} model takes this relation into account, \cite{KramersKronig}. It describes the relation of absorption and dispersion of light in a material by combining the real and the imaginary components into one complex model, Fig. \ref{Pic:Kramers_Kronig}.
This relation has a general validity for all materials. For spectral ranges with negligible absorption of the specific material ($k^\prime \rightarrow 0$), the resonance frequency tends to differ significantly from the frequency of the incident light. For this case the relation can be simplified and expressed in terms of the wavelength \glssymbol{Lambda}/\glssymbol{LambdaJ}
\begin{equation}\label{material_dispersion_zero_absorption}
n^2 - 1 =  \frac{Ne^2}{\varepsilon_0m}\sum_{j}^{}\frac{F_j}{\omega_j^2-\omega^2 } = \frac{Ne^2}{4\pi^2 c^2\varepsilon_0m}\sum_{j}^{}\frac{F_j\lambda_j^2\lambda^2}{\lambda^2-\lambda_j^2 },
\end{equation}
where one can abbreviate the material specific terms with \glssymbol{MatIndAbrrev}
\begin{equation}\label{dispersion_coefficiant}
A_j =  \frac{Ne^2F_j\lambda_j^2}{4\pi^2 c^2\varepsilon_0m},
\end{equation}
which leads to the simplified equation
\begin{equation}\label{material_dispersion_praxis}
n^2 - 1 =  \sum_{j}^{}\frac{A_j\lambda^2}{\lambda^2-\lambda_j^2 }.
\end{equation}
When analyzing real materials like glasses or polymers, the absorption characteristics differ very much from the simplified model. Most notably, multiple absorption peaks exist in contrast to only one single absorption peak at one resonant frequency. Furthermore, these peaks inherit a distinct structure which leads to a complex dispersion characteristic. For this reason, a number of approximation relations exist. These relations are always valid in a well defined spectral range which is most likely far away from a strong absorption band. An extrapolation outside the valid spectral range is not considered valid. Over the years a broad range of approximation relations have emerged for different materials , spectral ranges and achievable accuracies, Tab. \ref{table_dispersion_equations}.
\begingroup
\renewcommand{\arraystretch}{1.5}
\begin {table}[ht]
\caption[Models for the description of material dispersion]{Models for the description of material dispersion according to \cite{BachNeuroth}} \label{table_dispersion_equations}
\begin{tabular}{p{2.5cm}p{2.5cm}p{4cm}p{5cm}}
	Model & Properties & Equations & Remarks \\
	\hline \hline
	\textsc{Cauchy} & precision in VIS $10^{-4}$ & $n(\lambda) = a + \frac{b}{\lambda^2} + \frac{c}{\lambda^4} $ & first model from 1830 \\
	\hline
	\textsc{Hartmann} & precision in VIS $10^{-4} - 10^{-4}$ & $n(\lambda) = n_0 + \frac{A}{(\lambda - \lambda_{0})^{B}}$ & constants have to be found empirically, $0.5 \leq B \leq 2$ \\
	\hline
	\textsc{Sellmeier} & precision  $10^{-5} - 5 \cdot 10^{-6}$ & $n(\lambda)^2 = 1 + \sum_{j=1}^{N}\frac{a_j \lambda^2}{(\lambda^2 - \lambda_j^2)}$ & based on physical model, very good fit characteristics \\
	\hline
	\textsc{Helmholtz-Ketteler-Drude} & based on \textsc{Sellmeier}, precision similar & $n(\lambda)^2 = a_0 + \sum_{j=1}^{N}\frac{a_j}{(\lambda^2 - \lambda_j^2)}$ & see also \cite{Born} \\
	\hline
	\textsc{Schott} & precision $10^{-5} - 10^{-6}$ & $n^2(\lambda) = A_0 + A_1 \lambda^2 + A_2 \lambda^{-2} + A_3 \lambda^{-4} + A_4 \lambda^{-6} + A_5 \lambda^{-8}$ & derived from \textsc{Sellmeier} equation; valid over large spectral range and for nearly all glasses\\
	\hline
	\textsc{Herzberger} & precision $10^{-4}$ (VIS) & $n(\lambda) = A_0 + A_1\lambda^2 + \frac{A_2}{\lambda^2 - \lambda_0^2} + \frac{A_3}{(\lambda^2 - \lambda_0^2)^2}$ & based on analytical investigation; $\lambda_0$\,=\,168\,nm\\
	\hline
	\textsc{Geffcken} & precision $10^{-6}$ & $n(\lambda) = 1 + \lbrack 1-D(\lambda) \rbrack \lbrack n(\lambda_1) - 1 \rbrack + B(\lambda) \lbrack n(\lambda_2) - n(\lambda_1)\rbrack + D(\lambda)\delta(\lambda) $ & combination of two functional relations (spectral behavior of a normal glass and abnormal dispersion from that reference) \\
	\hline
	\textsc{Buchdahl} & precision $1-2 \cdot 10^{-4}$ & $n(\omega) = n_0 + \nu_1\omega + \nu_2\omega^2 + \nu_3\omega^3 \dots  + \nu_n\omega^n$ with $\omega(\lambda) = \frac{\lambda - \lambda_{0}}{1 + \alpha(\lambda - \lambda_{0})}$ & reference $\lambda_{0}$\,=\,587.6\,nm (VIS); $\alpha$\,=\,2.5\,=\,const. for glasses; fast conversion for $\omega$ of $j = 2$ or $j = 3$
\end{tabular}
\end{table}
\endgroup
The measurement of the refractive index of a medium is usually defined in relation to the refractive index of air, Eq. (\ref{relative_refractive_index}). For this reason, the conditions of the surrounding air should be well defined in terms of temperature, pressure and humidity. Calculations are based on typical values for laboratory conditions, Tab. \ref{table_measurement_conditions}.
\begingroup
\renewcommand{\arraystretch}{1.5}
\begin {table}[ht]
\caption[Typical environmental conditions for the calculation of refractive index influences]{Typical environmental conditions for the calculation of influences regarding the refractive index according to \cite{BachNeuroth}}
\centering
\begin{tabular}{cc}\label{table_measurement_conditions}
parameter & value \\
\hline \hline
temperature $T$ &  20 $^\circ C$\\
thermal expansion coefficient $\alpha$ &  0.00367 $K^{-1}$ \\
wavelength $\lambda$ &  589.592 nm \\
air pressure  $p$ & 760 Torr \\
partial pressure of vapor $w$ & 10 Torr
\end{tabular}
\end{table}
\endgroup
In order to perform these measurements for refractive indices at defined single wavelengths with high accuracy, methods like refractometry (accuracy $\pm$\,\num{e-6}) and interferometry (accuracy $\pm$\,\num{e-7}) are preferred.
Measurements are typically performed at one or few defined wavelengths of a gas discharge lamp such as the sodium line at \glssymbol{LambdaD}\,=\,589.592\,nm. The influence of the temperature \glssymbol{Temp}, pressure \glssymbol{Pressure} and humidity \glssymbol{ParVapor} on the refractive index of air and therefore on the relative refractive index measurements of materials can be described as follows
\begin{eqnarray}
\dv{n_{air}}{p} &=& \frac{n_{air}-1}{p} = +0.268\times 10^{-6}\text{ hPa}^{-1}\label{refractive_index_to_pressure_change}\\
\dv{n_{air}}{T} &=& -\alpha\frac{n_{air}-1}{1+\alpha T} = -1.071\times 10^{-6}\text{ K}^{-1}\label{refractive_index_to_temperature_change}\\
\dv{n_{air}}{w} &=& \frac{41 \times 10^{-9}hPa^{-1}}{1+\alpha T} = -0.039 \times 10^{-6}hPa^{-1}.\label{refractive_index_to_partial_pressure}
\end{eqnarray}
Where the thermal expansion coefficient is introduced as \glssymbol{AlphaTemp}. From these equations, it can be calculated that minor changes in the environmental conditions such as a temperature change of \glssymbol{DeltaTemp}\,=\,$\pm$\,5\,K or a pressure change of \glssymbol{DeltaPressure}\,=\,$\pm$\,20\,hPa can have a significant influence on $n_{air}$. For comparability, the joint commission for spectroscopy has developed an engineering equation for the determination of the refractive index of dry air (with 0.03\,\%\,$CO_2$ volume content) $n_{air}$ at $T$\,=\,\ang{15}C and the reference pressure \glssymbol{BasePressure}\,=\,760\,Torr in the spectral range between (200\,-\,1350)\,nm, as well as for other temperatures of air under the same pressure in the visible spectral range, \cite{BachNeuroth}.

\subsection{Thermo-optic coefficient}
Analogous to the behavior of air, the refractive index of a material is dependent on its temperature where the discrete form can be noted
\begin{equation}\label{thermo-optic_coefficient}
\frac{n\left(\lambda_k,T_{i+1}\right)-n\left(\lambda_k,T_{i}\right)}{T_{i+1}-T_i} = \frac{\Delta n\left(\lambda_k,T_{i,i+1}\right)}{\Delta T_{i,i+1}}.
\end{equation}
For practical reasons this dependency is often evaluated by measuring the refractive index at different \textsc{Fraunhofer} lines \glssymbol{LambdaK} at different temperatures of interest \glssymbol{TempI}, \cite{BaakThermal, SchottTemperature, ToyodaTemperature}. 
An appropriate fit of this equation has to determine six parameters when calculating a single temperature. When incorporating the correct thermal dependency, that number of necessary fit parameters increases to eighteen. In order to simplify the calculation, some boundary conditions have to be implemented. As the influence of temperatures on the refractive index is relatively low, it can be described by using a simplified \textsc{Sellmeier} model. Hoffmann et al., \cite{Hoffmann}, have shown, that a reduction to a one-term model ($i=1$) is sufficient  where measurements are performed at only one wavelength \glssymbol{Lambda} in reference to a base wavelength \glssymbol{Lambda0}
\begin{equation}\label{thermo-optic_dispersion_simplified}
 \begin{split}
\frac{\text{d}n(\lambda,T)}{\text{d}T} = \frac{n^2(\lambda,T_0)-1}{2n(\lambda,T_0)}\times \\
 \left(D_0 + 2D_1(T-T_0) + 3D_2(T-T_0)^2 + \frac{E_0+2E_1(T-T_0)}{\lambda^2-\lambda_{0}^2}\right),
\end{split}
\end{equation}
which results in
\begin{equation}\label{thermo-optic_dispersion_simplified_integrated}
 \begin{split}
\Delta n(\lambda,T-T_0) = \frac{n^2(\lambda,T_0)-1}{2n(\lambda,T_0)} \times \\
\left(D_0(T-T_0) + D_1(T-T_0)^2 + D_2(T-T_0)^3 + \frac{E_0(T-T_0)+E_1(T-T_0)^2}{\lambda^2-\lambda_{0}^2}\right),
\end{split}
\end{equation}
after integration. Usually, calculations are performed with a reference Temperature \glssymbol{Temp0} and a single measurement temperature \glssymbol{Temp} in order to fit the values for the thermo-optic coefficients \glssymbol{ThermOpCoeffs}. For a commonly used glasses these coefficients can be found in tables, e.g. in \cite{SchottTemperature}.

\subsection{Photo-elastic influences}
A second important influence on the refractive index and therefore on the dispersion is (mechanical) stress acting on a material. While materials like glasses or polymers are usually showing isotropic behavior, the application of mechanical stress \glssymbol{Stress} can lead to anisotropic behavior of the refractive index. The reason for this is that the refractive index is dependent on the electric field vector in relation to the stress plane in a sample. For a calculation, the relation of the refractive index to the stress plane can be described in parallel, \glssymbol{RefIndPar} and perpendicular, \glssymbol{RefIndPerp} orientation, \cite{BachNeuroth},
\begin{equation}\label{refractive_index_parallel_plane}
n_{\parallel} = n + \frac{\text{d}n_\parallel}{\text{d}\sigma}\sigma = n + K_{\parallel}
\end{equation}
and
\begin{equation}\label{refractive_index_perpendicular_plane}
n_{\perp} = n + \frac{\text{d}n_\perp}{\text{d}\sigma}\sigma = n + K_{\perp}.
\end{equation}
If the deformation as a result of stress is elastic, \textsc{Hooke}s law can be applied. In this case $\text{d}n/\text{d}\sigma$ can be written as the stress optical coefficients \glssymbol{KPar} and \glssymbol{KPerp}.
These coefficients are known for most common materials and are to be determined for example in a 4-point-bending test.
In materials like glasses, the refractive index typically changes equally for both directions in relation to the stress plane. In the case of hydrostatic pressure \glssymbol{Pressure}, not only the refractive index of the material itself but also of its surrounding medium will change. In this case the change is determined as
\begin{equation}\label{refractive-index_hydrostatic-pressure}
\frac{\text{d}n}{\text{d}p} = K_{\parallel} + 2K_{\perp}.
\end{equation}
Acousto-optical modulators are applications that make use of this effect. The stress-optical constant \glssymbol{K} can be found for a variety of materials in tables such as in \cite{RohrbachSpOp}. In some cases, it might be necessary to determine it experimentally for a specific material. An experimental setup would consist of a sample under uni-axial load such as pressure or tension where a polarized beam of light illuminates the sample. The angle of the polarization will be \ang{45} towards the main stress axis. The polarized beam of light can be denoted as a superposition of one component perpendicular to the main stress axis and one component parallel to this axis. By applying stress on the sample, the two components will experience an optical path difference \glssymbol{SOPathDiff} dependent on the sample length \glssymbol{SOSampLen} which can be described with
\begin{equation}\label{photoelastic_path-difference}
\Delta s = \left(n + \frac{\text{d}n_\parallel}{\text{d}\sigma}\sigma - n - \frac{\text{d}n_\perp}{\text{d}\sigma}\sigma \right) l = (K_{\parallel} - K_{\perp})\sigma l.
\end{equation}
This can also be written as a phase difference \glssymbol{SOPhaseDiff} between the two components with
\begin{eqnarray}\label{photoelastic_path-difference-phase}
\Delta \Phi = \frac{2\pi}{\lambda}(K_{\parallel} - K_{\perp})\sigma l = \frac{2\pi}{\lambda}K\sigma l \\
\text{with }K = K_{\parallel} - K_{\perp}.\label{photoelastic_coefficient_combined}
\end{eqnarray}
Clearly, a material's stress optical constant is also dependent on the dispersion of the material, \cite{Hoffmann1992stress}. Furthermore, the stress optical coefficient $K$ shows a temperature sensitivity. Hoffmann et al. \cite{Hoffmann1994temperature} showed that usually only the stress-thermo-optical coefficients  \glssymbol{StressThermOpCoeffs} of the equation:
\begin{eqnarray}\label{photoelastic_coefficient_temperature-dependence}
K(\lambda,\Delta T) = A_{0} + A_{1}\cdot \Delta T + A_{2}\cdot \Delta T^2 + \frac{B}{\lambda^2 - \lambda_{0}^2}\\
\text{with } \Delta T = T- T_0
\end{eqnarray}
are relevant for glasses. For most materials, the temperature tends to have a minor influence until the glass transition temperature is reached, \cite{Berger, Manns}.

\subsection{Characterization of dispersion}
The knowledge of the dispersion of materials and optical components plays a significant role in several areas of photonics. For example, the slope of dispersion in optical fibers in communications, due to different mechanisms like waveguide and material dispersion, determines the bandwidth and range of a transmission system, \cite{Liu}. During the construction of laser sources with ultra short pulses the knowledge and control of the dispersion of components is crucial in order to ensure the generation of short pulses with high energy in a determined spectral range, \cite{Chong}. Components such as fibers and mirrors have high demands regarding the determination of their dispersion behavior, \cite{Dombi}.
According to the specific requirements of each application, different metrology approaches are common.

\subsubsection{Time-of-flight measurement and phase-shift method}
In the field of optical communications, methods which utilize the direct measurement of the propagation time of an optical pulse as a measure for the dispersion characteristics are common and known as time-of-flight measurements. As communication lengths are typically very long (tens to hundreds of kilometers), dispersion has a heavy impact on the propagation characteristics. Dispersion can lead to time delays of pulses and to spectral broadening of the optical pulses, \cite{Abas}. In order to increase the transmission speed and data rates, modern systems implement approaches such as \gls{WDM} where information is sent as spectrally fine separated pulses. The occurrence of dispersion influences the ability to clearly separate the pulses and therefore the data sent. Additional compensation mechanism have to be implemented,  \cite{Abas}. As a method of characterization, the direct measurement of time delays between sent pulses is typically used in optical communications to characterize the dispersion behavior. A major advantage of such methods is that it can be used also in already existing fiber transmission installments, \cite{Page}. In this way, not only the dispersion of the fiber, but also of the system as a whole can be determined. In the case that fiber sections are 4\,km or longer, a significant amount of dispersion is present, so conventional metrology can be used to determine the time delay between pulses. The measured delay is a superposition of different dispersion mechanisms such as material and waveguide dispersion.\\
Typical light sources used in this context are \textsc{Raman} lasers, \glspl{SLD}, erbium-doped fiber lasers and semiconductor lasers. According to \cite{Page}, the main requirements for these light sources are a broad spectral range, a good tunability of a certain center wavelength and its full width half maximum as well as a high spectral power density.\\
One major disadvantage of this method is the requirement of appropriate hardware in order to achieve the necessary temporal resolution. In the case of samples with relatively low dispersion (circa 100\,\nicefrac{fs}{nm$\cdot$km}) it is necessary to have samples a substantial length in order to characterize them with standard equipment. Fast photo detectors are able to work at rise times of a few picoseconds (circa 15 ps according to \cite{Dematos}). Although \cite{Dematos} could implement an autocorrelator in order to realize the measurement of very small phase differences, the shortest measurable fiber length was not shorter than 100\,m. An application of the time-of-flight method to characterize samples with very low dispersion or very short lengths in the range of a few mm or \textmugreek m is not possible.\\
A further development of the time-of-flight method is a phase-sensitive detection. In this approach, light sent through a sample is modulated with a sine signal which can be detected by an appropriately designed phase-selected amplifier circuit, \cite{Costa}. This technique ,in contrast, allows to delay measurements, the detection of much smaller differences in the dispersion related delays (with a resolution of about 10\,ps). Problems arise from low light intensities and a bad \gls{SNR}, \cite{Thevenaz}. Despite these disadvantages, this method is widely used as a resolution enhancement of time-of-flight measurements with rather long fiber lengths in the telecommunications industry.

\subsubsection{Time-domain interferometric measurements}
A method to cope with the requirements regarding the temporal resolution in time-of-flight measurements is interferometry in the time domain. For this approach, an interferometer (e.g. of the Michelson type) with one fixed and one movable arm is set up, \cite{Spittel}. Usually broadband light sources with low coherence lengths, such as \glspl{SLD} and gas-discharge lamps filtered using a monochromator, are used in these setups. In a typical experiment, the movable, reference arm is translated in a continuous fashion while the corresponding, combined intensity signal of both arms is recorded with a photo diode. The shape of the recorded interferogram depends on the coherence properties of the light source. This method is similar to coherence scanning interferometry used for profilometry, see subsection \ref{Lit:SubSec:CSI}. The execution of similar experiments using different center wavelengths is the basis for the dispersion measurements. After these experiments in a dispersion-free interferometer, a sample of interest can be introduced in one of the interferometer arms. After the recording of several new interferograms at interesting wavelengths, a temporal delay of the maxima of the interferograms in relation to the dispersion-free measurements can be done, \cite{Knox,Diddams,Boeswetter}. A fit, using an appropriate dispersion model such as the Sellmeier model, can be performed using the maxima of the recorded interferograms. The temporal resolution in this approach is strongly dependent on the mechanical resolution of the translation stage. The step width and its repeatability define the possible measurable temporal differences between different interferograms.\\
The main disadvantage of this approach is the effort necessary to perform the measurements. Although broadband light sources can be used, the dispersion characterization over a large spectral range with a high spectral resolution takes time. Also, the appropriate filtering for each measurement has to be carried out. Furthermore, the accuracy of common translation stages limits the temporal resolution of the approach. A characterization of small dispersion, i.e. of thin structures (\textmugreek m to mm) is not possible due to instrumental constraints. Therefore, the approach is mostly used for the characterization of optical fibers with at least a few centimeters of length, \cite{Thevenaz2,Levchenko,Thomann}.

\subsubsection{Frequency-domain interferometric measurements}
Analogous to time-domain measurements, interferometry in the frequency domain uses broadband light sources as well. In contrast, these sources are not spectrally constricted. Furthermore, the interferometer is usually built using two fixed arms. The detection of the signal is performed on the complete spectral bandwidth of the light sources using a spectrometer, \cite{Kardas}.
A typical signal of the recombined intensity of both arms leads to a modulated signal over a certain spectral bandwidth with a characteristic frequency. The frequency is dependent on the difference in the arm lengths of the interferometer as well as on the dispersion of the system, \cite{Gosteva}. The insertion of an dispersive element leads to a characteristic change in the modulation and its frequency.
The determination of dispersion can be done by different methods in this approach. On the one hand, the modulation can be analyzed using a \gls{FFT}. On the other hand, the analysis can be based on the determination of the so called stationary phase point.\\
The usage of an \gls{FFT}-based approach enables the fast determination of very small dispersion which can occur due to material characteristics or small amounts of media. The analysis is similar to \gls{FD-OCT}, although the broadening of the Fourier-peaks due to dispersion can be neglected if dispersion is very low. Otherwise, complex dispersion compensation methods have to be implemented. Liebermann et al. \cite{Liebermann} showed an experiment to determine the dispersion characteristics of distilled water using a Michelson interferometer in a free-space configuration, utilizing a supercontinuum light source as well as two combined spectrometers, (350\,-\,1100)\,nm and (900\,-\,1780)\,nm, to cover a large measurement range. For the analysis, two FFT-based approaches were compared. The so called indirect approach transforms the signal initially from the frequency- into the time-domain using zero-padding in order to achieve an even distribution of the signal after back transformation. The transformed signal is filtered afterwards and Fourier transformed for a second time in order to extract the spectral phase. A Taylor expansion enables the calculation of the dispersion coefficients. The second analysis method discussed in \cite{Liebermann} is the so called direct approach. In this approach, the spectral data is Fourier transformed into the non-even distributed time-domain and filtered using a Heaviside-Filter. After the back transformation and a Taylor expansion the dispersion can be determined. According to the authors, the indirect approach is computationally more time-consuming but also more suitable to characterize samples with large dispersion. Opposing to this, the direct approach is relatively fast and suitable for samples with small dispersion in process-accompanying problems.\\
Another approach to characterize refractive indices, is to fit dispersive interferometer data from experiments, \cite{Delbarre}. The authors assumed that the frequency chirp of the spectral intensity was a measure for the dispersion. By fitting this data the fringe periodicity was analyzed for different sample configurations. Under the utilization of a mathematical model (e.g. Sellmeier) the wavelength-dependent refractive index was estimated. In order to increase the fitting quality, more complex fit methods or correction algorithms could be used, \cite{Hlubina,Borzsonyi}.

\chapter{Surface profilometry}\label{ChapterProfilometry}

\section{Experimental setup}
\begin{figure}[h]
	\begin{center}
		\begin{tabular}{c}
		\includegraphics[scale=0.65]{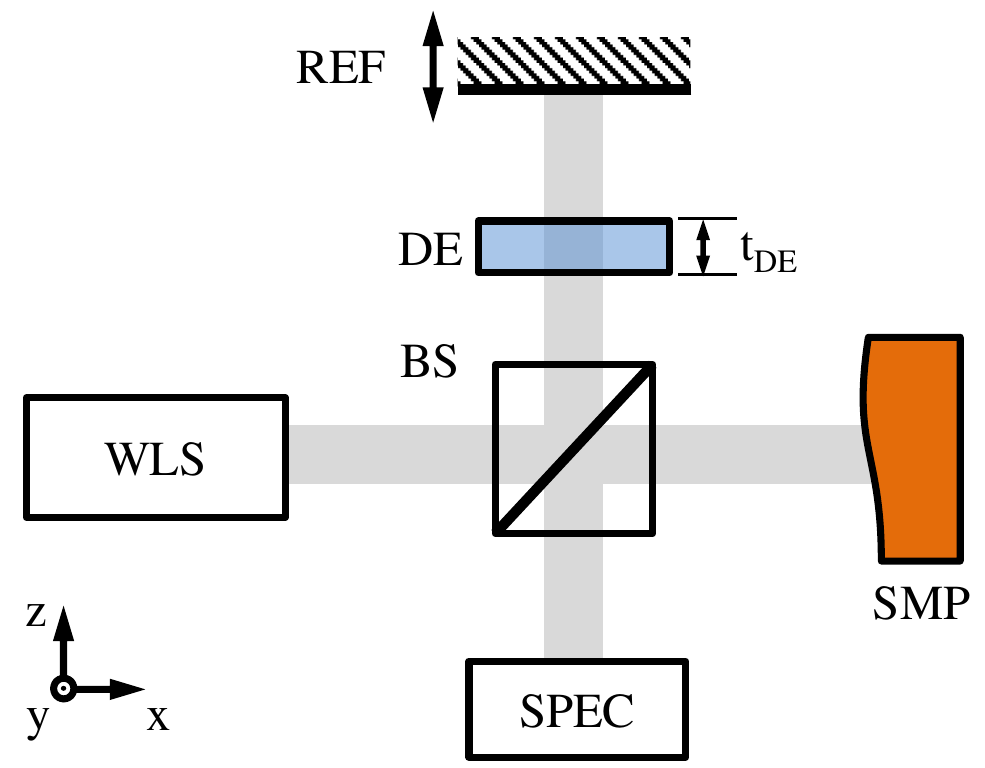}
		\end{tabular}
	\end{center}
	\caption[Principle of a dispersion-enhanced low-coherence interferometry setup]{Principle of a profilometry setup with WLS - white light source, BS - beamsplitter, DE - dispersive element (with thickness $t_{DE}$), REF - reference mirror (translatable in one dimension for adjustment purposes), SMP - sample with a surface profile and SPEC - spectrometer.}\label{ProfiloPic:BasicSetup}
\end{figure}

As outlined in the previous chapter, existing technologies for surface profilometry show certain drawbacks in terms of resolution, dynamic measurement range, three-dimensional measurement capabilities and speed. The following chapter introduces a novel approach to surface profilometry which aims to provide solutions to the problems named. The basic setup for all experiments is centered around a two-beam interferometer of the Michelson type, Fig. \ref{ProfiloPic:BasicSetup}.
A WLS -white-light source emits a collimated beam which is typically split in a 50:50 ratio by a BS - cube beamsplitter. The reference arm typically consists of a DE - dispersive element with the thickness $t_{DE}$. Following the transmission through the \gls{DE}, the light is reflected off of a REF reference mirror which causes the light to transmit through the DE a second time before it is guided by the beamsplitter. In the sample arm, the light is reflected on a sample surface before it is recombined with the reference arm light by the beamsplitter. The optical path difference of both arms $\delta$ is fixed as any change in $\delta$ is usually measured as a change in the samples surface profile\footnote{In practical implementations, one of the arms can be equipped with a translation stage for adjustment purposes.}. The recombined signal is detected by a grating spectrometer.
As the dispersive element plays a significant role in gathering relevant information on the topography of a sample, the term dispersion-encoded low coherence interferometry (DE-LCI) is used throughout the work for this approach.

\section{Measurement range and resolution}\label{Profilo:Sec:Meas_range}
\begin{figure}[h]
	\begin{center}
		\begin{tabular}{c}
			\begin{overpic}[scale=0.45]{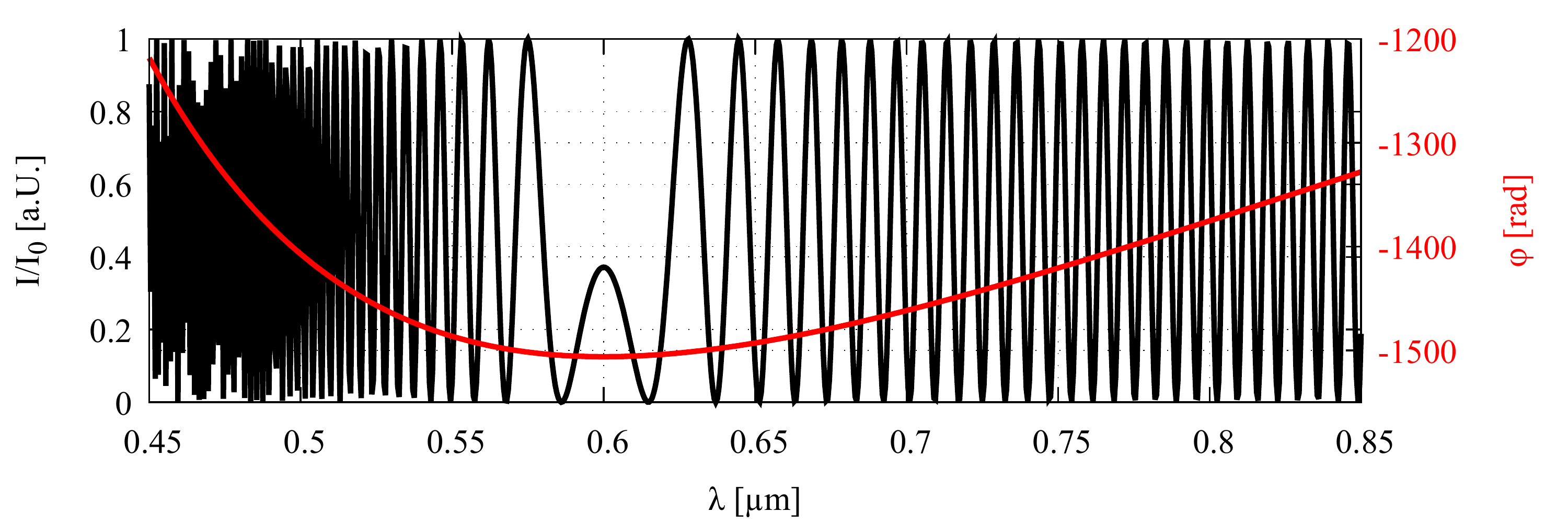}
			\end{overpic}  
		\end{tabular}
	\end{center}
	\caption[Simulation of a typical spectral interference signal with spectral intensity $I(\lambda)$ and phase $\varphi(\lambda)$]{Simulation of a typical spectral interference signal $I(\lambda)$ from a single point having the \gls{OPD} $\delta$ with a equalization wavelength $\lambda_{eq}$ = 600 nm plotted in black and the corresponding phase signal $\varphi(\lambda)$ plotted in red.}\label{ProfiloPic:Simple_dispersive_signal}
\end{figure}

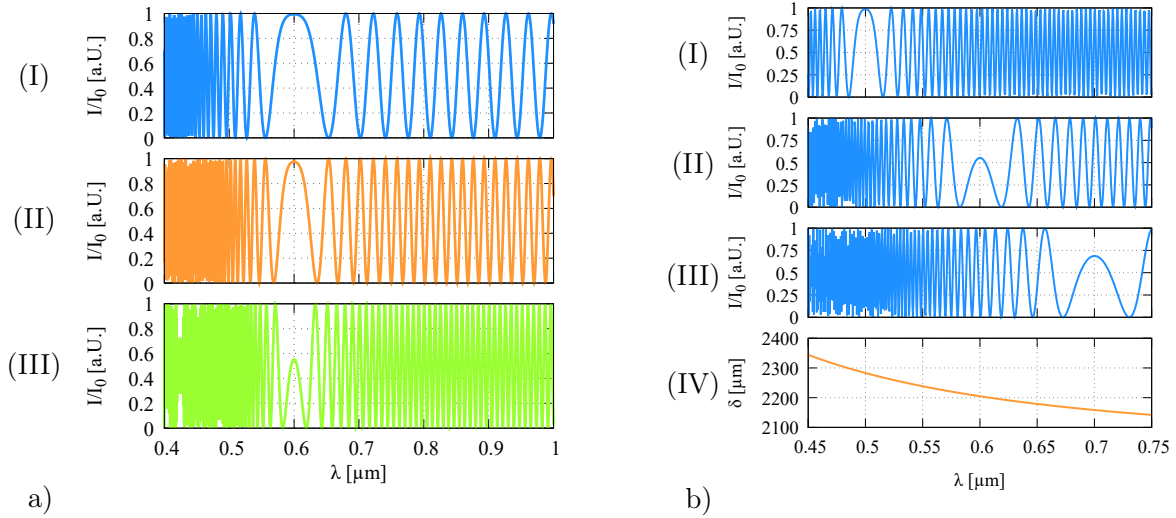
\begin{figure}[h]
	\begin{center}
		\begin{tabular}{c}
			\begin{overpic}[scale=0.34, grid = false]{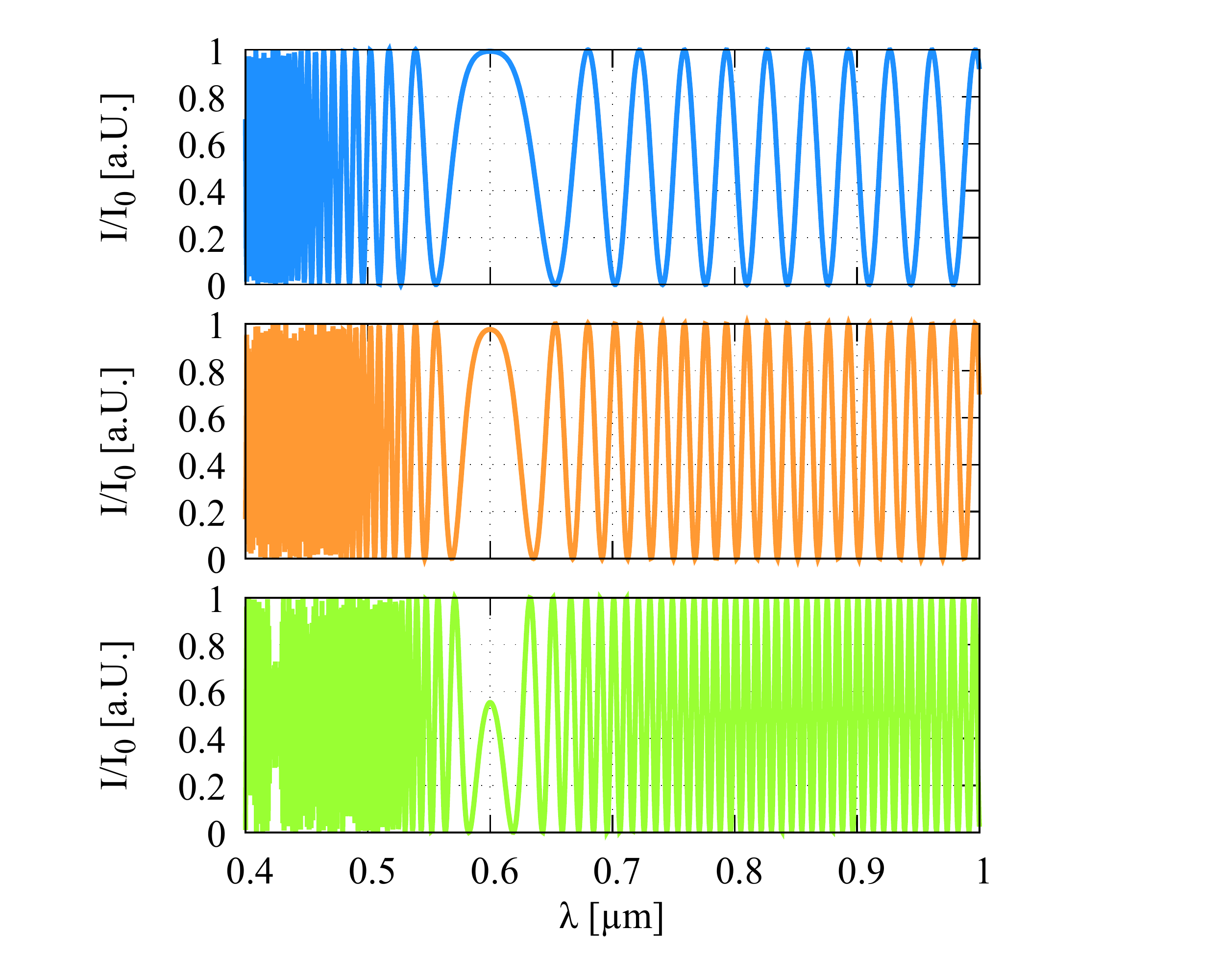}
				\put(1,1){\makebox(0,0){a)}}
				\put(0,66){\makebox(0,0){(I)}}
				\put(0,44){\makebox(0,0){(II)}}
				\put(0,21.5){\makebox(0,0){(III)}}
			\end{overpic}  
			\begin{overpic}[scale=0.25,grid = false]{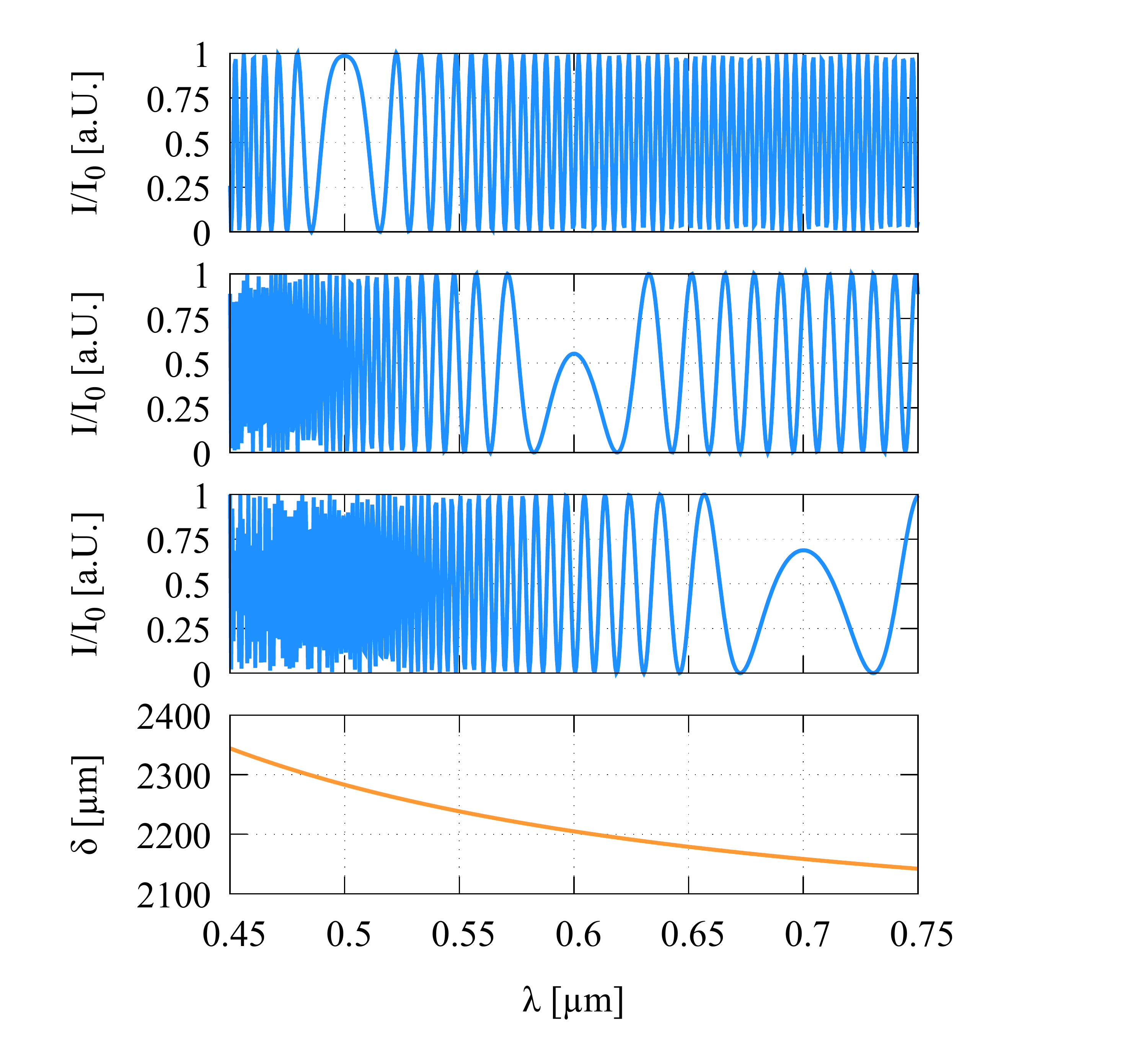}
				\put(1,1){\makebox(0,0){b)}}
				\put(0,78.75){\makebox(0,0){(I)}}
				\put(0,59.5){\makebox(0,0){(II)}}
				\put(0,40){\makebox(0,0){(III)}}
				\put(0,21){\makebox(0,0){(IV)}}
			\end{overpic}
		\end{tabular}
	\end{center}
	\caption[Simulation of the basic, dispersion-related dependencies of the interferometric signal]{Simulation of the basic dispersion-related dependencies of the interferometric signal with a) dispersion-dependent behavior where (I) shows the signal with a \gls{DE} of N-BK7 having a $t_{DE}$\,=\,1\,mm, (II) $t_{DE}$\,=\,2\,mm and (III) $t_{DE}$\,=\,4\,mm as well as b) the dependency regarding the \gls{OPD} for a N-BK7 \gls{DE} with $t_{DE}$\,=\,4\,mm where (I) $\lambda_{eq}$\,=\,0.5\,\textmugreek m, (II) $\lambda_{eq}$\,=\,0.6\,\textmugreek m, (III) $\lambda_{eq}$\,=\,0.7\,\textmugreek m and (IV) representing the slope of $\delta$ which is a function of the refractive index of the dispersive element $n^{DE}$.}\label{ProfiloPic:Dependencies}
\end{figure}

\begin{figure}[h]
	\begin{center}
		\begin{tabular}{c}
			\begin{overpic}[scale=0.28]{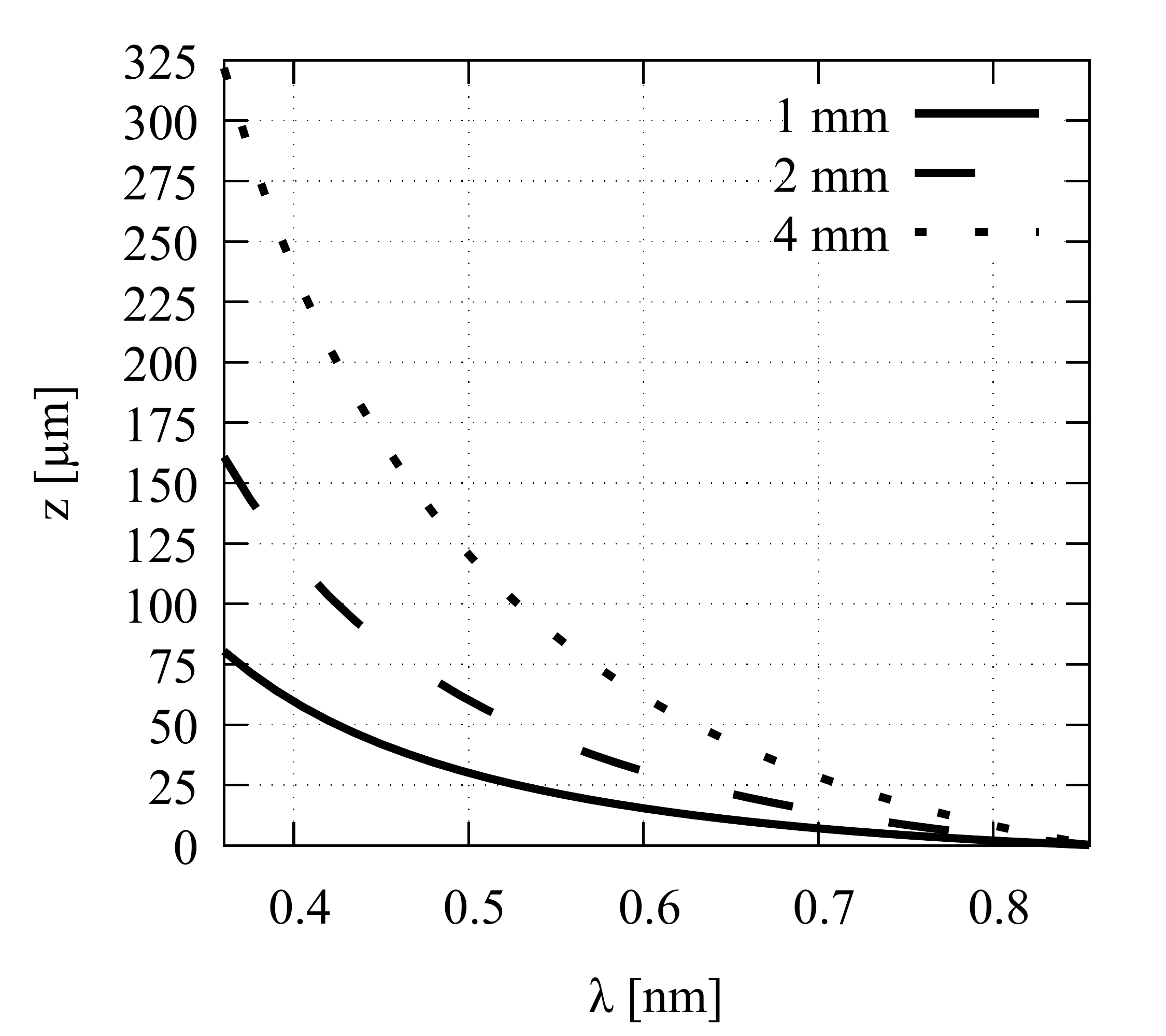}
				\put(1,1){\makebox(0,0){a)}}
			\end{overpic}
			\begin{overpic}[scale=0.28]{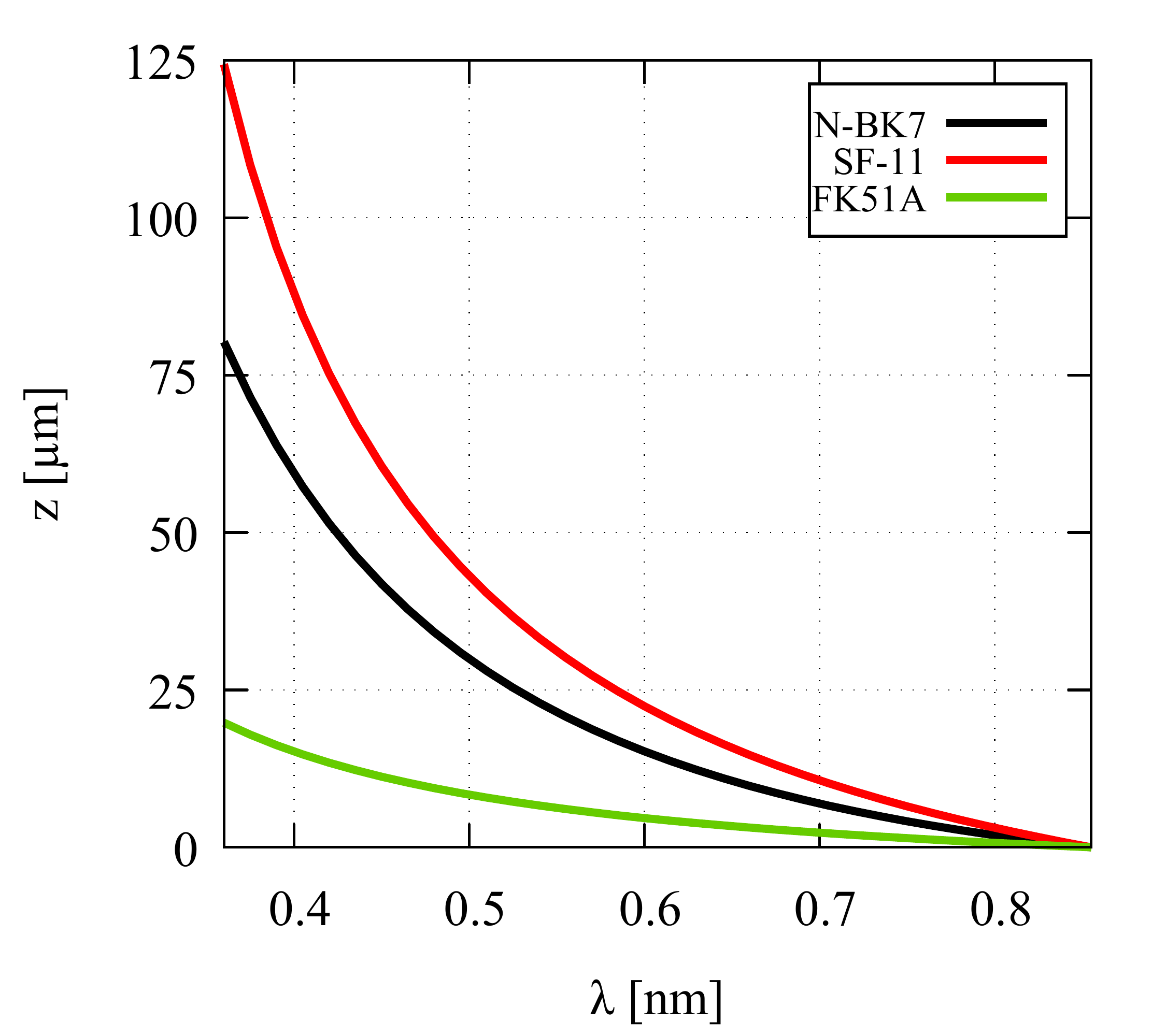}
				\put(1,1){\makebox(0,0){b)}}
			\end{overpic}
		\end{tabular}
	\end{center}
	\caption[Simulation of possible material-dependent height measurement ranges in DE-LCI]{a) Simulation of measurable height ranges in dependence of the spectral range of the setup $\Delta z$ for different thicknesses $t_{DE}$ of N-BK7 and b) simulated measurement ranges for different materials with a $t_{DE}$\,=\,1\,mm.}\label{simu_materials_thickness}
\end{figure}

\begin{figure}[h]
	\centering
		\begin{tabular}{c}
			\begin{overpic}[scale=0.32]{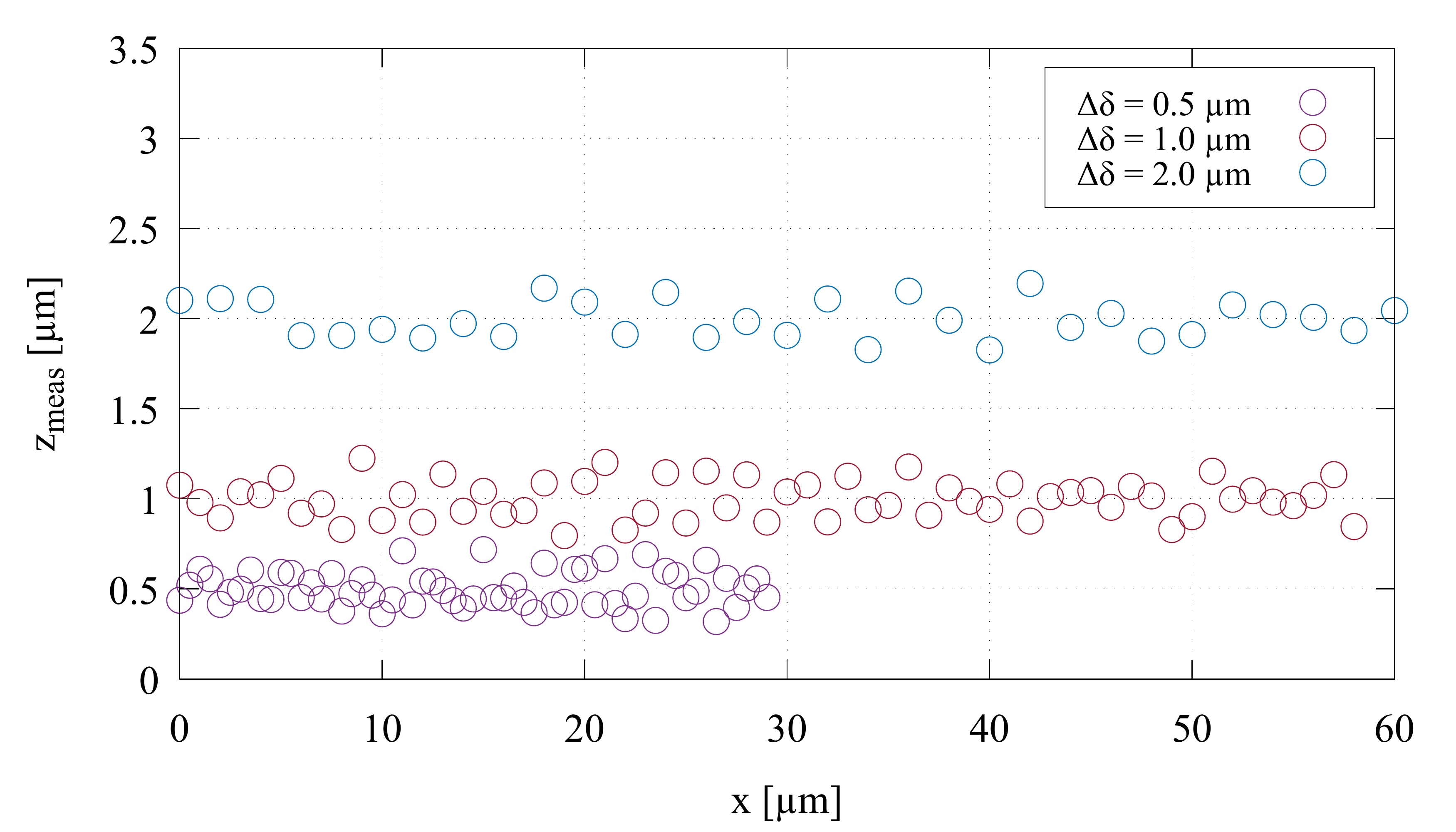}

			\end{overpic}
		\end{tabular}
	\caption[Results for the measured emulated height steps $z_{meas}$]{Results for the measured emulated height steps $z_{meas}$ for the targeted steps of $\Delta \delta$\,=\,0.5, 1 and 2\,\textmugreek m respectively over a given travel range of a piezo stage in the reference arm.}\label{Profilo:Pic:Results_simple_height_emulation}
\end{figure}

\begin{figure}[h]
	\begin{center}
		\begin{tabular}{c}
			\begin{overpic}[scale=0.3]{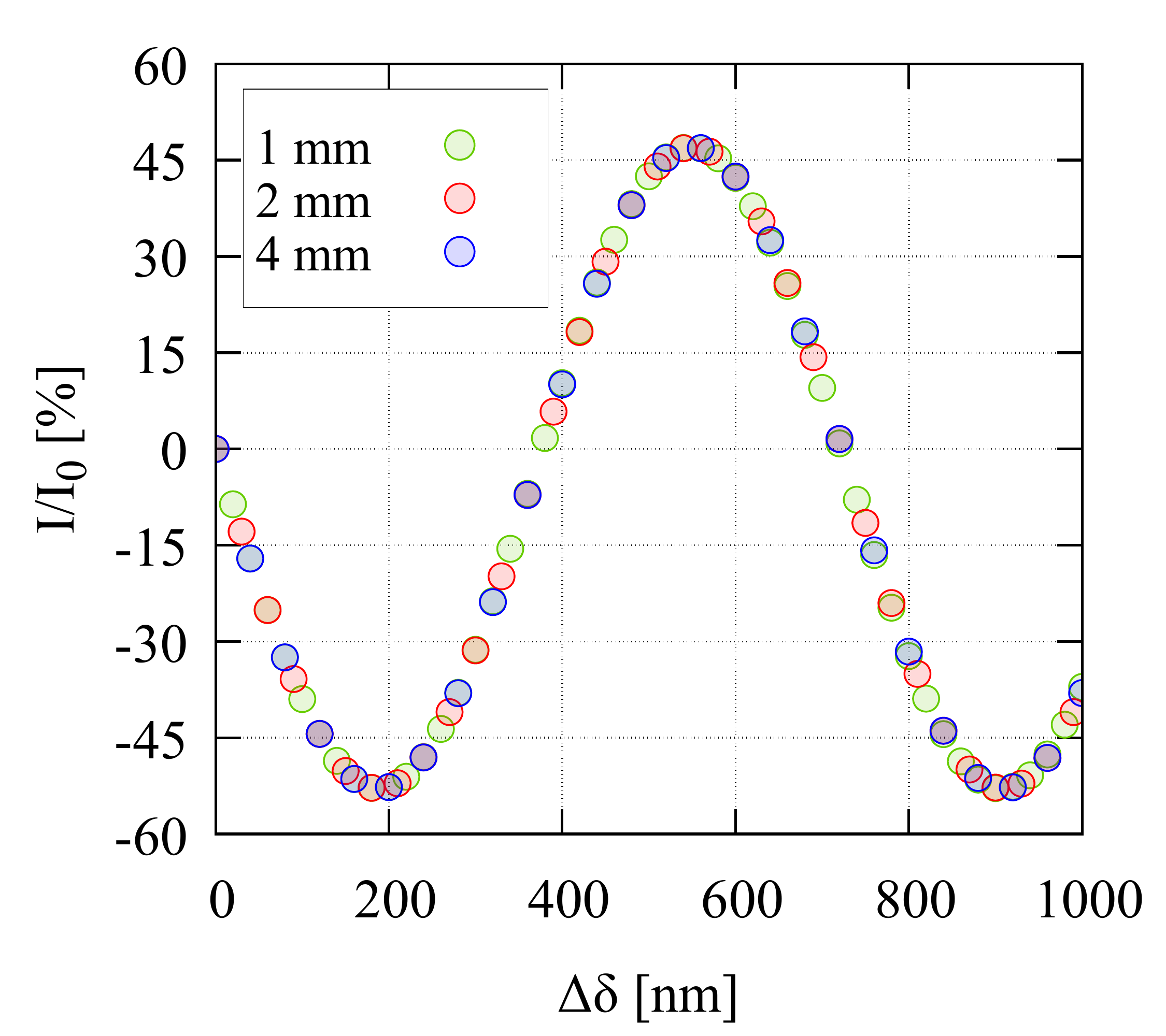}
				\put(1,1){\makebox(0,0){a)}}
			\end{overpic}
				\begin{overpic}[scale=0.3]{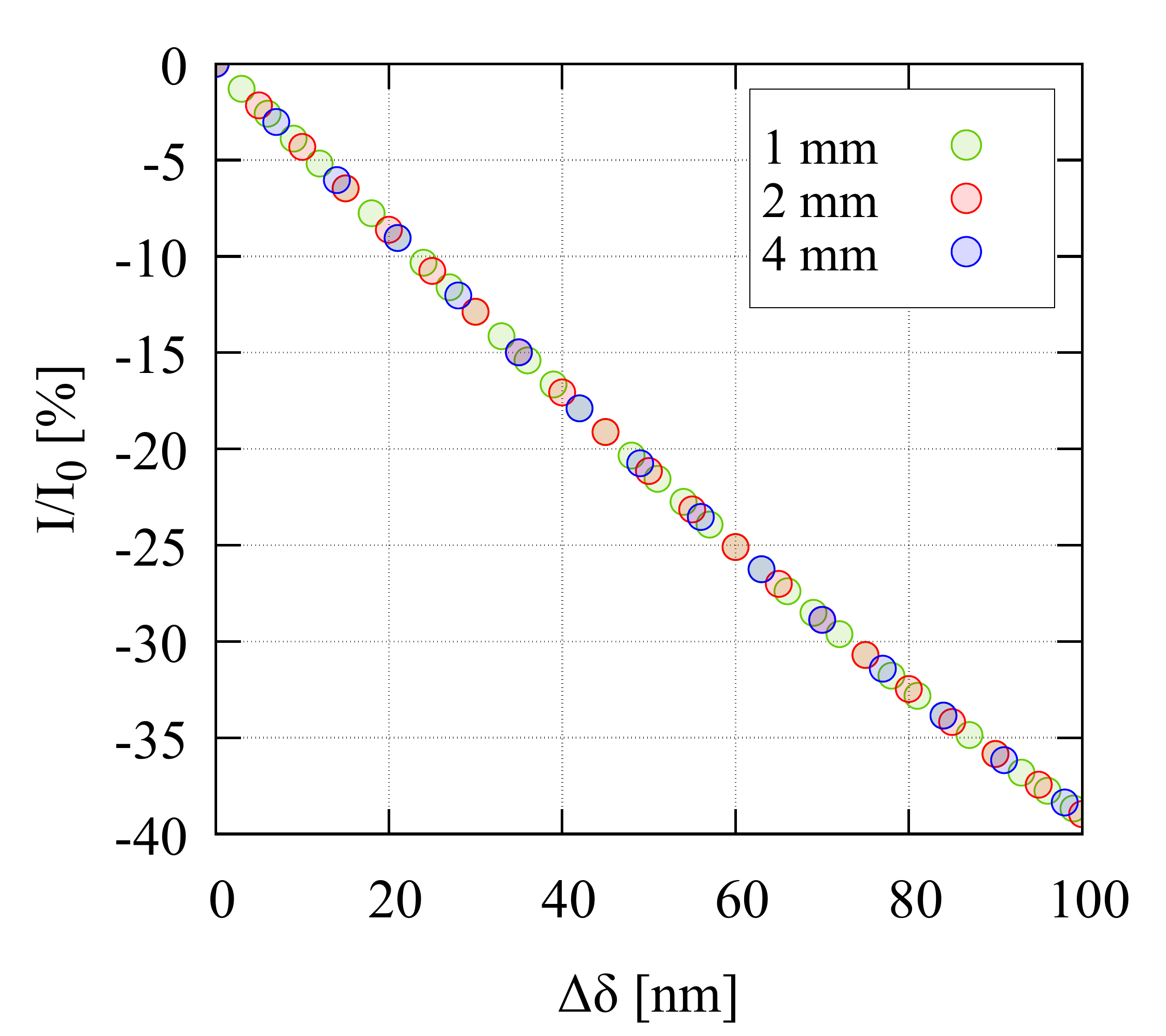}
				\put(1,1){\makebox(0,0){b)}}
			\end{overpic}
		\end{tabular}
	\end{center}
	\caption[Simulated intensity change in relation to a height change $\Delta \delta$ for different thicknesses of the \gls{DE} $t_{DE}$ of N-BK7]{Simulated intensity change in relation to a height change $\Delta \delta$ for different thicknesses of the \gls{DE} $t_{DE}$ of N-BK7 with a) oscillations visible over a $\Delta\delta$-range of 1\,\textmugreek m and b) reduced region of interest for $\Delta\delta$ where the slope is nearly linear and the maximum difference between different \glspl{DE} is about 0.012\,\%.}\label{simu_intensity_fit}
\end{figure}

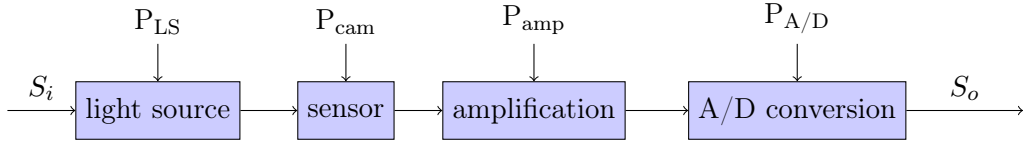
\begin{figure}[h]
	\begin{center}
		\begin{tabular}{c}
			\tikzstyle{int}=[draw, fill=blue!20, minimum size=2em]
			\tikzstyle{init} = [pin edge={to-,thin,black}]
		\begin{tikzpicture}[node distance=2.5cm,auto]
			\node [int, pin={[init]above: P\textsubscript{LS}}] (a) {light source};
			\node (b) [left of=a,node distance=2cm, coordinate] {a};
			\node [int, pin={[init]above: P\textsubscript{cam}}] (c) [right of=a] {sensor};
			\node [coordinate] (end) [right of=c, node distance=4cm]{};
			\node [int, pin={[init]above: P\textsubscript{amp}}] (e) [right of=c] {amplification};
			\node [coordinate] (end) [right of=e, node distance=4cm]{};			
			\node [int, pin={[init]above: P\textsubscript{A/D}}] (f) [right of=e, node distance=3.5cm] {A/D conversion};
			\node [coordinate] (end) [right of=f, node distance=3cm]{};
			\path[->] (b) edge node {$S_i$} (a);
			\path[->] (a) edge node {} (c);
			\path[->] (c) edge node {} (e);
			\path[->] (e) edge node {} (f);
			\path[->] (f) edge node {$S_{o}$} (end);
		\end{tikzpicture}
		\end{tabular}
	\end{center}
	\caption[Noise-affected components of the setup]{Noise-affected components of the setup with the initial spectrum $S_i$ which is manipulated by the noise sources  P\textsubscript{LS}, P\textsubscript{cam}, P\textsubscript{amp}, P\textsubscript{A/D} resulting in the measured spectrum $S_{o}$. }\label{Proflo:Pic:MeasRange_block-diagram}
\end{figure}

\begin{figure}[h]
	\begin{center}
		\begin{tabular}{c}
			\begin{overpic}[scale=0.3]{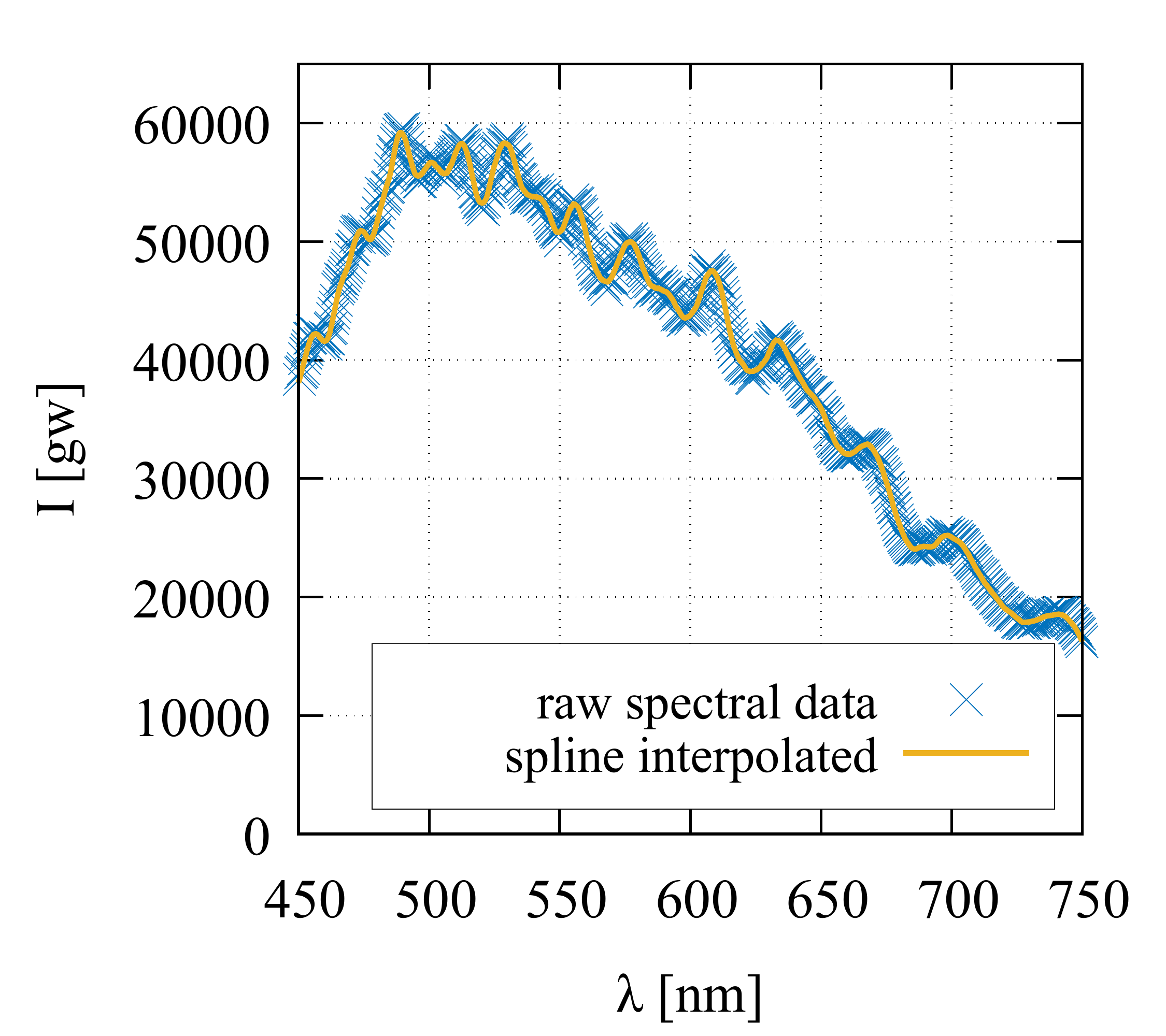}
				\put(1,1){\makebox(0,0){a)}}
			\end{overpic}
			\begin{overpic}[scale=0.3]{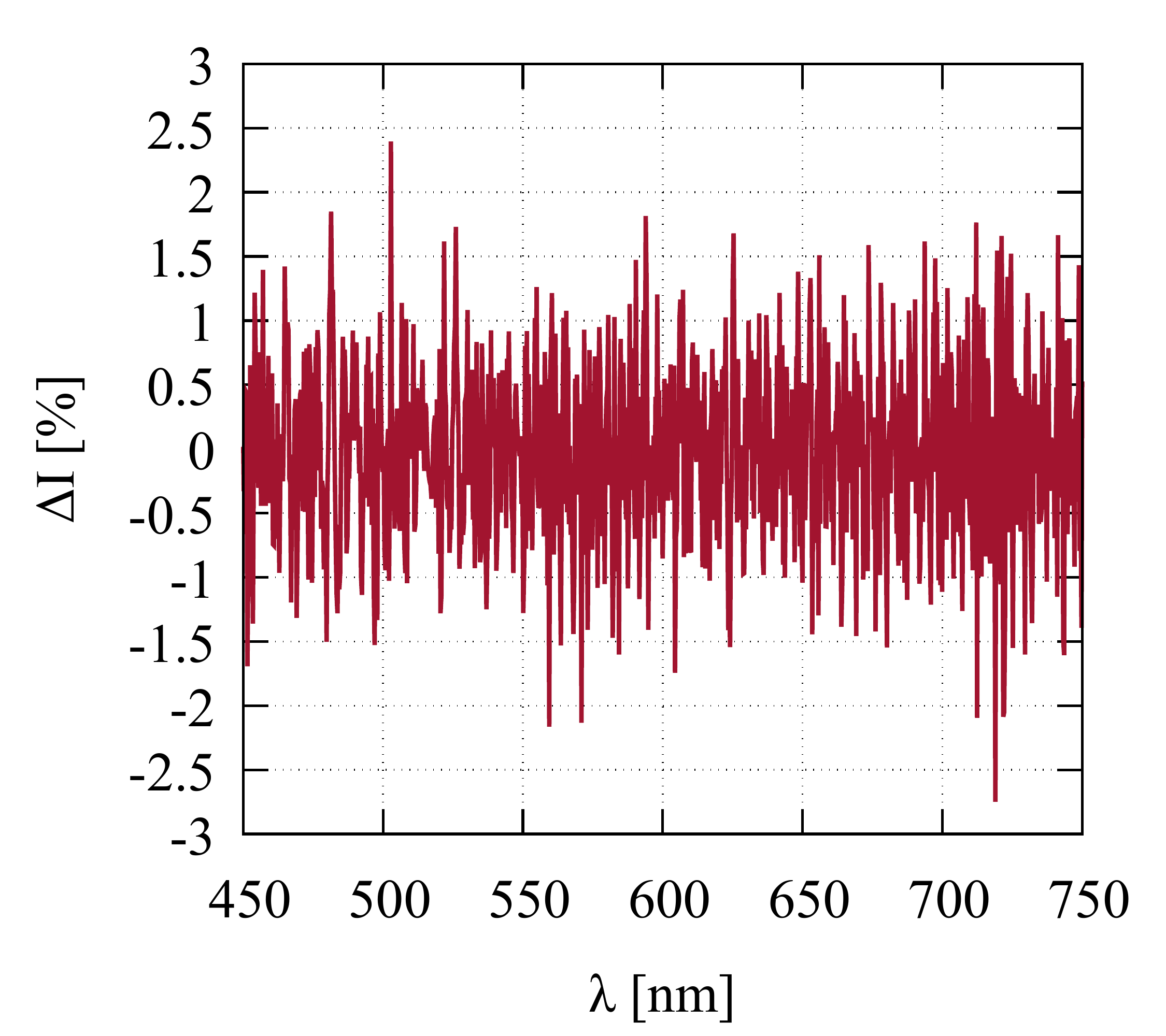}
				\put(1,1){\makebox(0,0){b)}}
			\end{overpic}
		\end{tabular}
	\end{center}
	\caption[Analysis of the spectral intensity noise of the setup ]{Analysis of the noise of the setup with a) raw measured spectral intensity in gray-values (abbrev. \textit{gw}) and spline interpolation as well as b) relative spectral noise $\Delta I$ with respect to the spectral range.}\label{Proflo:Pic:MeasRange_spline}
\end{figure}

\begin{figure}
	\centering
		\begin{tabular}{c}
			\begin{overpic}[scale=0.28]{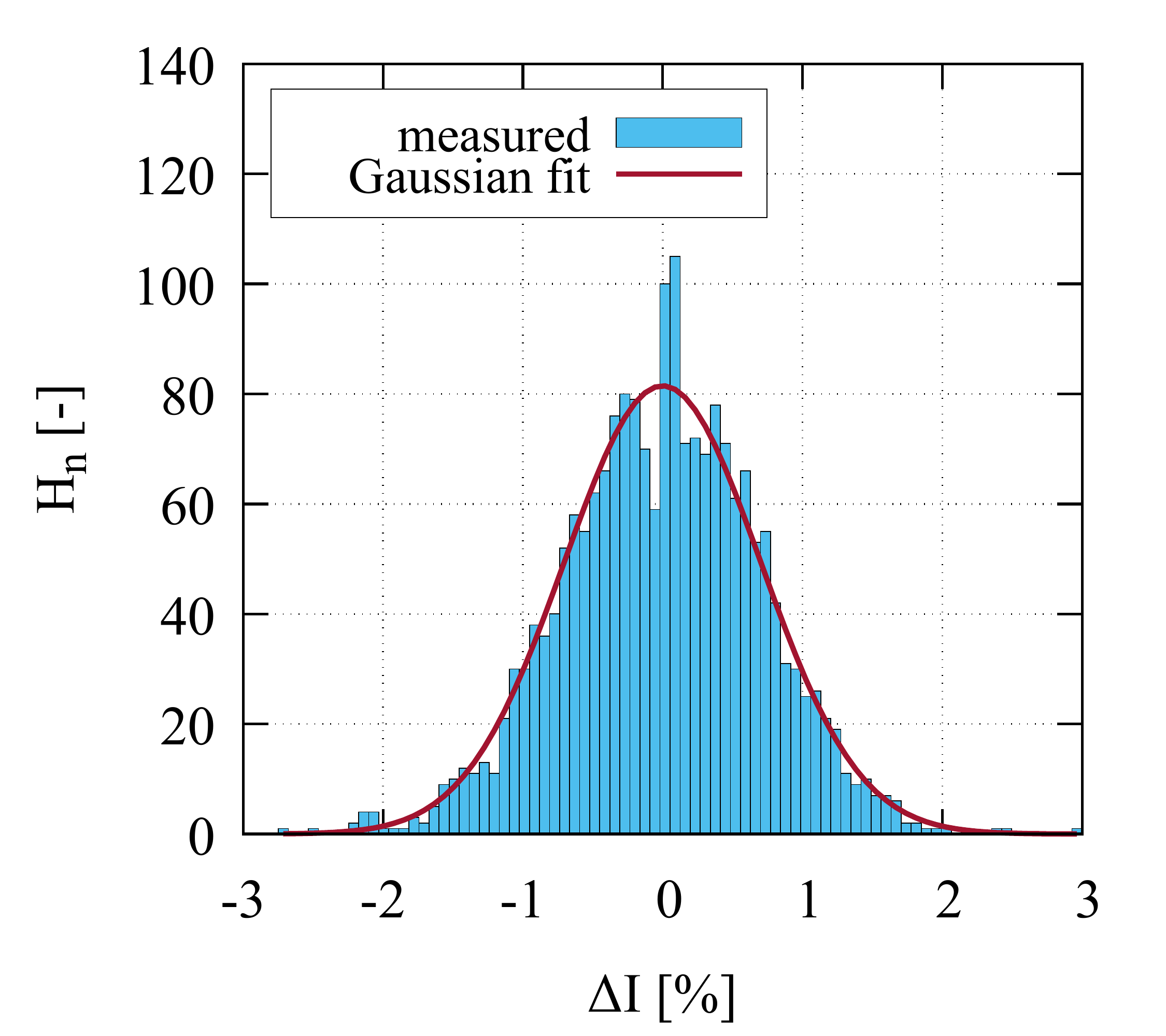}
				\put(1,1){\makebox(0,0){a)}}
			\end{overpic}
			\begin{overpic}[scale=0.28]{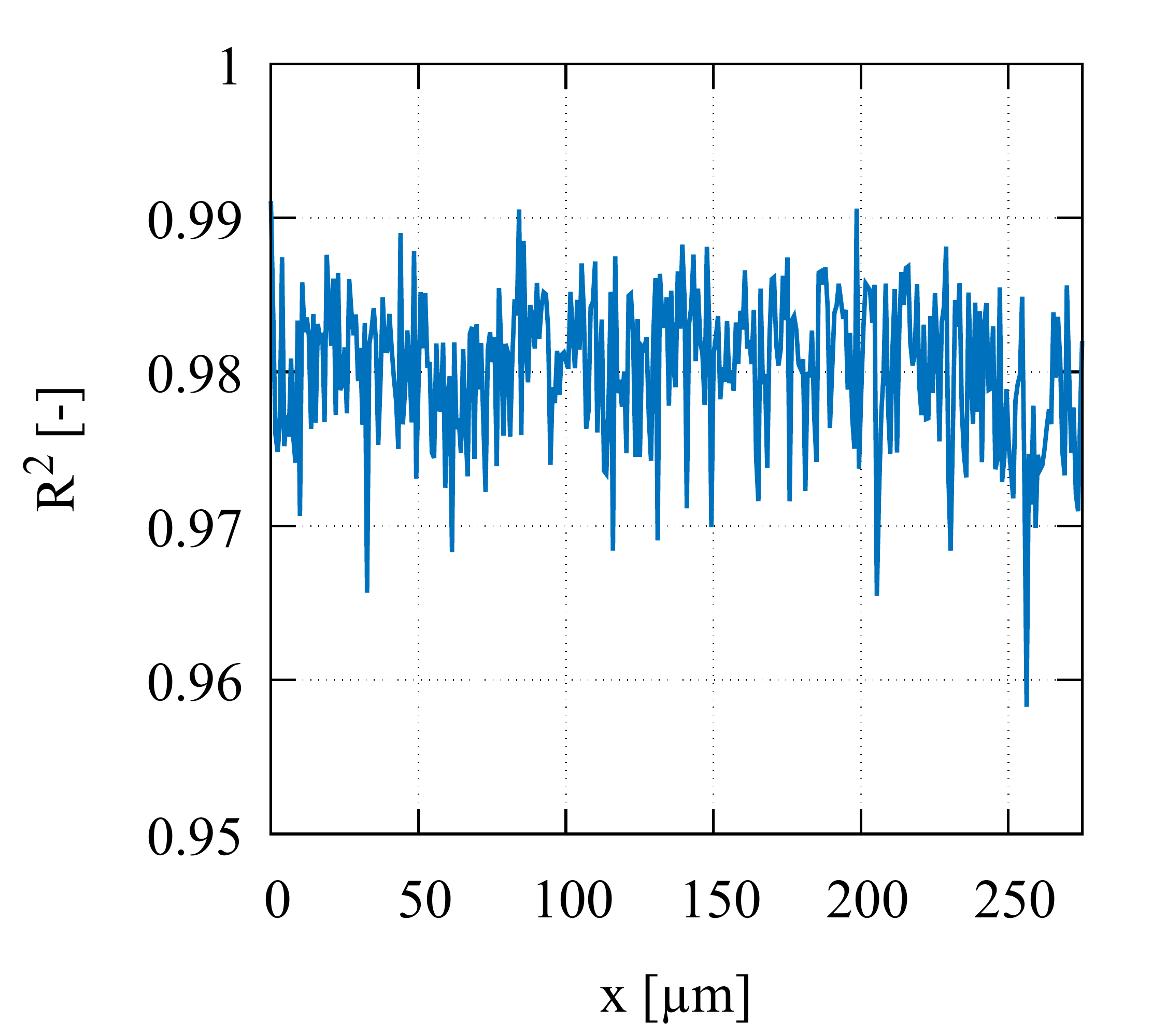}
				\put(1,1){\makebox(0,0){b)}}
			\end{overpic}
		\end{tabular}
	\caption[Analysis of the statistical behavior of the captured intensity noise]{Analysis of the statistical behavior of the captured intensity noise with a) Distribution of the measured, relative intensity noise of all positions in the $x$-dimension for a gain of 0\,dB and a relative signal magnitude of 100\,\% and Gaussian fit of the same having a mean $R_{s}^2$\,=\,0.979 where $\Delta I$\,=\,0.75\,\% (21.25\,dB). could be measured as the averaged intensity noise and b) spatially-dependent plot of coefficient of determination R\textsuperscript{2} for Gaussian fits of the relative noise distribution.} \label{Proflo:Pic:R2_distribution}
\end{figure}

In general, the mathematical model of a two-beam interferometer applies in the case of the setup described in Fig. \ref{ProfiloPic:BasicSetup} according to \cite{SPIEfieldguide}
\begin{equation}
E_{out} = E_0 \cdot e^{i(\omega t \pm \varphi_0)} = E_0 \cdot e^{i2\pi (ft \pm \frac{\delta n}{\lambda})},
\end{equation}
where the resulting electric field \glssymbol{Eout} is composed of the initial electric field \glssymbol{E0} in combination with the oscillating portion defined by the angular optical frequency \glssymbol{Omega} or the optical frequency \glssymbol{Frequency}, the time \glssymbol{Time} and the phase \glssymbol{PhaseDispFree} which consists of the optical path difference \glssymbol{DeltaInter}, the refractive index of the surrounding medium \glssymbol{RealRefInd} and the wavelength \glssymbol{Lambda}. In an experimental setup, where only the time averaged signal of the electric field can be detected the intensity \glssymbol{I} is of interest it can be formulated
\begin{eqnarray}
&& E_{out}^2 = \frac{c \cdot \varepsilon_0}{2} \left( |E_1|^2 + |E_2|^2 + 2 \cdot E_1 \cdot E_2 \right) \\
&& I = \frac{c \cdot \varepsilon_0}{2T}\int{\overline{E}_{out}^2 dt},
\end{eqnarray}
which is composed of the velocity of light \glssymbol{VelLight}, the vacuum permittivity constant \glssymbol{VacPerm} as well as the two electric field components of the interferometer arms \glssymbol{E1} and \glssymbol{E2} and is integrated over a given time.
As the relative change between the two arms is of interest when measuring surface heights, the phase term holds the relevant information. Therefore, the intensity \glssymbol{ILambda} can be written as the spectral dependency of the phase with  
\begin{eqnarray}\label{Profilo:EQ:basic_interferometer}
&& I(\lambda) = I_0 \cdot (1 + \gamma(\lambda) \cdot \cos\varphi_0) \\
&& \varphi_0 = \frac{2\pi \cdot \delta \cdot n_{air}}{\lambda},
\end{eqnarray}
where \glssymbol{I0} is the initial spectral intensity and \glssymbol{InterContrast} is the spectral contrast of the interference fringes. It is dependent of the \gls{OPD} between the both arms denoted with $\delta$, the wavelength \glssymbol{Lambda} and the refractive index of air $n_{air}$ which was assumed to equal one for simplification. These equations are suitable under the consideration that the interferometer is dispersion-free. In this general case, the phase is changing proportionally with the wavelength. Through the introduction of a dispersive medium in one arm of the interferometer, Fig. \ref{ProfiloPic:BasicSetup}, the phase term \glssymbol{PhaseDispInter} extends to \cite{Delbarre, SPIEfieldguide}
\begin{equation}
\varphi = 2\pi \frac{(n^{DE}(\lambda)-1)t_{DE}-\delta}{\lambda}.
\label{disp_phase}
\end{equation}
Assuming that pure material dispersion is the only effective mechanism\footnote{In the cases considered within this work, only material dispersion was used as a mechanism to encode surface information as the setup was implemented in free-space. By using a fiber-based setup, other mechanisms of dispersion such as waveguide dispersion might be utilized.}, the transformation of the interference signal is dependent on the  wavelength-dependent refractive index \glssymbol{RefIndDE} and the material's thickness \glssymbol{ThickDE}. The periodicity of fringes tends to a minimum at the so called equalization wavelength \glssymbol{LambdaEQ}, which is dependent on $\delta$, Fig. \ref{ProfiloPic:Simple_dispersive_signal}.
Using the interferometer as a profilometer, every height change in the sample's surface changes the \gls{OPD} and thus leads to a different equalization wavelength. The equalization wavelength can be calculated from the intensity signal analytically as the minimum of the derivative of the phase signal with respect to the wavelength
\begin{eqnarray}
\left(\frac{\partial\varphi}{\partial\lambda}\right)_{\lambda_{eq}} = 0 = 2\pi \frac{\left[1-n_g^{DE}(\lambda_{eq})\right]t_{DE} + \delta}{\lambda^2}\label{Profilo:EQ:deriv}\\
\text{with } n_g^{DE}(\lambda) = n^{DE}(\lambda) - \lambda \cdot \dv{n}{\lambda}, 
\end{eqnarray} 
where \glssymbol{GroupRefIndDE} is the group refractive index of the dispersive element. It can be concluded from this equation that the dispersive element and the path difference $\delta$ have the most significant influence on the signal. In this setting, the \gls{DE} defines the phase slope of the signal and furthermore, the axial measurement range where the equalization wavelength can be observed, Fig. \ref{ProfiloPic:Dependencies} a).
With increasing thickness of the \gls{DE} of the same $n^{DE}$, the phase gradient around the equalization wavelength increases which results in a higher frequency intensity signal around $\lambda_{eq}$. In practical terms, this signal modification is responsible for the effective axial measurement range. This measurement range \glssymbol{MeasRange} can be estimated within a spectral range of interest
\begin{equation}
\Delta z(\lambda)  =  \left[n_g^{DE}(\lambda) -1\right] \cdot t_{DE}.
\label{zRange}
\end{equation}
This characteristic of the \gls{DE} defines the response of $\lambda_{eq}$ due to change of the optical path difference, Fig. \ref{ProfiloPic:Dependencies} b) which therefore can be used as a measure for surface profile changes. During the design of a \gls{DE-LCI} setup, the detector size and its resolution as well as its dynamic range in the desired spectral range set the initial boundaries for the axial measurement range and resolution. The characteristics of the \gls{DE} enable fine tuning of the axial measurement range and resolution even after the initial design of the system.\\ 
In a system like this, the axial measurement range and resolution are determined by the spectral probing range \glssymbol{WaveLenRange}, the center wavelength \glssymbol{CenWaveLen}, the dispersive element and the detecting spectrometer configuration. In order to evaluate the possible axial measurement range as well as the corresponding resolution, the interconnection of the detector properties with the material properties of the \gls{DE} have been studied. For this purpose, simulations for a system with a defined set of parameters have been performed, Tab. \ref{Profilo:Tab:parameters}.
\begin{table}[h]
	\captionabove[Spectrometric system properties for initial simulations regarding measurement range and resolution]{Spectrometric system properties for initial simulations regarding measurement range and resolution of the designed DE-LCI.} \label{Profilo:Tab:parameters}
	\begin{center}       
		\begin{tabular}{cc} 
			\hline
			\rule[-1ex]{0pt}{3.5ex} component & value  \\
			\hline
			\rule[-1ex]{0pt}{3.5ex}  number of spectrometer pixels \glssymbol{NumPix}&  3648\\
			\rule[-1ex]{0pt}{3.5ex}  spectral range $\Delta \lambda$ &  500 nm  \\
			\rule[-1ex]{0pt}{3.5ex}  center wavelength $\lambda_c$ & 610 nm \\
			\rule[-1ex]{0pt}{3.5ex}  spectral resolution \glssymbol{ResSpectro} & 0.3 nm \\
			\rule[-1ex]{0pt}{3.5ex}  dispersive element (N-BK7) $t_{DE}$ & 1, 2 and 4  mm \\
			\hline
		\end{tabular}
	\end{center}
\end{table}
As stated before, the center wavelength in relation to the refractive index of the \gls{DE} and the spectral range have a characteristic influence on the axial height measurement range which follows an e.g. \textsc{Sellmeier}-like slope. This behavior can be used to manipulate the measurement range in a pre-designed setup. According to Eq. (\ref{zRange}) the measurement range $\Delta z(\lambda)$ = $\Delta z$ can be calculated over a given spectral range $\Delta \lambda$ as a function of the group refractive index and the thickness of the \gls{DE}, Fig. \ref{simu_materials_thickness} a). 
The behavior, which is in most cases \textsc{Sellmeier}-like, increases the measurement range for a \gls{DE}, following this characteristic especially towards shorter wavelengths. With the aid of different dispersive characteristics from e.g. glasses, polymers or thin films, this behavior can be controlled in a wide range while the setup is kept constant, Fig.\,\ref{simu_materials_thickness}\,b). \\
Light sources such as \glspl{SLD} or swept-source lasers  are commonly used in other \gls{LCI} approaches. With regard to the system design of a \gls{DE-LCI}, these light sources are considered suboptimal as the spectral bandwidth is usually about (80\,-\,120)\,nm with center wavelengths in the range of (800\,-\,1300)\,nm, \cite{ZhuSingleShot,RuizHSI,PavlicekHaeussler2005}. The results of the simulation, Fig.\,\ref{simu_materials_thickness} a), show that such a spectral range with a center wavelength at 1050\,nm would significantly limit the height measurement range. In fact, \gls{DE-LCI} benefits from broadband light sources such as supercontinuum or laser-driven plasma light sources which provide a spectral power output from (300\,-\,2000)\,nm, \cite{BaseltIRSC}.\\
With regard to the \gls{DE}, a spectrally defined resolution parameter, \glssymbol{ResDE}, can be calculated
\begin{equation}\label{EQ:reso_dis}
r_{DE} = \frac{\Delta z}{\Delta \lambda}.
\end{equation}
Clearly, the resolution can change due to dispersion as a function of the center wavelength while the nominal spectral range is constant. Nonetheless, the center wavelength influences $\Delta z$, see also Eq. (\ref{zRange}). As the detection of spectral interference is usually performed in a spectrometer, its specific resolution properties have to be considered as well. Therefore, the resolution of the system, \glssymbol{ResSys}, can be calculated with  
\begin{equation}
r_{sys} = r_{DE} \cdot \Delta r_{spec}
\end{equation}
Due to the construction of e.g. a grating spectrometer, its resolution $r_{spec}$ is defined by the combination of the slit size, the grating constant, the detector size and its specific pixel size. In consequence, a set of typical values for the measurement range and axial resolution was simulated for three different dispersive elements, Tab. \ref{Profilo:Tab:resolution}.  
\begin{table}[h]
	\captionabove[Calculated system properties regarding measurement range and resolution]{Calculated system properties regarding measurement range and resolution of the designed dispersion-enhanced low-coherence interferometer for investigations in the spectral range of $\Delta \lambda$\,=\,360\,-\,860\,nm.} \label{Profilo:Tab:resolution}
	\begin{center}       
		\begin{tabular}{ccc} 
			\hline
			\rule[-1ex]{0pt}{3.5ex} $t_{DE}$ [mm] & $\Delta z$ [\textmugreek m]  & $r_{sys}$ [nm]\\
			\hline
			\rule[-1ex]{0pt}{3.5ex}  1  &  80.45 & 48\\
			\rule[-1ex]{0pt}{3.5ex}  2 & 160.9 & 97 \\
			\rule[-1ex]{0pt}{3.5ex}  4  &  321.7 & 193  \\
			\hline
		\end{tabular}
	\end{center}
\end{table}
From this simulation, it can be deduced that the axial height resolution is tied to the measurement range and scales linearly with the thickness of the dispersive element. Additionally, it becomes clear that the dispersive element holds a large potential to tune both measurement range and resolution in a setup where the remaining components are already selected and fixed as shown on Fig. \ref{simu_materials_thickness}.\\ 
In order to evaluate the simulations, a temporally controlled experiment was carried out with the setup seen in Fig. \ref{ProfiloPic:BasicSetup}. In contrast to classical temporal \gls{LCI}, the experiment was designed to be controlled temporally but detected spectrally. For this purpose, a spectrometer according to the specifications of Tab. \ref{Profilo:Tab:parameters} (AvaSpec-ULS3648 VB, Avantes BV, Apeldoorn, The Netherlands) was used to capture interference data. The plain mirror in the reference arm acted as a sample which was translated to different \glspl{OPD} in order to emulate height changes of a sample. The sample arm, which is also equipped with a plain mirror, was kept at a constant position. Aluminum coated mirrors with a flatness of $\lambda$/20 (EO Partno. 34360; Edmund Optics Inc., Barrington, NJ, USA), having a scratch-dig of 20-10, served as sample and reference reflectors. In the experiment, a dispersive element of $t_{DE}$\,=\,2\,mm (EO Partno. 49121; Edmund Optics Inc., Barrington, NJ, USA) was used. The height emulation was performed with a piezo-driven precision stage (SLC 2412, SmarAct GmbH, Germany).
As a common reference, the setup was adjusted to an equalization wavelength of $\lambda_{eq}$\,=\,500\,nm. Subsequently, the translation stage was used to move the reference mirror in steps of \glssymbol{DeltaDelta}\,=\,0.5, 1 and 2\,\textmugreek m along a maximum measurement range of $\Delta z$\,=\,60\,\textmugreek m. The spectral interference at the spectrometer was recorded and the equalization wavelength was determined for every emulated height step. Based on $\lambda_{eq}$, the analysis of the measured height step, \glssymbol{MeasHeight}, was performed relative to the previous position with Eq.\,(\ref{zRange}), Fig.\,\ref{Profilo:Pic:Results_simple_height_emulation}.
The results show that the emulated height steps could be measured with some statistical deviations. Specifically, heights of $z_{meas}$\,=\,(0.497\,$\pm$\,0.098), (0.998\,$\pm$\,0.106) and (1.997\,$\pm$\,0.108)\,\textmugreek m could be measured for the nominal emulated steps of $\Delta \delta$\,=\,0.5,\,1 and 2\,\textmugreek m respectively. The detected deviations were only slightly larger than the calculated resolution of 0.097\,\textmugreek m for the dispersive element of $t_{DE}$\,=\,2\,mm, Tab. \ref{Profilo:Tab:resolution}. In reference to the work of Ruiz et al. \cite{RuizHSI}, the \gls{DR} can be calculated as the quotient of the measurement range $\Delta z$ and the resolution $r_{sys}$. The simulations as well as the experiments show that the \glssymbol{DynRange} of about 1667 is constant for the different measurement ranges since it is limited by the hardware. Due to the limitations in available detector sizes and trade offs between measurement range and resolution, other analysis schemes have to be considered in order to enable a higher dynamic range.\\
It can be noted that the signal not only shows a dependency of $\delta$ towards a change in equalization wavelength but also towards the signal amplitude at $\lambda_{eq}$. In a more advanced analyzing scheme this behavior is the primary component of a two-part process in order to determine height profiles with high-dynamic range. When analyzing the intensity behavior of the signal at the equalization wavelength in a significantly smaller $\delta$-range such as $10 \cdot r_{sys}\approx$ 1\,\textmugreek m, an oscillatory behavior becomes visible, Fig.\,\ref{simu_intensity_fit}\,a).
This simulation reveals that the differences in the height dependent intensity behavior are negligibly small for the different dispersive elements. By deceasing the $\delta$-range of interest even further towards the maximum hardware resolution of $t_{DE}$\,=\,2\,mm to $\Delta \delta$\,=\,97\,nm, the intensity dependency becomes nearly linear and unambiguous, Fig. \ref{simu_intensity_fit} b). This effect can be exploited in a more advanced analysis scheme where the first step consists of the determination of the equalization wavelength as a rough measure for height changes. During a second step, a fit of the intensity amplitude around this point can contribute to a fine-scaled analysis with resolutions beyond the hardware limit. Furthermore, the behavior can be considered independent of the thickness of the \gls{DE} and hence of the measurement range. The simulation revealed that the intensity difference between the signals of  $t_{DE}$\,=\,1\,mm and $t_{DE}$\,=\,4\,mm are about 0.012\,\% at maximum. Once the analysis becomes not only dependent on the wavelength accuracy but also on the intensity signal measured, the noise of this signal determines the height resolution. In the setup presented within this work, Fig. \ref{ProfiloPic:BasicSetup}, the actual measured spectrum, \glssymbol{NoiseSpecSigIn}, is affected by four main noise components, of the light source, \glssymbol{NoiseLightSource}, the cameras chip, \glssymbol{NoiseCam}, the cameras amplifier circuit, \glssymbol{NoiseAmp}, as well as of the A/D converter, \glssymbol{NoiseAD}, which contribute to the measured signal, \glssymbol{NoiseSpecSigOut}, Fig. \ref{Proflo:Pic:MeasRange_block-diagram}.
The impact of the light source as well as the camera sensor together with its amplification are highly influenced by the experimental conditions such as integration time and gain of the camera. On a more detailed level, the four noise components represent more fundamental noise sources. The photon flux of the light source is the main source of statistical variation which leads to photon noise of this component. While receiving this fluctuating source of energy, the camera sensor converts incoming photons into electron-hole pairs with a given quantum efficiency, which in itself is a statistical process known as photo-electron noise. The camera is also source of photo-current noise, which arises when electron-hole pairs are converted to pulses of electric current. In addition to the before mentioned noise sources, the process is dependent on the area of the detector, its quantum efficiency and the integration time, \cite{DissCimalla}. If the incoming photons are Poisson distributed, this noise source is known as shot noise. Additionally, the amplification of the photon-induced electrical current will be statistically dependent as well as the quantization of the detector current in A/D conversion. Furthermore, thermal variations will cause deviations and random signal contributions in the electronic circuitry of the sensor, the amplification and the A/D conversion which are also known as receiver circuit noise.\\
It should also be noted that thermal influences will affect the setup as a whole and lead to geometrical deformations of critical components. Due to the slow rate of change of these fluctuations and the expected short integration times of the detector, it can be assumed that they have a neglectable influence on the signals. In determining a combined noise level of a typical measurement situation, these influences will be captured during a noise characterization measurement.\\
In order to qualify the combined influence, an experiment was conducted to characterize the induced noise. For this purpose, the spectral intensity response of the system with a mirror in the reference arm (Thorlabs PF10-03-P01) and a typical sample surface (silicon height standard Simetrics VS, Simetrics GmbH, Germany) was recorded. The recording was performed at gain levels ranging from 0 to 12\,dB in steps of 3\,dB and with five equally spaced integration times per gain level. The integration time was used to control the relative magnitude of the intensity signal in steps of 20\,\% starting from 100\,\%\footnote{In this context, the 100\,\% reference was defined as the integration time where the recorded intensity maximum was just not provoking overexposure. While this section only discusses the behavior of the camera for a gain of 0\,dB, the appendix \ref{ANPDX:Profilo_noise} presents an analysis for the other gain levels.}. The recorded spectral intensity was spline interpolated along the spectral range, which was recorded in order to gather the moving average of the data, Fig. \ref{Proflo:Pic:MeasRange_spline} a).
Subsequently, the spline interpolated data was subtracted from the raw, gray-valued data. A normalized signal was computed to get information on the relative spectral noise, \glssymbol{DeltaI}, Fig.\,\ref{Proflo:Pic:MeasRange_spline}\,b). For one particular spatial position, a constant fluctuation over the complete spectral range is visible. The analysis of the distribution of this data reveals that it can be modeled using a Gaussian function which fits the data set with a coefficient of determination of \glssymbol{RSquareS}\,=\,0.979\,, Fig.\,\ref{Proflo:Pic:R2_distribution}\,a).
By analyzing the $\sigma$ of the Gaussian distribution, the averaged intensity noise was found to be $\Delta I$\,=\,$\pm$\,0.75\,\% (21.25\,dB). To gain a better insight in the spatial dependency of the noise, the distribution of noise for a number of points along the spatial axis was evaluated. The coefficient of determination $R_{s}^2$ for a Gaussian function describing the noise at the individual point was calculated, Fig.\,\ref{Proflo:Pic:R2_distribution}\,b). It can be seen, that no spatial dependency is present. For this reason, the spatial domain was not investigated in further detail and all data was integrated over along this domain.\\
In order to estimate the impact of this noise on the determination of the height of a surface, an analysis of the initial interferometric equation Eq. (\ref{Profilo:EQ:basic_interferometer}) was performed. The equation was solved for the path length difference $\delta$,
\begin{equation}
\delta = \left(n^{DE}(\lambda)-1\right)t_{DE} - \frac{\lambda}{2\pi} \cdot \cos^{-1}\left[\frac{I}{I_0} - 1\right].
\end{equation}
Where the derivative of the path length difference $\delta$ with respect to the intensity variation $dI$ is given by
\begin{equation}
\frac{d \delta}{dI} = \frac{d}{dI}\left[\left(n^{DE}(\lambda)-1\right)t_{DE}  \right] - \frac{\lambda}{2\pi} \cdot  \frac{d}{dI}\left[\cos^{-1}\left(\frac{I}{I_0} - 1\right) \right]
\end{equation}
and the path length uncertainty $\Delta \delta$ is
\begin{eqnarray}\label{noise_resolution}
&& \Delta \delta (I,\lambda) =  \frac{\lambda}{2\pi} \frac{1}{\sqrt{1 - \frac{(I - I_0)^2}{I_0^2}}} \cdot \Delta I.
\end{eqnarray}
This relation describes the limits for the described profilometry approach, which is also dependent on the spectral range as well as on the relative intensity, Fig. \ref{Proflo:Pic:MeasRange_resolution_estimation} a).
\begin{figure}[h]
	\begin{center}
		\begin{tabular}{c}
			\begin{overpic}[scale=0.27]{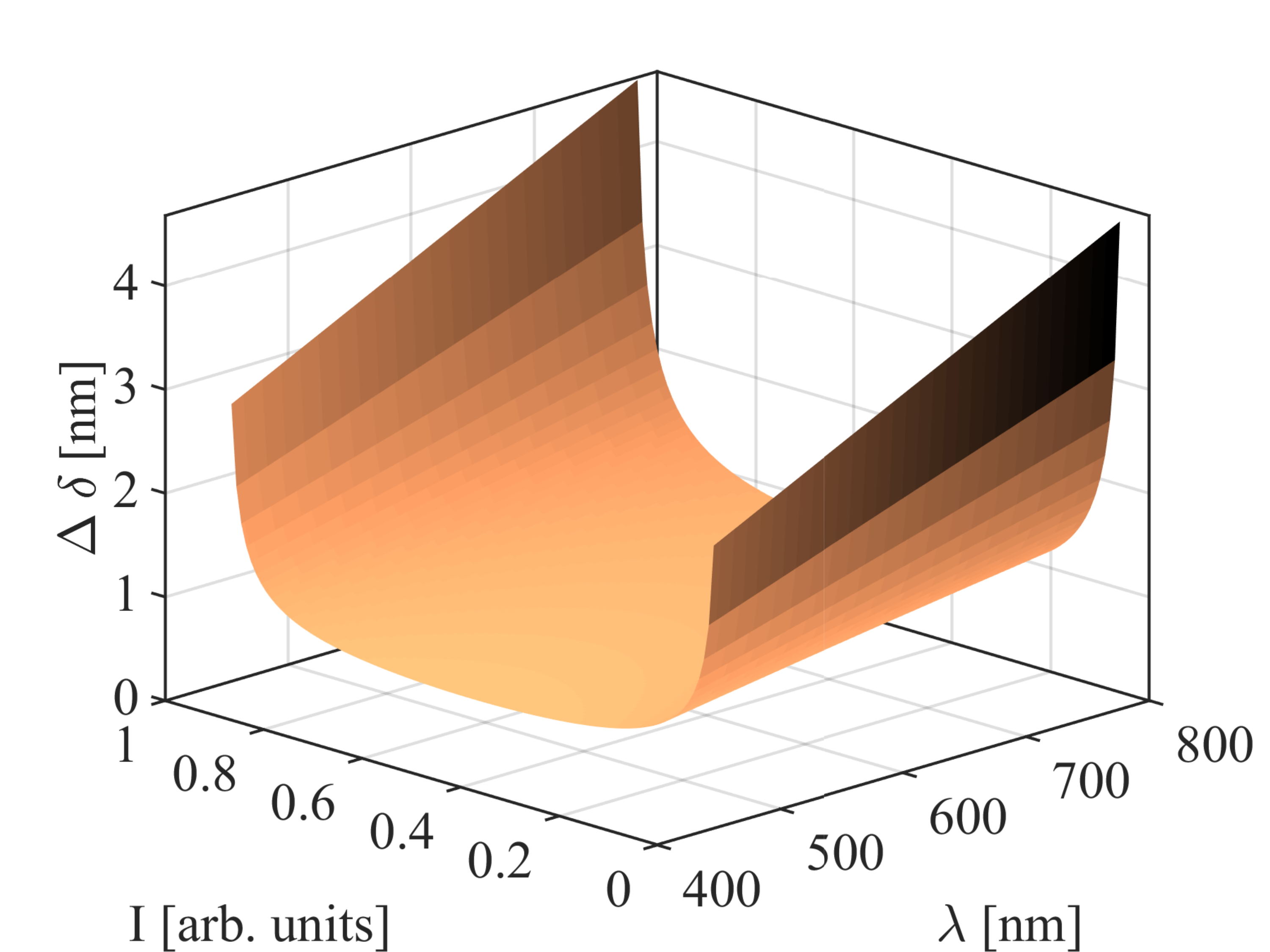}
				\put(1,1){\makebox(0,0){a)}}
			\end{overpic}
			\begin{overpic}[scale=0.28]{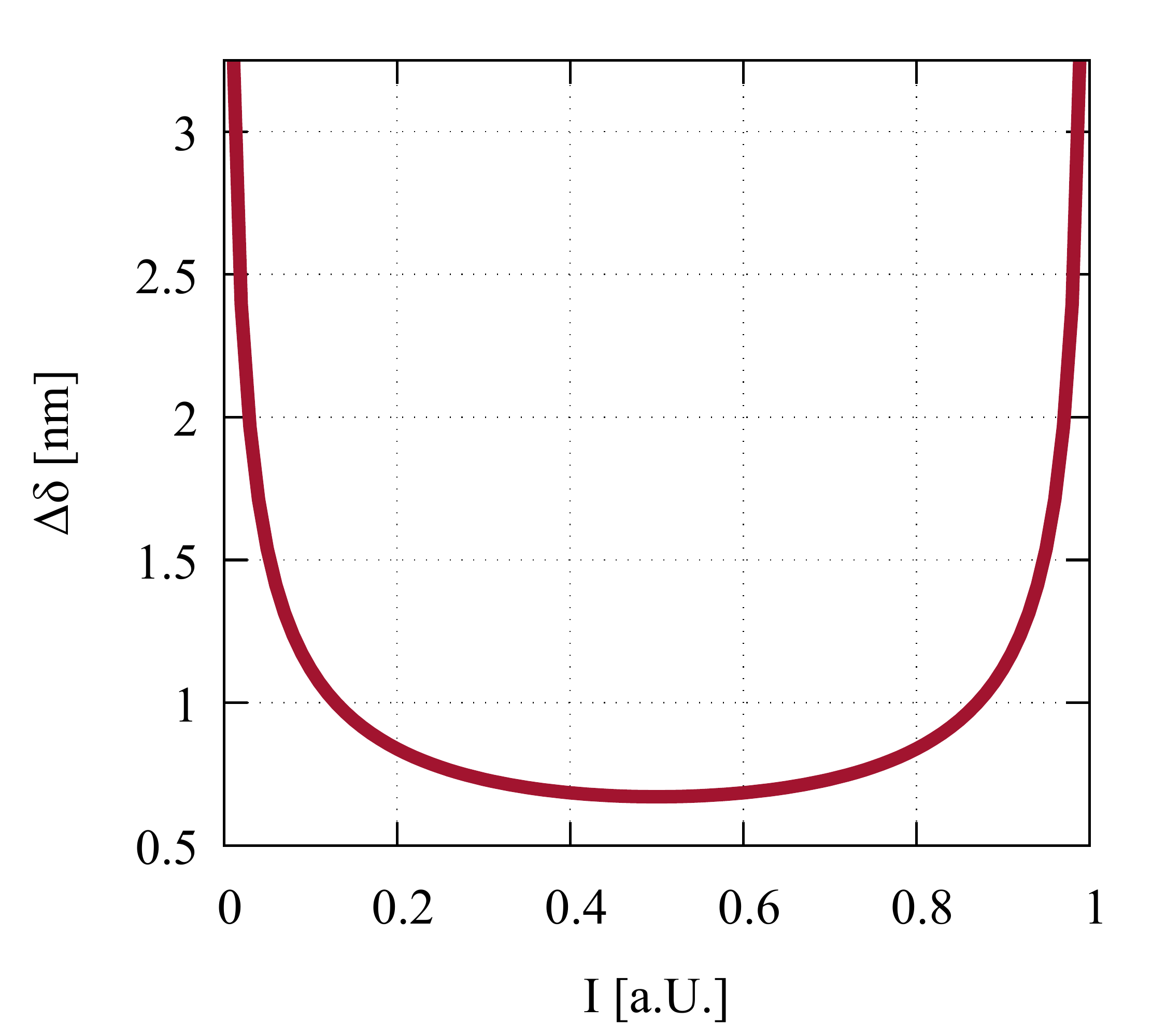}
				\put(1,1){\makebox(0,0){b)}}
			\end{overpic}
		\end{tabular}
	\end{center}
	\caption[Estimation of the resolution based on the measured intensity noise]{Estimation of the resolution based on the measured intensity noise with a) three-dimensional dependency related to the measured spectral and intensity range according to Eq. (\ref{noise_resolution}) as well as b) detailed plot at a equalization wavelength of $\lambda_{eq}$ = 562 nm.}\label{Proflo:Pic:MeasRange_resolution_estimation}
\end{figure}
As Eq.\,(\ref{noise_resolution}) describes, the resolution depends linearly on the wavelength. It could be determined that it has a slope of 0.0012\,nm per nm at $I_0$\,=\,0.5\,arb.\,units. Furthermore, the resolution strongly depends on the probing intensity in a more complicated relation, Fig.\,\ref{Proflo:Pic:MeasRange_resolution_estimation}\,b). In order to qualify this dependency, the intensity range in which the resolution increase is $<$\,10\,\% of the minimal value, was calculated. It was found that this equals an intensity range of $I$\,=\,0.3\,-\,0.7\,arb.\,units. For practical purposes, this means, that the intensity range should be chosen together with an equalization wavelength as low as possible in order to achieve high resolutions. This can be ensured by adjusting the path length difference of the interferometer.\\
In all experiments, the setup was adjusted to $\lambda_{eq}$\,=\,562\,nm. As the spectrometers detection range started at 447\,nm, the chosen equalization wavelength enabled the acquisition of enough data for fitting in proximity of $\lambda_{eq}$. This helped to minimize imaging distortions which are typically present close to the border of the detector. Using the values for $I_{0}$ and $\lambda_{eq}$ as well as the intensity noise $\Delta I$ in Eq.\,(\ref{noise_resolution}), a typical resolution of $\Delta \delta(\lambda_{eq})$\,=\,0.67\,$\pm$\,0.05\,nm was calculated. The calculated detection limit according to Eq. (\ref{noise_resolution}) is valid for the analysis at one spectral position. The data analysis of the presented experiments utilized the fit of spectra in a \gls{ROI} around $\lambda_{eq}$ which is explained in more detail in subsection \ref{Profilo:SubSec_fitting}. Specifically, the measured intensity data was fitted with Eq. (\ref{Profilo:EQ:basic_interferometer}) and (\ref{disp_phase}), where the thickness of the dispersive element $t_{DE}$ and its refractive index $n^{DE}(\lambda)$ were assumed to be known and $\delta(x,y)$ as well as $\gamma(\lambda)$ were approximated. In order to account for utilization of the \gls{ROI} in fitting, the single point detection limit $\Delta \delta(\lambda_{eq})$ was used to calculate the resolution of the fitted data, \glssymbol{ResFit}, with the aid of an \gls{RMS} approach as
\begin{equation}\label{EQ:resolution_fit}
r_{fit} = \sqrt{\frac{\Delta \delta(\lambda_{eq})^2}{n}}
\end{equation}
where \glssymbol{NumFitPoints} is the number of spectral data points used for fitting. Within this work, $n$\,=\,530 spectral data points were used so that the theoretical resolution was estimated as $r_{fit}$\,=\,0.029\,nm. With a measurement range of $\Delta z$\,=\,79.91\,\textmugreek m and the above calculated minimal resolution, the \gls{DR} is \num{2.75e6}. This value is significantly higher than the initial estimation which was solely based on the evaluation of $\lambda_{eq}$ which was \gls{DR}\,=\,1667.\\
As this value is based on the model expressed through Eq. (\ref{noise_resolution}), the experimentally achievable \gls{DR} might additionally be limited by other influences such as thermal fluctuations or the data processing routines which are not included in the model. Utilizing this analyzing scheme, the limitation on the axial resolution imposed by the thickness of the dispersive element was minimized.

\section{Signal formation and analysis}\label{SecProfiloSigAn}
As shown in the previous examinations, signal analysis in dispersion-encoded low-coherence profilometry needs to be based on a combined evaluation of the equalization wavelength $\lambda_{eq}$ and the amplitude at this wavelength in order to achieve sufficient resolution. This holds especially true for comparatively large measurement ranges. Different approaches to analyze the information in recorded spectra have been developed and assessed within this work.

\subsection{Fitting of oscillating data}\label{Chap:Profilo_Sec:SigAn_SubSec:brute_force} 
\begin{figure}[h]
	\begin{center}
		\begin{tabular}{c}
			\begin{overpic}[scale=.32]{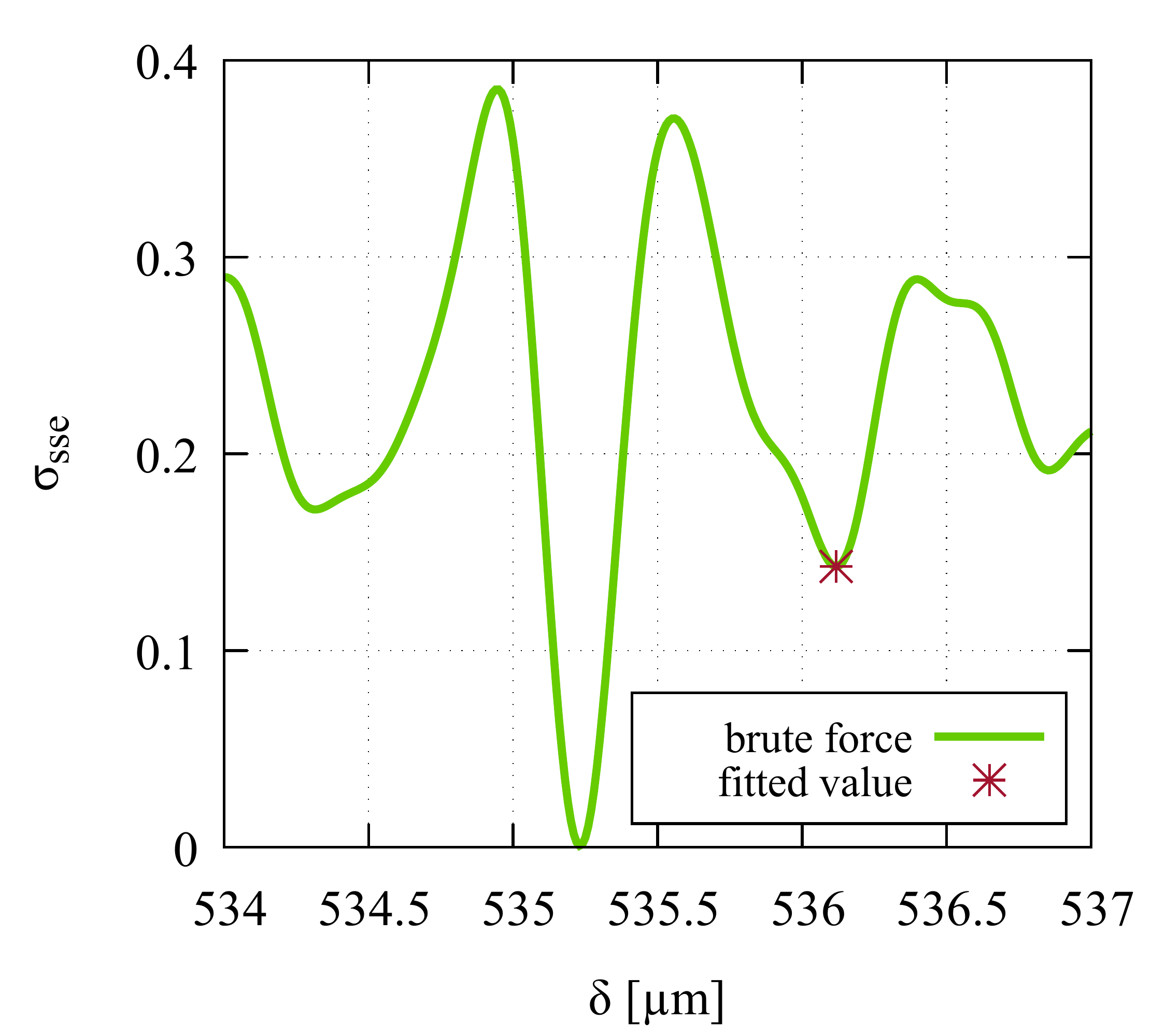}
				\put(1,1){\makebox(0,0){a)}}
			\end{overpic}
			\begin{overpic}[scale=.32]{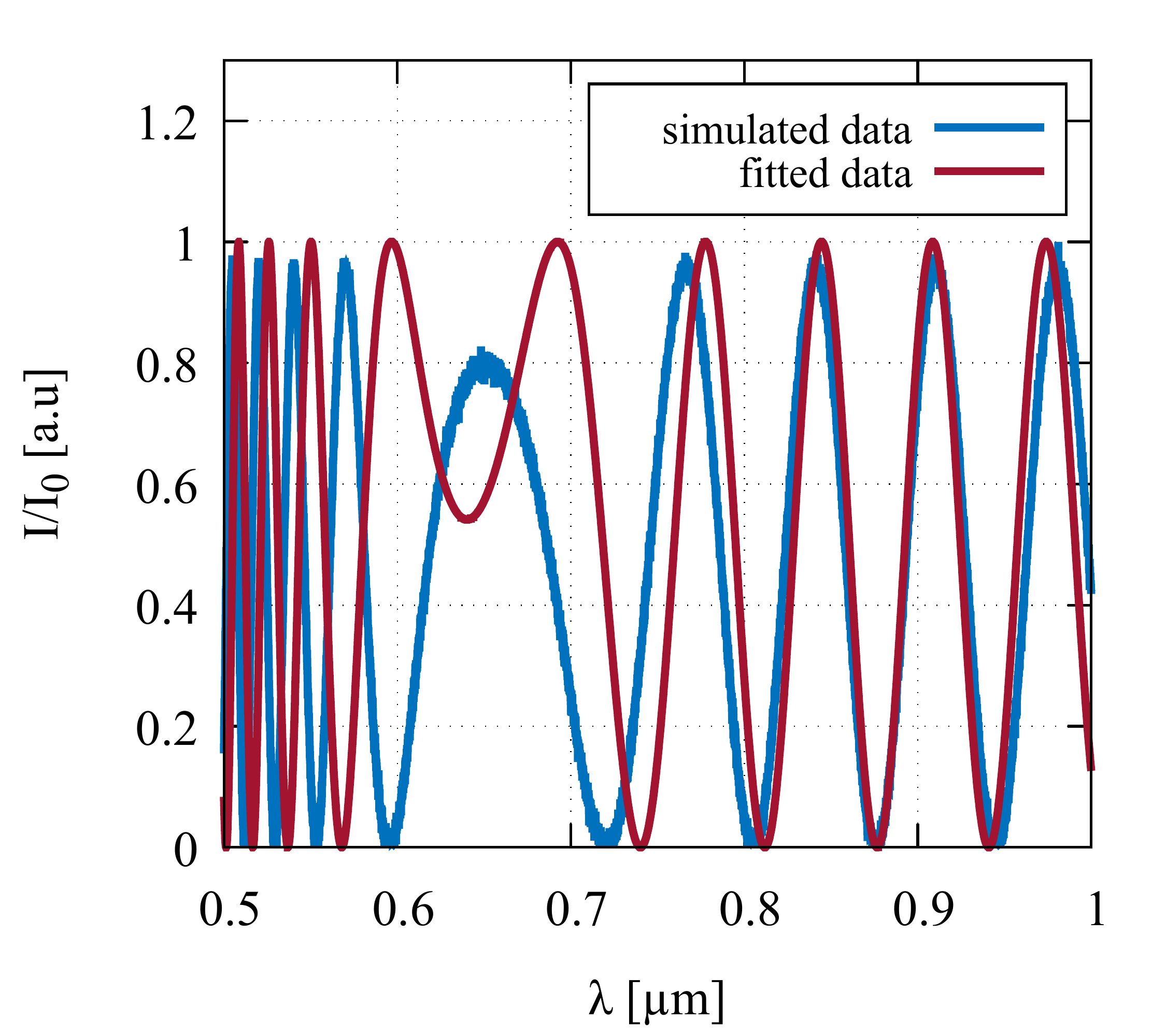}
				\put(1,1){\makebox(0,0){b)}}
			\end{overpic}
		\end{tabular}
	\end{center}
	\caption[Evaluation of classical Levenberg-Marquardt-based fitting approach versus brute force calculation]{Evaluation of classical Levenberg-Marquardt-based fitting approach versus brute force calculation for dispersion-encoded interferometric data with a) error progression of spectra using different path differences $\delta$ showing several local and a global minima where fitting is likely to fall on a local minimum if initial fit parameters are not carefully chosen and b) resulting spectrum of the fit routine which converged on a value of  $\delta_{fit}$\,=\,536.118\,\textmugreek m for an initial guess of $\delta_{guess}$\,=\,536\,\textmugreek m in relation to the simulated spectrum.}\label{Profilo:Pic:Local_minimum}
\end{figure}

Conventional fitting approaches such as the Levenberg-Marquardt algorithm, converge typically fast for periodical signals with constant phases. In the case where the phase of the data is varying over one dimension as in DE-LCI, fits can converge fast as well but most likely on a local minimum rather than on the global minimum, Fig. \ref{Profilo:Pic:Local_minimum} a).
In case of this simulation the actual OPD was \glssymbol{DeltSim}\,=\,535.233\,\textmugreek m. A brute force calculation of spectra according to the model described with Eq.\,(\ref{Profilo:EQ:basic_interferometer}) and (\ref{disp_phase}) and the respective squared sum of errors \glssymbol{SigmaSSE} was calculated in a range of \glssymbol{DeltRange}\,=\,(534\,-\,537)\,\textmugreek m. It can be seen that the $\sigma_{sse}$ value oscillates and shows two distinct local minima at 534.3 and 536.1\,\textmugreek m apart from the global minimum. An approximation with the Levenberg-Marquardt method using an initial guess of \glssymbol{DeltGuess}\,=\,536\,\textmugreek m converged in \textless\,10 iterations to a value of \glssymbol{DeltFit}\,=\,536.118\,\textmugreek m which falls on one of the local minima. It is clearly visible that the resulting spectrum is significantly different from the one which was actually present, Fig.\,\ref{Profilo:Pic:Local_minimum}\,b).  
One way to circumvent this problem is to compute spectra within a very large range of possible values for the fit parameters. As this approach can be time and memory consuming, strategies like downsampling of the measured data or the introduction of more advanced fitting approaches have to be taken into consideration, \cite{Kulkarni2020,Rajshekhar2013,Gurov2012}. Here, Monte-Carlo-based methods can be used in order to guess different start parameters and perform fitting in the parameter range with the highest likelihood of convergence on the global minimum, \cite{Ermolaev2015}. Furthermore, approximating spectral data in multiple stages with coarse variations of the fit parameters to determine the area of interest and finer variations to exactly converge on the final parameters is another strategy. In either way, it is desirable to determine the start value $\delta_{guess}$ with high precision. It should be emphasized that under the consideration of a precise determination of good start values any established fitting approach such as Levenberg-Marquardt can be used instead of the brute-force approach.

\subsection{Frequency analysis}\label{Profilo:SubSec:FreqAna}
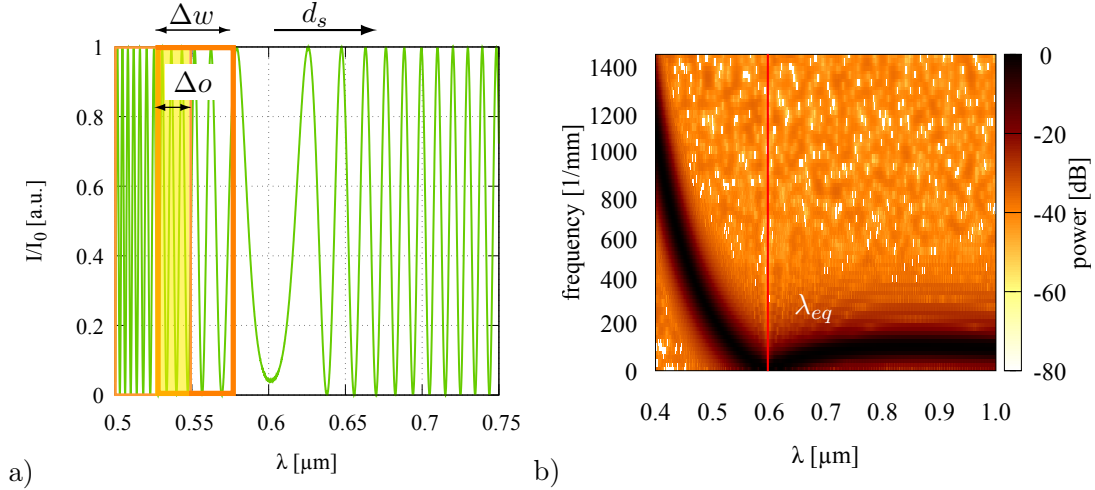
\begin{figure}[h]
\centering
		\begin{tabular}{c}
			\begin{overpic}[scale=.30, grid=false]{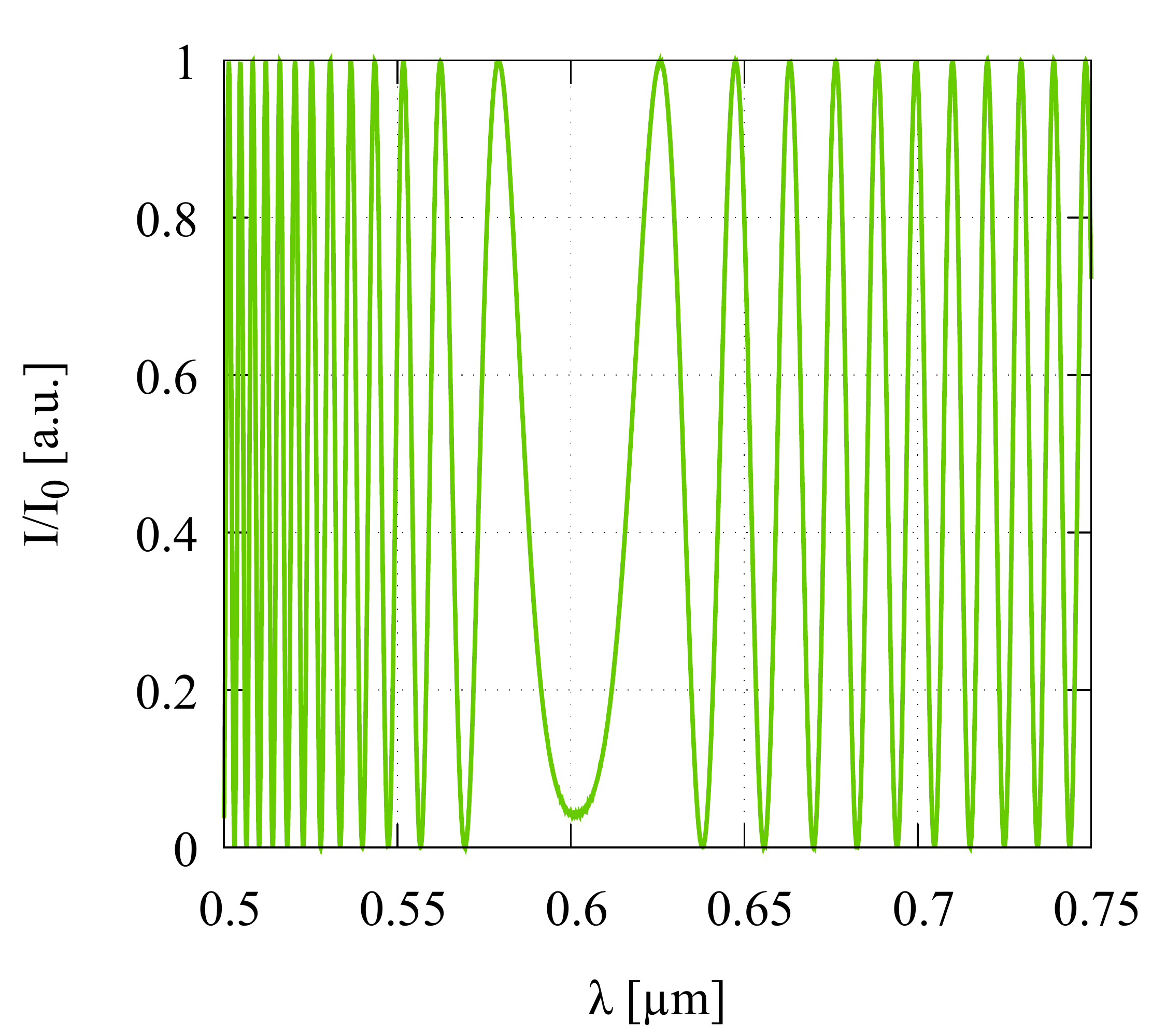}
				\put(1,1){\makebox(0,0){a)}}
					\put(27,16){\tikz \fill[color = yellow, opacity=0.3] (0,0) rectangle (.45,4.62);}
					\put(19,16){\tikz \draw[line width=1pt, color = orange!80] (0,0) rectangle (1,4.62);}
					\put(27,16){\tikz \draw[line width=2pt, color = orange] (0,0) rectangle (1,4.60);}
					\put(27,16){\tikz \fill[color = yellow, opacity=0.4] (0,0) rectangle (.45,4.60);}
					\put(27,87){\color{black}\vector(1,0){14.5}}
					\put(41.5,87){\color{black}\vector(-1,0){14.5}}
					\put(34,90){\makebox(0,0){$\Delta w$}}
					\put(33.75,72){\color{black}\vector(-1,0){7}}
					\put(28,72){\color{black}\vector(1,0){6}}
					\put(29,75){\colorbox{white}{\textcolor{black}{$\Delta o$}}}
					\thicklines \put(50,87){\color{black}\vector(1,0){20}}
					\put(58,90){\makebox(0,0){$d_s$}}
			\end{overpic}
		\begin{overpic}[scale=0.33]{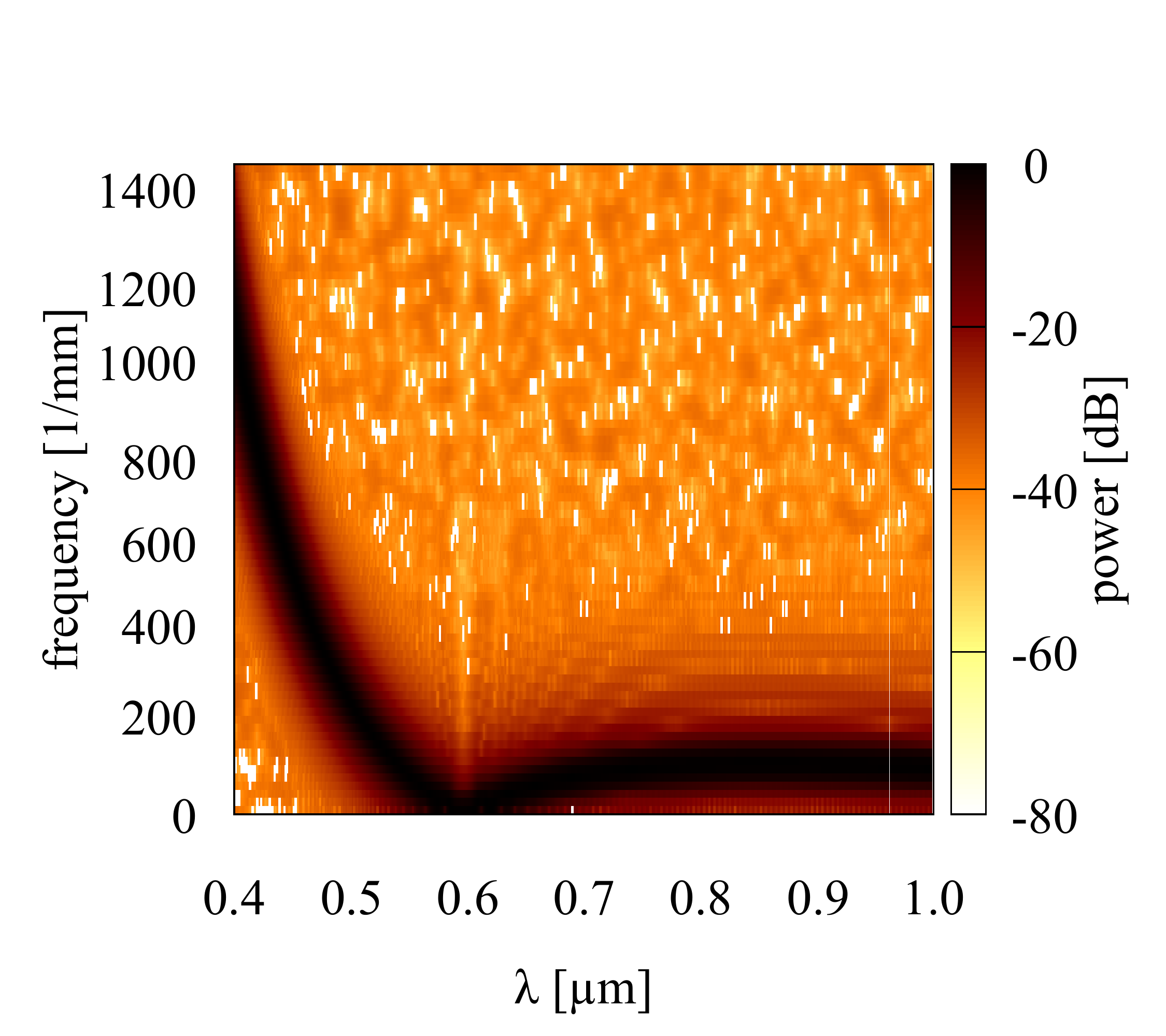}
			\put(1,1){\makebox(0,0){b)}}
			\put(48,30){\makebox(0,0){\textcolor{white}{$\lambda_{eq}$}}}
			\put(38,18.75){
				\begin{tikzpicture}
				\draw[thick,red] (0,0) -- (0,4.22);
				\end{tikzpicture}}
		\end{overpic}
		\end{tabular}
	\caption[Principle depiction of the \gls{STFT}]{Principle depiction of the \gls{STFT} where a window of the width $\Delta w$ is continuously slid over the signal in the direction \glssymbol{DirSTFT} to form spectral slices \glssymbol{SpecSliceSTFT} with an overlap of $\Delta o$  and b) resulting spectrogram of the stacked, Fourier-transformed slices $\mathfrak{F}(S_n^\prime)$.}\label{Profilo:Pic:STFT}
\end{figure}

In order to evaluate $\delta_{guess}$ from measured data, the equalization wavelength $\lambda_{eq}$ can be used in conjunction with Eq. (\ref{zRange}) which follows from $\frac{\partial\varphi}{\partial\lambda} = 0$, Eq. (\ref{Profilo:EQ:deriv})
\begin{eqnarray}
\delta_{guess} =  \left[n_g^{DE}(\lambda_{eq}) -1\right] \cdot t_{DE}
\end{eqnarray}
where the value for the group refractive index of the dispersive element $n_g^{DE}(\lambda_{eq})$ as well as its thickness $t_{DE}$ are used to calculate the initial estimate for the path length difference $\delta_{guess}$. However, this approach relies on the correct estimation of $\lambda_{eq}$ from measured data. Although this method describes a theoretical way to gather the necessary data, it is problematic in its implementation for measured data. In order to analyze the phase signal, a $\cos^{-1}$-operation can be performed which results in wrapped data ranging in $[-\pi,\pi]$. To avoid phase wrapping, the data processing established within this work relied on a frequency-based analysis. The analysis of periodic signals is typically performed by frequency analysis approaches such as Fourier transforms like \gls{FFT}. This is the standard method in e.g. \gls{FD-OCT}, where interference at different axial positions can be separated as different frequency components from the spectral domain signal, \cite{octBook}. As a number of publications demonstrated, dispersion in the experimental setup or in samples, which might be composed of different materials, leads to measurement uncertainties. For this reason, a variety of dispersion compensation approaches have been researched, \cite{Banaszek2007,Lippok2012,Wojtkowski2004}.\\
In distinction to approaches performing dispersion compensation of the gathered spectral data, \gls{DE-LCI} aims to make use of the dispersion within the system. In order to gain information on the phase and its wavelength-dependent slope, a Fourier-based analysis was performed on small slices $S_n^\prime$ of the signal. Known as windowed or short-time Fourier-transform, this method assumes that the phase is constant in a small slice of the signal, \cite{STFT}. It composes a resulting spectrogram by sliding an observation window in steps over the signal having a fixed length and overlap, Fig. \ref{Profilo:Pic:STFT} a).
By changing the window shape, window width \glssymbol{DeltaW} as well as the overlap between subsequent windows \glssymbol{DeltaO}, the resolution with regard to the phase minimum can be controlled. In consequence, a spectrogram of the stacked and Fourier-transformed slices \glssymbol{FourierSpecSliceSTFT} is used to analyze the phase minimum and the resulting equalization wavelength $\lambda_{eq}$, Fig. \ref{Profilo:Pic:STFT} b). The approach can also be used in analyzing signals composed of multiple frequencies from different axial positions. In the context of \gls{DE-LCI}, the approximation of the equalization wavelength $\lambda_{eq}$ was used to calculate an initial value of $\delta_{guess}$ in the interferometer in order to enable the fit to converge fast on a global minimum.

\subsection{Two-stage fitting}\label{Profilo:SubSec_fitting}
\begin{figure}[h]
	\centering
	\begin{tabular}{c}
		\begin{overpic}[scale=0.3, grid = false]{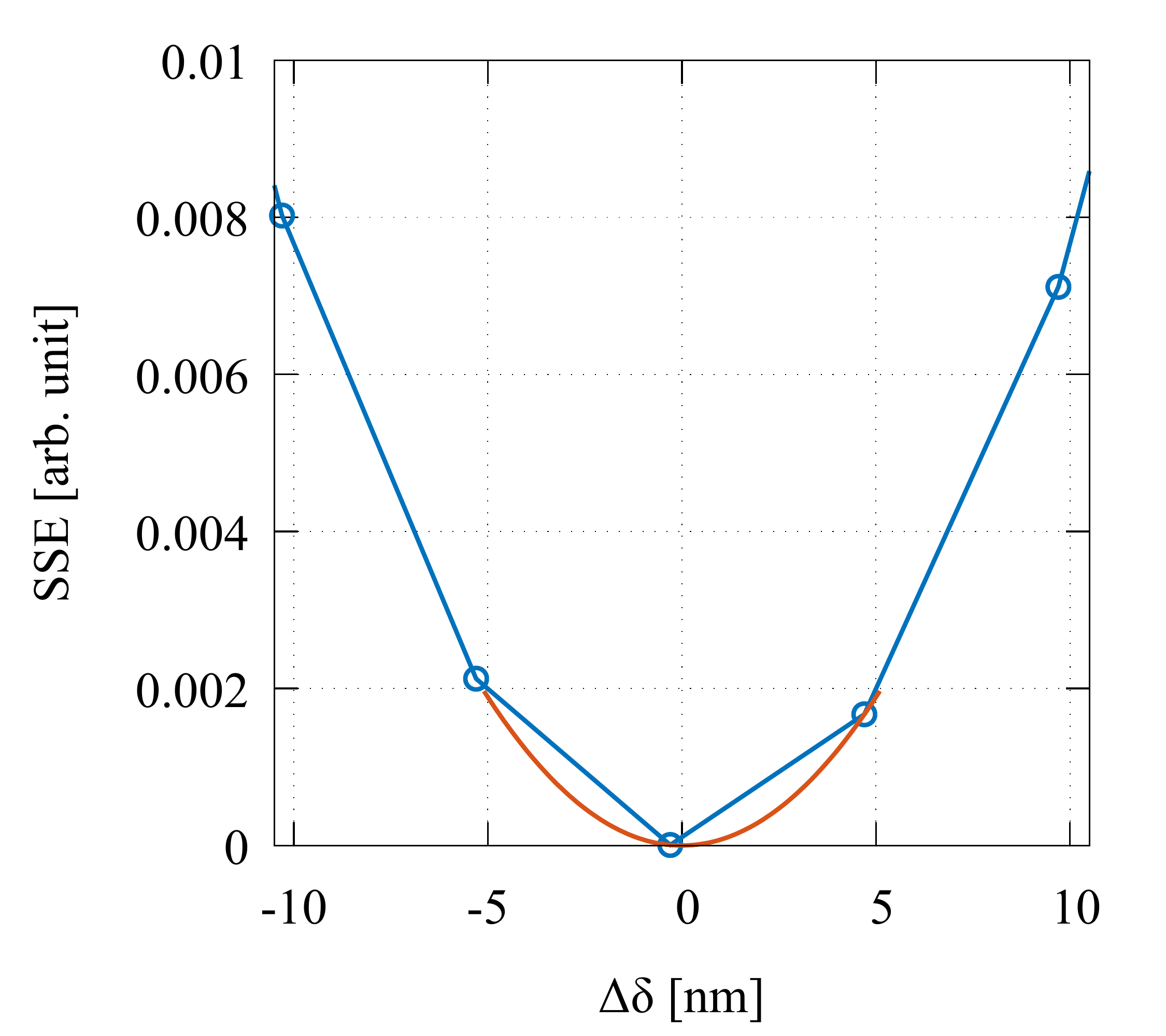}
			\definecolor{blue_plot}{HTML}{0072BD}
			\definecolor{orange_plot}{HTML}{D95319}
			\put(59,16.2){\tikz \draw[line width=0.2mm,orange_plot] (0,0) -- (0,4.17);}
			\put(57.5,16.2){\tikz \draw[line width=0.2mm,blue_plot] (0,0) -- (0,4.17);}
			\put(64.5,12){\makebox(0,0){\textcolor{orange_plot}{$\delta_2$}}}
			\put(53,12){\makebox(0,0){\textcolor{blue_plot}{$\delta_1$}}}
			\put(23,72){\tikz \draw [line width=0.5mm,<->,>=stealth] (0,.5) -- (4.6,.5);}
			\put(53,76){\colorbox{white}{\textcolor{black}{$\Delta \delta_1$}}}
			\put(42,30){\tikz \draw [line width=0.5mm,<->,>=stealth] (0,.5) -- (2.2,.5);}
			\put(53,34){\colorbox{white}{\textcolor{black}{$\Delta \delta_2$}}}
		\end{overpic}
	\end{tabular}
	\caption[Visualization of the simplified two-step fitting process]{Visualization of the simplified two-step fitting process based on \gls{SSE} determination for simulated data sets in two ranges ($\Delta \delta_1$ and $\Delta \delta_2$) for the path length difference $\delta$.}\label{APNDX_fit}
\end{figure}
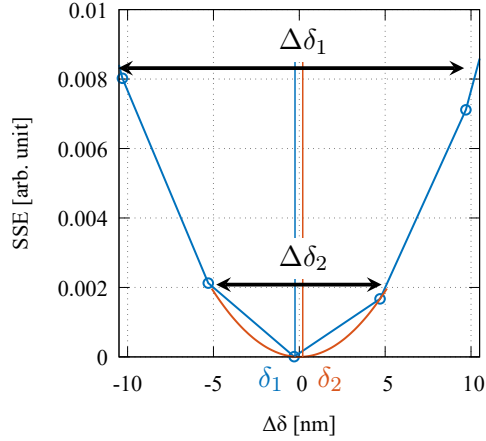

\begin{figure}[h]
\centering
		\begin{tabular}{c}
			\begin{overpic}[scale=0.27]{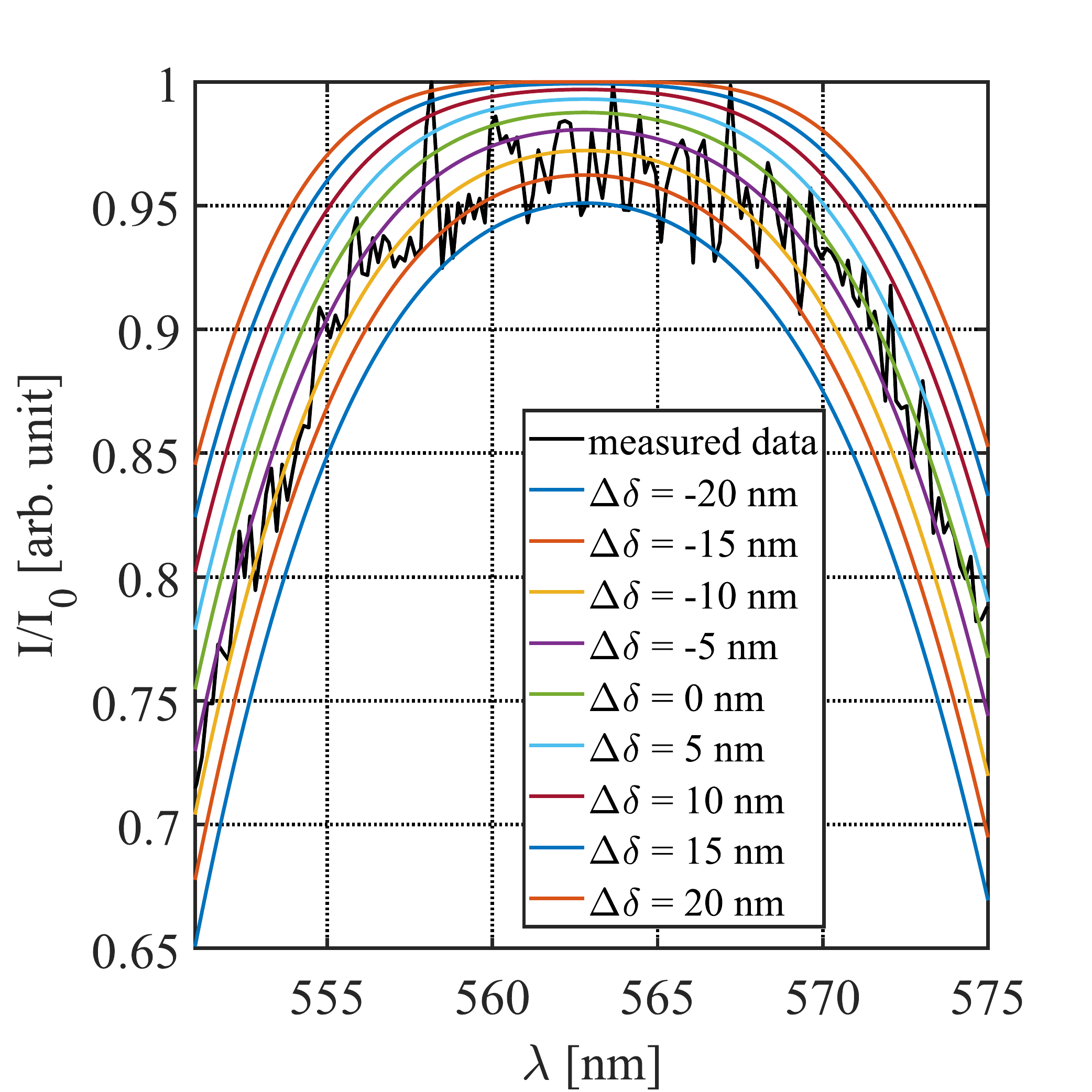}
				\put(1,1){\makebox(0,0){a)}}
			\end{overpic}
			\begin{overpic}[scale=0.27]{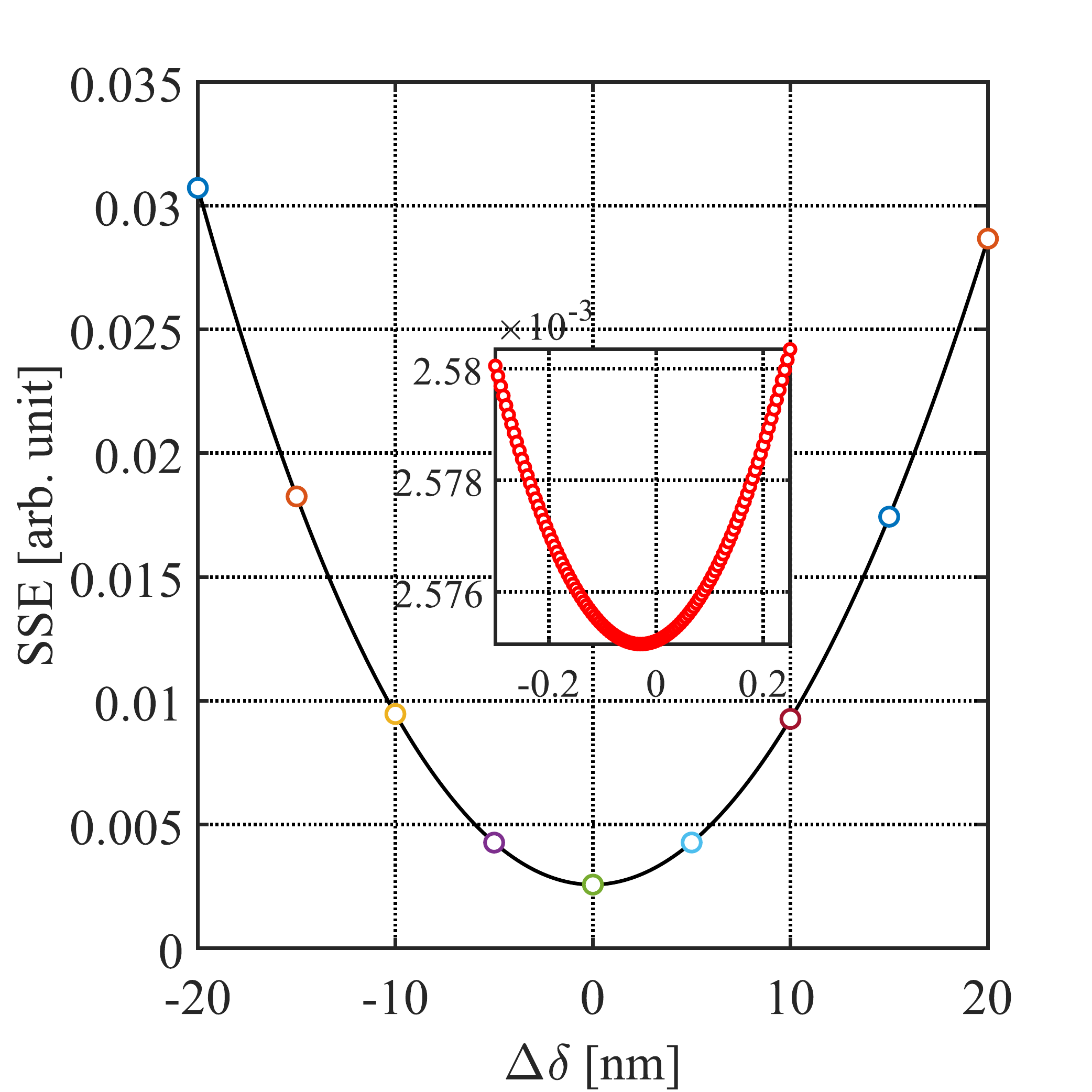}
				\put(1,1){\makebox(0,0){b)}}
				\put(54.25,20){\tikz \draw [line width=0.5mm,->,>=stealth] (-1.5,.5) -- (-0.8,1.25);}
				\put(49,20.5){\tikz \draw [line width=0.5mm,->,>=stealth] (-1.5,.5) -- (-1.75,1.25);}
			\end{overpic}
		\end{tabular}
		\caption[Example of the two-stage fitting routine based on a single-point measurement]{Example of the two-stage fitting routine with a) plot of a measured intensity signal at $\lambda_{eq}$ with a selection of fit curves where the \gls{OPD} is separated by $\Delta \delta$\,=\,5\,nm for each iteration and b) corresponding error sum of squares (\gls{SSE}) for the different $\Delta\delta$ and a interpolating curve plotted in black where the arrows indicate a magnified plot of the area in close proximity of the minimum \gls{SSE}.}\label{Proflo:Pic:SSE_estimation}
\end{figure}
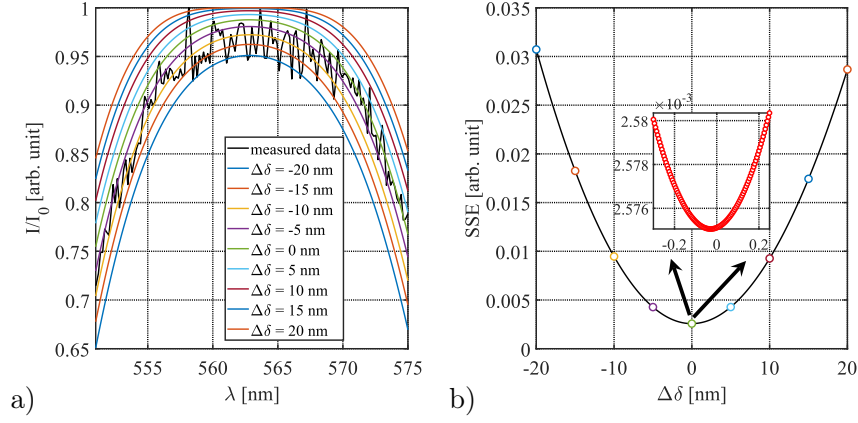

As described in subsection \ref{Chap:Profilo_Sec:SigAn_SubSec:brute_force}, the usage of conventional fitting routines on oscillatory data with varying phase can lead to problems. For this reason, the here developed fitting routine was constructed as a two-step process. Using a range of $\Delta \delta_1$\,=\,$\pm$\,1\,\textmugreek m with a step size of 2\,nm centered around the previously calculated $\delta_{guess}$, a set of simulated spectra based on Eq.\,(\ref{Profilo:EQ:basic_interferometer}) and (\ref{disp_phase}) was calculated in a brute-force fashion. The determination of the \gls{SSE} of these calculated spectra with respect to the measured spectrum enabled the estimation of a more precise value for the path length difference $\delta_1$ at the minimum of the \gls{SSE} curve, Fig. \ref{APNDX_fit}.
The calculated $\delta_1$ was used in a second iteration of the routine to calculate another set of spectra with a finer spacing in $\Delta \delta_2$\,=\,$\pm$\,140\,nm with steps of 0.02\,nm\footnote{These values were typically used throughout this work. They are dependent on the used DE as well as on the sample to be measured.}. Comparably, the \gls{SSE} of the calculated spectra was evaluated with respect to the measured spectrum. The minimum \gls{SSE} indicates the path length difference $\delta_2$ which can be used to compute the height at a point of the sample, see also Fig.\,\ref{Proflo:Pic:SSE_estimation}\,b).\\
The described method was chosen in instead of other established fitting algorithms to ensure the convergence on the global minimum rather than a local minimum which can be the case due to the oscillating nature of the data. The iterative fitting approach bears further potential for optimizations regarding the processing time.\\
Opposing to doing a fit of the whole spectrum, fitting in a \gls{ROI} was carried in order to reduce processing time, as only about 25 \% of the gathered data had to be processed. The processing time of a whole profile with this spectral \gls{ROI} was about 2 seconds\footnote{This value is strongly dependent on the used hardware and the optimization performed with regard towards parallelization. Within this work neither specialized hardware nor software optimizations have been used. This is up to future work.}. The \gls{ROI} was selected as a fixed set of 530 data points distributed symmetrically around the equalization wavelength. The size of the \gls{ROI} was determined in preliminary experiments in order to include at least one spectral modulation to each side of $\lambda_{eq}$ and is dependent on the used \gls{DE}. It is not expected that the size of the \gls{ROI} influences the resolution of the setup.\\
As described above, the fit of the measured data in close proximity of the equalization wavelength significantly enhances the resolution, see Eq.\,(\ref{noise_resolution}). In order to prove the resolution limit experimentally, a sample data set was evaluated regarding its regression error towards simulated data sets within a range of path length differences $\Delta \delta$, Fig.\,\ref{Proflo:Pic:SSE_estimation}\,a).
Due to the ability to perform a search for the regression minimum based on a \gls{SSE} approach, the best fitting data set can be used to determine the path length difference as a basis for the height calculation. This method assumes that the calculated minimum value of regression \glssymbol{SSEmin} is the center of a confidence interval which has a variance \glssymbol{Variance}. The variance can be computed using the number of fitted parameters \glssymbol{FitParams}\,=\,3 and the number of points in the fit interval n\,=\,530 with, \cite{SachsStatistik} p. 287
\begin{equation}
\sigma^2 = \frac{SSE_{min}}{(n-m)} = \num{4.89e-6}
\end{equation} 
By using to the slope calculated in Fig. \ref{Proflo:Pic:SSE_estimation} b), the variance can be used to determine the interval of path length difference with $\Delta \delta$ = $\pm$ 0.27 nm. Using typically N\,=\,10 consecutive measurements, a mean height profile deviation can be calculated with $\Delta \delta$ and Eq. (\ref{EQ:resolution_fit}) to \glssymbol{ResFitExp}\,=\,0.085\,nm. The intensity of the data at the equalization wavelength influences the resolution as shown in Fig. \ref{Proflo:Pic:MeasRange_resolution_estimation}. Taking this into account, a further analysis can be done where the data set of Fig.\,\ref{Proflo:Pic:SSE_estimation}\,a) showed a relative intensity of I\,=\,0.97\,arb.\,units. From the theoretical calculation of Eq. (\ref{noise_resolution})/Fig. \ref{Proflo:Pic:MeasRange_resolution_estimation} it can be deduced that this intensity corresponds to a theoretical resolution of $r_{fit}$ = 0.087 nm. This is well aligned with the expected experimental value. Further experimental evaluation in the spatial domain was performed in subsection \ref{Profilo:Sec:2D_approach}.

\subsection{Error estimation of the data processing}\label{Profilo:SigAn:SubSec:error_data_proc}
\begin{figure}[h]
	\centering
		\begin{tabular}{c}
			\begin{overpic}[scale=0.50]{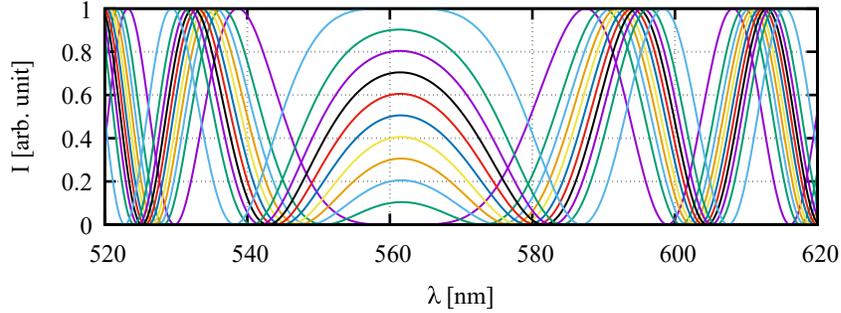}
			\end{overpic}
		\end{tabular}
	\caption{Example of the simulated data set for the characterization of the influence due to noise from the data processing routines where the curves are separated by $\Delta I$\,=\,0.05 in a range of $I_{0}$\,=\,0\,-\,1.}\label{Profilo:Pic:simulated_spectra_for_noise}
\end{figure}

\begin{figure}
	\begin{tabular}{c}
		\begin{overpic}[scale=.32]{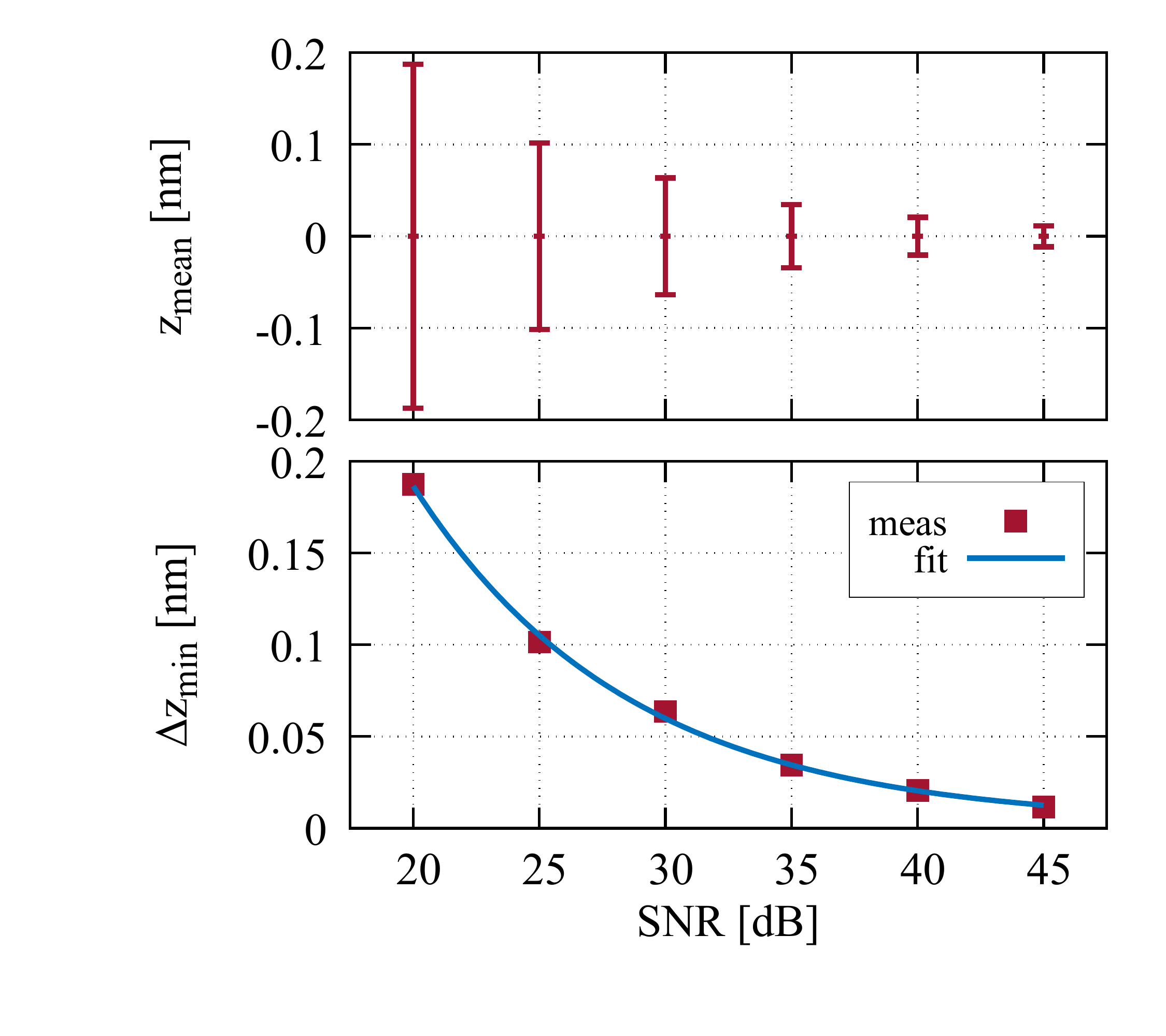}
			\put(1,1){\makebox(0,0){a)}}
		\end{overpic}
		\begin{overpic}[scale=.32]{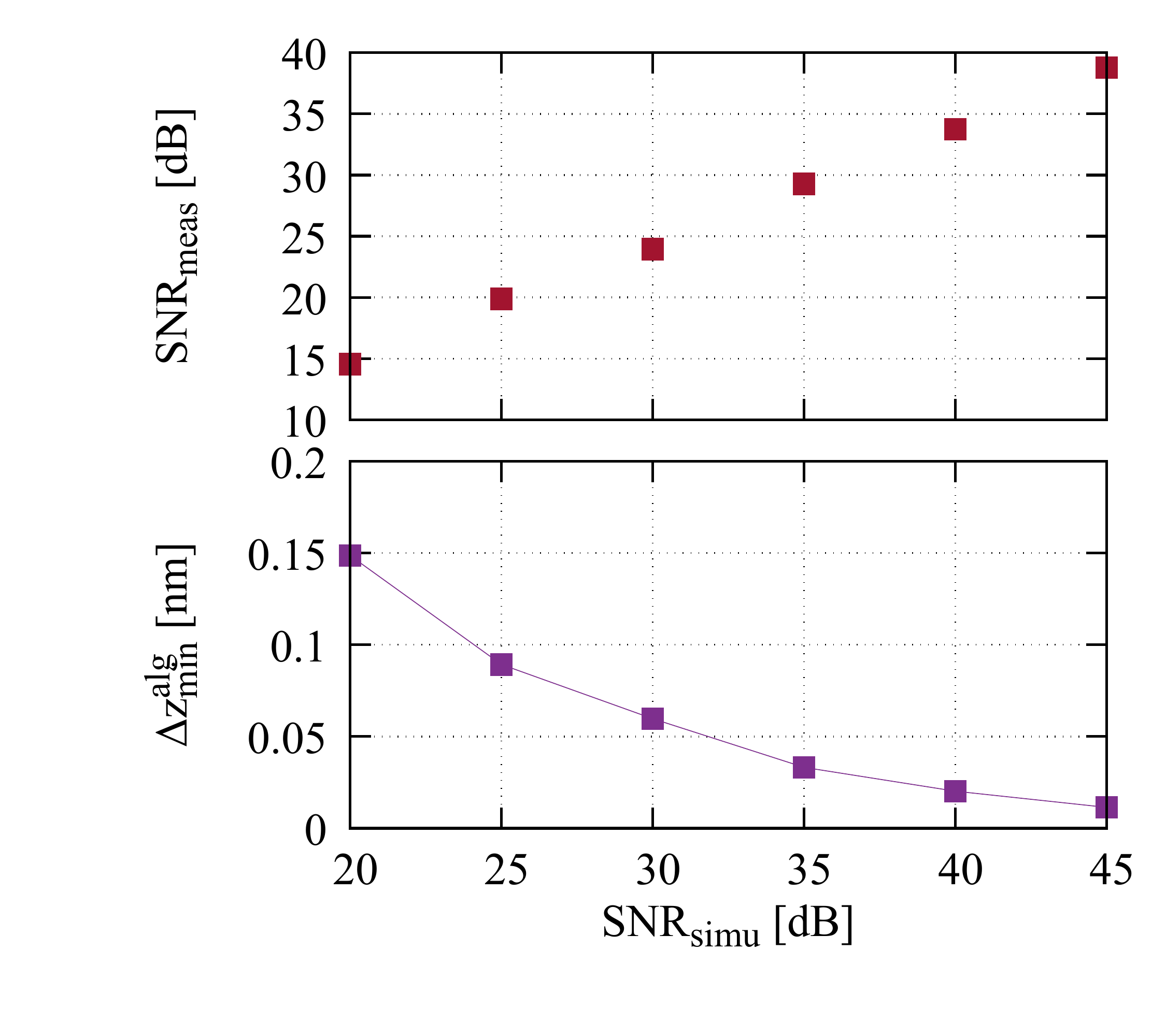}
			\put(1,1){\makebox(0,0){b)}}
		\end{overpic}
	\end{tabular}
{\caption[Depiction of the analyzed simulated data in relation to the different noise levels]{Depiction of the analyzed simulated data in relation to the different noise levels at a relative intensity of $I_{0}$\,=\,0.5\,arb.\,units with a) mean height data of averaged profiles $z_{mean}$ and related standard deviation $\Delta z_{min}$ where the nominal value was $z_{nom}$ = 0 as well as the slope of $\Delta z_{min}$ with a fitted inverse logarithmic relationship with respect to the added noise; b) Plot of the measured SNR from the power of $z_{mean}$ over the length of 500 analyzed spectra per data set with respect to the simulated noise levels displayed on top of a plot of the effective, noise-dependent influence of the algorithm on the resolution $\Delta z^{alg}_{min}$.} \label{Proflo:Pic:noise_simu_vs_SNR}}
\end{figure}

\begin{figure}[h]
\centering
		\begin{tabular}{c}
			\begin{overpic}[scale=0.31]{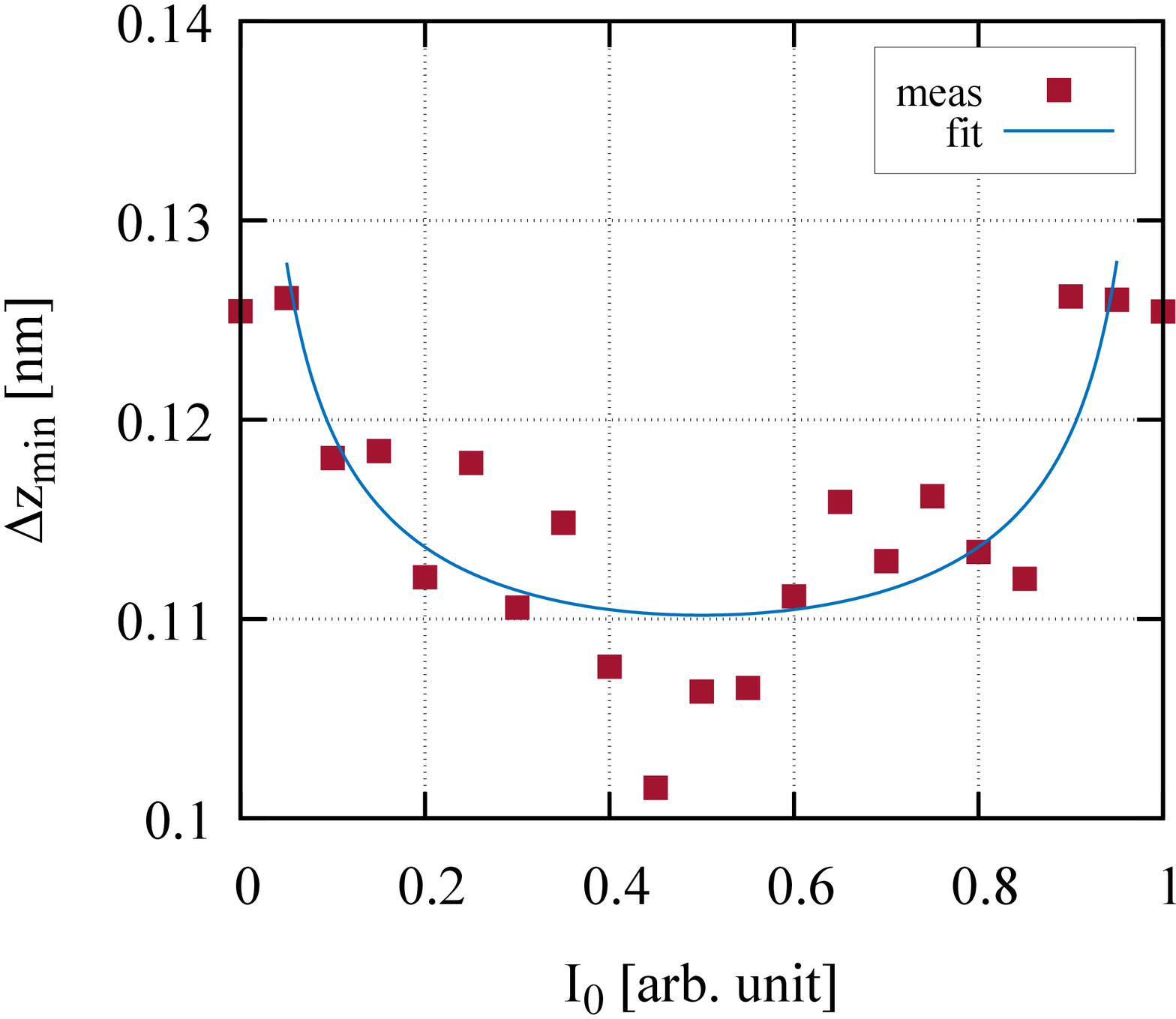}
				\put(1,1){\makebox(0,0){a)}}
			\end{overpic}
			\begin{overpic}[scale=0.32]{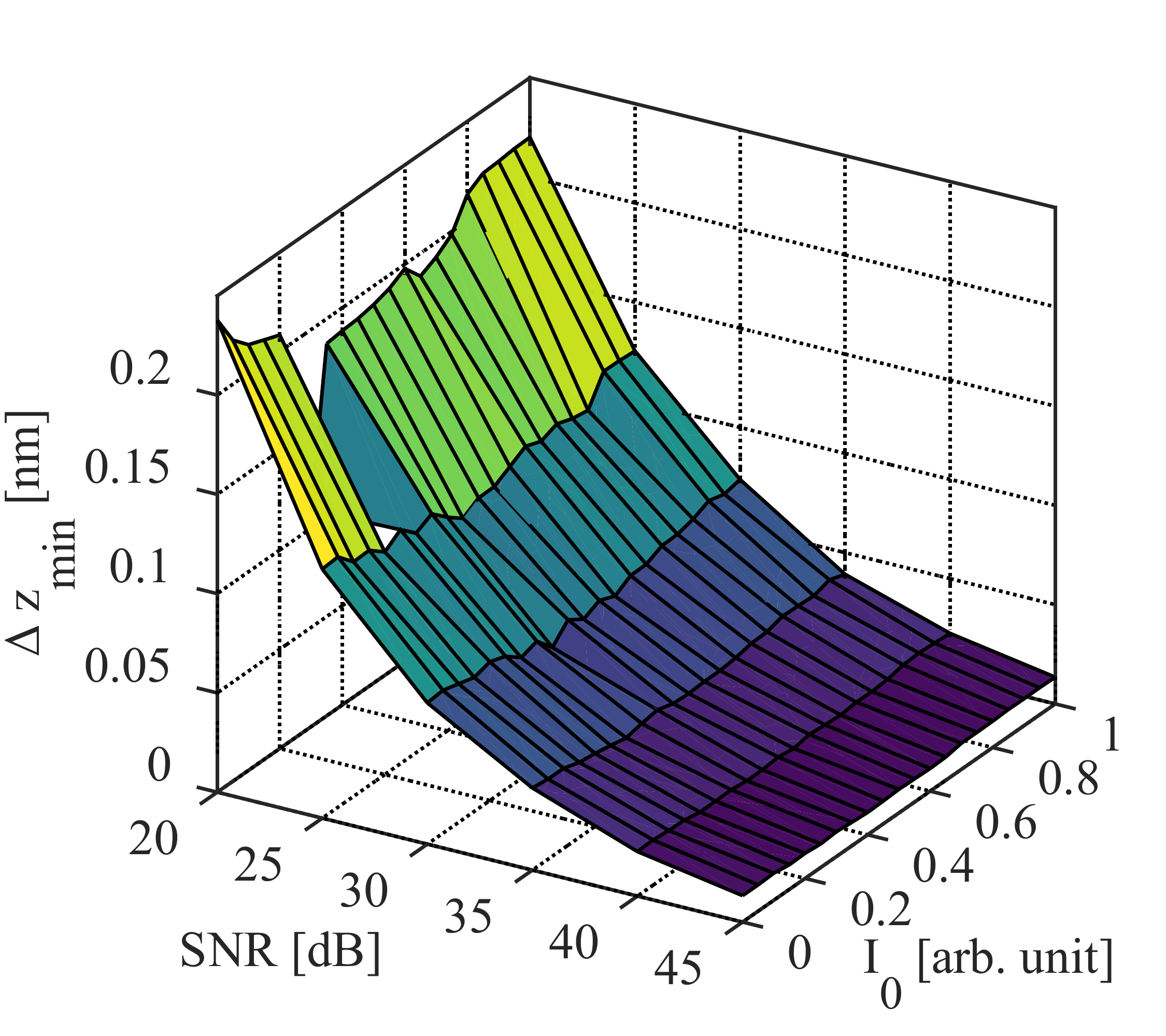}
				\put(1,1){\makebox(0,0){b)}}
			\end{overpic}
		\end{tabular}
	\caption[Comparison of all noise influences induced by the fitting algorithm]{a) Plot of the measured resolution $\Delta z_{min}$ with respect to the relative intensity at the equalization wavelength $\Delta I$ at a simulated noise level of 25\,dB and  b) three-dimensional representation of the influence of simulated noise and change in relative intensity on the height resolution.}\label{Proflo:Pic:noise_simu_3D}
\end{figure}

\begin{figure}[h]
	\begin{center}
		\begin{tabular}{c}
			\begin{overpic}[scale=0.65]{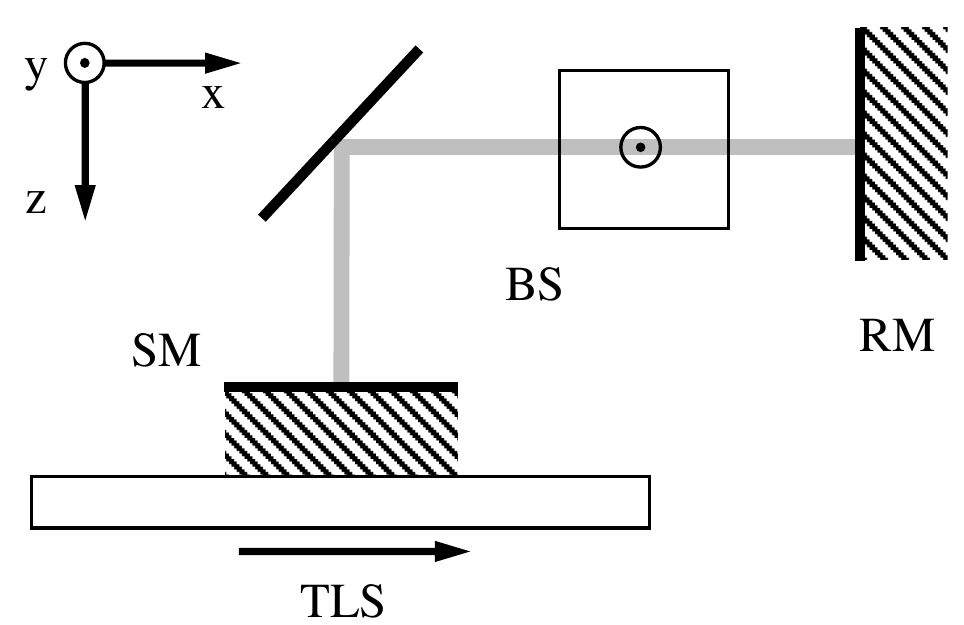}
				\put(1,1){\makebox(0,0){a)}}
			\end{overpic}
			\begin{overpic}[scale=0.25]{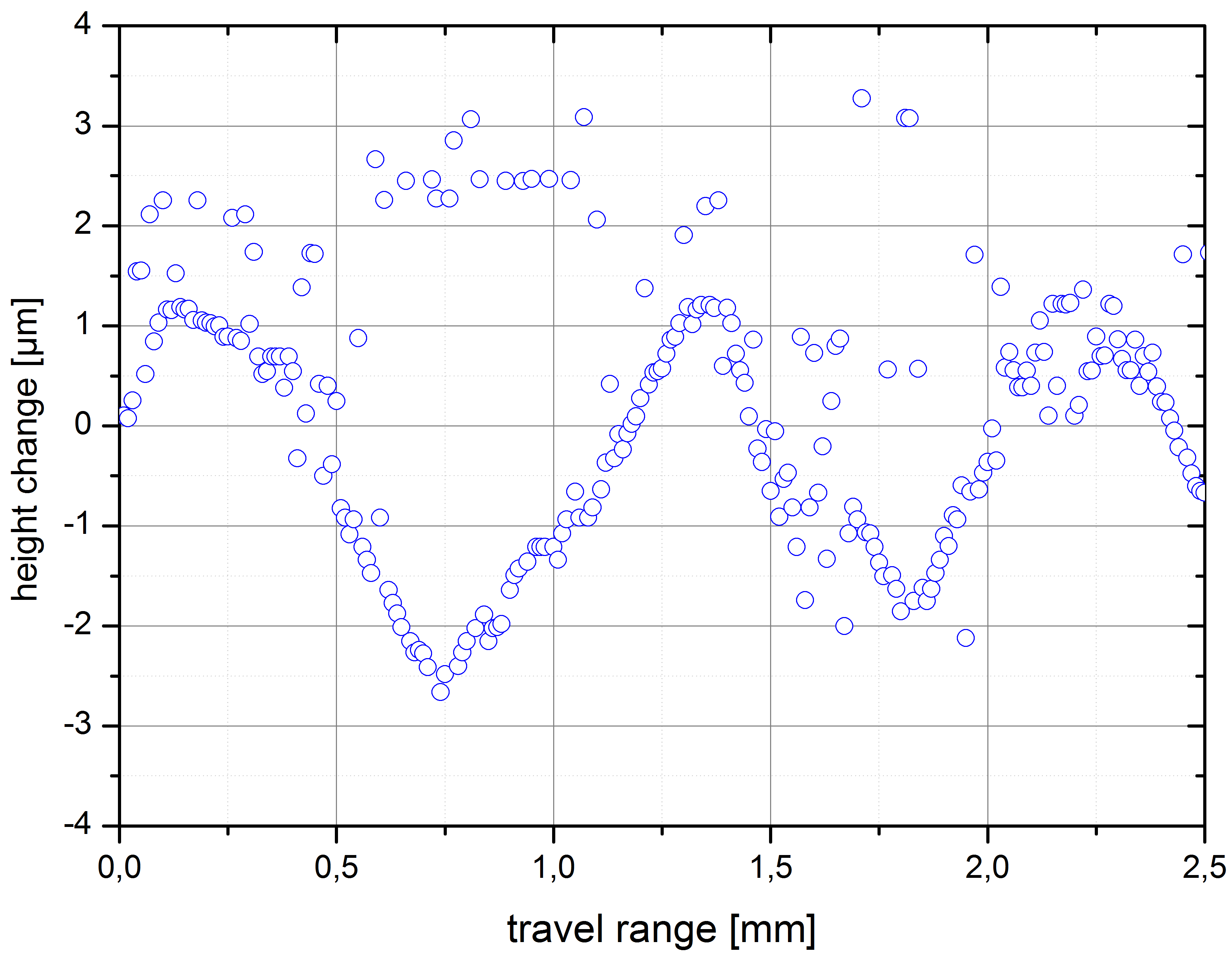}
				\put(1,1){\makebox(0,0){b)}}
			\end{overpic}
		\end{tabular}
	\end{center}
	\caption[Setup and results of the evaluation of influences through lateral movement]{ a) Principle of a setup for the characterization of axial influences from lateral movement where a TLS - translation stage moves a SM - sample mirror in the $x$-dimension. The sample arm beam is recombined with the beam from the RM - reference mirror using a BS - beamsplitter in order to characterize the errors introduced as height changes in the $z$-dimension interferometrically and b) Result of the translation stage evaluation showing the introduced, oscillating height error over a scanning range in the $x$-dimension.}\label{Profilo:Pics:Surface_Scan_etup_and_data}
\end{figure}

Besides the previously discussed error influences due to the optical setup, the implementation of the analysis algorithm introduces further noise. In order to quantify its influence, a simulation and subsequent analysis were conducted. For this purpose, a simulated height profile with a nominal height of \glssymbol{HeightNom}\,=\,0 was constructed. It was represented by 500 spectra per profile and 10 repetitions. Based on a typical equalization wavelength of $\lambda_{eq}$\,=\,562\,nm, a set of spectra in close proximity to $\lambda_{eq}$ was calculated having relative intensities of $I_{0}(\lambda=\lambda_{eq})$\,=\,0\,-\,1\,arb.\,units with $\Delta I$\,=\,0.05\,arb.\,units, Fig. \ref{Profilo:Pic:simulated_spectra_for_noise}. 
The dispersive element was assumed to be of N-BK7 with $t_{DE}$\,=\,2\,mm within this simulation. Every calculated spectrum was subsequently obstructed by white, Gaussian noise having a \gls{SNR} \glssymbol{SNRsimu}\,=\,20\,-\,45\,dB in steps of $\Delta$SNR\,=\,5\,dB. In total, this resulted in 630,000 spectra processed by the analysis algorithm.\\
The analysis was concentrated on two important features, the averaged height \glssymbol{HeightMean} as well as its corresponding standard deviation over the length of the 500 spectra per data set \glssymbol{HeightMin} as an indicator for the resolution. By analyzing the mean height and its standard deviation for the spectra at $I_{0}$\,=\,0.5\,arb.\,units with respect to the added noise levels, some initial insight can be drawn, Fig. \ref{Proflo:Pic:noise_simu_vs_SNR} a).
As expected, the mean height is close to zero while the standard deviation is dependent on the noise level. A mean value $z_{mean}$\,=\,\num{-2.9e-15}\,nm of across all noise levels could be measured. On a closer look, the standard deviation of the height measurement, which can be regarded as a measure for the resolution, can be fitted by an inverse logarithmic model, which corresponds well with the logarithmic scale of the noise levels. Using this fitted equation, the influence of the used algorithms could be estimated.\\
It is known from the analysis of the spectral intensity signal of the light source, Fig.\,\ref{Proflo:Pic:MeasRange_spline}, that the typical noise is 21.25\,dB. By evaluating the fitted curve of the standard deviation of $z_{mean}$, a value of $\Delta z_{min}(21.25dB)$\,=\,0.16\,nm can be found. As this value is significantly apart from the calculated minimal resolution for this noise level of $r_{fit}$\,=\,0.029\,nm, a further investigation was performed. The power \glssymbol{NoisePower} of the averaged height over the range of \glssymbol{NumSpectra}\,=\,500 spectra was used to calculate the signal-to-noise ration \glssymbol{SNRmeas} of the height profile \glssymbol{SurfProfile} using 
\begin{eqnarray}
P = \frac{1}{S} \cdot \sum_{0}^{S}|z|^2\\
SNR_{meas} = 10*\log{P},
\end{eqnarray}
where it was compared to the initially simulated noise $SNR_{simu}$, Fig.\,\ref{Proflo:Pic:noise_simu_vs_SNR}\,b). It can be seen that the relation follows a linear slope as expected, but a general offset of about 5\,dB is present. Under consideration of this offset, the resolution for the typical noise of the system in an experimental situation was re-calculated as $\Delta z_{min}(21.25 dB)$\,=\,0.088\,nm. As value takes the complete processing of data into account, it is supposed to be the expected minimal resolution of the system at this noise level. A subtraction of the noise-related resolution according to Eq.(\ref{noise_resolution}) from the resolution calculated in this simulation, see bottom plot of Fig.\,\ref{Proflo:Pic:noise_simu_vs_SNR}\,a), results in the effective influence of the algorithm on the resolution  $\Delta z^{alg}_{min}$, bottom plot of Fig.\,\ref{Proflo:Pic:noise_simu_vs_SNR}\,b). Obviously, it is dependent on the effective noise of the system.\\
According to Eq. (\ref{noise_resolution}), the resolution $\Delta z_{min}$ of the \gls{DE-LCI} setup is also dependent on the relative intensity at the equalization wavelength $\Delta I$. Consequently, the analysis of the height measurement, the standard deviation and the \gls{SSE} in relation to the relative intensity. In correspondence to the finding of Fig.\,\ref{Proflo:Pic:noise_simu_vs_SNR}\,b), this investigation was performed at a noise level of 25\,dB, Fig.\,\ref{Proflo:Pic:noise_simu_3D}\,a).
From the height measurement it can be derived that the measured values are normally distributed around the nominal value $z_{nom}$\,=\,0. More interestingly, the values for the resolution with respect to the relative intensity can be fitted using the already derived relationship from Eq. (\ref{noise_resolution}). This fit supports the finding that the algorithm has a significant influence on the resolution which prevents it from achieving a minimal theoretical resolution of $r_{fit}$\,=\,0.029\,nm.\\
As a result of the simulations, the noise behavior of the data processing routines can be plotted against all three influences, the relative intensity, the original \gls{SNR} level as well as the standard deviation of the measured average height, Fig. \ref{Proflo:Pic:noise_simu_3D} b). From this plot, the described trends can be analyzed by direct comparison. The mean standard deviation for the height estimation follows an inverse logarithmic relationship with an increase in \gls{SNR}. Simultaneously, the expected relationship to the relative intensity $\Delta I$ according to Eq. (\ref{noise_resolution}) is visible.\\
In typical measurement scenarios, the characterization of height profiles is desired in one lateral dimension or even in an areal fashion. As other approaches have shown, gathering line profile data either by scanning the beam or by moving the sample relative to the beam is common. In either case, the accuracy of the moving parts has an influence on the accuracy of measurement. In order to evaluate this influence, an experiment with a plane mirror on a translation stage (Z812B, Thorlabs Inc., USA) was conducted, Fig. \ref{Profilo:Pics:Surface_Scan_etup_and_data} a).
The translation stage was moved continuously over a range of 2.5\,mm during the experiment along the $x$-axis. Simultaneously, data was collected with the interferometer. As the same mirrors with high flatness were used on the translation stage as well as reference mirror, the result should ideally show no height differences between both arms, Fig. \ref{Profilo:Pics:Surface_Scan_etup_and_data} b). It is visible in the data that the translation stage introduces a significant amount of height differences to the measurements. Over a scanning range of 2.5\,mm the stage oscillates in $z$-direction up to $\pm$\,1.5\,\textmugreek m. In order to achieve a resolution in the nanometer range, such deviations have to be diminished or calibrated.

\section{Two-dimensional approach and characterization}\label{Profilo:Sec:2D_approach}
\begin{figure}[h]
	\begin{center}
		\begin{tabular}{c}
			\begin{overpic}[scale=0.7]{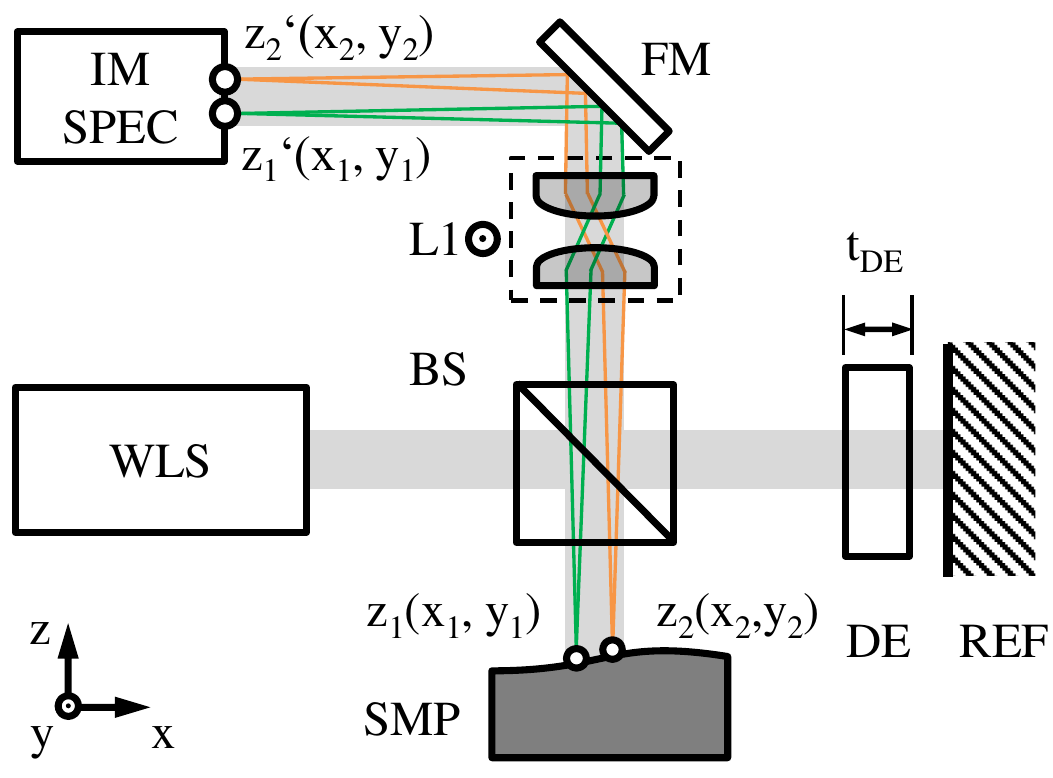}
				\put(0,0){\makebox(0,0){a)}}
			\end{overpic}
		\begin{overpic}[scale=0.7]{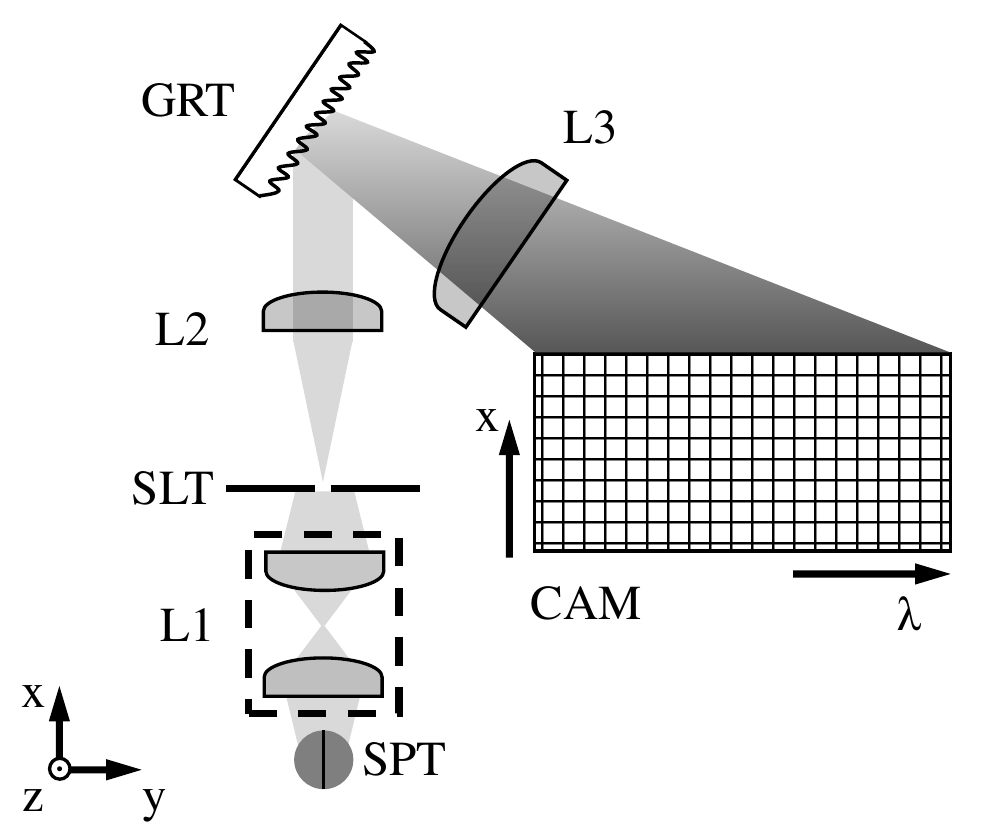}
			\put(0,0){\makebox(0,0){b)}}
		\end{overpic}
		\end{tabular}
	\end{center}
	\caption[Experimental setup for scan-free two-dimensional profilometry]{a) Experimental setup with WLS - white light source, BS - beamsplitter, DE - dispersive element (having the thickness $t_{DE}$ and the refractive index $n^{DE}(\lambda)$), REF - reference mirror, SMP - sample profile including the points $z_1(x_1,y_1)$ and $z_2(x_2,y_2)$ which are imaged with a given magnification M (typically M\,=\,1.3 or 4) by the L1 - imaging configuration, relayed by a FM - folding mirror onto the slit of the IMSPEC - imaging spectrometer as magnified points $z_1'(x_1,y_1)$ and $z_2'(x_2,y_2)$ and a detailed view of the same in b) with SPT - measurement spot, SLT - slit, L2 - collimating lens, GRT - grating, L3 - imaging optics which is used to realize an internal magification (refer also to appendix \ref{Profilo:APNDX_Calc_ImagSpec})  and CAM - camera where the spectral information for every point on the line in $x$-dimension is recorded.}\label{ChaProf:Sec2D:Pic-setup}
\end{figure}

The estimations of section \ref{Profilo:Sec:Meas_range} demonstrated the capability to perform high-resolution profilometry with the dispersion-encoded low-coherence approach. But they also demonstrated the scanning mechanism's influence on deviations of the result. In order to avoid these and other unwanted deviations from e.g. thermal or vibrational influences, an imaging approach was developed to gather two-dimensional data without scanning, Fig. \ref{ChaProf:Sec2D:Pic-setup} a).
In this configuration the reference arm is composed of an element with known dispersion (here Schott N-BK7, $t_{DE}$\,=\,2000\,\textmugreek m) and a plain mirror. The sample arm holds a sample with a varying height profile along the $x$-$y$ plane noted with $z(x,y)$. The recombined collimated light from sample and reference arm is imaged on the slit of an imaging spectrometer. Within this spectrometer the light is spectrally decomposed and imaged onto the two-dimensional \gls{CMOS} array of a camera, Fig. \ref{ChaProf:Sec2D:Pic-setup} b). In contrast to a single-line detector of a standard spectrometer, this configuration enables the detection of spectra at every point on a line in the $x$-dimension of the measurement spot. The information is only selected from one position in the $y$-direction which means that the acquisition of height profiles along a single line at once becomes possible. Following this approach, the recorded signal enables the detection of spectral interferograms along a line of the $x$-dimension which can be described analogous to Eq. (\ref{Profilo:EQ:basic_interferometer}) and (\ref{disp_phase}) with
\begin{eqnarray}\label{Profilo:EQ:2d_int_equation}
& I(x,\lambda) = I_0(\lambda) \cdot \left[ 1 + \cos \varphi(x,\lambda)  \right]\\
& \textrm{with } \varphi(x,\lambda) = 2\pi \frac{ \left[ \left( n_{DE}(\lambda)  - 1 \right)t_{DE} \right] -\delta(x) }{\lambda},
\end{eqnarray}
where $I_0(\lambda)$ is the initial spectral intensity before the beamsplitter and $\varphi(\lambda,x)$ is the absolute phase of the signal at every point in the $x$-dimension which is dependent on the \gls{OPD} between both arms denoted with $\delta(x)$ and the wavelength $\lambda$. The signal detected by the camera of the imaging spectrometer is composed of stacked spectral interferograms where the resolution of the $x$-dimension is dependent on the magnification,\glssymbol{Mag}, of the interferometer and the spectrometer construction, see Fig. \ref{ChaProf:Sec2D:Pic-setup}. Typically, magnifications M\,=\,1.3 and 4 where used within this work. For all investigations within this work, an imaging spectrometer with the following parameters was designed and built, Tab. \ref{ChaProf:Sec2D:Tab-specs_imaging_spectrometer}.
\begin{table}[h]
	\captionabove{Calculated parameters and components for the designed imaging spectrometer.} \label{ChaProf:Sec2D:Tab-specs_imaging_spectrometer}
	\begin{center}       
		\begin{tabular}{p{5cm}p{3cm}p{5cm}}
			\hline
			Name & Value & Component  \\
			\hline
			Wavelength range $\Delta \lambda$ & 447 - 780 nm& -   \\
			Slit & 10 \textmugreek m & Thorlabs S10RD \\
			Collimation lens f\textsubscript{3} & 25 mm & Thorlabs AC127-025-A-ML  \\
		 	Grating & 300 lines / mm, 500 nm blaze & Thorlabs GR25-0305  \\
			Off-axis parabolic mirror f\textsubscript{4} & 101.6 mm & Thorlabs MPD149-G01  \\
		 	\gls{CMOS} camera & 2048 x 2048 px / 11.8 x 11.8 mm & Basler acA2040-90um-NIR  \\
			\hline
		\end{tabular}
	\end{center}
\end{table}
A detailed calculation including ray-tracing and optimization of optical components for the designed imaging spectrometer as well as of the calibration methods can be found in appendix \ref{Profilo:APNDX:Sec:imaging_spectrometer}.
\subsection{Height standard evaluation}
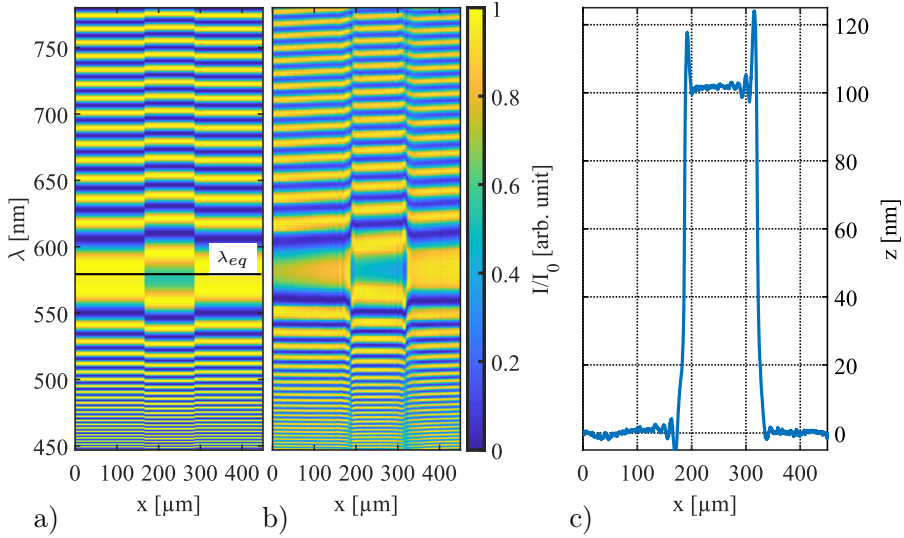
\begin{figure}[h]
	\centering
	\begin{overpic}[scale=0.4, grid = false]{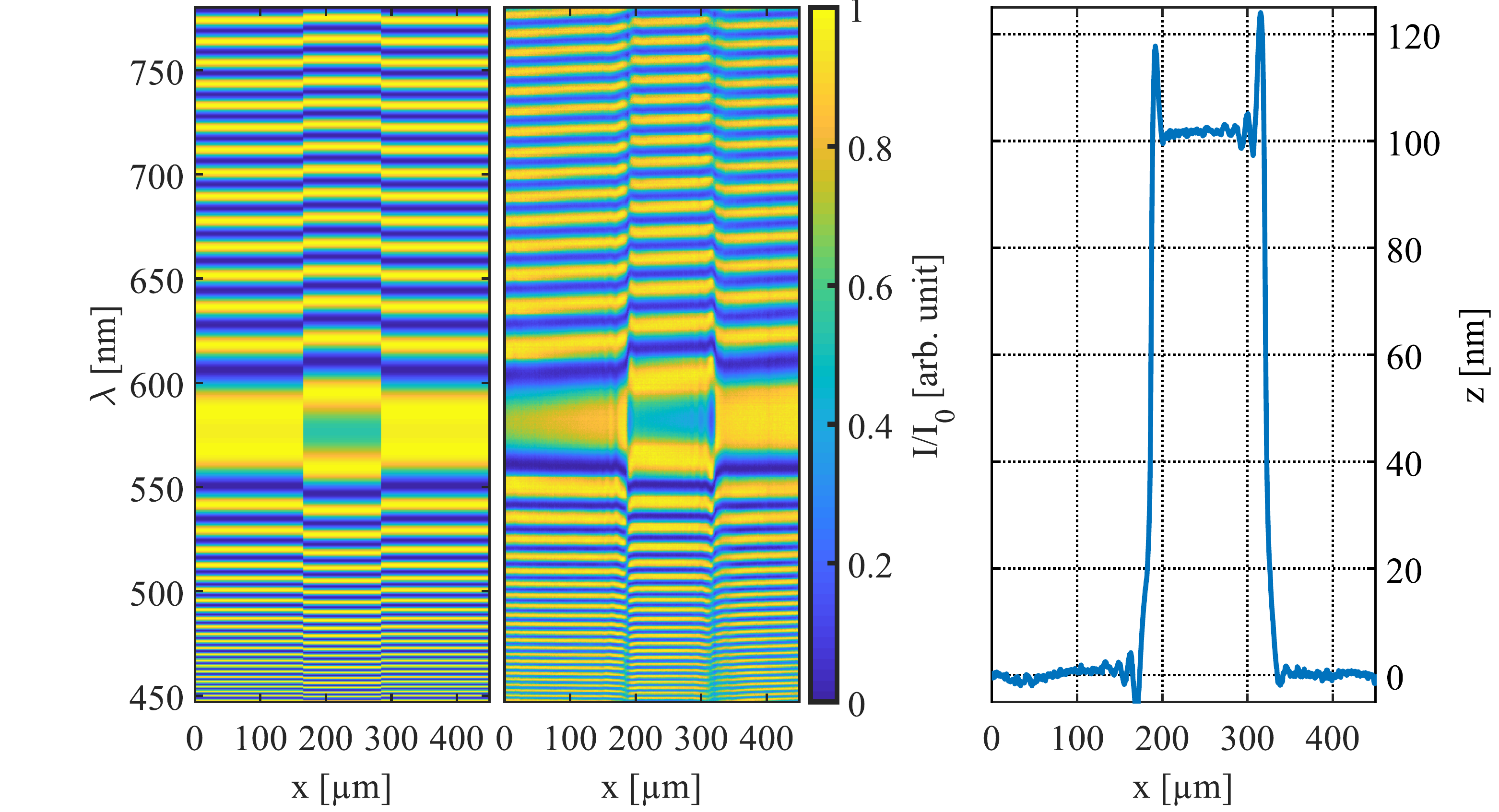}
		\put(10,0){\makebox(0,0){a)}}
		\put(34,0){\makebox(0,0){b)}}
		\put(66,0){\makebox(0,0){c)}}
		\put(27,27){\colorbox{white}{\textcolor{black}{$\scriptscriptstyle \lambda_{eq}$}}}
		\thicklines
		\put(13,25.75){\color{black}\line(1,0){19.5}}
	\end{overpic}
	\caption[Results of the measurement of a (100\,$\pm$\,7)\,nm nominal height \gls{Si}-standard]{Results of the measurement of a (100\,$\pm$\,7)\,nm nominal height \gls{Si}-standard with a) simulated spectral interference signal over a spectral range of 333\,nm and lateral dimension of 450\,\textmugreek m with the equalization wavelength $\lambda_{eq}$ marked and b) corresponding measured spectral interferences data with visible intensity modulations due to diffraction (marked with red ellipses) and c) calculated mean height profile from the raw data with diffraction-induced bat-wing effects at the sharp edges (marked with red ellipses) having a measured height of (101.8\,$\pm$\,0.1)\,nm.}\label{Profilo:Pic:result_2d_standard_nm}
\end{figure}

\begin{figure}[h]
\centering
		\begin{tabular}{c}
			\begin{overpic}[scale=.36]{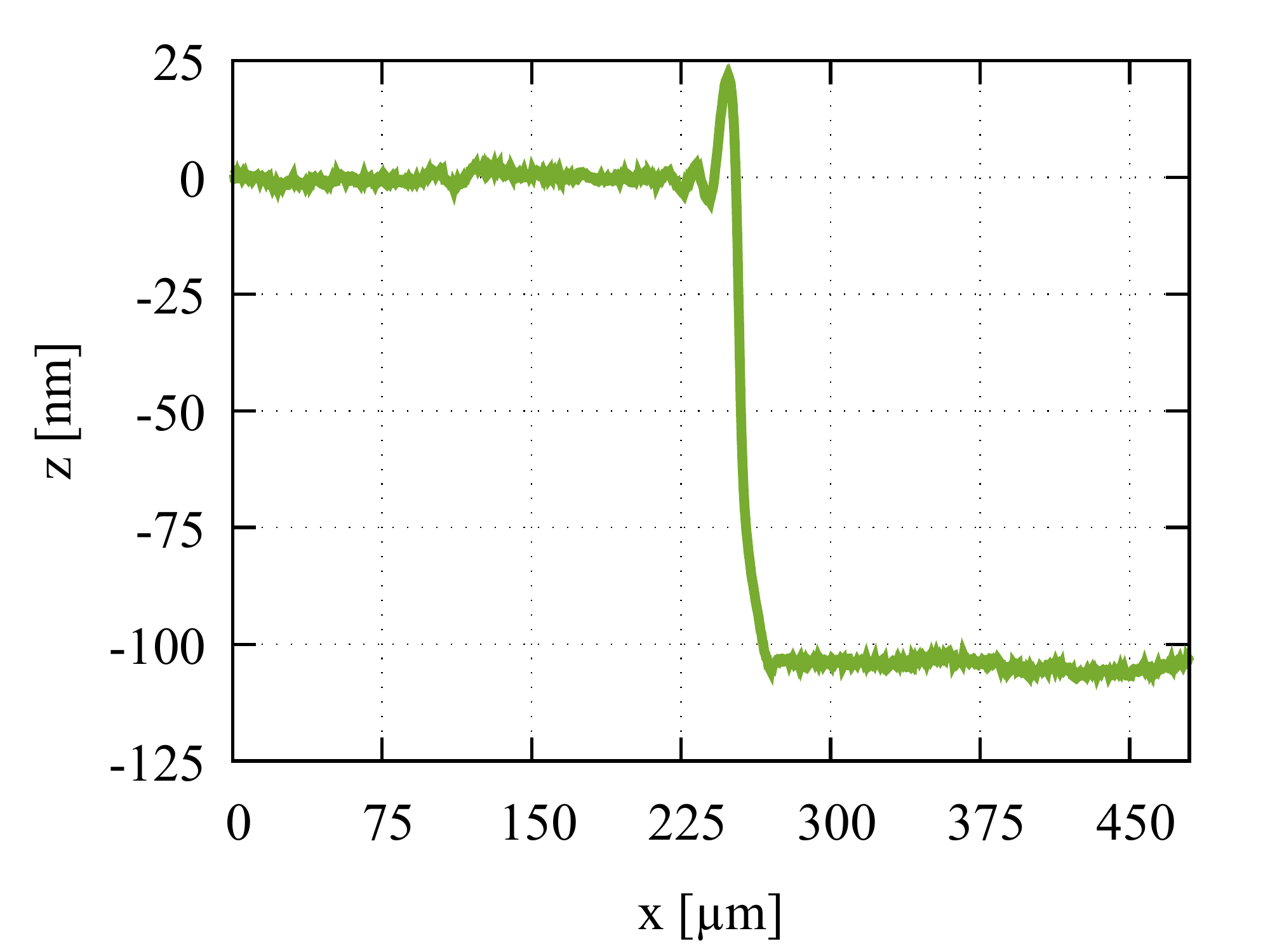}
				\put(1,1){\makebox(0,0){a)}}
			\end{overpic}
			\begin{overpic}[scale=.36]{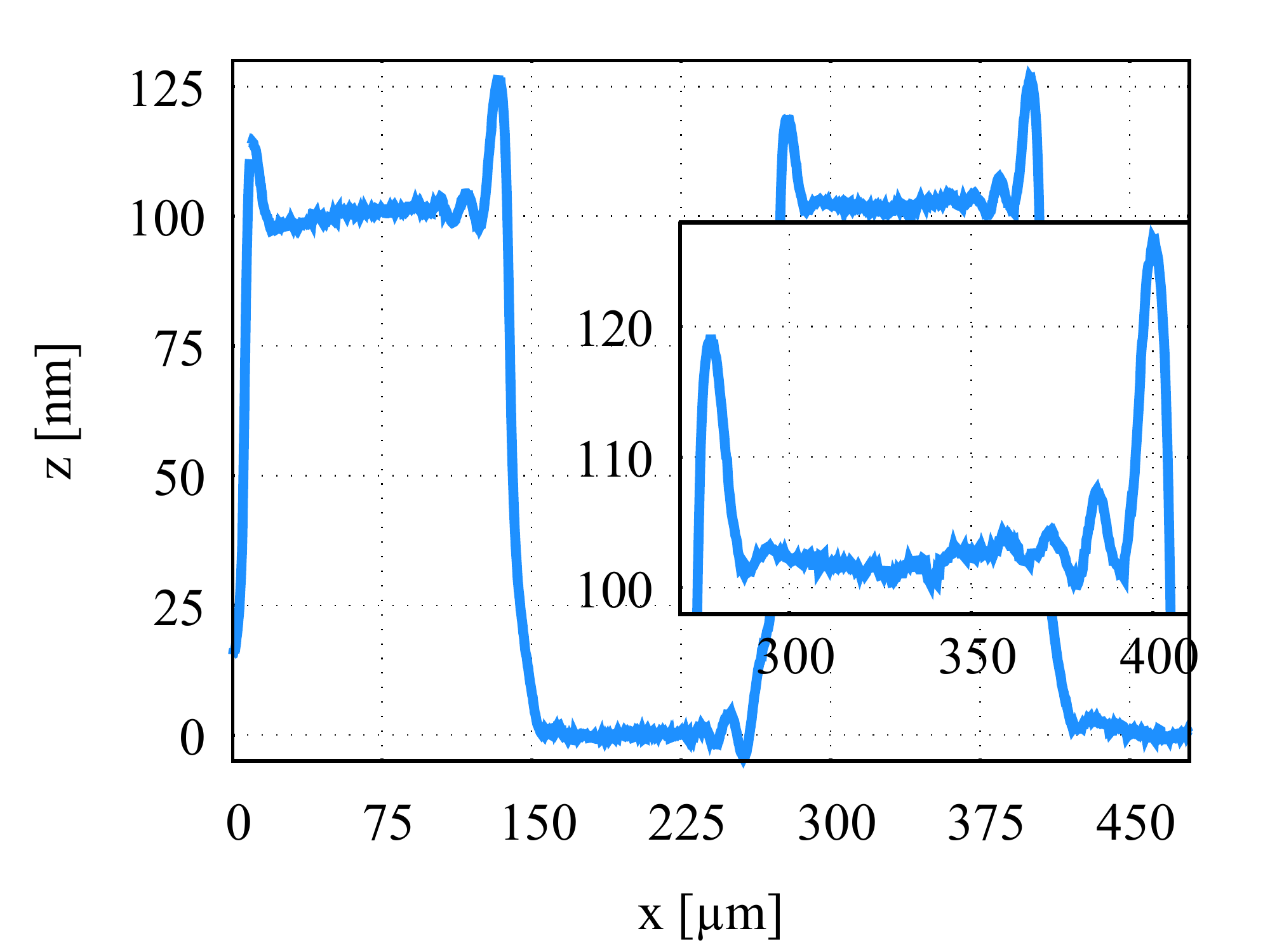}
				\put(1,1){\makebox(0,0){b)}}
			\end{overpic}
		\end{tabular}
	\caption[Plot of measured structures of the Si-based height standard showing diffraction effects]{Plot of measured structures of the Si-based height standard showing diffraction effects with a) single edge having a mean height of (104.35\,$\pm$\,0.11)\,nm and b) series of steps with a mean height of (99.88\,$\pm$\,0.11)\,nm and diffraction effects of up to 50\,\textmugreek m lateral size from each edge which influence the step width significantly, see inset.}\label{Profilo:Pic:result_2d_standard_nm_edge_multi}
\end{figure}

Additionally, experiments have been conducted to evaluate the accuracy of the developed system for the determination of small height steps. For this purpose, a \gls{Si}-based step standard (VS 0.10, Simetrics GmbH, Germany) was examined, Fig.\,\ref{Profilo:Pic:result_2d_standard_nm}.
The results of this examination revealed a good ability to resolve nm-sized height steps with a measured height of (101.8\,$\pm$\,0.1)\,nm which is in good agreement to the nominal value of (100\,$\pm$\,7)\,nm quoted by the manufacturer. The corresponding \gls{RMS} error with regard to the nominal value was 1.1\,nm. The roughness, \glssymbol{RoughnessRa}, of the \gls{Si} surface could be measured with 0.8 nm which scales with a factor of about 8.7 to the roughness measure \glssymbol{RoughnessRt}\,=\,7.0\,nm, \cite{conversion_roughness}. This is within the range of 6\,-\,10\,nm quoted by the manufacturer, \cite{SimetricsVSdatasheet}. The recorded and measured profiles show bat-wing effects at the sharp edges, \cite{Xie_PE16}. The measurement error increases in the regions of these effects due to diffraction and deflections. It was visible that deviations of up to 20\,nm occur, marked with red ellipses in Fig.\,\ref{Profilo:Pic:result_2d_standard_nm}\,b) and c). These deviations were attributed to diffraction effects visible as additional intensity modulation in the spectral interference raw data, Fig. \ref{Profilo:Pic:result_2d_standard_nm}\,b). For calibration purposes, the oscillations of the diffraction can be modeled as Fourier filtering by the aperture of the capturing optical system, \cite{Xie_PE16}. In relation to the simulated raw data, Fig. \ref{Profilo:Pic:result_2d_standard_nm}\,a), it was visible that not only diffraction occurs, but other distortions as well. In case of a flat, properly aligned sample, the spatial distribution of the maxima and minima is parallel to the x-axis of the plot, see Fig. \ref{Profilo:Pic:result_2d_standard_nm}\,a). It can be seen in the actual measured data, that this was not the case, Fig.\,\ref{Profilo:Pic:result_2d_standard_nm}\,b). This was the result of a slight tilt of the sample (about 0.11\,nm/\textmugreek m) in relation to the sample arm. It was corrected during post-processing of the final measured profiles assuming a linear tilt.\\
On the same standard, a single edge as well as a series of steps having a pitch of 250\,\textmugreek m were studied, Fig.\,\ref{Profilo:Pic:result_2d_standard_nm_edge_multi}.
For both sample positions heights of (104.35\,$\pm$\,0.11)\,nm and (99.88\,$\pm$\,0.11)\,nm were measured respectively. It is clearly visible that both of these features show the same diffraction effect, which confirms that it is independent of the sample position but dependent on the feature size and slope. Fig. \ref{Profilo:Pic:result_2d_standard_nm_edge_multi}\,a) leads to the note, that the effect has a length of influence of about $l_i$\,=\,50\,\textmugreek m into the profile. When measuring structures with lateral feature sizes smaller than $2 \cdot l_i$ information of these structures can be obscured. It can be seen in the measured profiles of multiple successive steps with a width of only 125\,\textmugreek m that an evaluation of e.g. roughness is influenced by this effect, see inset Fig. \ref{Profilo:Pic:result_2d_standard_nm_edge_multi}\,b).


\subsection{Repeatability and resolution characterization}\label{Profilo:SubSec_repeatability_resolutiuon}
\begin{figure}[h]
	\centering
	\begin{tabular}{c}
		\begin{overpic}[scale=0.78, grid = false]{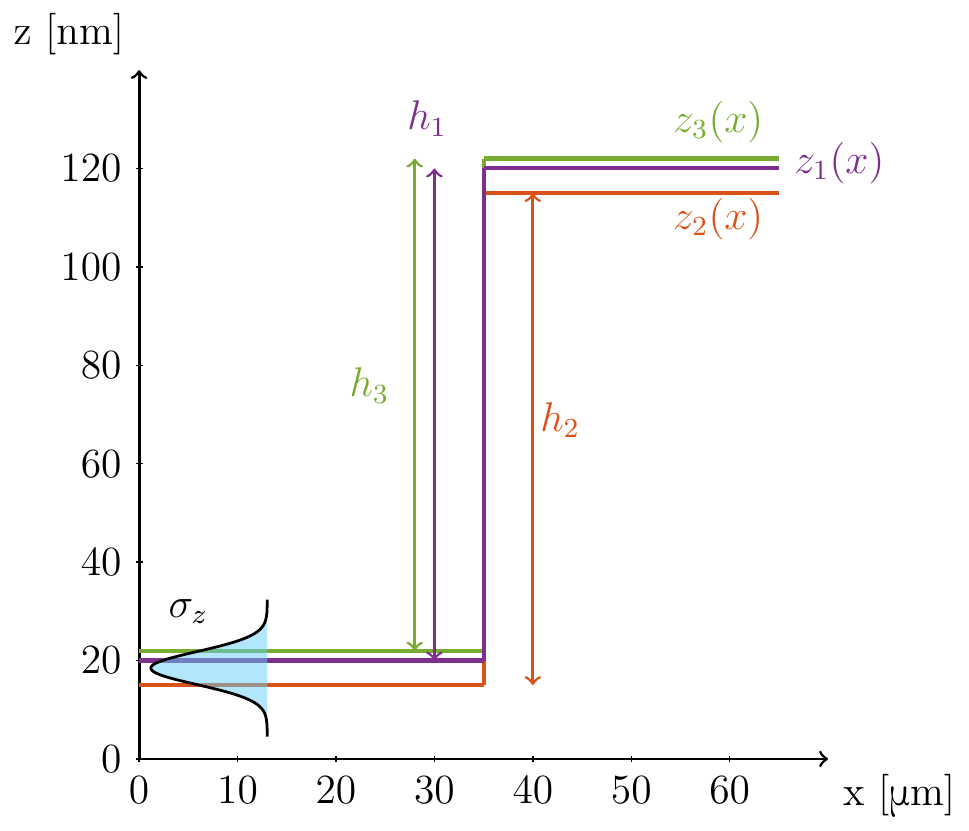}
			\put(1,1){\makebox(0,0){a)}}
		\end{overpic}
		
		\begin{overpic}[scale=0.32, grid = false]{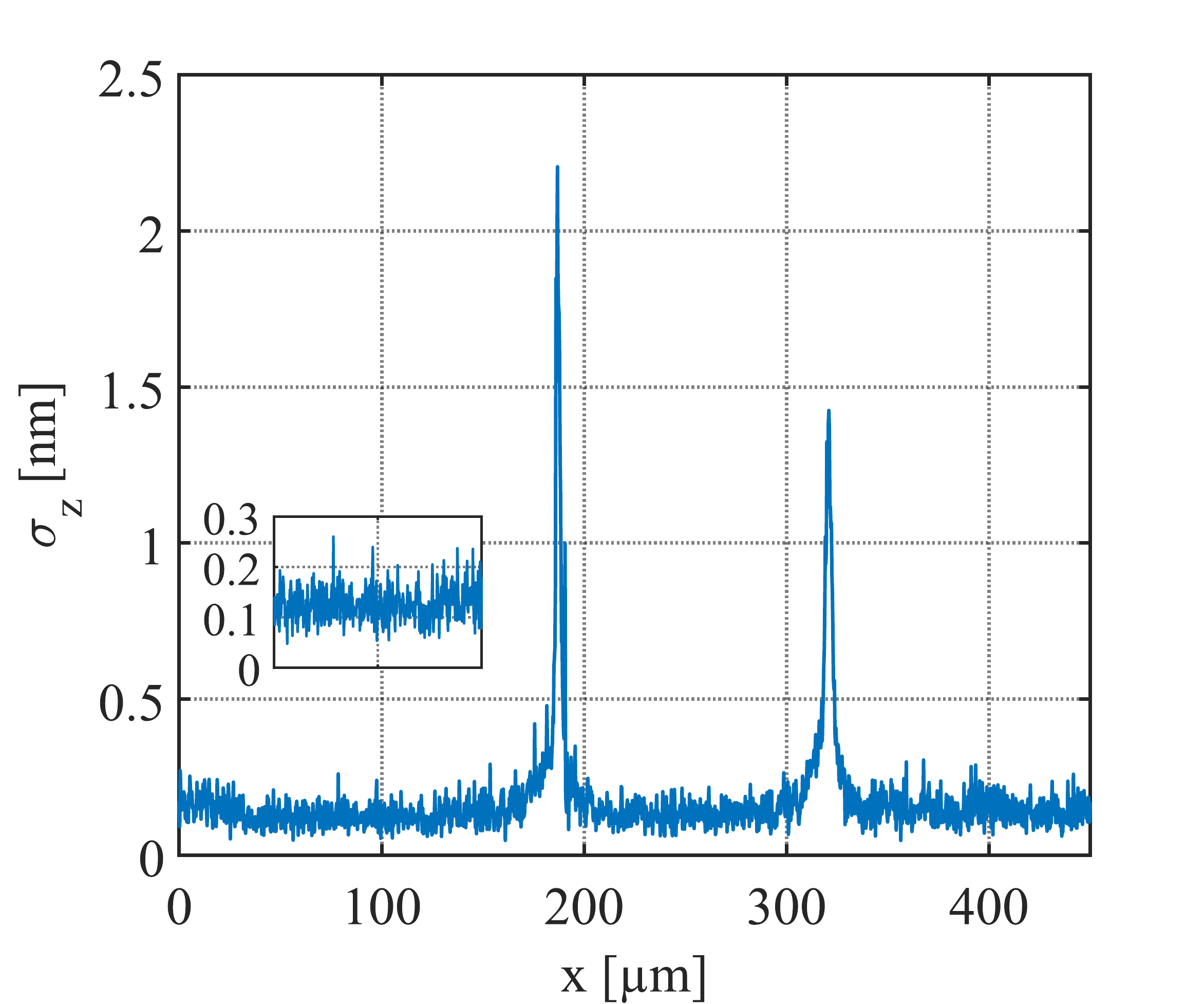}
			\put(1,1){\makebox(0,0){b)}}
			\put(40,18.0){\color{red}\vector(0,1){10.}}
			\put(22.8,18.0){\color{red}\vector(0,1){10.}}
		\end{overpic}
	\end{tabular}
	\caption[Depiction of repeatability and resolution measures for the two-dimensional DE-LCI setup]{a) Depiction of the calculation of repeatability as the standard deviation $\sigma_z(x)$ of multiple profiles $z_i(x)$ according to Eq. (\ref{EQ:repeatability}) as well as the resolution $\Delta z_{min}$ as the standard deviation of the feature height $h_i$ according to Eq. (\ref{EQ:resolution}) and b) plot of the spatially resolved repeatability of the nm-sized height standard of Fig. \ref{Profilo:Pic:result_2d_standard_nm} c) where the impact of diffraction is visible at $x =$\,150\,-\,210\,\textmugreek m and $x =$ 300\,-\,350\,\textmugreek m with an inset to visualize the magnitude of $\sigma_z(x)$ between $x =$ 50\,-\,150\,\textmugreek m.}\label{result_std_si}
\end{figure}

\begin{figure}[h]
	\centering
	\begin{tabular}{c}
		\begin{overpic}[scale=0.75, grid = false]{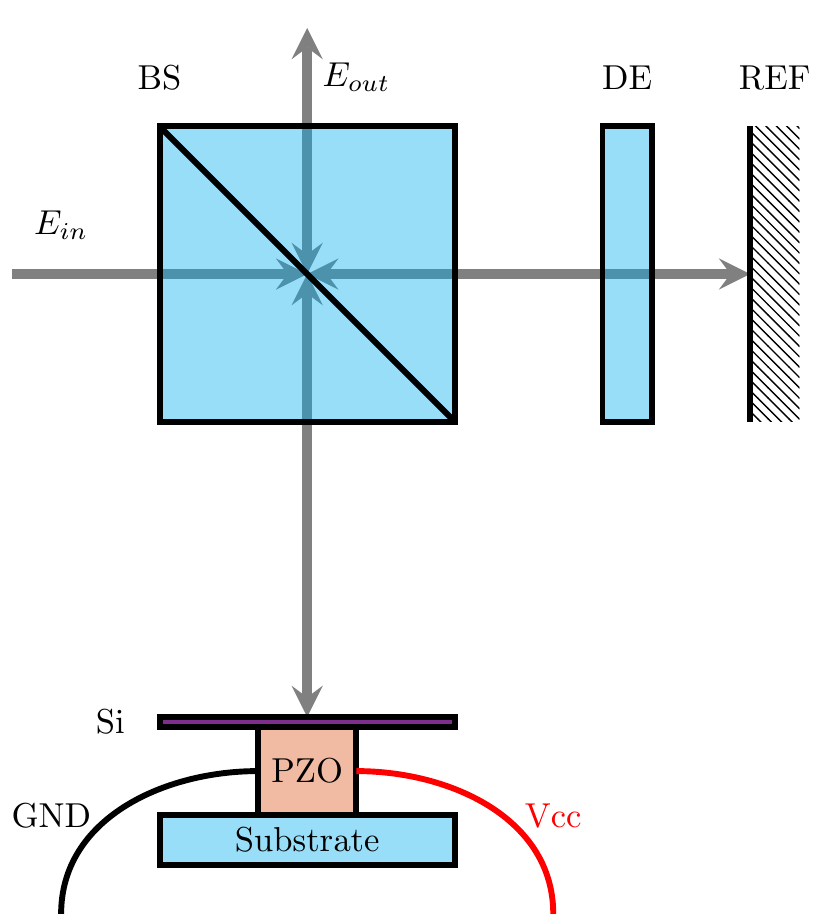}
			\put(1,1){\makebox(0,0){a)}}
		\end{overpic}
		
		\begin{overpic}[scale=0.35, grid = false]{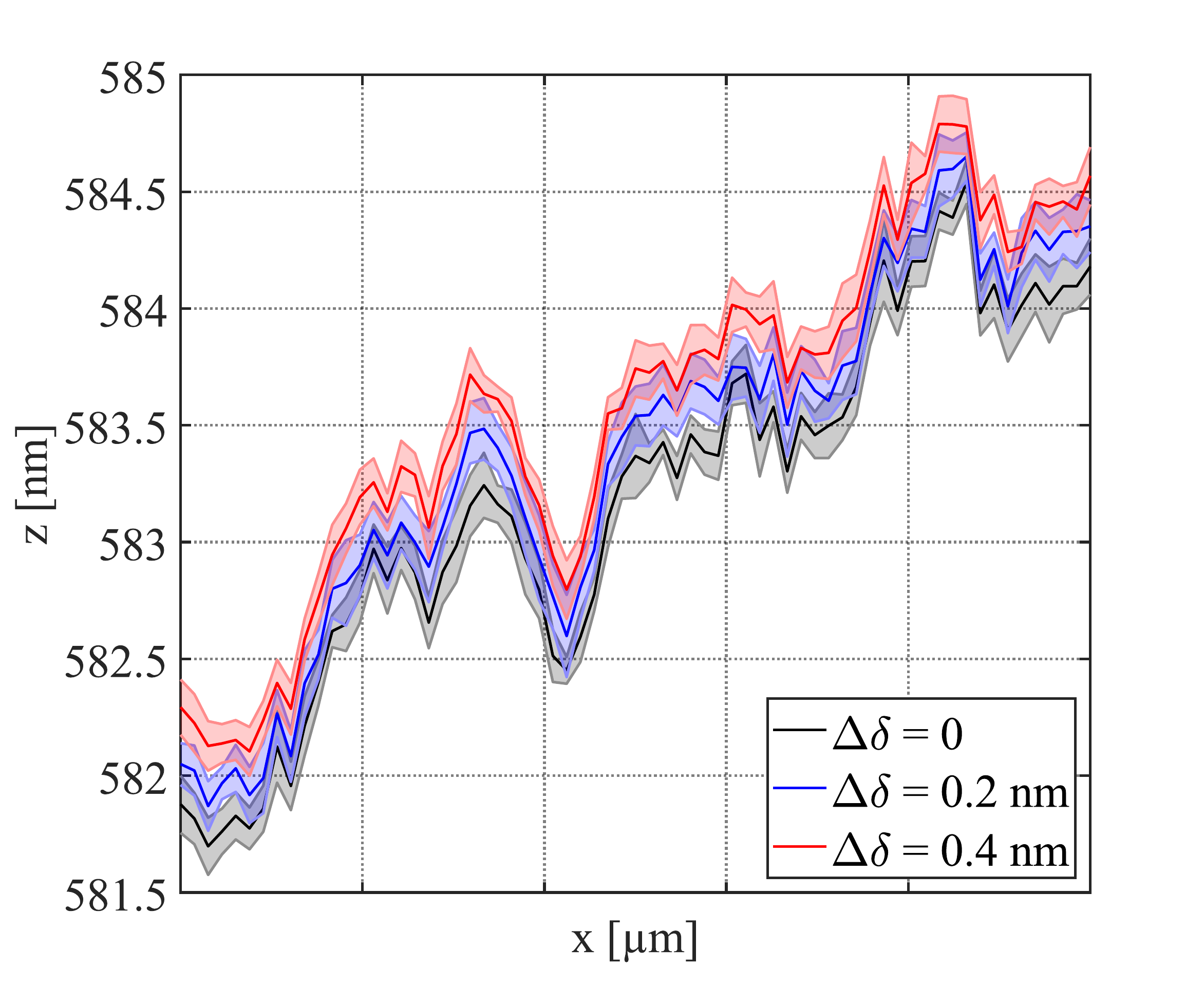}
			\put(1,1){\makebox(0,0){b)}}
		\end{overpic}
	\end{tabular}
	\caption[Setup and result for the repeatability determination using a piezo stack]{a) Depiction of the modified setup to characterize the repeatability where the incoming electric field $E_{in}$ is split by a BS - cube beamsplitter, while 50\,\% of the light passes the DE - dispersive element (N-BK7, $t_{DE}$\,=\,2\,mm) and gets reflected on the REF - reference mirror. In the second arm, light gets reflected of the designed sample which was a piece of a Si - silicon wafer attached to a PZO - piezo stack ($\Delta$\,=\,1.8\,\textmugreek m per 100\,V, PA3JEW, Thorlabs Inc., USA) which was mounted on a bulk glass substrate. The voltage of the PZO was remotely controlled in order to change the path length difference in steps of $\Delta \delta$\,=\,0.2\,nm where b) plot of the recorded averaged surface profiles along the $x$-dimension and the corresponding standard deviation shaded around the slopes for three different positions of the piezo stack.}\label{Profilo:Pic:setup_result_piezo_translation}
\end{figure}

\begin{figure}[h]
	\begin{center}
		\begin{tabular}{c}
			\begin{overpic}[scale=.8, grid = false]{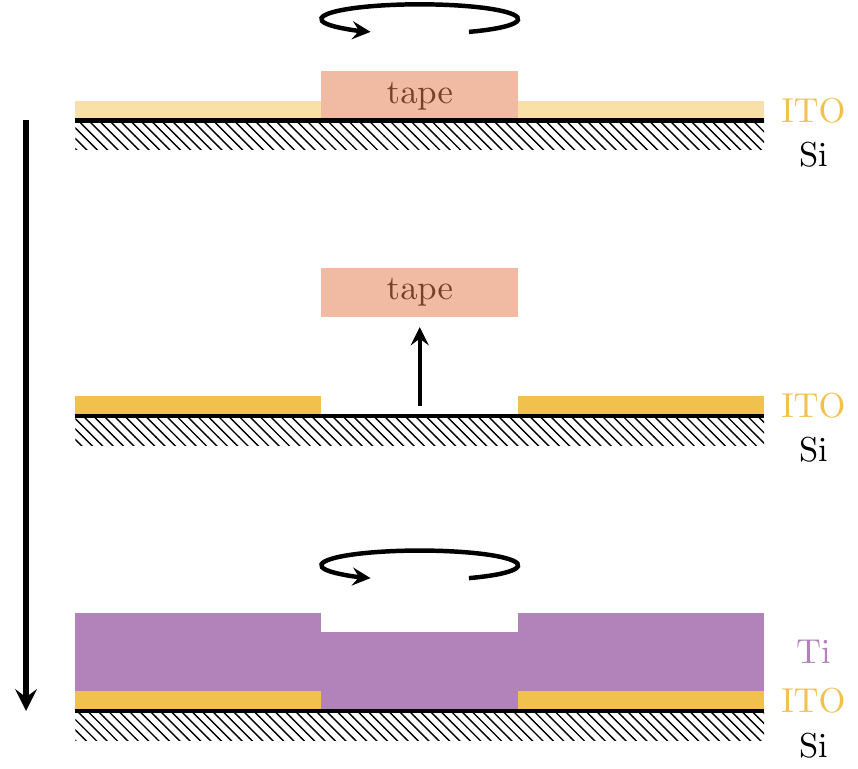}
				\put(1,1){\makebox(0,0){a)}}
				\put(15,22.){\makebox(0,0){(III)}}
				\put(15,47){\makebox(0,0){(II)}}
				\put(15,82){\makebox(0,0){(I)}}
			\end{overpic}
			\begin{overpic}[scale=.32]{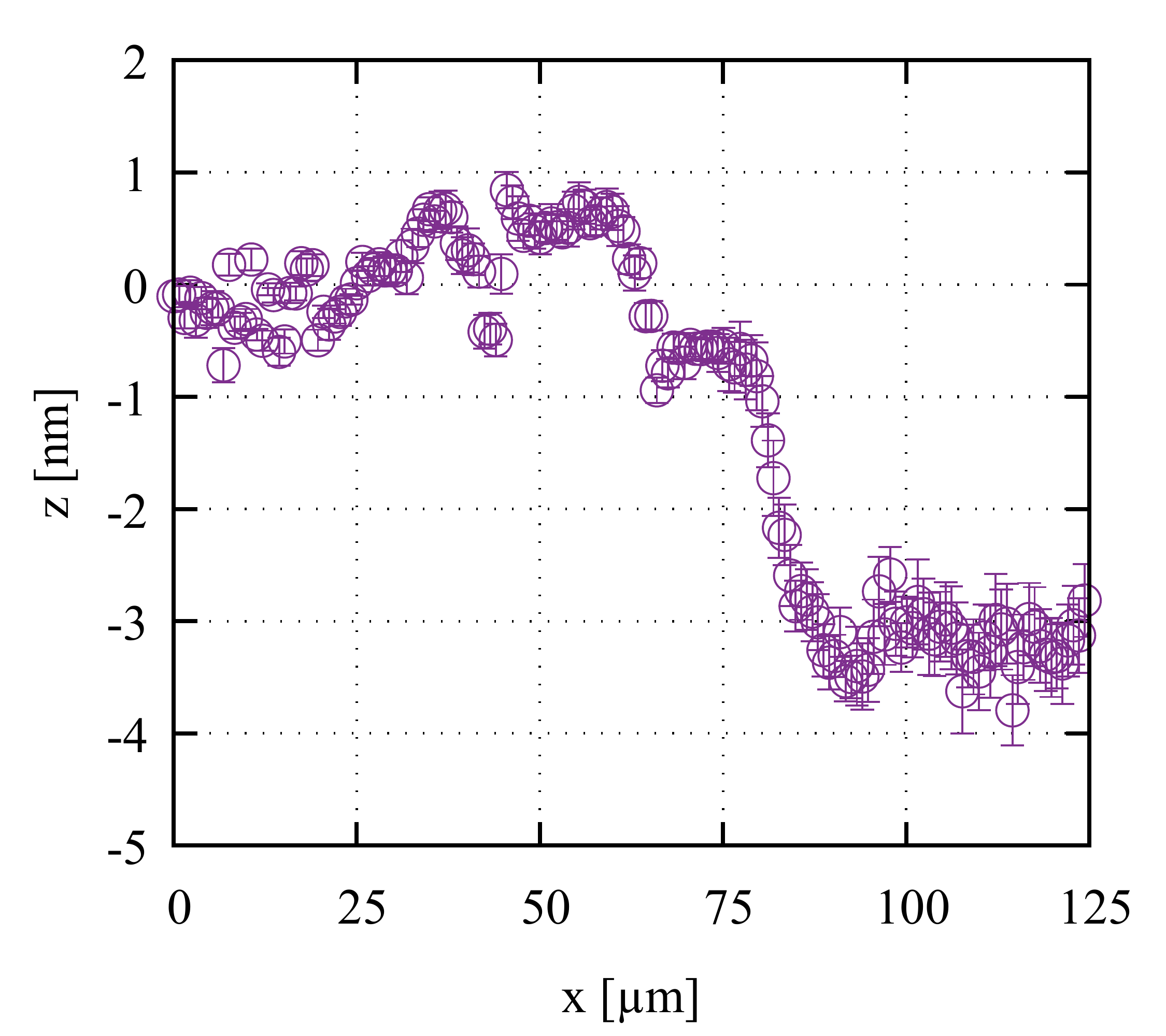}
				\put(1,1){\makebox(0,0){b)}}
			\end{overpic}
		\end{tabular}
	\end{center}
	\caption[Sample processing and result for the resolution characterization using a 3\,nm step]{a) Depiction of the processing chain to fabricate a sample with a step of 3\,nm nominal height where (I) polyimide tape is placed as mask on a \gls{Si}-substrate and a layer of \gls{ITO} is sputtered on the substrate having the desired thickness. (II) The removal of the tape finishes the step formation. (III) In order to generate a sample with a uniformly reflecting surface, a layer of 40\,nm \gls{Ti} is sputtered onto the sample which maintains the step. b) Plot of the averaged height profile of the step with $h_{stp}$\,=\,3.45\,$\pm$\,0.19\,nm using the \gls{DE-LCI} setup.}\label{Profilo:Pic:result_3nm_step}
\end{figure}

The error of the system can be analyzed by utilizing two measures where one is the repeatability, defined by the standard deviation $\sigma_z(x)$ of multiple profiles $z_i(x)$ gathered in a short time frame. The second measure is the resolution, calculated as the standard deviation $\Delta z_{min}$ of a feature such as the height \glssymbol{FeatureHeight}, Fig. \ref{result_std_si}. a).
In order to analyze the repeatability, the structure presented in Fig. \ref{Profilo:Pic:result_2d_standard_nm} c) was measured \glssymbol{NumSurfProfiles}\,=\,10 times in a row without any other delay than the acquisition and data transfer time. The analysis of the standard deviation of the profiles with respect to their mean, \glssymbol{MeanSurfProfiles}, allows one to conclude on the repeatability, 
\begin{equation}\label{EQ:repeatability}
\sigma_z(x) = \sqrt{\frac{1}{N-1} \sum_{i=1}^{N}\left( z_i(x) - \overline{z(x)} \right)^2 }.
\end{equation}
It can be noticed by analyzing the standard deviation in relation to the lateral dimension \glssymbol{Repeatability} that the error significantly increased due to the sharp edges and the diffraction effects, Fig.\,\ref{result_std_si}\,b). In order to estimate the repeatability of the setup in this configuration, the standard deviation was evaluated without the data points that were affected by defraction ($x =$\,150\,-\,210\,\textmugreek m and $x =$ 300\,-\,350\,\textmugreek m). A mean value of \glssymbol{MeanRepeatability}\,=\,0.13\,nm was calculated. The value of $\overline{\sigma_z}$\,=\,0.13\,nm, measured on a low scattering nm-height standard, is expected to be the lower limit of the setup, as sections of the sample with disturbing influences were excluded from the calculation. In order to characterize the repeatability of the setup further, an experiment was designed where the path length difference was altered with a translatable sample, Fig. \ref{Profilo:Pic:setup_result_piezo_translation} a).
The designed sample was a flat piece of a \gls{Si}-wafer (15\,x\,15 mm\textsuperscript{2}) diced and mounted on a piezo stack ($\Delta$\,=\,1.8\,\textmugreek m per 100\,V, PA3JEW, Thorlabs Inc., USA) which was attached to a bulk glass substrate. During the experiment, the voltage of the piezo was discretely controlled (power supply QL355TP, Aim and Thurlby Thandar Instruments, United Kingdom) and monitored (digital multimeter DM3068, RIGOL Technology Co., Ltd, China) to adjust the path length difference in defined steps of $\Delta \delta$\,=\,0.2\,nm. For every adjusted step, N\,=\,10 consecutive measurements were taken, Fig. \ref{Profilo:Pic:setup_result_piezo_translation} b). The data for the three pictured positions demonstrates that the setup is capable of resolving steps of 0.2\,nm as the standard deviations hardly overlap. The mean standard deviation of every position was calculated as $\overline{\sigma(\Delta \delta=0)}$\,=\,0.11\,nm, $\overline{\sigma(\Delta \delta=0.2)}$\,=\,0.13\,nm, $\overline{\sigma(\Delta \delta=0.4)}$\,=\,0.10\,nm. These values correspond well with the measurements on the \gls{Si} height standard shown previously and demonstrate the limit of the current setup with regard to stability.\\
The gathered data was further analyzed using a two-sample Student's \textit{t}-test in order to statistically evaluate if the measured averaged slopes are significantly different from each other, \cite{SachsStatistik}. A mean value of $\hat{t}_{12}$\,=\,3.282 and $\hat{t}_{23}$\,=\,3.619 was calculated between the profiles with $\Delta \delta$\,=\,0 / $\Delta \delta$\,=\,0.2\,nm and $\Delta \delta$\,=\,0.2\,nm / $\Delta \delta$\,=\,0.4\,nm respectively. In contrast to the value of $t$\,=\,3.250 (with $n$\,=\,9 at 99\,\% probability) it was found that $\hat{t} > t$ which rejects the null hypothesis. Consequently, the averaged profiles are significantly different in this experiment. Furthermore, this test can be used to estimate the minimal repeatability for the case where \glssymbol{TtestHat}\,=\,\glssymbol{Ttest}. It was found to be $\Delta \delta_{min}$\,=\,0.12\,nm for a 95\,\% probability and $\Delta \delta_{min}$\,=\,0.18\,nm for a 99\,\% probability.\\
The creation and measurement of samples which are close to the proposed resolution of the setup is complex. For the calibration of \gls{AFM} instruments, height step samples exist which have heights in the size of one atomic layer of silicon, \cite{Homma1990,Suzuki1996}. However, the steps on these samples usually have widths below 1\,\textmugreek m which are not resolvable with the presented approach. While the repeatability is a measure for temporal fluctuations that occur from one measurement to the other, the ability to resolve structures along the spatial domain (here denoted as the $x$-coordinate) is independent from these fluctuations. A measure for the resolution can be found in the standard deviation $\Delta z_{min}$ of feature sizes such as the height of structures $h_i$ relative to the mean height of multiple measured features, \glssymbol{MeanHeight}. It can be assumed that in-between short time frames of the acquisition time of single data sets the sample does not change,
\begin{equation}\label{EQ:resolution}
\Delta z_{min} = \sqrt{\frac{1}{N-1} \sum_{i=1}^{N}\left( h_i - \overline{h} \right)^2 }.
\end{equation}
In case of the nm-sized, \gls{Si}-height standard, the height was measured as the difference between the two base levels, $x_1 =$\,100\,-\,150\,\textmugreek m and $x_2 =$ 350\,-\,400\,\textmugreek m, and the top plateau of the step at $x_3 =$ 225\,-\,275\,\textmugreek m. The quadratic mean of $\Delta z_{min}$ for 20 measured heights, and therefore the resolution, was found to be 0.1\,nm. 
The standard deviation of the feature size represents a cumulative measure for the resolution which includes influences of the optical setup, the electronics, the calibration routines and the data processing alike. During the data analysis of the presented results it became obvious that the difference between the calculated minimal resolution of 0.088\,nm, see bottom plot of  Fig.\,\ref{Proflo:Pic:noise_simu_vs_SNR} a), and the minimal measured resolution of 0.1\,nm are partly due to data processing routines. As the recorded profiles usually were tilted by a minor degree, an appropriate tilt correction was performed based on the linear fit of every captured surface profile. Although the tilt correction was optimized, a minor influence on the standard deviation cannot be excluded.\\  
As a supplement, a step with a nominal height of \glssymbol{FeatureHeightNom}\,=\,3\,nm was created using standard semiconductor procedures, Fig.\,\ref{Profilo:Pic:result_3nm_step}\,a).
A silicon substrate was prepared with polyimide tape and sputtered with \gls{ITO} of about 3\,nm thickness. Afterwards, the tape was removed and a second sputtering step using titanium was performed. Due to the initial application of tape, the sample has a defined height difference which is maintained after subsequent generation of the titanium layer. This layer was applied in order to generate a uniformly, high reflecting surface and prevent any thin-film interferences which a sole layer of \gls{ITO} would have caused.\\ 
The created sample was analyzed using the setup similar to the measurements on the height standard where N\,=\,10 consecutive measurements were taken and averaged, Fig.\,\ref{Profilo:Pic:result_3nm_step}\,b). Using the same scheme as before, the height of the step was determined as \glssymbol{FeatureHeightStep}\,=\,3.45\,$\pm$\,0.19\,nm. The result fortifies the claim that the proposed \gls{DE-LCI} method is capable of measuring surface profiles with sub-nm resolution. In particular, the repeatability of $\overline{\sigma}$\,=\,0.12\,nm and the resolution of the height measurement $\Delta z_{min}$\,=\,0.19\,nm support this.\\
On the basis of the calculated resolution, the \gls{DR} of the setup was calculated with the measurement range of $\Delta z$\,=\,79.91 \textmugreek m and the resolution of $\Delta z_{min}$\,=\,0.1\,nm as DR\,=\,\num{7.99e5}. Compared to the latest findings of other areal profilometer approaches such as of Reichold et al., \cite{Reichold2019}, the achieved dynamic range is about 5.8 times higher.\\


\subsection{Edge effects}
\begin{figure}[h]
	\begin{center}
		\begin{tabular}{c}
			\begin{overpic}[scale=.3, grid = false]{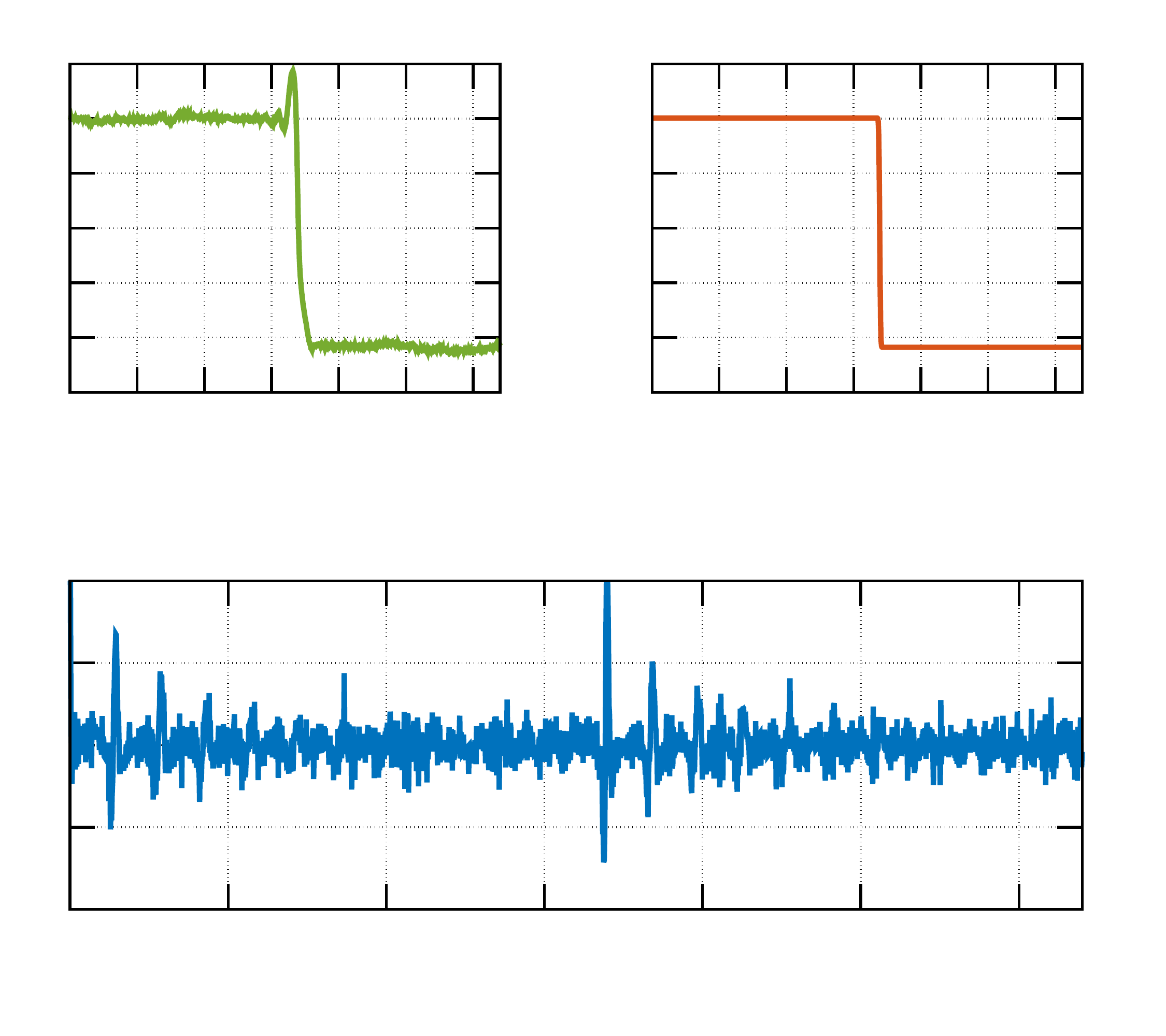}
				\put(1,1){\makebox(0,0){a)}}
				\put(85,35){\makebox(0,0){$H$}}
				\put(85,75){\makebox(0,0){$Z$}}
				\put(35,75){\makebox(0,0){$U$}}
			\end{overpic}
			\begin{overpic}[scale=.3]{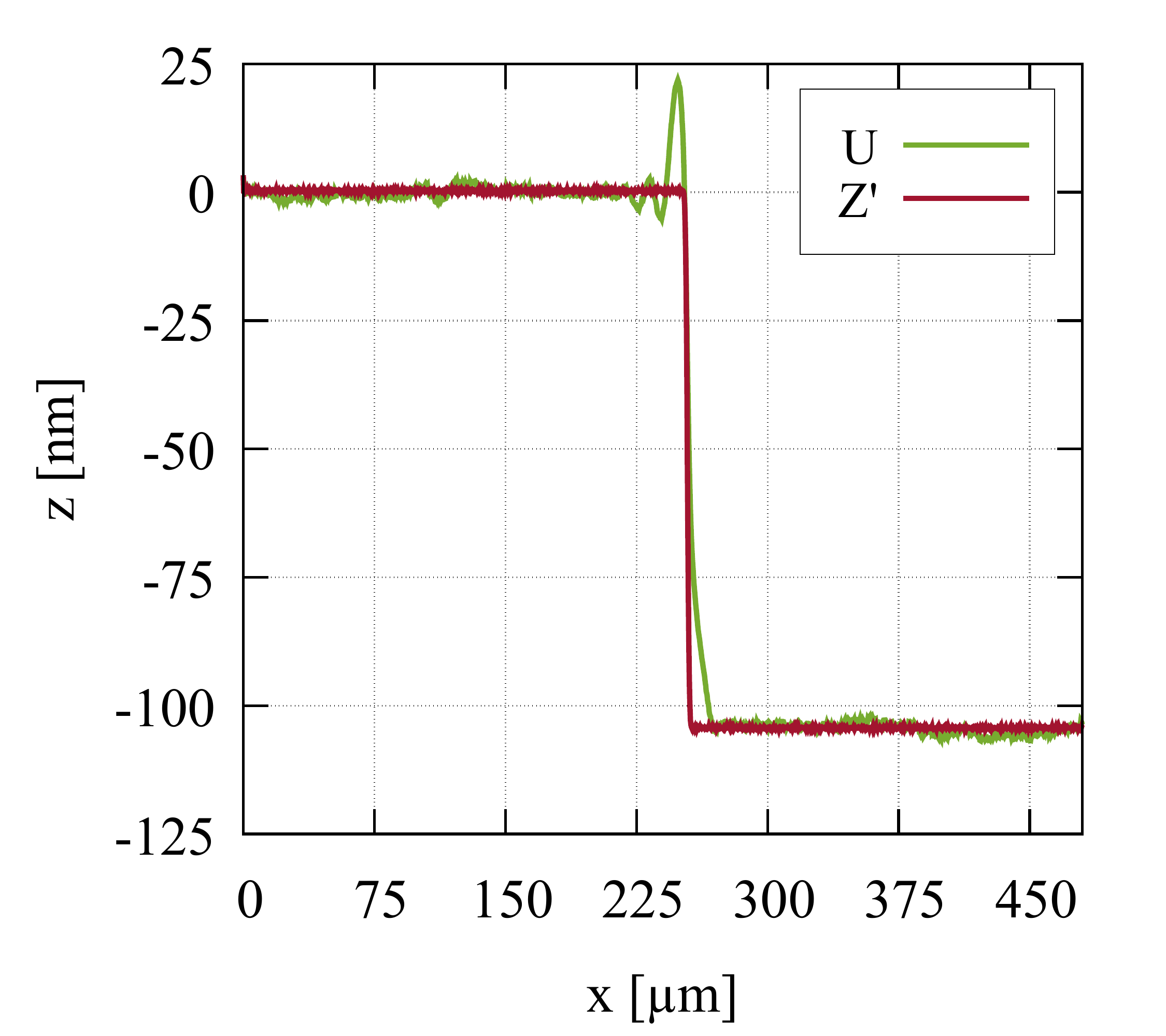}
			\put(1,1){\makebox(0,0){b)}}	
			\end{overpic}
		\end{tabular}
	\end{center}
	\caption[Principle description of the correction of edge effects on steep slopes using a \textsc{Wiener} filter]{Depiction of the \textsc{Wiener}-based deconvolution routine with a) components of the process where $U$ represents the measured signal (see Fig.\,\ref{Profilo:Pic:result_2d_standard_nm_edge_multi}\,a), $Z$ is the ideal profile and $H$ is the estimated impulse response function as well as b) comparison of the measured profile $U$ with the deconvoluted profile $Z^\prime$.}\label{Profilo:Pic:wiener_filter_batwing_approach}
\end{figure}
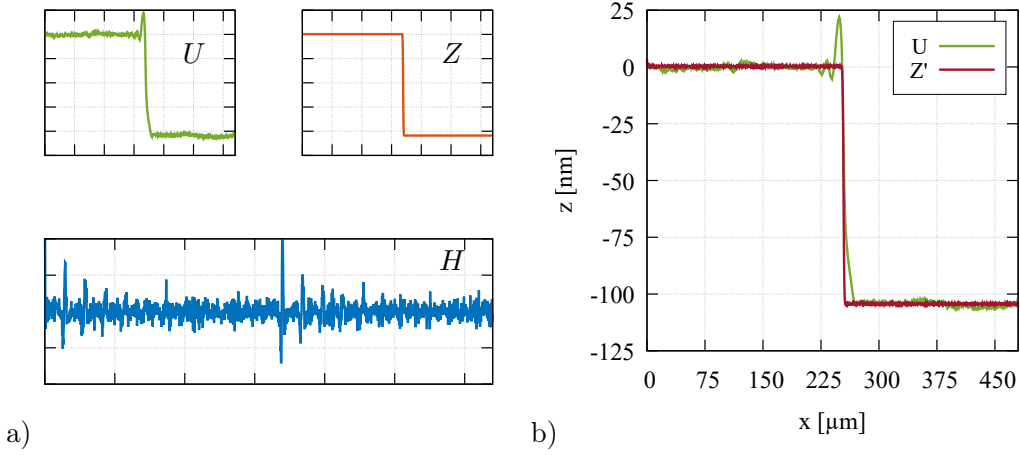

\begin{figure}[h]
\centering
		\begin{tabular}{c}
			\begin{overpic}[scale=.35]{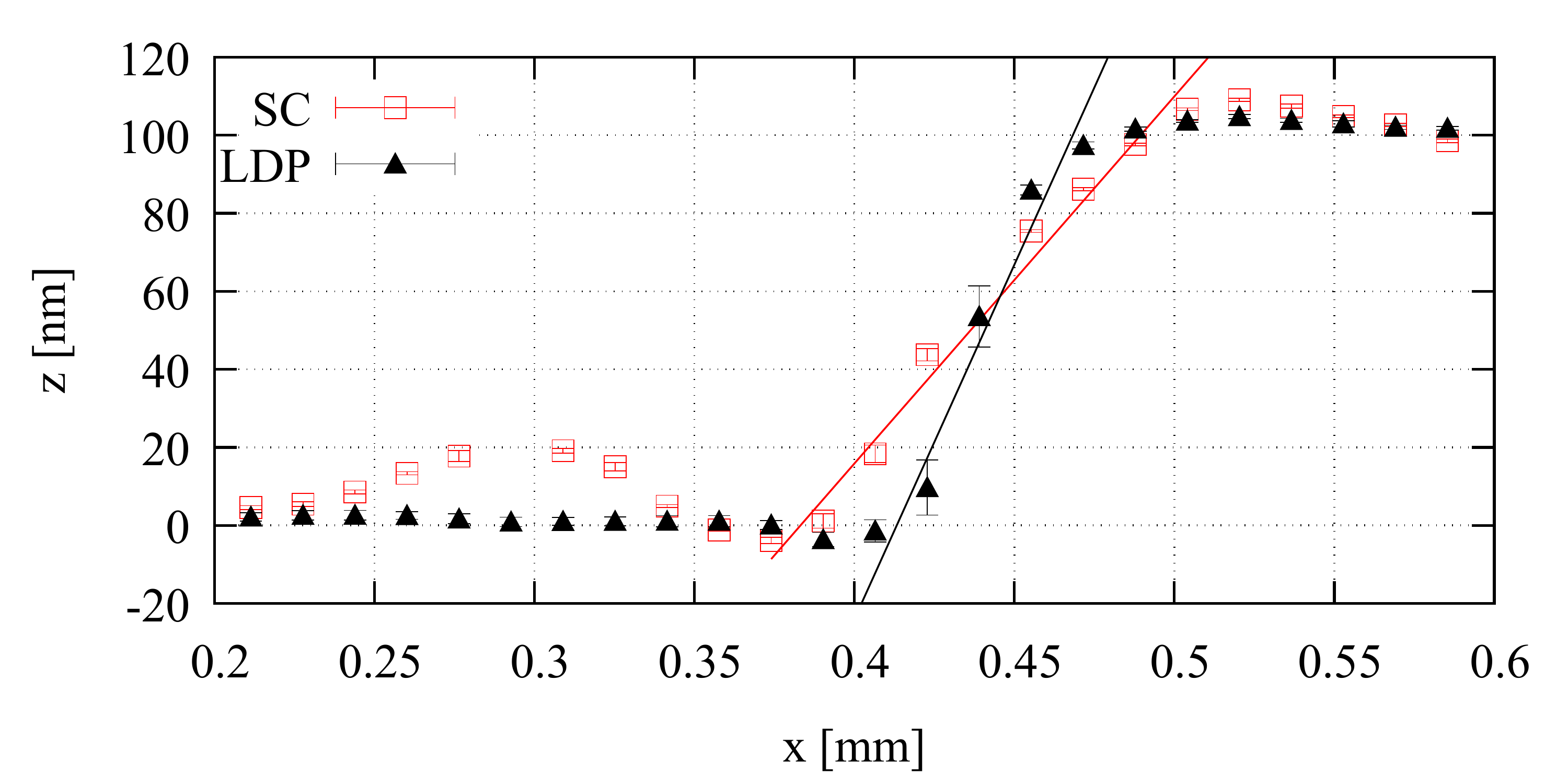}
			\end{overpic}
		\end{tabular}
	\caption[Result of the measured slopes using a SC versus a LDP light source]{Result of the measured slopes using a SC versus a LDP light source.}\label{Profilo:Pic:edge_comparison}
\end{figure}

The occurrence of edge effect on sharp edges is caused by different sources. While diffraction and scattering play an important role, also filtering effects of the aperture, shadowing from the sample's steep slopes as well as interferometric mixing of components of the different height levels due to the lateral resolution are relevant.
A rough estimation of the influence of the aperture revealed that its contribution is neglectable, such that it is assumed that diffraction due to the spatial coherence properties of the light source are dominant. As these influences are mixed with in an unknown ratio, no single model can be used to filter the signal appropriately. For this reason, a deconvolution according to the \textsc{Wiener} approach was found to be suitable, \cite{BookImageProc}, Fig.\,\ref{Profilo:Pic:wiener_filter_batwing_approach}\,a). 
This approach typically assumes that besides the a measured profile \glssymbol{WienerU}, a response function of the system \glssymbol{WienerH} is known. Based on these functions, a deconvolution for the ideal, filtered profile \glssymbol{WienerZ} can be performed. Using this notation, a measured profile is the convolution of the ideal profile with the response function\footnote{These functions are assumed to be handled in Fourier-space.} 
\begin{equation}
U = H \cdot Z.
\end{equation}
Following \textsc{Wieners} idea, a wild-card function \glssymbol{WienerG} acting on the measured signal is utilized to minimize the error between the ideal, but unknown profile $Z$ and the deconvoluted signal \glssymbol{WienerZPrime}
\begin{eqnarray}
Z^\prime = G \cdot U \label{WienerDeConv}\\
\min (Z^\prime - Z).
\end{eqnarray}
As the response function in the typical use case of the \textsc{Wiener} deconvolution is known, the wild-card function can be constructed of $H$
\begin{equation}\label{EQ:WienerG}
G = \frac{H^*}{|H|^2 + \left(\frac{1}{SNR}\right)}.
\end{equation}
In the case investigated here, a modification has to be performed as $H$ is unknown but $Z$ is known. Hence, in order to perform the deconvolution, a wild-card function \glssymbol{WienerQ} has to be introduced based on $Z$
\begin{equation}
Q = \frac{Z^*}{|Z|^2 + \left(\frac{1}{SNR}\right)}
\end{equation} 
which can be used to compute the response function
\begin{equation}
H = Q \cdot U.
\end{equation}
With the aid of this response function, finally the deconvolution of the original measured signal $U$ can be done using Eq. (\ref{EQ:WienerG}) and (\ref{WienerDeConv}) resulting in the filtered signal \glssymbol{WienerZPrime}. This computation was performed on the measured data of a single silicon edge, originally presented in Fig. \ref{Profilo:Pic:result_2d_standard_nm_edge_multi}\,a), Fig.\,\ref{Profilo:Pic:wiener_filter_batwing_approach}\,b). It is obvious that the edge effects can be filtered well. The outlined procedure can be made part of a calibration routine when measuring similar structures repeatedly.\\
Additionally, the dependency of edge effects with regards to the used light source was evaluated. For this purpose, a \gls{SC} ($\Delta \lambda$\,=\,380\,-\,1100\,nm) and a \gls{LDP} ($\Delta \lambda$\,=\,200\,-\,1100\,nm) were used in comparison. Due to the limited coherence of the broadband light sources the interference contrast of the wavelength-integrated signal forms a considerable envelope function which can be detected by using a spectrometer. This data is the basis for the calculation of the coherence length. The coherence length $l_c$ can be determined exactly as the integral of the area under normalized degree of coherence. Usually, an approximation can be found by estimating the width of the interferogram at an intensity of \nicefrac{1}{e}, \cite{octBook}. The respective coherence lengths were determined as $l_c$\,=\,1.81\,\textmugreek m for the LDP light source and $l_c$\,=\,1.58\,\textmugreek m for the SC light source. Consequently, the same step standard of 100\,nm nominal height was measured with both light sources. While the measured height was comparable with both light sources, the behavior on the edge of the standard showed differences, Fig. \ref{Profilo:Pic:edge_comparison}.
The analysis reveals that using a LDP source, a much steeper slope can be measured. A linear approximation results in a slope of 0.94\,\nicefrac{\textmugreek m}{mm} for the SC source and a slope of 1.82\,\nicefrac{\textmugreek m}{mm} for the LDP source. Furthermore, it can be seen that the formation of edge effects is stronger in the measurements of the SC light source. As the coherence length is very similar for both light sources, a possible explanation is that the spatial coherence of SC light source is higher. Due to its operation principle, highly spatially coherent light of a microchip laser is used to generate a broadened spectrum . In contrast, the LDP light sources generates its broad spectrum from a random process both temporally and spatially.\\
Further knowledge of the precise spatial coherence properties of the used measurement light source as well as of possible sample geometries can help to develop appropriate filter models for the correction methods demonstrated above. The spatial coherence properties of the light source and resulting effects such as diffraction and scattering are the main limiting factors for measurements requiring high lateral resolutions.


\subsection{Roughness evaluation}\label{SubSec_roughness_eval}
\begin{figure}[h]
\centering		\begin{tabular}{c}
			\begin{overpic}[scale=.5]{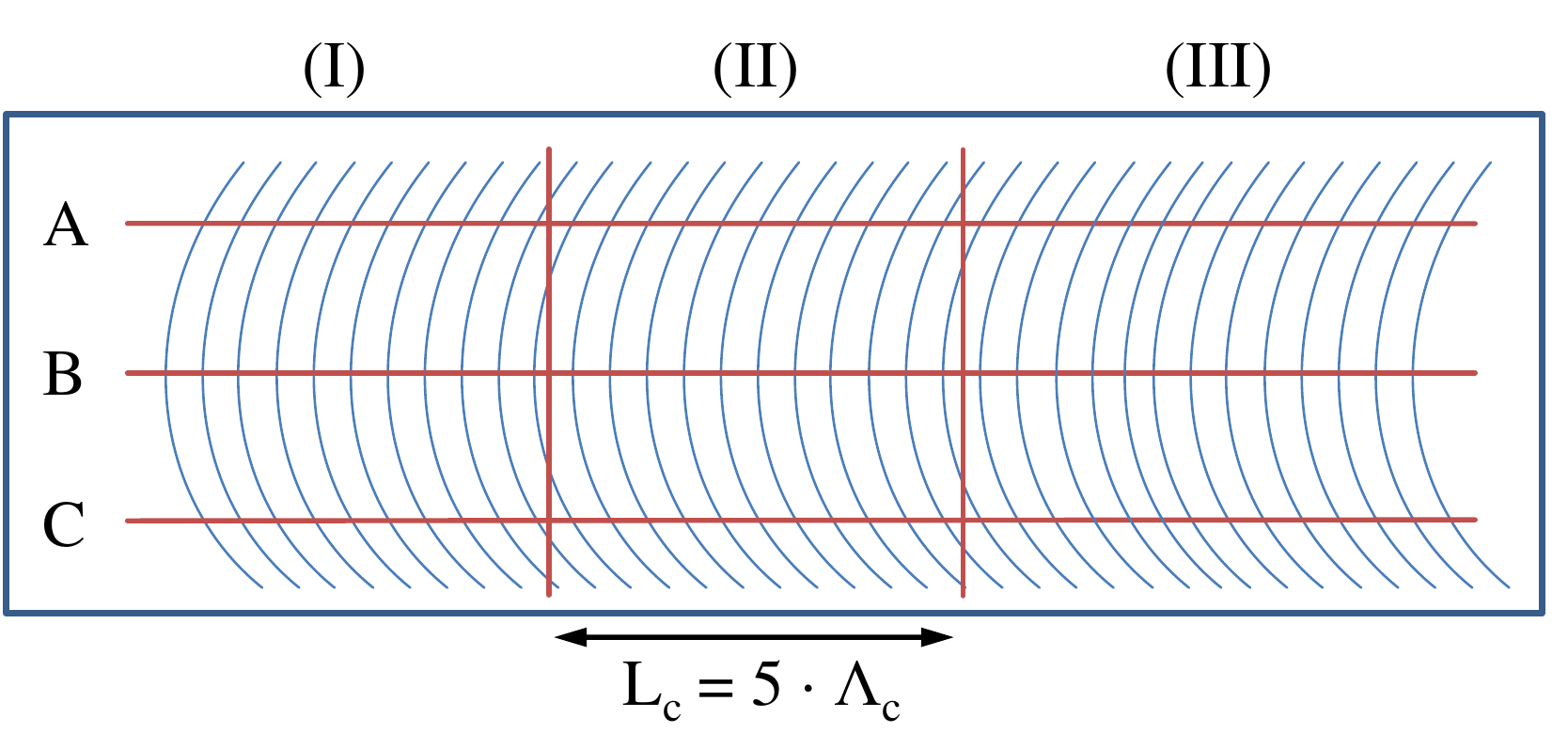}
			\end{overpic}
		\end{tabular}
	\caption[Schema of the measured \gls{PTB}-traceable surface roughness standard]{Schema of the measured \gls{PTB}-traceable surface roughness standard where measurements are taken in the slices A, B, and C with segments of (I), (II) and (III) all having the same evaluation length of $L_c$\,=\,5\,$\cdot \Lambda_c$.}\label{Profilo:Pic:Result_roughness_meas_scheme}
\end{figure}

\begin{figure}[h]
	\begin{center}
		\begin{tabular}{c}
			\begin{overpic}[scale=.5]{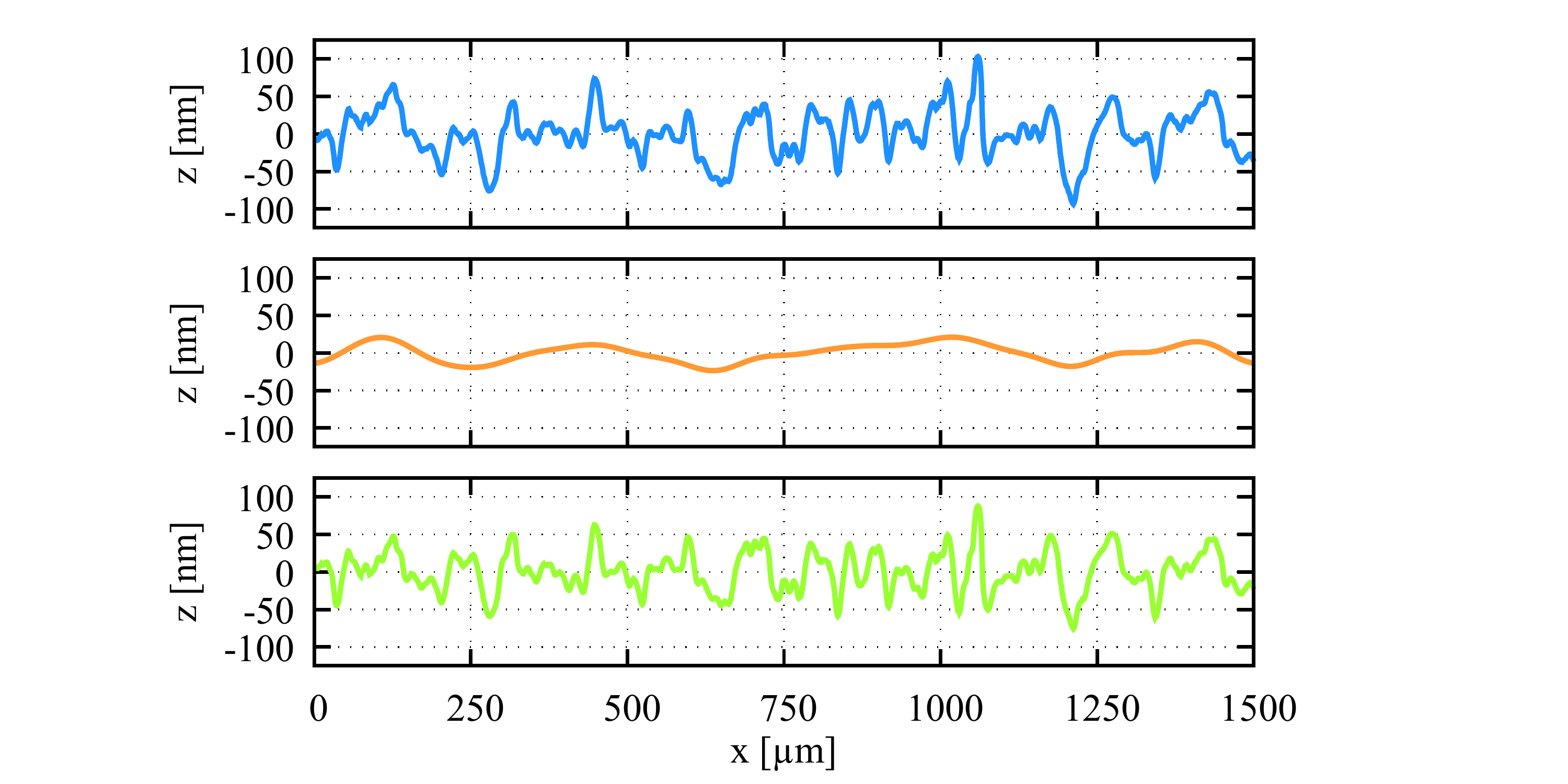}
				\put(1,42){\makebox(0,0){a)}}
				\put(1,28){\makebox(0,0){b)}}
				\put(1,14){\makebox(0,0){c)}}
			\end{overpic}
		\end{tabular}
	\end{center}
	\caption[Result of the measurements on a \gls{PTB}-traceable surface roughness standard]{Result of the measurements on a \gls{PTB}-traceable surface roughness standard with a) primary, form corrected profile, b) waviness profile (gaussian low-pass filtration of a)) and c) roughness profile (Gaussian high-pass filtration of a) ) both using a cut-off wavelength of $\Lambda_c$\,=\,250\,\textmugreek m.}\label{Profilo:Pic:Result_roughness_PTB}
\end{figure}

\begin{figure}[h]
	\begin{center}
		\begin{tabular}{c}
			\begin{overpic}[scale=.32]{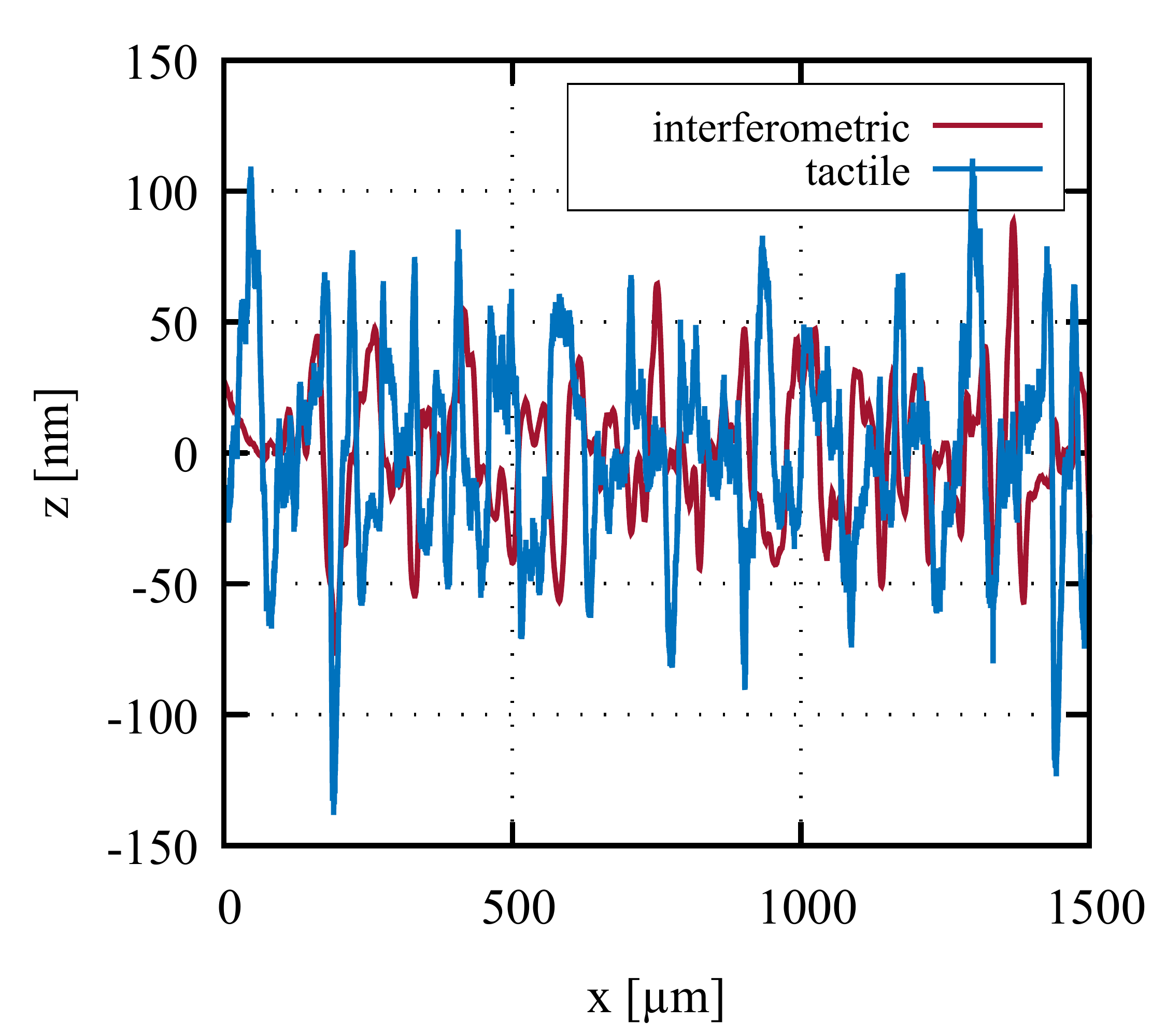}
				\put(1,1){\makebox(0,0){a)}} 
			\end{overpic}
			\begin{overpic}[scale=.32]{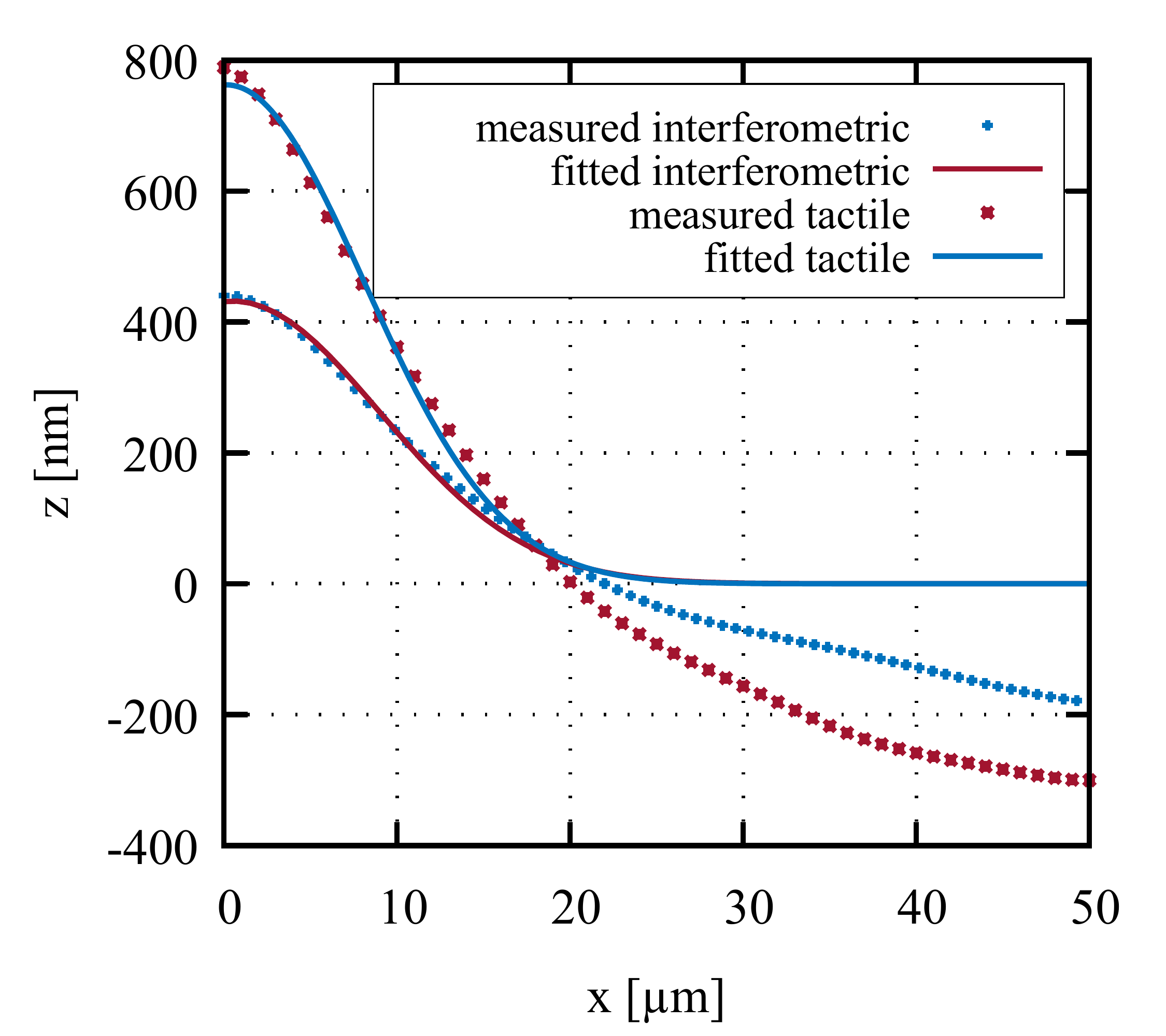}
				\put(1,1){\makebox(0,0){b)}}
			\end{overpic}
		\end{tabular}
	\end{center}
	\caption[Results of the comparative roughness evaluation using the ACF method]{Results of the comparative roughness evaluation with a) roughness profiles of the \gls{PTB}-calibrated height standard from an interferometric and tactile measurement as well as b) calculated auto-correlation functions and appropriate fits using a Gaussian approximation.}\label{Profilo:Pic:Result_roughness_ACF}
\end{figure}

\begin{figure}[h]
	\centering
			\begin{overpic}[scale=.36]{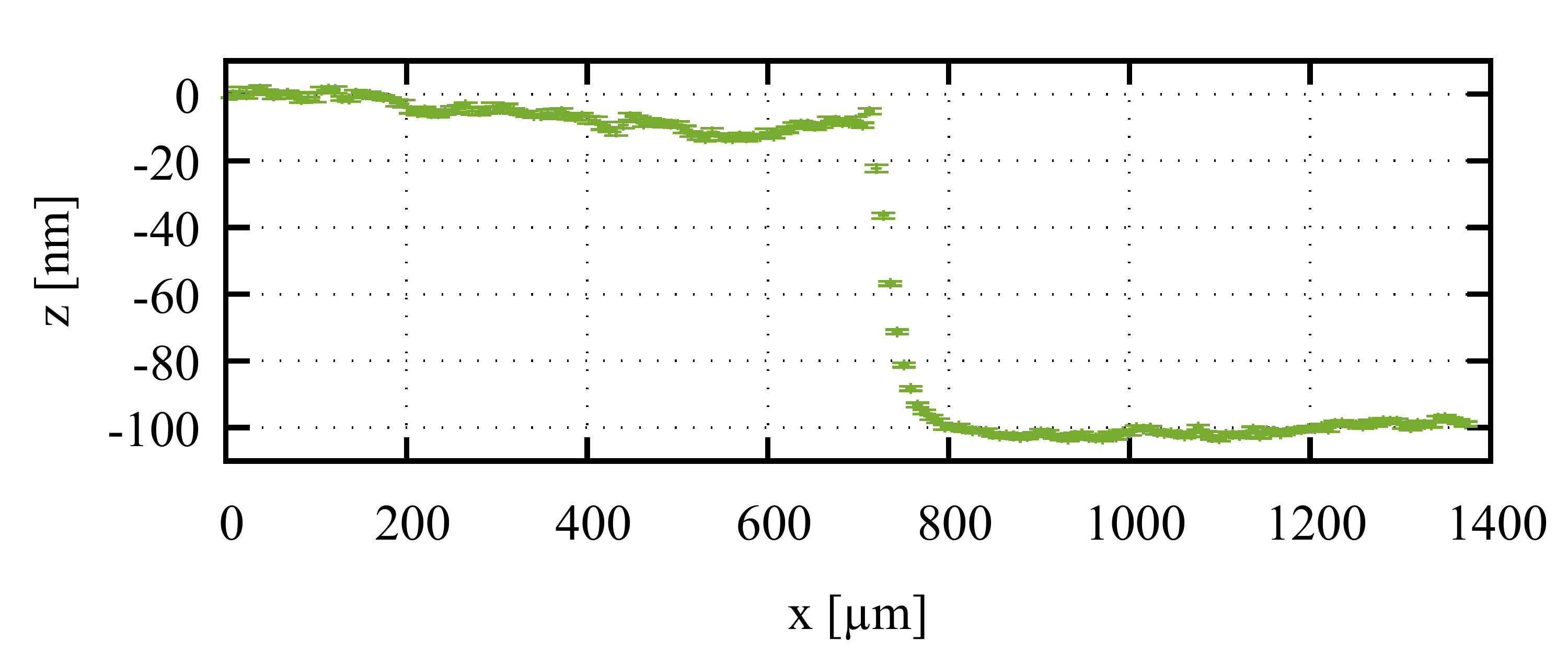}
			\end{overpic}
	\caption[Plot of a measured aluminum mirror edge on a float glass substrate]{Plot of a measured aluminum mirror edge on a float glass substrate with height profile as mean value of ten measurements on one position along the edge having a height of $h_{AL}$\,=\,(99.47\,$\pm$\,0.12)\,nm.}\label{Profilo:Pic:result_layertec_edge}
\end{figure}

\begin{figure}[h]
	\begin{center}
		\begin{tabular}{c}
			\begin{overpic}[scale=.32]{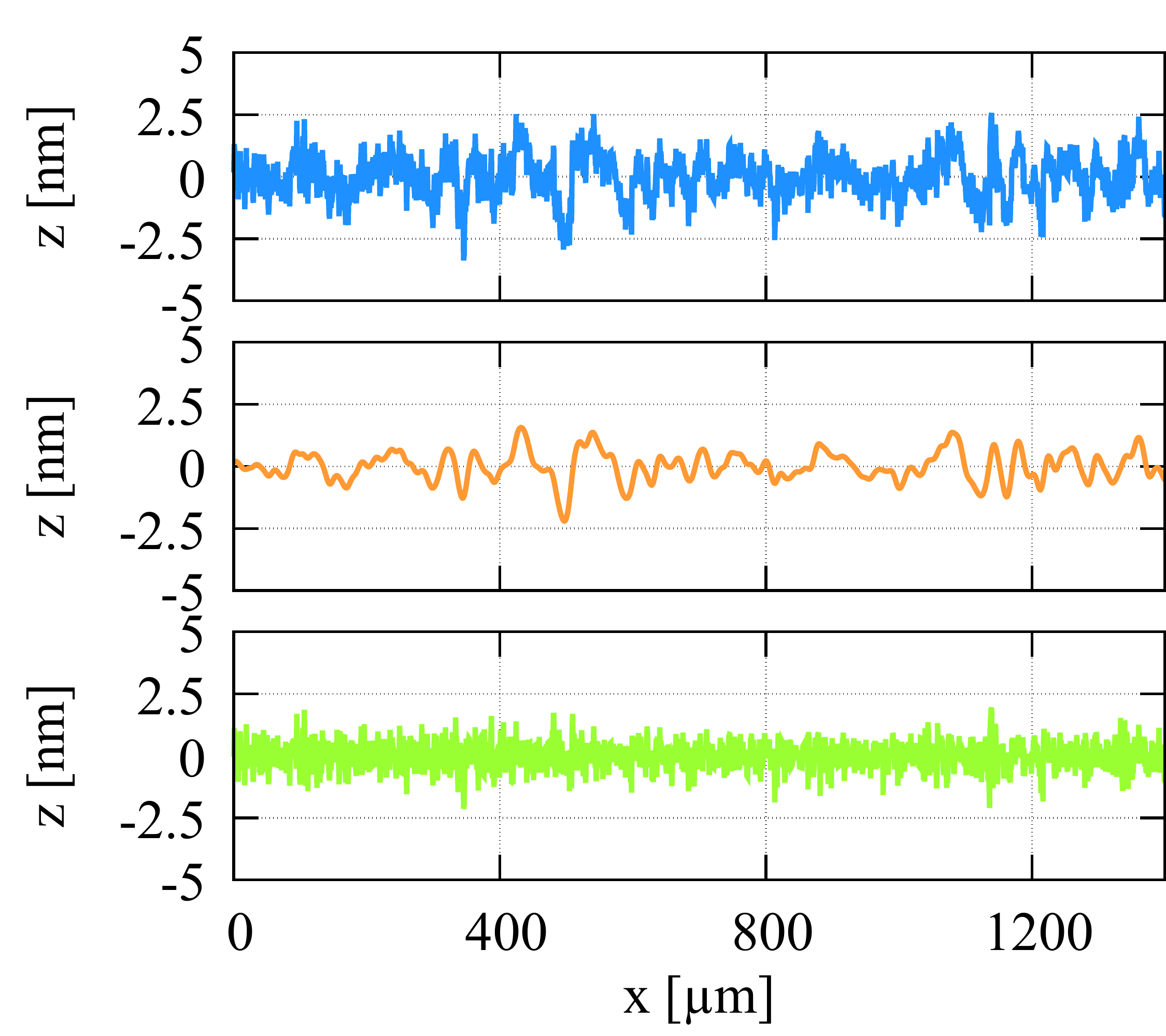}
				\put(1,1){\makebox(0,0){a)}}
				\put(-6,74){\makebox(0,0){\small{(I)}}}
				\put(-6,47){\makebox(0,0){\small{(II)}}}
				\put(-6,25){\makebox(0,0){\small{(III)}}}
			\end{overpic}
			\begin{overpic}[scale=.32]{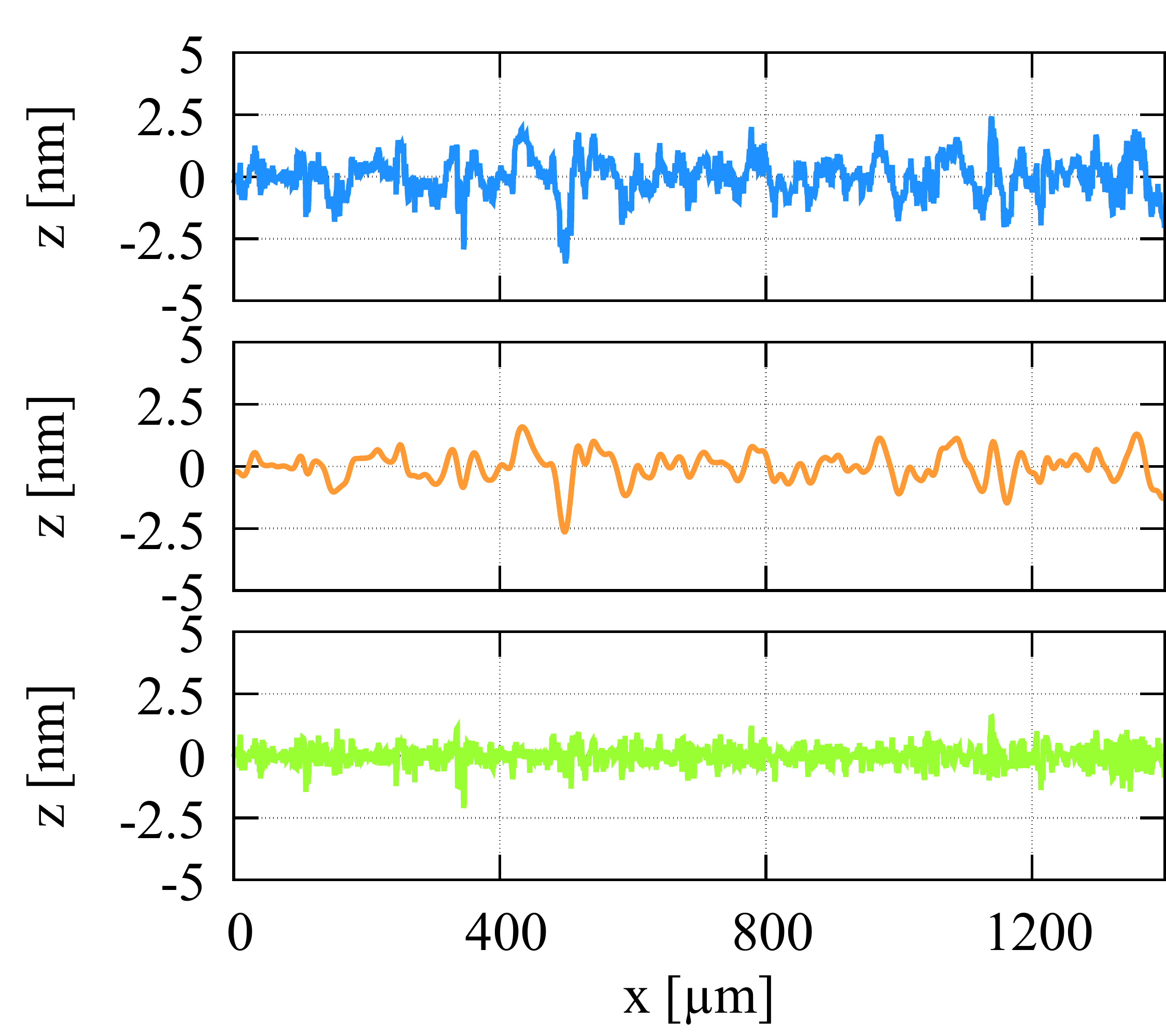}
				\put(1,1){\makebox(0,0){b)}}
			\end{overpic}
		\end{tabular}
	\end{center}
	\caption[Result of roughness measurements of an aluminum mirror edge on a float glass substrate]{Result of roughness measurements on a) an aluminum mirror surface and b) a float glass substrate with (I) primary, form corrected profile, (II) waviness profile (Gaussian low-pass filtration of (I)) and (III) roughness profile (Gaussian high-pass filtration of (I)) using a cut-off wavelength of $\Lambda_c$\,=\,25\,\textmugreek m (mean of ten measurements).}\label{Profilo:Pic:Result_roughness_layertec_edge}
\end{figure}

As surface quality and roughness in particular can be essential for the function of technical products and components, their in-line assessment is of high interest, \cite{Leach,Leach2008,bennett_roughness,influence_roughness_1,influence_roughness_2}. Various norms and guidelines exist for a number of established measurement technologies such as tactile profilometers, confocal microscopes and \gls{AFM}, \cite{ISO4288,ISO4287,ISO25178}. Most commonly, quantification methods based on the distribution of heights like the averaged roughness Ra and the root-mean-square roughness \glssymbol{RoughnessRq} are determined.\\
As the demonstrated \gls{DE-LCI} approach is capable of capturing precise height profiles over a large lateral measurement range of several hundred micrometers, an application for roughness evaluation of samples is possible. For the initial qualification, a \gls{PTB}-traceable surface roughness standard (KNT 20170/3 \textit{superfine}, Halle GmbH, Germany) was analyzed, Fig. \ref{Profilo:Pic:Result_roughness_meas_scheme}.
According to the specifications of \gls{PTB}, surface profiles were captured on the slices A, B and C in segments of (I), (II) and (III) in each slice. Finally, the results of the roughness measurement of all segments were averaged. According to ISO 4288-1996, surface quality is assessed by the application of appropriate processing steps and filters, \cite{ISO4288}. Following these processing steps, the captured surface profiles were form corrected (\glssymbol{LambSFilter}-filtering according to ISO 3274-1996 \cite{ISO3274}) and referenced to their respective mean values in order to generate the so called primary profile, Fig.\,\ref{Profilo:Pic:Result_roughness_PTB}\,a).
Based on this profile, a Gaussian filter was applied to calculate the waviness and the roughness of the primary profile, Fig.\ref{Profilo:Pic:Result_roughness_PTB} b) and c) respectively. Here, the application of the low-pass filter results in the waviness profile, while the application of the high-pass filter results in the roughness profile. The cut-off wavelength \glssymbol{LambCFilter} is determined in correspondence with the evaluation length of the profile which should be approximately 5\,$\cdot\Lambda_c$. In case of the roughness standard, the captured evaluation length was \glssymbol{EvaLenRough}\,=\,1.25\,mm as the filter length was $\Lambda_c$\,=\,250\,\textmugreek m. In order to exclude possible deviations on the roughness calculation that are due to filtering effects, the actual measured length was expanded equally by \nicefrac{1}{2}\,$\cdot\Lambda_c$ at the beginning and the end of the profile. By applying this methodology to the data captured from the roughness standard, values of Ra\,=\,(21.15\,$\pm$\,0.8)\,nm and Rq\,=\,(26.58\,$\pm$\,1.0)\,nm were calculated. The value of Ra is within the specifications given by the \gls{PTB} calibration which measured a mean value of Ra\,=\,(22.4\,$\pm$\,0.5)\,nm with an uncertainty of $\pm$\,5\,\%. As a second measure of comparison, a roughness evaluation using a confocal microscope was performed. Using the same probing scheme highlighted in Fig. \ref{Profilo:Pic:Result_roughness_meas_scheme}, data was captured in stitching mode as the 50x magnification objective was needed to achieve sufficient axial resolution, is restricted to a lateral measurement range of 220\,\textmugreek m. The captured data was filtered for noise using a $\Lambda_c$\,=\,8\,\textmugreek m Gaussian low-pass filter and analyzed in the same way as the interferometric data. The roughness parameters were calculated with Ra\,=\,(21.42\,$\pm$\,0.6)\,nm and Rq\,=\,(26.81\,$\pm$\,0.7)\,nm. These values support the interferometric values within the respective errors.\\
It is known that roughness parameters which are based on amplitude values of the height distribution, such as Ra and Rq, are not directly comparable in different measurement techniques, \cite{bennett_roughness}. This is particularly due to the different spatial bandwidth limitations of the individual techniques, \cite{Duparre_roughness}. For the purpose of enhancing comparability, alternative methods to determine the \gls{RMS} roughness were established. The most commonly used ones are based on the determination of the integral of either the \gls{ACF} or the \gls{PSDf} of a profile, \cite{Walsh:99_PSD, NECAS_ACFPSD,Scheer_mroughness_standard}. In order to perform a comparison of optically measured surfaces with tactile measured ones, data from the same standard sample was captured using both methods, Fig. \ref{Profilo:Pic:Result_roughness_ACF} a).
It can be seen from the roughness profiles that both data sets show similar amplitudes. However, the data from the tactile measurement shows significant noise which could be measured with $\pm$\,2.6\,nm. This noise can be attributed to the typical noise of a tactile measurement system consisting of mechanical and electronic components, \cite{TH_iSeries}. With the expected roughness levels of about 20 nm, this noise has an influence on the roughness measurements. For the comparison of both methods, the respective auto-correlation functions were determined using a Fourier-based approach, Fig. \ref{Profilo:Pic:Result_roughness_ACF} b). The acquired data was subsequently fitted using a Gaussian approximation in order to perform further analysis, \cite{DavidNecasTalk2012}. A fundamental analysis is the determination of the correlation length \glssymbol{TauX}. Most commonly, it is measured as the distance $x$ where the approximated \gls{ACF} reaches a value of $g(x) = \nicefrac{1}{e}$, \cite{Panda2017}. While the analysis of the tactile measurement yielded in a value of $\tau_x$\,=\,11.37\,\textmugreek m, the nine analyzed measurements of the interferometric evaluation resulted in a value of $\tau_x$\,=\,(11.69\,$\pm$\,1.0)\,\textmugreek m. Obviously, the measurements of the different technical approaches show a high similarity within the standard deviation of the measurement. Furthermore, the amplitude of the fitted \gls{ACF} can be interpreted as the squared \gls{RMS} roughness of the measured data. The comparison of these properties reveals the influence of noise on the tactile measurement. A value of \glssymbol{SigmaRMS}\,=\,27.6\,nm was calculated for the tactile measurement, while a mean value of $\sigma_{rms}$\,=\,(21.46\,$\pm$\,1.8)\,nm was calculated for the interferometric measurements. Taking the noise of the tactile measurements into account, both measurements are very similar within the respective error bars.\\
In addition to measurements on a standard, the characterization of an industry-relevant configuration, in particular an aluminum mirror coating on a float glass substrate (Layertec GmbH, Mellingen, Germany), was performed. The coating was applied using magnetron sputtering on one half of the circular substrate for evaluation purposes, Fig. \ref{Profilo:Pic:result_layertec_edge}.
Using the averaged data of three measurements on different positions along the coating edge with ten measurements at every position, a mean height of $z_{AL}$\,=\,(99.47\,$\pm$\,0.12)\,nm was measured. In comparison with the data analyzed for the height standard, the edge of the mirror was less steep which resulted in significantly lower edge effects. Furthermore, the evaluation of sub-nm roughness differences as a part of production accompanying characterization is of interest. For this purpose, data from the float glass substrate as well as from the the aluminum coated part of the mirror was analyzed with the described roughness methodology, Fig.\,\ref{Profilo:Pic:Result_roughness_layertec_edge}.
The separation of roughness\,(III) and waviness\,(II) from the primary, form-corrected profile\,(I) reveals a distinct difference. While the waviness of both profiles is in the same order of magnitude, the roughness of the aluminum coated part of the sample is larger. The mathematical analysis resulted in values of  Ra\,=\,(0.27\,$\pm$\,0.01)\,nm and Rq=\,(0.35\,$\pm$\,0.01)\,nm for the substrate area. The roughness of the aluminum coated part was  Ra\,=\,(0.38\,$\pm$\,0.02)\,nm and Rq=\,(0.47\,$\pm$\,0.02)\,nm. Compared to  Ra\,=\,0.31\,nm for the substrate and Ra=\,0.40\,nm for the coating, which are quoted by the manufacturer, the measured values correspond well.\\
In contrast to other technologies such as \gls{AFM}, scanning white-light interferometric microscopy and confocal microscopy, \gls{DE-LCI} is able to capture data for roughness evaluation on a large lateral measurement range of a few mm in one single data acquisition. The lack of the necessity to scan a sample eliminates problems of stitching, vibration and speed.


\subsection{High-dynamic range measurements}\label{Profilo:SubSec:HDR}
\begin{figure}[h]
	\begin{center}
		\begin{tabular}{c}
			\begin{overpic}[scale=0.315]{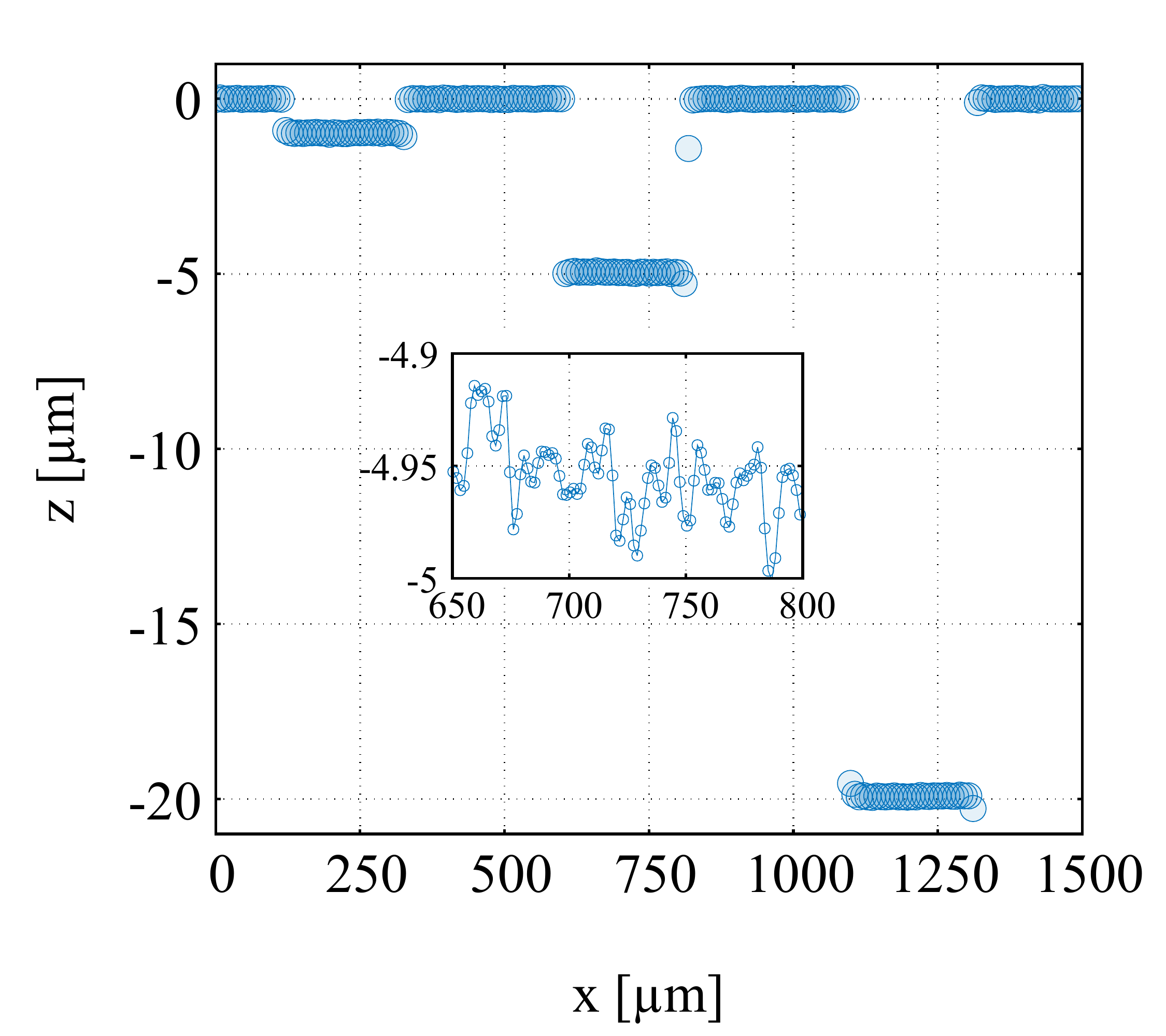}
				\put(1,1){\makebox(0,0){a)}}
				\put(39,58.5){\color{black}\vector(1,.9){7.5}}
				\put(69,58.5){\color{black}\vector(-1,.7){8.25}}
			\end{overpic}
			\begin{overpic}[scale=0.315]{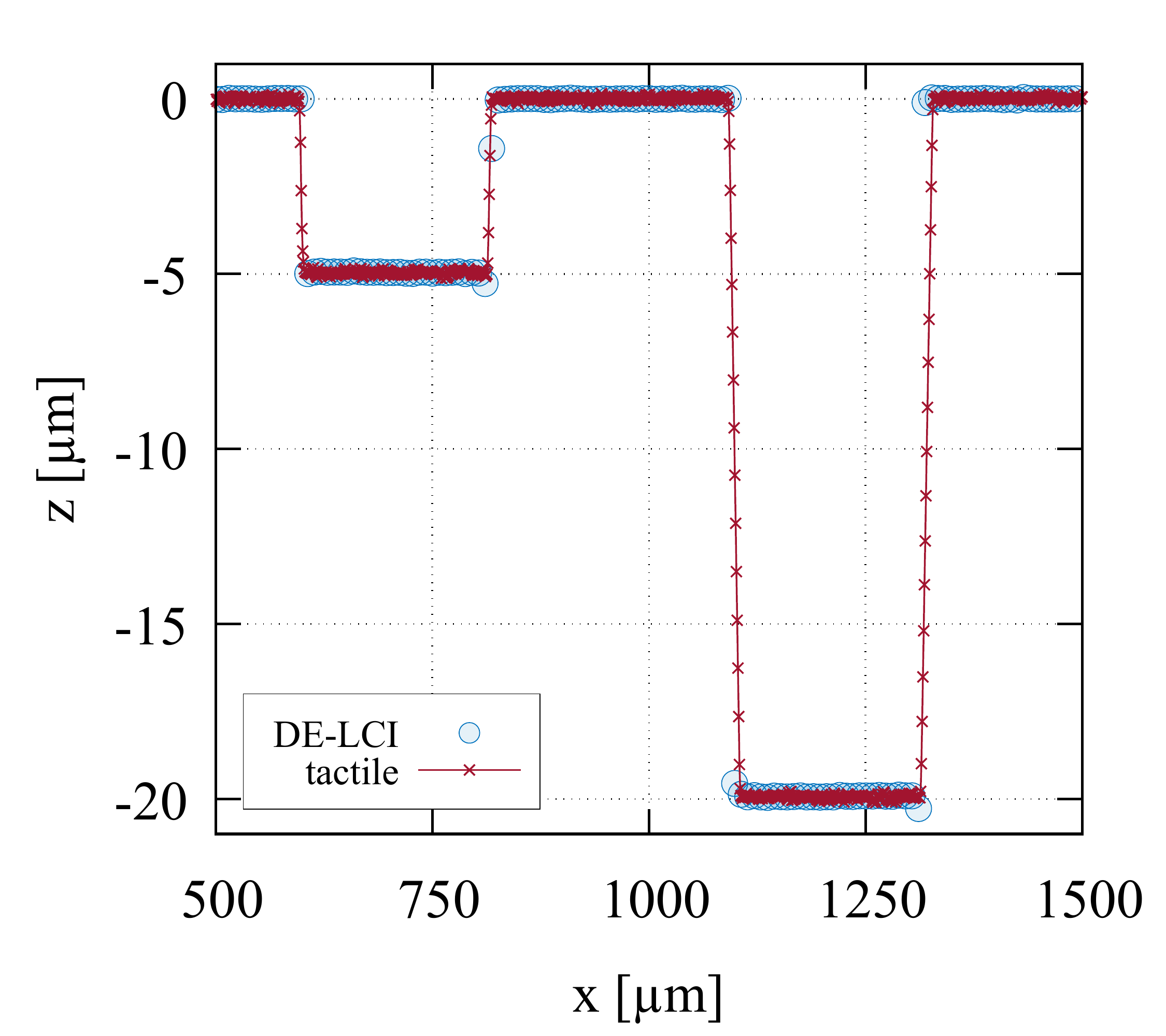}
				\put(1,1){\makebox(0,0){b)}}
			\end{overpic}
	\end{tabular}
	\end{center}
	\caption[Plot of an averaged line profile of a precision turned height standard with nominal height steps of 1, 5 and 20 \textmugreek m]{Plot of an averaged line profile where recorded depths of (971.26\,$\pm$\,0.31), (4951.40\,$\pm$\,0.28) and (19924.00\,$\pm$\,0.36)\,nm could be measured with a mean \gls{RMS} error of 26.9\,nm with respect to a measurement on a tactile profilometer while having the ability to capture roughness information which is shown in the inset where a value of Rq\,=\,26.7\,nm was calculated as well as  b) overlay of a dataset from the same sample taken with a tactile profilometer which shows significantly better capabilities to capture data on steep edges.}\label{Profilo:Pic:result_2d_standard_mum}
\end{figure}

In order to measure the performance of the setup with a high-dynamic range where the measurement range is $>$\,10\,\textmugreek m while the achievable height resolution should still be in the nm-range, a precision-turned height standard (EN14-3, \gls{PTB}, Germany) was examined. The standard provides grooves of defined heights with steps of 1, 5 and 20\,\textmugreek m which were subject to a series of measurements, Fig. \ref{Profilo:Pic:result_2d_standard_mum}.
The recorded data includes measured steps of (971.26\,$\pm$\,0.31), (4951.40\,$\pm$\,0.28) and (19924.00\,$\pm$\,0.36)\,nm. This results in an overall averaged \gls{RMS} error of 26.9 nm with regard to a measurement on a tactile profilometer. According to the calibration certificate of the standard, these values are within the quoted uncertainty for the nominal height steps of $\pm$  33 nm. Furthermore, the high axial resolution leads to the ability to capture roughness data in the nm-range on all height steps, see inset in Fig.\,\ref{Profilo:Pic:result_2d_standard_mum}\,a). A \gls{RMS} value of Rq\,=\,26.7\,nm was calculated. Some edge effects and noise occur in slopes of the steps. In the current optical design, that uses a NA\,=\,0.06 imaging system, a large lateral measurement range could be covered but data on the slopes with an 70\si{\degree} angle could not be gathered reliably. Depending on the application, the setup can be optimized to increase the sensitivity on these parts of the sample. Reference measurements with a tactile profilometer confirmed these heights but emphasized the fact that the transitions between the different levels are formed by segments of 70\si{\degree}, Fig. \ref{Profilo:Pic:result_2d_standard_mum} b). In contrast, the tactile profilometer is able to gather a much higher number of data points in these areas while having a lower overall resolution.\\ 
Additionally, the repeatability according to Eq.\,(\ref{EQ:repeatability}) was analyzed by investigating N\,=\,10 consecutive measurements of the profile. The sample showed a slightly increased averaged standard deviation of $\overline{\sigma_z}$\,=\,0.52\,nm with respect to the measurements on a low-scattering silicon sample, Fig.\,\ref{result_std_si} b). This can be attributed to influences of noise due to diffraction, scattering and other effects affecting the measured repeatability.
In a comparative manner to the analysis of the silicon height standard, see subsection \ref{Profilo:SubSec_repeatability_resolutiuon}, the heights on the \textmugreek m-sized standard, Fig. \ref{Profilo:Pic:result_2d_standard_mum} a), were evaluated for 10 measurements as the difference of the two closest base levels (z = 0) to the particular step. From the data, quadratic means of $\Delta z_{min}$ for the three steps with nominal heights of 1, 5 and 20 \textmugreek m were calculated as $\Delta z_{min1}$\,=\,0.31\,nm, $\Delta z_{min5}$\,=\,0.28\,nm and $\Delta z_{min20}$\,=\,0.36\,nm respectively. The result shows that the resolution is not dependent on the size of the measured step.\\
In contrast to the silicon standard, the \textmugreek m-sized height standard has a significantly higher roughness which leads to scattering. In consequence, the measurements on this sample were affected by noise which led to a number of outlier data points. The implemented post-processing routines took these outliers into account and corrected them. The outlier correction was performed by the detection of rising edges using data of the first derivative of the profile with respect to the $x$-coordinate in combination with the correction of the difference between the outlier and the mean value of five previous data points. In the analysis of the step heights and its standard deviation it could be detected that the outlier correction scheme influences the profile on a sub-nm level. As outliers occur on different spatial positions for consecutive measurements, a higher standard deviation was measured opposing to measurements where less outliers occurred, as on the silicon height standard.


\subsection{Dual-channel approach}
\begin{figure}[h]
	\begin{center}
		\begin{tabular}{c}
			\begin{overpic}[scale=.58]{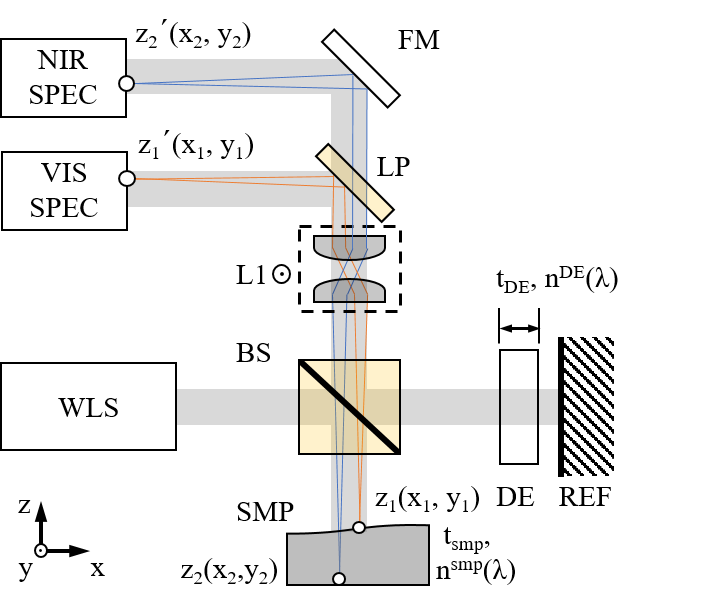}
				\put(1,1){\makebox(0,0){a)}}
			\end{overpic}
			\begin{overpic}[scale=.3]{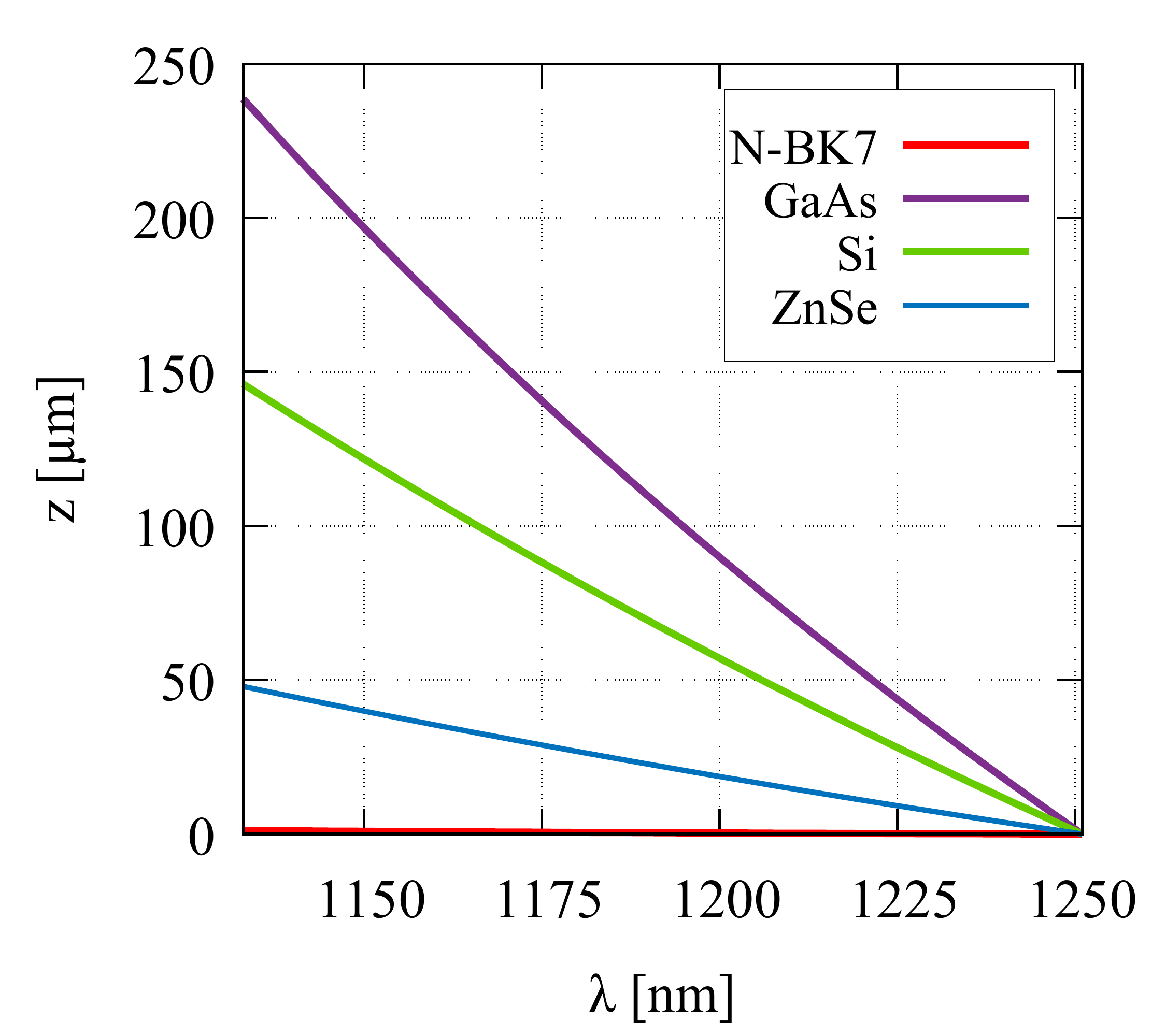}
				\put(1,1){\makebox(0,0){b)}}
			\end{overpic}
		\end{tabular}
	\end{center}
	\caption[Schema of the extended optical setup for NIR-enhanced profilometry]{a) Schema of the extended optical setup for dual-channel interferometry with WLS - white-light source, BS - beam splitter, SMP - sample having a  thickness $t_{smp}$ and a refractive index $n^{smp}(\lambda)$ where  $z_1(x_1,y_1)$ and  $z_2(x_2,y_2)$ are two points, one on the surface, one on the back side, REF - fixed reference mirror, DE - dispersive element with the thickness $t_{DE}$ and $n^{DE}(\lambda)$, L1 - lens to image the sample with a given magnification M (typically M = 1.3 or 4), HP - high-pass filter @800\,nm to reflect the \gls{VIS} part of the spectrum where VISSPEC - \gls{VIS} imaging spectrometer detects the surface information of the magnified point $z_1'(x_1,y_1)$ and FM - folding mirror relays the \gls{NIR} part of the spectrum to NIRSPEC - \gls{NIR} imaging spectrometer which detects the depth information of the magnified point $z_2'(x_2,y_2)$ as well as b) simulation of possible measurement ranges of materials suitable as dispersive elements for \gls{NIR} investigations.}\label{Profilo:Pic:IR_meas_setup}
\end{figure}

\begin{figure}[h]
	\begin{center}
		\begin{tabular}{c}
			\begin{overpic}[scale=.95]{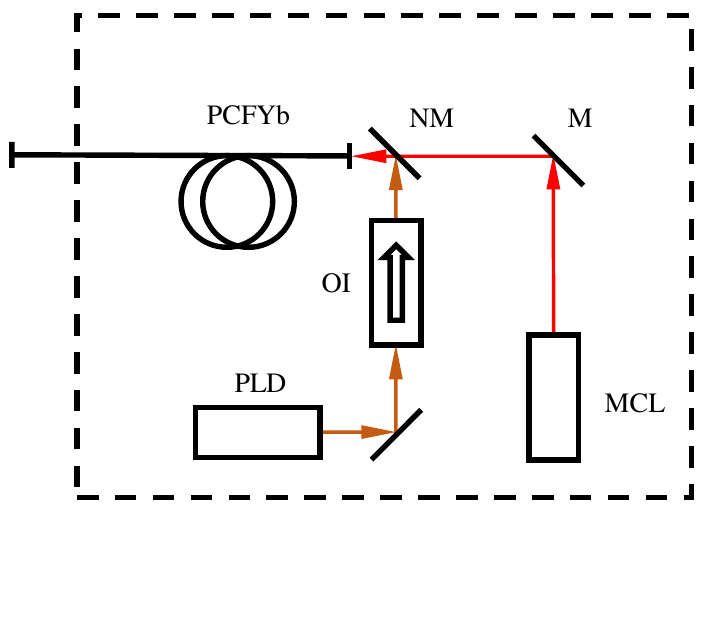}
				\put(1,1){\makebox(0,0){a)}}
			\end{overpic}
			\begin{overpic}[scale=.32]{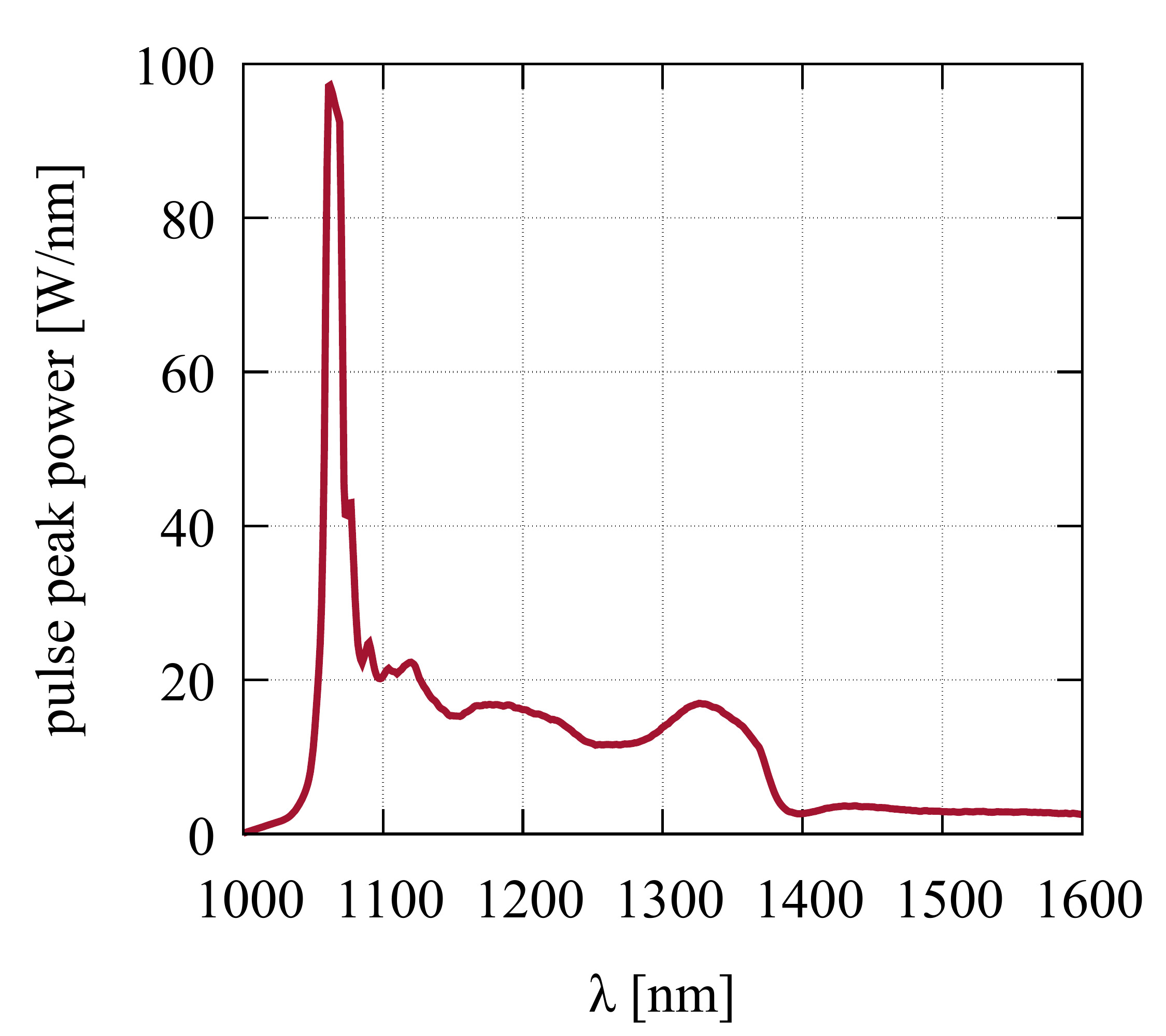}
				\put(1,1){\makebox(0,0){b)}}
			\end{overpic}
		\end{tabular}
	\end{center}
	\caption[Schema of the optical setup for the generation of amplified supercontinuum]{Schema of the optical setup for the generation of amplified supercontinuum (ASC) in the \gls{NIR} range with MCL - microchip Laser, PLD - pump laser diode ($\lambda_{pmp}$ = 976 nm), OI - optical isolator, M - mirror, NM - notch mirror, PCFYb - Yb:doped PCF – fiber as well as b) the optical spectral pulse peak power density of the light source (both adapted from \cite{BaseltIRSC}).}\label{Profilo:Pic:IR_SC_setup}
\end{figure}

As surface profile evaluation is crucial at different industrial processing steps, tomographic evaluation of structures becomes also interesting. For this purpose, the \gls{DE-LCI} approach was adapted to the \gls{NIR} to perform tomographic examinations of \gls{Si}-based structures, Fig.\,\ref{Profilo:Pic:IR_meas_setup}\,a).
The setup was designed to work in a dual-channel configuration where a broadband light source illuminates the sample and a dichroic mirror at 800\,nm separates the recombined light for the two analysis channels. In contrast to conventional \gls{DE-LCI}, the light of the \gls{NIR} spectral range holds information from inside the sample while light of the \gls{VIS} spectral range holds only surface information. This is of course valid for samples like e.g. silicon which are transmissive in the \gls{NIR} range but not in the \gls{VIS}. Consequently, an appropriate imaging spectrometer was calculated and designed for a spectral range of $\Delta \lambda$\,=\,(1133\,-\,1251)\,nm based on a \gls{InGaAs} camera (Bobcat 640, Xenics Ltd., Belgium). A detailed description of assumptions, parameters and components is given in appendix \ref{Profilo:APNDX:Sec:imaging_spectrometer}.\\ 
In order to generate spectral power densities in the spectral range of (1000\,-\,1400)\,nm, an \gls{ASC} was used, \cite{BaseltIRSC}. The source utilizes an Yb\textsuperscript{3+} doped photonic crystal fiber as medium for non-linear spectral broadening and as gain medium in a fiber amplifier configuration, Fig. \ref{Profilo:Pic:IR_SC_setup} a).
The configuration utilizes a pump diode laser (\glssymbol{WavLenPMP}\,=\,976\,nm) to core pump the fiber while a passively Q-switched microchip laser (\glssymbol{WavLenSeed}\,=1064\,nm) with pulse durations of 1.3\,ns and a variable repetition rate of up to 20\,kHz is used to seed the system. The amplification enables several \nicefrac{W}{nm} pulse peak power in the desired spectral range and beyond, Fig. \ref{Profilo:Pic:IR_SC_setup} b).
The high pulse peak power as well as the short pulse duration make the light source interesting for dynamic measurements to e.g. observe \gls{MEMS} movements. Stroboscopic illumination can be envisioned to achieve high penetration depths and high temporal resolution\footnote{These aspects were not studied within this work and are subject to further research.}.

\subsubsection{Measurement range and resolution in NIR evaluation}
With its dispersion characteristics, N-BK7 is a very suitable material to be used for \gls{DE-LCI} in the \gls{VIS} where typical measurement ranges of 79.91 \textmugreek m are achieved ($t_{DE}$\,=\,2\,mm). Due to lower spatial resolutions of cameras used for imaging spectrometers in the \gls{NIR} the covered spectral range is usually low. Furthermore, most materials have a rather flat $n(\lambda)$ slope in this spectral region. Both factors limit the possible axial measurement range. Using N-BK7 with $t_{DE}$\,=\,2 mm in the range of $\Delta \lambda$\,=\,(1133\,-\,1251)\,nm, yields in an axial measurement range of $\Delta z$\,=\,1.12\,\textmugreek m. A possibility to increase the range is the utilization of the sample as dispersive material if its thickness is known. Under the assumption that a sample is a silicon wafer with a thickness of 100\,\textmugreek m, the resulting range would be 7.03\,\textmugreek m. Depending on the application, this range can be suitable to detect buried marks in a wafer. Similar measurement ranges are achievable by substituting N-BK7 for a higher refractive glass. The usage of FK51A would enable a range of $\Delta z$\,=\,1.8\,\textmugreek m while SF11 would lead to $\Delta z$\,=\,5.8\,\textmugreek m assuming a thickness of 2\,mm. Significantly higher measurement ranges, which are comparable to the \gls{VIS} approach, can only be achieved with non-glass materials, Fig. \ref{Profilo:Pic:IR_meas_setup} b). As already discussed, \gls{Si} can be used to extend the range as it can reach $\Delta z$\,=\,146.04\,\textmugreek m for $t_{DE}$\,=\,2 mm. Even higher ranges can be observed with materials like \gls{GaAs} where $\Delta z$\,=\,238.6\,\textmugreek m. However, it has to be noted that both \gls{Si} and \gls{GaAs} are non-transmissive in the visible spectral range. Therefore, these materials are not suitable in a dual-channel approach where it is desired to simultaneously gather data of both \gls{VIS} and \gls{NIR} channels. A possible measurement mode would incorporate a highly dispersive material which is non-transmissive in the \gls{VIS} as a reference mirror. In this way, a glass \gls{DE} can be used for measurements in the \gls{VIS} while the reference mirror acts as a \gls{DE} for the \gls{NIR} investigations. Alternatively, a transmission mode operation can be implemented with a material which is transmissive in the \gls{VIS} and \gls{NIR} range. A material which meets this requirement is \gls{ZnSe}. It has high optical transmission starting at 550 nm. In the \gls{NIR}, a measurement range of $\Delta z$\,=\,47.88\,\textmugreek m can be achieved by the application of a \gls{DE} with $t_{DE}$\,=\,2\,mm\footnote{In a real measurement setup, the measurement range will be extended by the dispersion on the sample. The amount of increase is dependent on the sample depth to be analyzed.}. Consequently, this leads to a maximal achievable measurement range in the \gls{VIS} of about $\Delta z_{VIS}$\,=\,605.84\,\textmugreek m, if the \gls{DE} is used in transmission mode. All presented measurements in this work have been performed using a \gls{ZnSe} element ($t_{DE}$\,=\,2\,mm) in transmission mode. Furthermore, it has to be noted that the measurement range only describes the ability to detect surface height changes on one particular surface. In a tomographic measurement, the separation of the equalization wavelengths for multiple surfaces is also important. A simulation showed that a sample has to have a minimum thickness of 60\,\textmu m of silicon in order to capture the equalization wavelengths in both spectral channels separately for the particular setup described here.\\
While the equations derived in section \ref{Profilo:Sec:Meas_range} for the estimation of the profile height are still applicable for the surface information, the back reflected data from structures within the sample follow a different relation. In this configuration, the phase is transformed in slight variations regarding Eq. (\ref{Profilo:EQ:2d_int_equation}) with an additional component for the samples refractive index \glssymbol{RefIndSMP}\,=\,$n^{smp}$ and its thickness \glssymbol{ThickSMP}, where $n^{smp}$ is supposed to be a known quantity
\begin{equation}
\varphi = 2\pi \frac{ \left[\left(n^{DE} -1 \right) \cdot t_{DE} - \left(n^{smp} -1 \right) \cdot t_{smp} - \delta \right]}{\lambda} .
\end{equation}
Consequently, all data captured from the inside of a sample is scaled by the depth-dependent refractive index which has to be corrected, if height information should be obtained. In order to perform this, blind dispersion compensation techniques known from \gls{OCT} can be utilized, \cite{Banaszek2007,Lippok2012,Wojtkowski2004}.\\
Analogous to the measurements and calculations performed previously, see section \ref{Profilo:Sec:Meas_range}, the resolution limit of the \gls{NIR} approach was characterized. A measurement of the noise yielded in an average value of $\Delta I$\,=\,16.0\,dB which was found to be normally distributed along the spatial and spectral dimension of the \gls{NIR} imaging spectrometer. The measured noise was used to calculate the single point resolution limit $\Delta \delta$\,=\,4.73\,nm. This calculation was based on Eq.\,(\ref{noise_resolution}) where the wavelength range was $\Delta \lambda$\,=\,(1133\,-\,1251)\,nm while the equalization wavelength was $\lambda_{eq}$\,=\,1189\,nm and the relative normalized intensity at this point was $I_0$\,=\,0.5\,arb. units. Under the assumption that n\,=\,300 points were used for fitting, a resolution of the \gls{NIR} system of $r_{fit}$\,=\,0.27\,nm was calculated, referring also to Eq. (\ref{EQ:resolution_fit}). By extrapolating the estimated influence of the algorithm for the measured noise $\Delta I$, see subsection \ref{Profilo:SigAn:SubSec:error_data_proc}, an influence of 0.22\,nm can be computed. This leads to an expected resolution of 0.49\,nm for this experiment in the NIR spectral range. In relation to the measurement range of $\Delta z$\,=\,95.76\,\textmugreek m, which can be achieved by using a dispersive element of ZnSe with $t_{DE}$\,=\,2\,mm, a dynamic-range of DR\,=\,\num{1.95e5} was calculated.

\subsubsection{Results of tomographic profilometry}
\begin{figure}[h]
\centering
		\begin{tabular}{c}
			\begin{overpic}[scale=.40, grid = false]{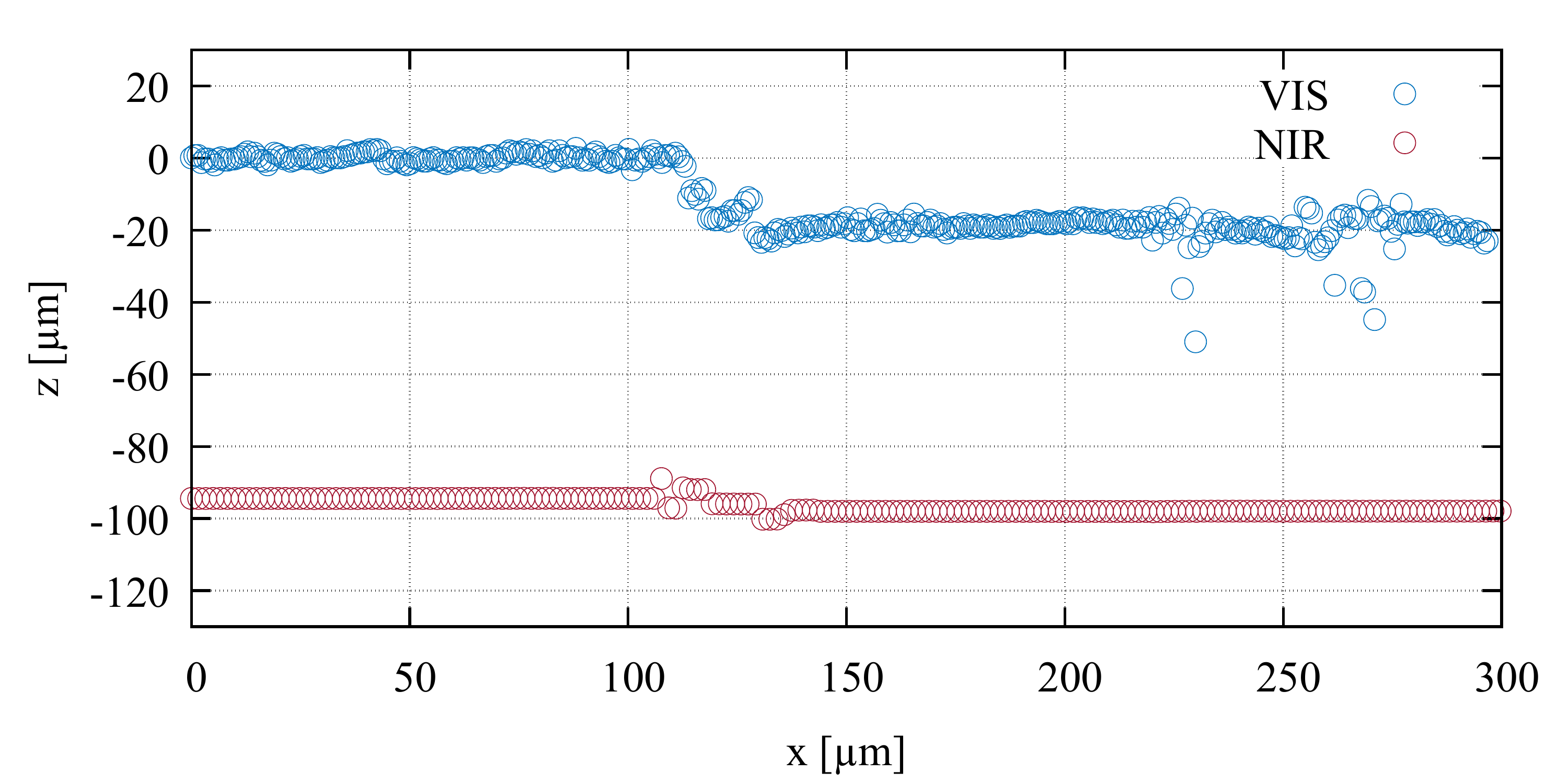}
				\put(12.2,18.){\tikz \fill[line width=1pt, color = orange!80, opacity=.2] (0,0) rectangle (10.2,2.1);}
				\put(14,21){\tikz \node[draw = none, color = orange!80,] at (0,0)   (a) {Si};}
				\put(12.2,35.3){\tikz \fill[line width=1pt, color = violet!80, opacity=.2] (0,0) rectangle (4,.65);}
				\put(14,34.5){\tikz \node[draw = none, color = violet!80,] at (0,0)   (a) {coating};}
				\put(14,41){\tikz \node[draw = none, color = black,] at (0,0)   (a) {top};}
				\put(14,12){\tikz \node[draw = none, color = black,] at (0,0)   (a) {bottom};}
			\end{overpic}
		\end{tabular}
	\caption[Result for the dual-channel measurement on a wafer]{Result of the profilometry measurement using a dual channel approach where the VIS channel shows the surface profile on top of a thinned Si - wafer with a coated area and a tomographic profile from the backside of the wafer using the NIR channel of the setup.}\label{Profilo:Pic:NIR_interferometer_wafer_meas}
\end{figure}
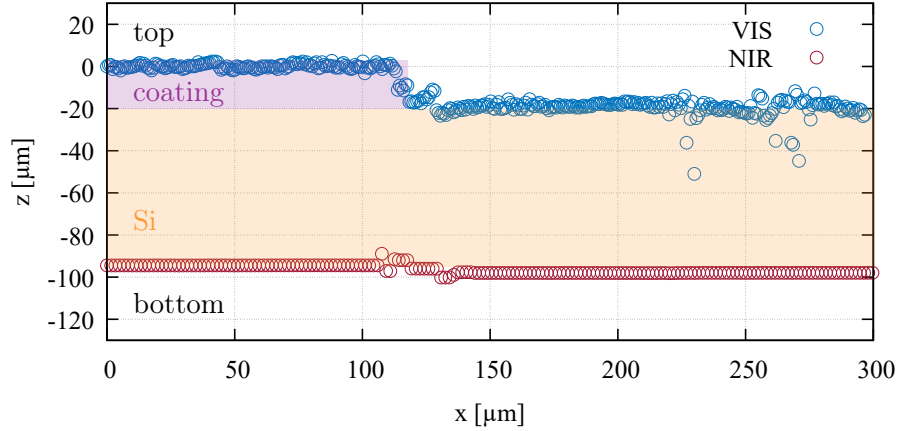

Using the above described setup and configuration of the \gls{DE}, an experiment for the tomographic imaging of a thinned wafer was conducted to capture the surface profile as well as the backside profile of the sample in a simultaneous measurement, Fig.\,\ref{Profilo:Pic:NIR_interferometer_wafer_meas}.
The sample was a Si-wafer with a partially coated area which was thinned previously. It was mounted on a thin glass substrate. The measurement shows the stepped surface profile where the coating height was determined with 18.70\,$\pm$\,1.42\,\textmugreek m. By analyzing the interferometric signal in the NIR channel and correcting the measured \gls{OPD} with the refractive index of silicon, the wafer thickness was determined with 84.71\,$\pm$\,0.38\,\textmugreek m. The edge of the coating can be identified in the VIS as well as in the NIR signal with some significant noise (x\,=\,110\,-\,140\,\textmugreek m). The refractive index of the coating introduced a dispersion-induced deviation in the NIR measurement. This deviation was not corrected in this measurement as the material composition of the coating was unknown. It is also visible that the VIS signal shows significant overall noise. This is due to the relatively small spectral power density of the used \gls{ASC} source in combination of the low optical transmission of \gls{ZnSe} in this spectral range. This led to a reduced \gls{SNR} of 16.5\,dB for the VIS measurement compared to the experiments presented before. Future developments will account for this and develop methods to increase the SNR. Furthermore, the development of automatic dispersion correction for tomographic data according to known approaches will be worked on.

\section{Areal measurement approaches}
In order to gather areal information, different approaches were developed. While two methods were developed only theoretically, one approach was implemented, characterized and tested.

\subsection{Translation-based areal information}
\begin{figure}[h]
\centering
		\begin{tabular}{c}
			\begin{overpic}[scale=0.28]{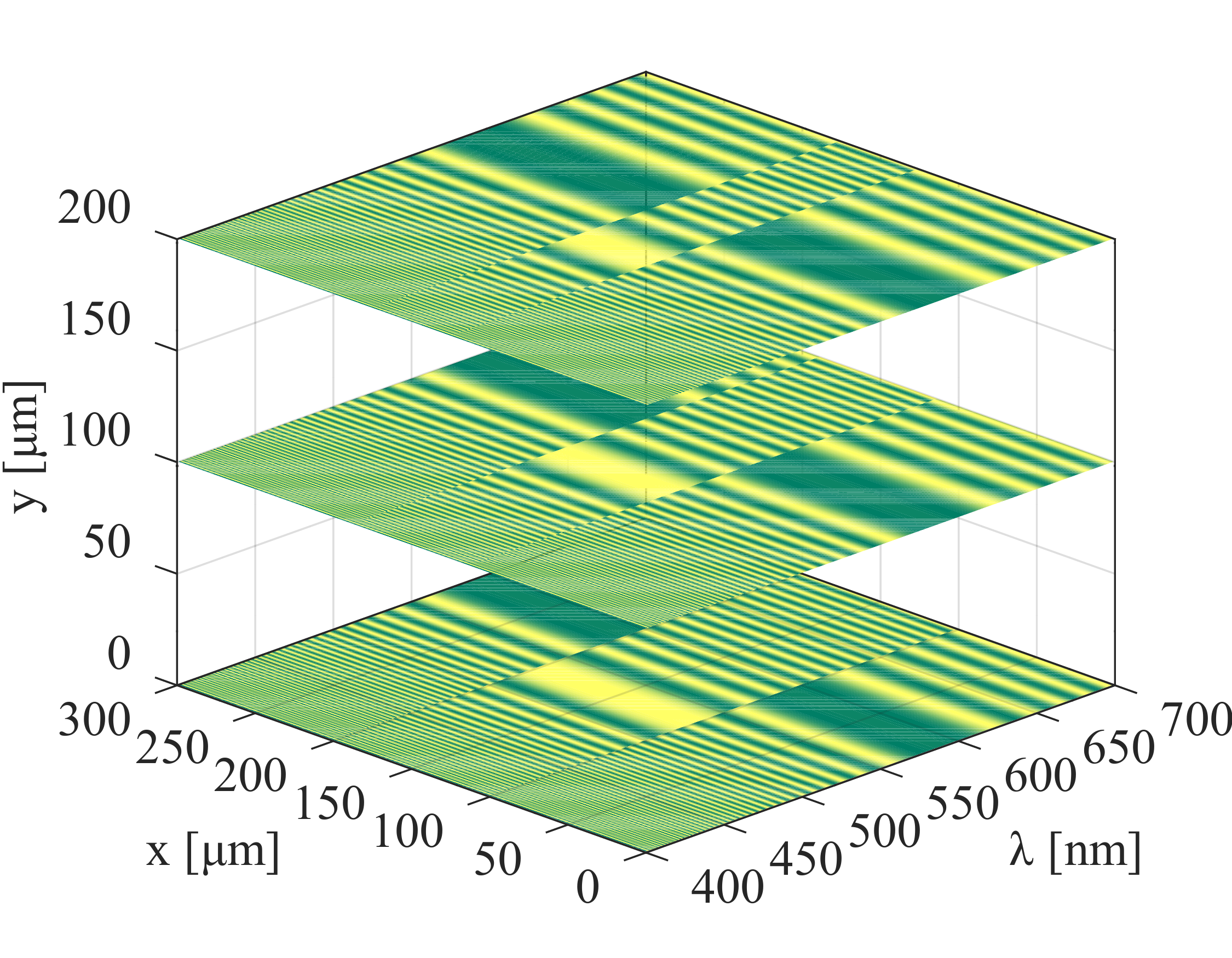}
				\put(1,1){\makebox(0,0){a)}}
			\end{overpic}
			\begin{overpic}[scale=0.21]{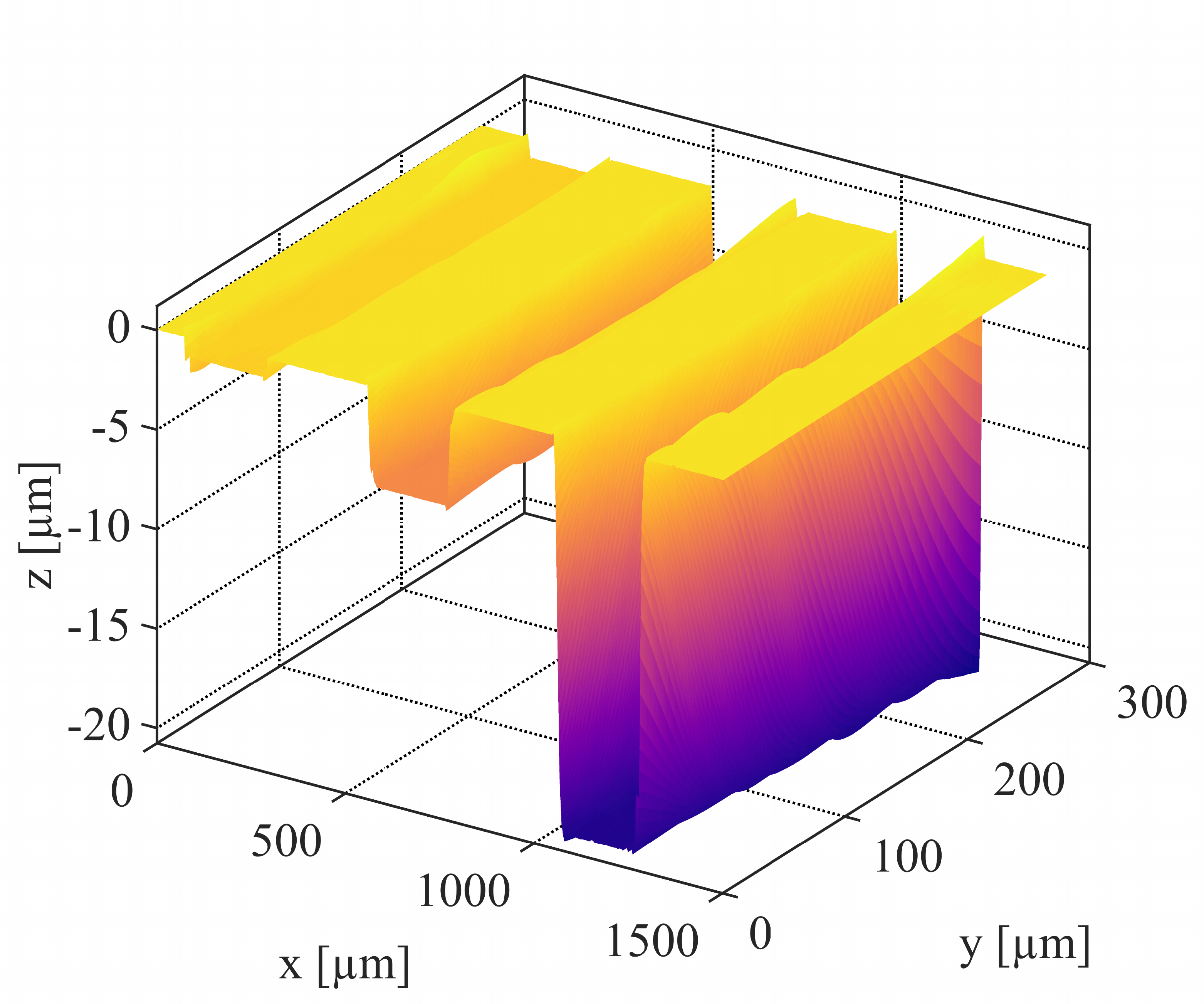}
				\put(1,1){\makebox(0,0){b)}}
			\end{overpic}
		\end{tabular}
	\caption[Plot of the measured three-dimensional surface of a precision-turned height standard]{a) Simulated stack of two-dimensional spectra gathered by the translation of the imaging lens L1 in the $y$-direction to capture 3D information and b) plot of the three-dimensional surface of a precision-turned groove standard (Gaussian filter applied to reduce edge effects for display purposes) with measured depths of (971.26\,$\pm$\,0.31), (4951.40\,$\pm$\,0.28) and (19924.00\,$\pm$\,0.36)\,nm.}\label{Profilo:Pic:Result_3D_standard_mum}
\end{figure}

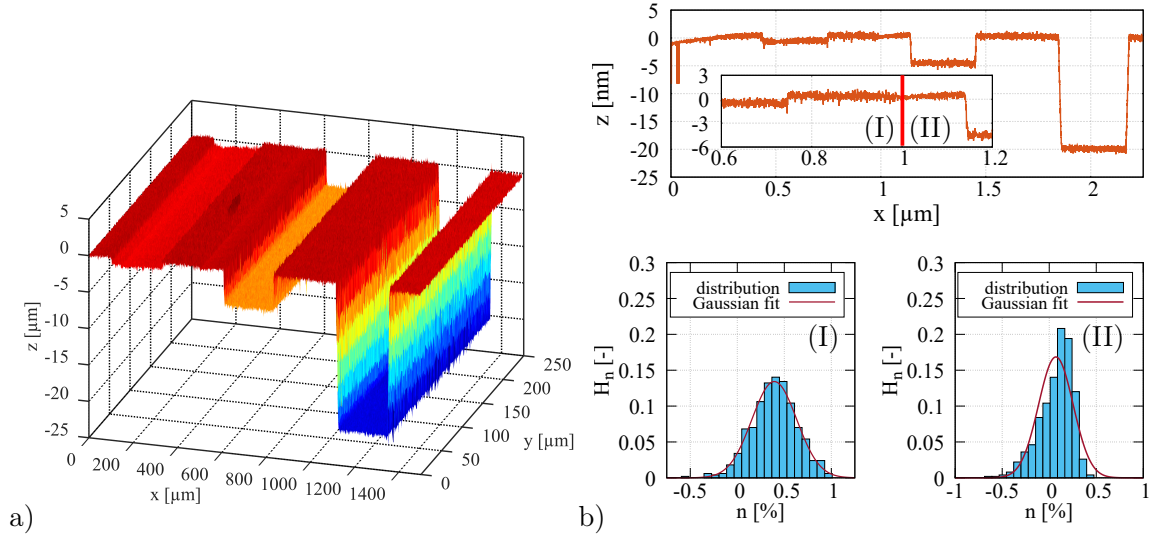
\begin{figure}[h]
\centering
		\begin{tabular}{c}
			\begin{overpic}[scale=.032]{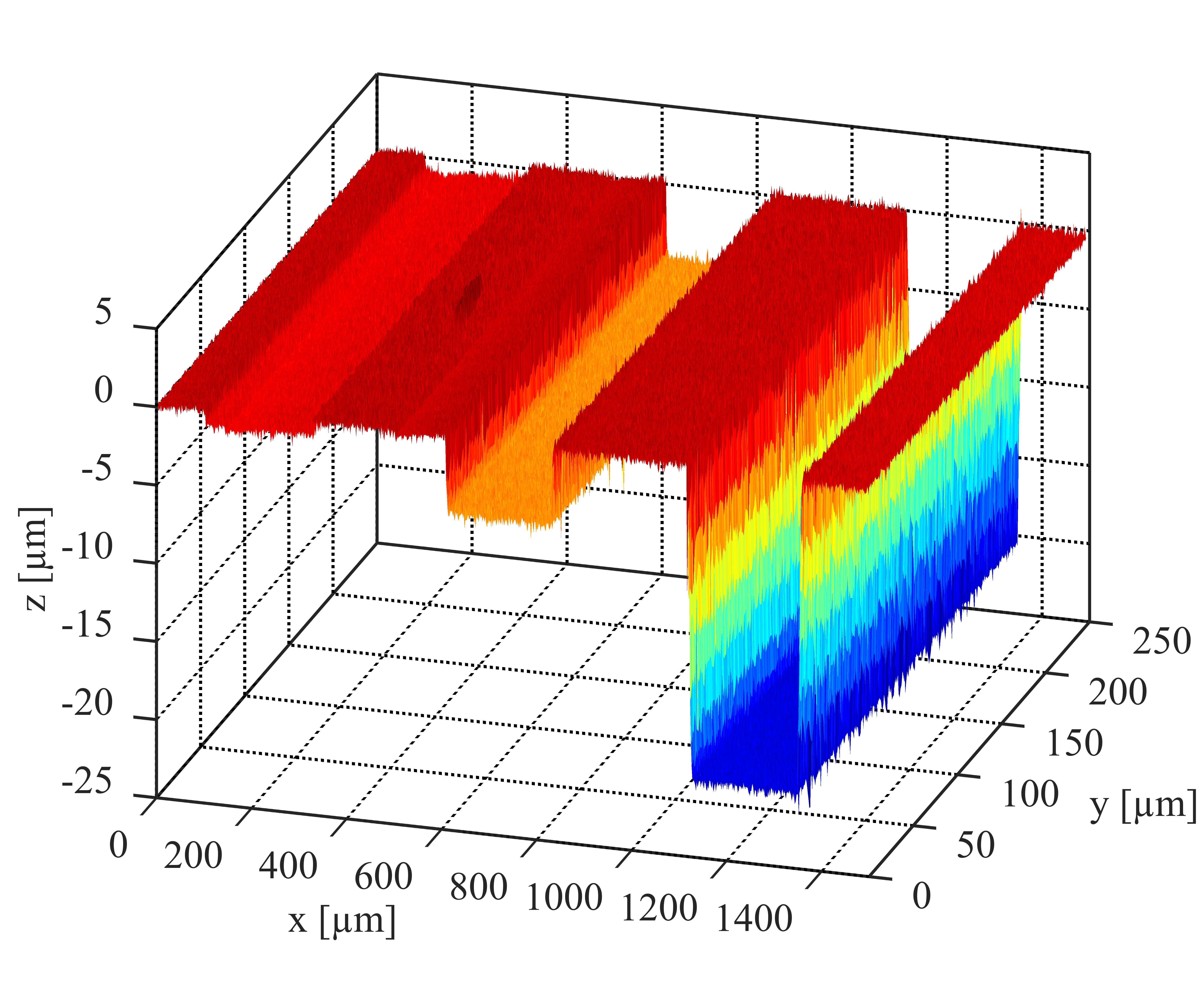}
				\put(1,1){\makebox(0,0){a)}}
			\end{overpic}
			\begin{overpic}[scale=1]{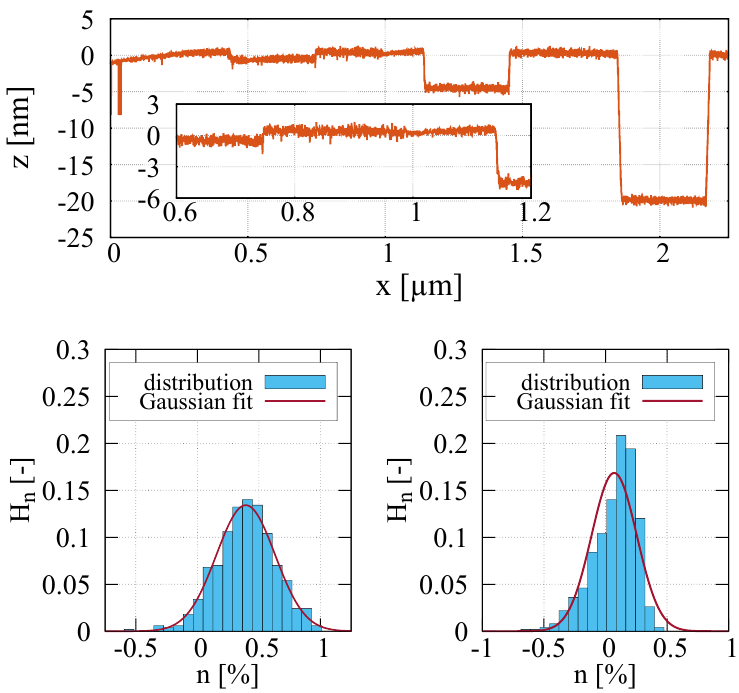}
				\put(1,1){\makebox(0,0){b)}}
				\put(52,70){\makebox(0,0){(I)}}
				\put(42,33){\makebox(0,0){(I)}}
				\put(61,70){\makebox(0,0){(II)}}
				\put(91.5,33){\makebox(0,0){(II)}}
				\put(56,67.25){\color{red}\linethickness{.5mm}\line(0,1){12.5}}
			\end{overpic}
		\end{tabular}
	\caption[Plot of the measurements on a PTB-traceable height standard using a confocal microscope]{Plot of the measurements on a PTB-traceable height standard using a confocal microscope with a) three-dimensional representation of the sample, b) profile plot in the middle of the data set with inset highlighting a typical stitching error with the segments (I) and (II) having different distributions of noise $n$ and Gaussian fits.}\label{Profilo:Pic:result_pizza_CLSM}
\end{figure}

\begin{figure}[h]
\centering
		\begin{tabular}{c}
			\begin{overpic}[scale=.21]{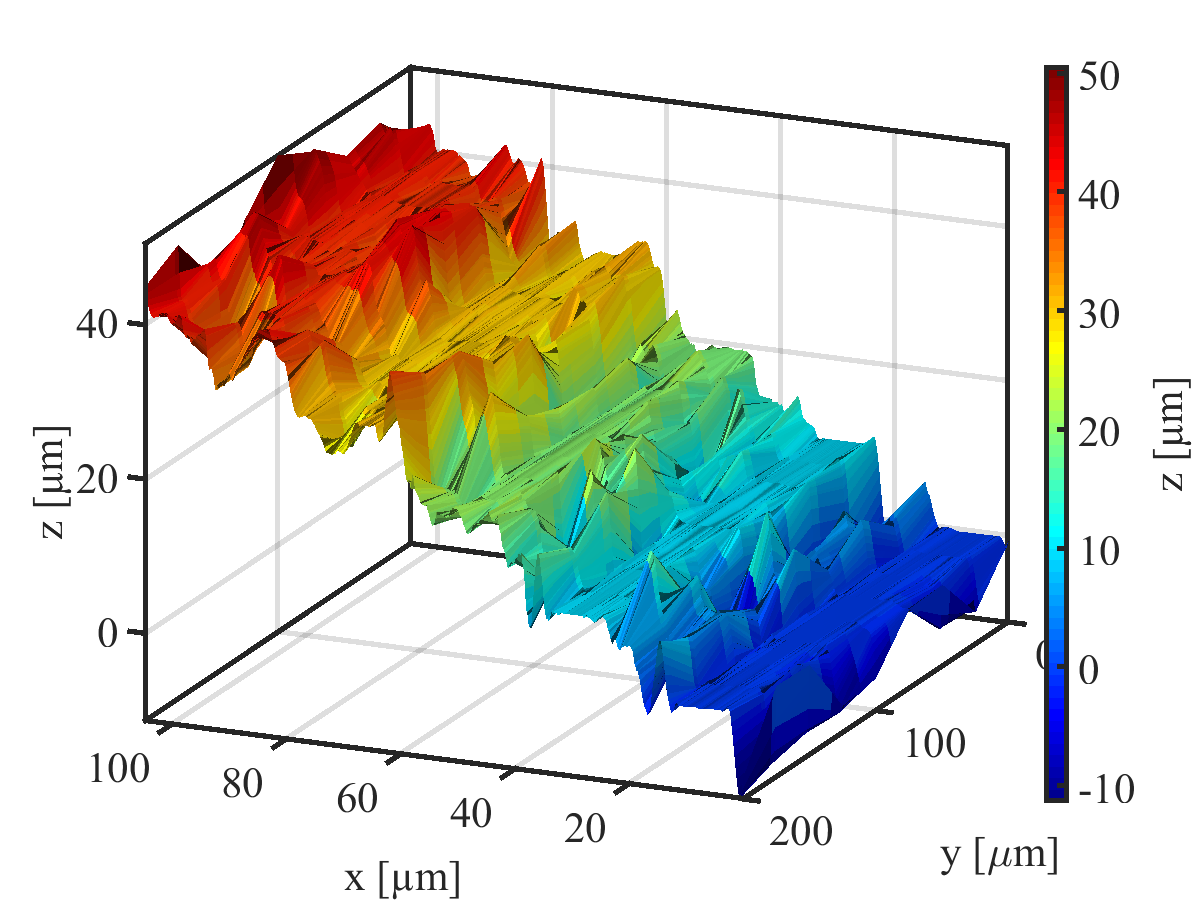}
				\put(1,1){\makebox(0,0){a)}}
				\put(64,45){\frame{\includegraphics[scale=.09]{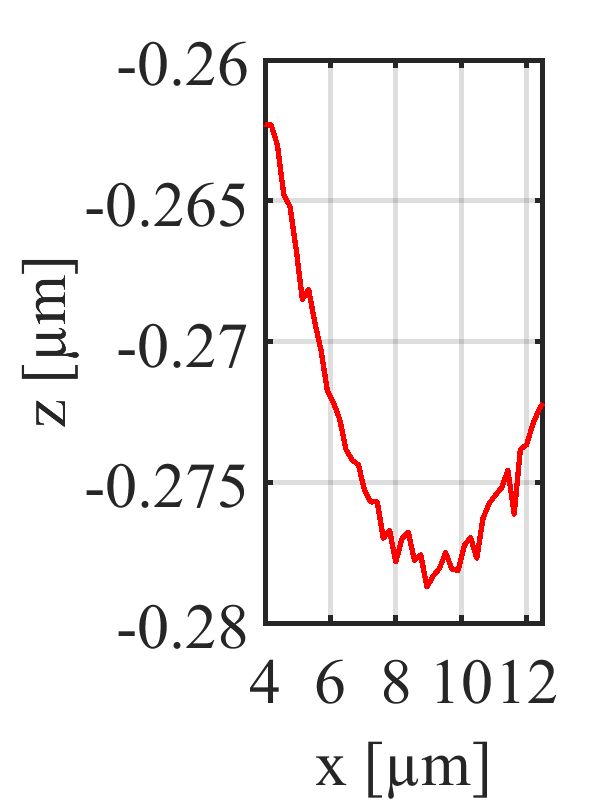}}}
				\put(80,45){\color{black}\vector(-1,-2){11}}
			\end{overpic}
			\begin{overpic}[scale=.23]{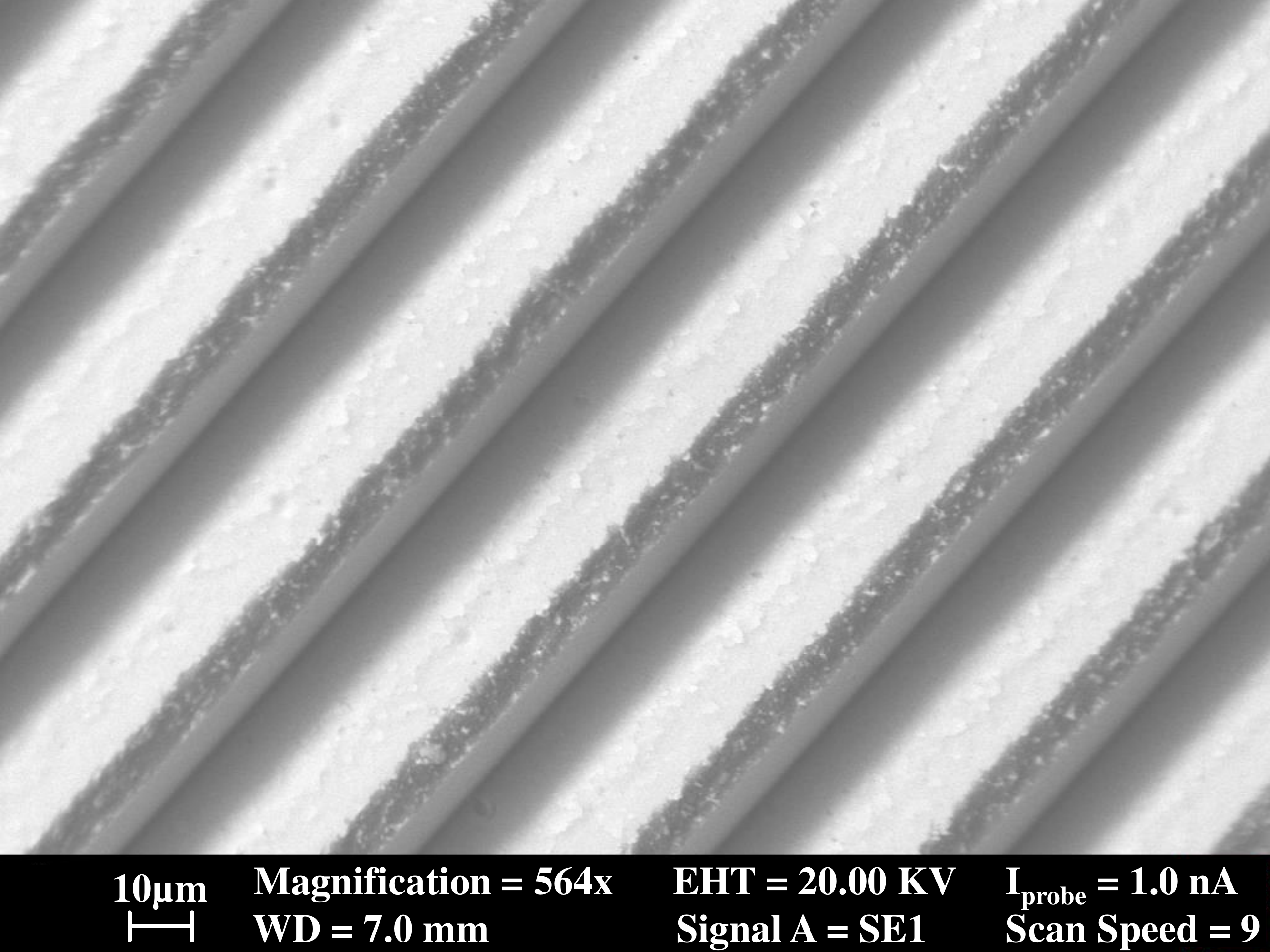}
				\put(0,1){\colorbox{white}{\textcolor{black}{b)}}}
			\end{overpic}
		\end{tabular}
	\caption[Result of the measured, three-dimensional surface of an echelle grating]{Result of the measured echelle grating with a) \gls{DE-LCI} measurement in Littrow configuration with five consecutive steps with a mean height of (9.66\,$\pm$\,0.4)\,\textmugreek m and an inset which shows the nm-fine structure of the first step at the position $y$\,=\,100\,\textmugreek m and b) \gls{SEM} image that captures the coarse structure of the individual step edges which are the reason for the noise in \gls{DE-LCI}.}\label{Profilo:Pic:Result_3D_echelle_grating}
\end{figure}

In this implemented approach, information was obtained by constantly translating the lens\,\textit{L1} in order to image different parts of the sample on the slit, as already depicted in Fig.\,\ref{ProfiloPic:BasicSetup}. In consequence, a stack of two-dimensional line profiles were gathered and analyzed in order to receive three-dimensional data. As the imaging lens was placed after the light of both interferometric arms was recombined and the sample was not moved between measurements, negative influences on the measurements were kept to a minimum.\\
Three-dimensional information of the precision-turned height standard used for the high-dynamic range evaluation, Fig.\,\ref{Profilo:Pic:result_2d_standard_mum}\,a), was gathered in steps of 25\,\textmugreek m along the $y$-direction and a rather small magnification to enable a lateral measurement range in the $x$-direction of 1.5\,mm, Fig.\,\ref{Profilo:Pic:Result_3D_standard_mum} b).
As noted before, see subsection \ref{Profilo:SubSec:HDR}, the high axial resolution leads to the ability to capture nanometer-sized roughness data on all height steps while maintaining a large axial measurement range of 79.91\,\textmugreek m. Furthermore, steps of (971.26\,$\pm$\,0.31), (4951.40\,$\pm$\,0.28) and (19924.00\,$\pm$\,0.36)\,nm were measured over an area of 1500\,x\,250\,\textmugreek m\textsuperscript{2} without the need for stitching, which distinguishes the approach clearly from other techniques such as confocal microscopy. In the current optical design, which utilizes an imaging system with a \gls{NA} of 0.06, a large lateral measurement range could be covered while data on the slopes with an 70\si{\degree} angle could not reliably be gathered. This is due to the comparatively low lateral resolution of 5\,\textmugreek m. Depending on the application, the setup can be optimized to increase the sensitivity on these parts of the sample. The results of this sample also highlight the capability of the approach to decouple the axial resolution from the lateral measurement range as nm features can be detected while measuring over a range of 1.5\,mm.\\
For comparison, the height standard was also analyzed using a confocal microscope (Smartproof 5, Carl Zeiss Microscopy GmbH, G\"ottingen, Germany), Fig.\,\ref{Profilo:Pic:result_pizza_CLSM}\,a).
It has to be noted that the capturing of a profile having the same length as the \gls{DE-LCI} measurement relied on stitching of multiple images, as a magnification of 20x with an lateral field of view of 450\,x\,450\,\textmugreek m\textsuperscript{2} was necessary to achieve a comparable resolution. This approach is not only time consuming (approx. 10 minutes per areal image) but also tends to be prone to errors as stitching inconsistencies exist, see inset Fig.\,\ref{Profilo:Pic:result_pizza_CLSM}\,b). It can be seen, that these errors on the nm-scale influence the representation of the surface topography in the stitched regions which in turn can have an effect on quantitative analysis. The roughness distribution in these areas is no longer a Gaussian one, which makes it unusable e.g. for roughness evaluation, Fig.\,\ref{Profilo:Pic:result_pizza_CLSM}\,c) and d). Furthermore, the \gls{RMS} value of the profile data was significantly larger compared to the other methods. This fact is usually addressed during roughness evaluation by filtering the signal with an appropriate low-pass filter, known as micro-roughness filtering, \cite{ISO3274}.\\
Further three-dimensional evaluation was performed by analyzing a commercially available echelle grating in the Littrow configuration, Fig.\,\ref{Profilo:Pic:Result_3D_echelle_grating}\,a).
A total of five facets of the grating could be imaged in a lateral range of 100\,x\,200\,\textmugreek m\textsuperscript{2}, having a mean height of (9.66\,$\pm$\,0.40)\,\textmugreek m. The data on the edges of the steps is notably noisy. A \gls{SEM} scan of the grating was performed to examine individual steps as a reference, Fig.\,\ref{Profilo:Pic:Result_3D_echelle_grating}\,b). From this image it can be seen that each edge has a very coarse structure in the size of about 6\,\textmugreek m. These lead to very low \glspl{SNR} during the \gls{DE-LCI} measurements which are the reason that a larger area of about 6.75\,\textmugreek m is obstructed on each side while only about 5.5\,\textmugreek m of the plateaus are visible. Apart from this, the center of the plateaus could be resolved clearly with sub-nm surface structures, see inset Fig.\,\ref{Profilo:Pic:Result_3D_echelle_grating}\,a).

\subsection{Alternative spectral encoding for areal measurements}
In order to gather full areal surface profile data, hence three-dimensional information, without any need for scanning two alternative approaches have been developed.
\subsubsection{Multi-slit approach}
\begin{figure}[h]
\centering
		\begin{tabular}{c}
			\begin{overpic}[scale=.47]{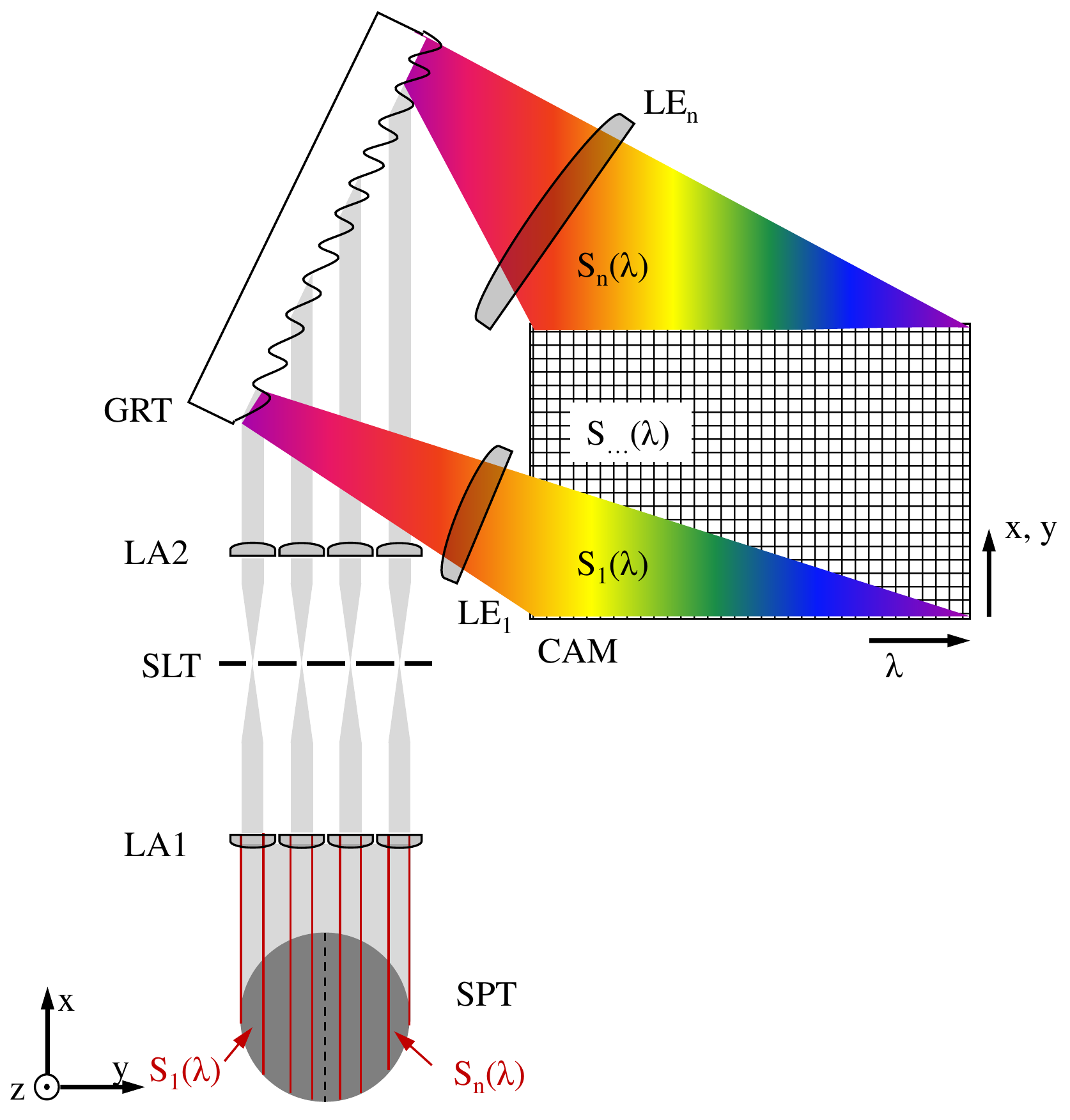}
				\put(-1,0){\makebox(0,0){a)}}
			\end{overpic}
			\begin{overpic}[scale=.47]{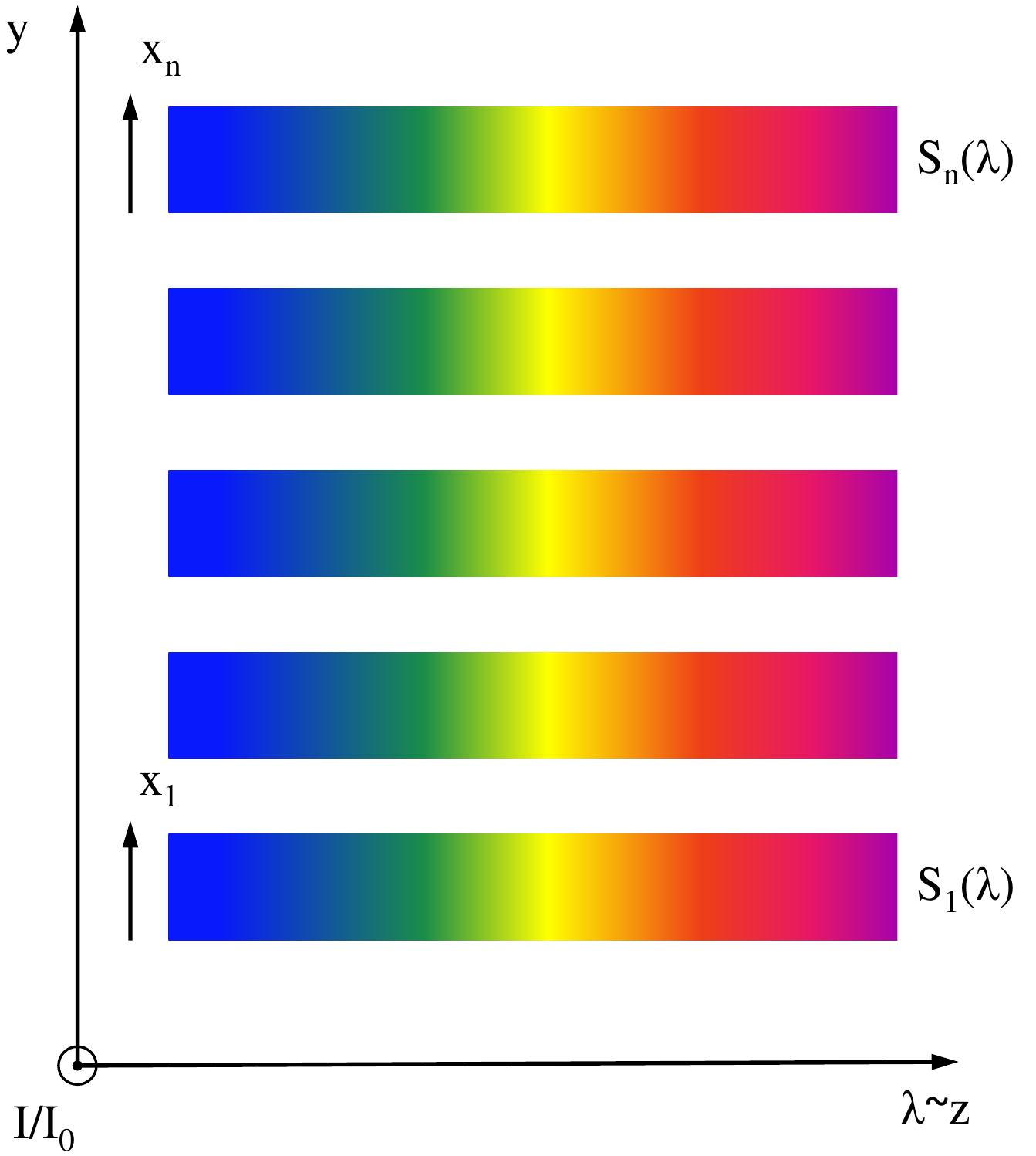}
				\put(-2,0){\makebox(0,0){b)}}
			\end{overpic}
		\end{tabular}
	\caption[Schematic representation of a modified imaging spectrometer for three-dimensional encoding variant 1]{a) Schematic representation of a modified imaging spectrometer for three-dimensional encoding with SPT - measurement spot which is spatially segmented by an LA1 - lens array onto the SLT - multi-slit arrangement. A second LA2 - lens array images the slits onto the grating which spectrally decomposes the light of every facette of LA2 while individual LE\textsubscript{n} - imaging elements image the components on the CAM - camera as well as b) simulation of a signal on the CAM where the axial information on the profile height $z$ is spectrally encoded in the cameras $x$-dimension, individual spectral slices $S_n(\lambda)$ are encoding information of one lateral dimension $x_n$ while the combination of these slices hold information of the second lateral dimension $y$.}\label{Profilo:Pic:SpecEncoding_setup1}
\end{figure}

In this approach, the measurement spot is spatially expanded and the single slit is substituted with a set of parallel slits in order to make use of a large area of the grating in the imaging spectrometer, Fig.\,\ref{Profilo:Pic:SpecEncoding_setup1}.
This arrangement allows the decomposition of the measured spot both spatially and spectrally. In consequence, several spatial parts of the measurement spot can be analyzed in the same fashion as described in Section \ref{Profilo:Sec:2D_approach} while the individual spectral slices are stacked in the y-dimension of the spectrometer, Fig.\,\ref{Profilo:Pic:SpecEncoding_setup1}\,b).\\
In a practical realization, the multi-slit approach can be implemented by using imaging fibers in a linear arrangement in order to simplify the setup and to avoid diffraction effects from tight fitted slits. The approach holds the potential for measurements with large axial resolution while the lateral resolution of both dimensions is dependent on the physical size of the camera used.

\subsubsection{Spatial combiner approach} 
\begin{figure}[h]
		\centering
		\begin{tabular}{c}
			\begin{overpic}[scale=.45]{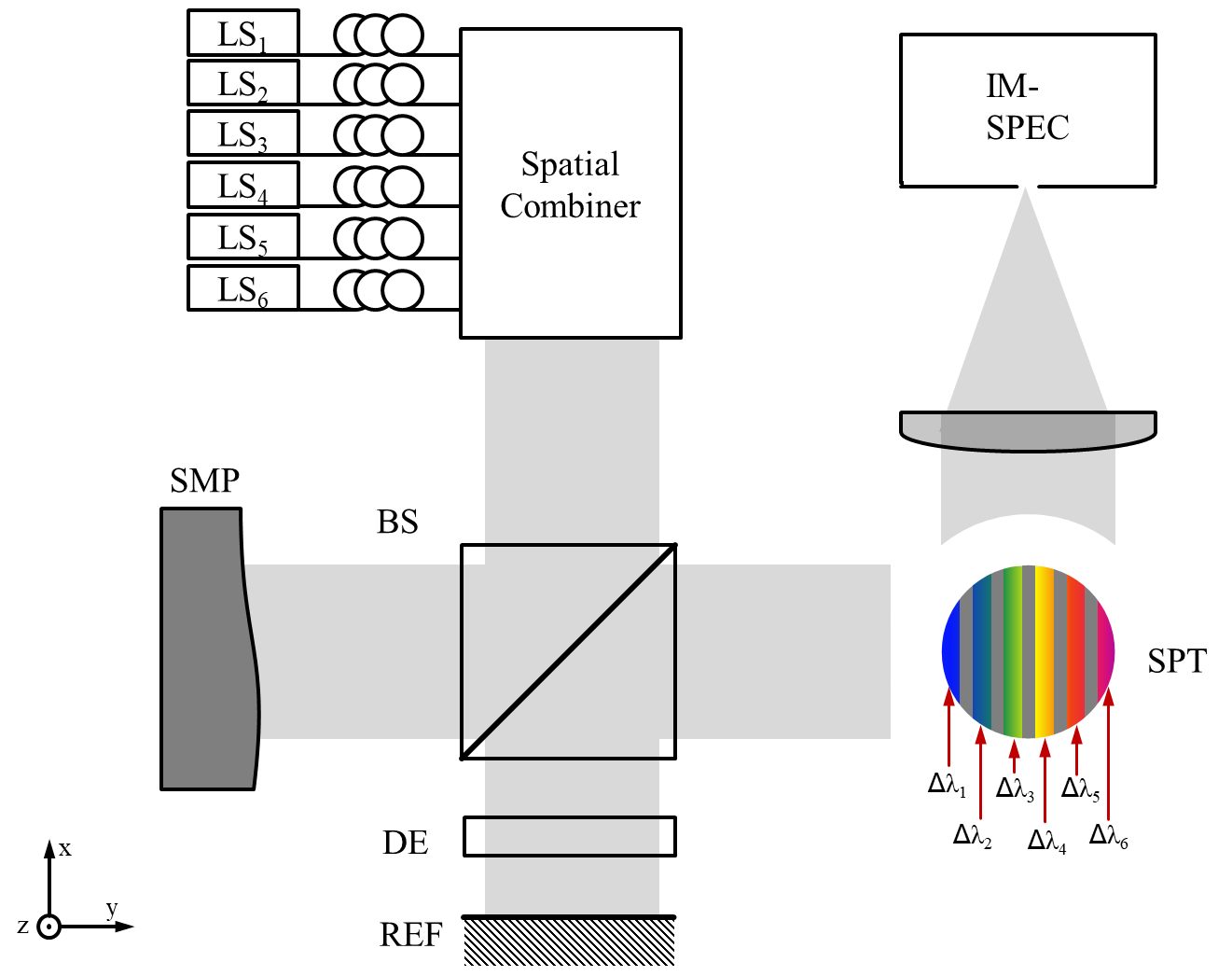}
			\end{overpic}
		\end{tabular}
	\caption[Schema of a modified imaging spectrometer for three-dimensional encoding variant 2]{Schema of a modified imaging spectrometer for three-dimensional encoding with $LS_n$ - low-coherent light sources which are coupled to the interferometer by a spatial combiner where a BS - beamsplitter delivers appropriate beams to the SMP - sample and REF - reference mirror where it is also manipulated by the DE - dispersive element. Finally, the SPT - measurement spot which consists spatially separated spectral ranges $\Delta \lambda_n$ which are imaged onto and analyzed by the IMSPEC - imaging spectrometer.}\label{Profilo:Pic:SpecEncoding_setup2}
\end{figure}

In a further approach, the encoding of a second lateral dimension is performed in the spectral domain. For this purpose, the measurement spot is composed of different spatial components which inhibit individual, discrete spectral ranges. These spectral slices, which are formed with low coherent light sources, are used to illuminate discrete regions of the sample, Fig.\,\ref{Profilo:Pic:SpecEncoding_setup2}.
The data evaluation in this approach is similar to the conventional, two-dimensional analysis. The $x$-dimension of the camera acquires spectral information while the $y$-dimension stores information of one lateral dimension. However, the data of certain spectral slices with ranges of $\Delta \lambda_1$ to $\Delta \lambda_i$ corresponds to the height of the respective lateral dimension $x_n$, so that each spectral slice has to be analyzed separately.
The resolution in the axial as well as in the lateral domain are controlled by the size of the spectral slices, the detector size as well as the imaging magnification of the measurement spot.

\chapter{Polymer characterization}\label{ChapterPolymer}
As discussed in section \ref{StateOfTheArtPolymer}, the characterization of cross-linking and especially its spatial distribution is crucial for the fabrication of micro optics, MEMS, and semiconductors. Some studies have found that the measurement of the refractive index over the cross-linking process can be used as an indicator for the degree of cross-linking of a sample. According to Kudo et al., \cite{Kudo}, the cross-linking of a polymeric material leads to a densification which can be directly related to an increase in refractive index using the Lorentz-Lorenz equation. The characterization of the refractive index is therefore a suitable measure for the degree of cross-linking, \cite{3D_laser_writing}.\\
Based on the spectral interferometric approach utilized for surface profilometry in section \ref{Profilo:Sec:Meas_range} of this work, a characterization method was developed, tested and evaluated. It is based on the fact that the wavelength-dependent refractive index \glssymbol{RefrIndSample} of a sample in correspondence with its thickness \glssymbol{SampThickness} determines the spectral output of a low-coherence interferometer, \cite{Taudt,Taudt2}. In contrast to dispersion-encoded profilometry, in cross-linking characterization, the sample itself is the dispersive element. Hence, the dispersion characteristic of \glssymbol{RefrInd} is the unknown quantity.\\
In the most basic configuration a two-beam interferometer with spectral detection is used to analyze a sample, Fig.\,\ref{poylmer_temporal_approach_simple_setup}\,a). In this case, the signal at the spectrometer $I(\lambda)$ can be described as an adaptation of Eq.\,(\ref{Profilo:EQ:2d_int_equation}),
\begin{eqnarray}
&& I(\lambda) = I_0(\lambda) \cdot \left[ 1 + cos \varphi(\lambda)  \right] \label{cross-linking_interferometer_equation}\\
&& \textrm{with } \varphi = 2\pi \frac{ \left[ n^{smp}(\lambda)  - 1 \right]t_{smp} -\delta }{\lambda}\label{phase_cross-linking_interferometer_equation},
\end{eqnarray}
where $I_0(\lambda)$ is the spectral profile of the light source and  $\varphi$ the phase. In the assumed simple case, the thickness of the sample $t_{smp}$ is a constant whereas the path difference $\delta$ can be altered with a translation stage.

\section{Temporal approach}\label{subsection_temporal_approach}
\begin{figure}[h]
	\begin{center}
		\begin{tabular}{c}
			\begin{overpic}[scale=0.62]{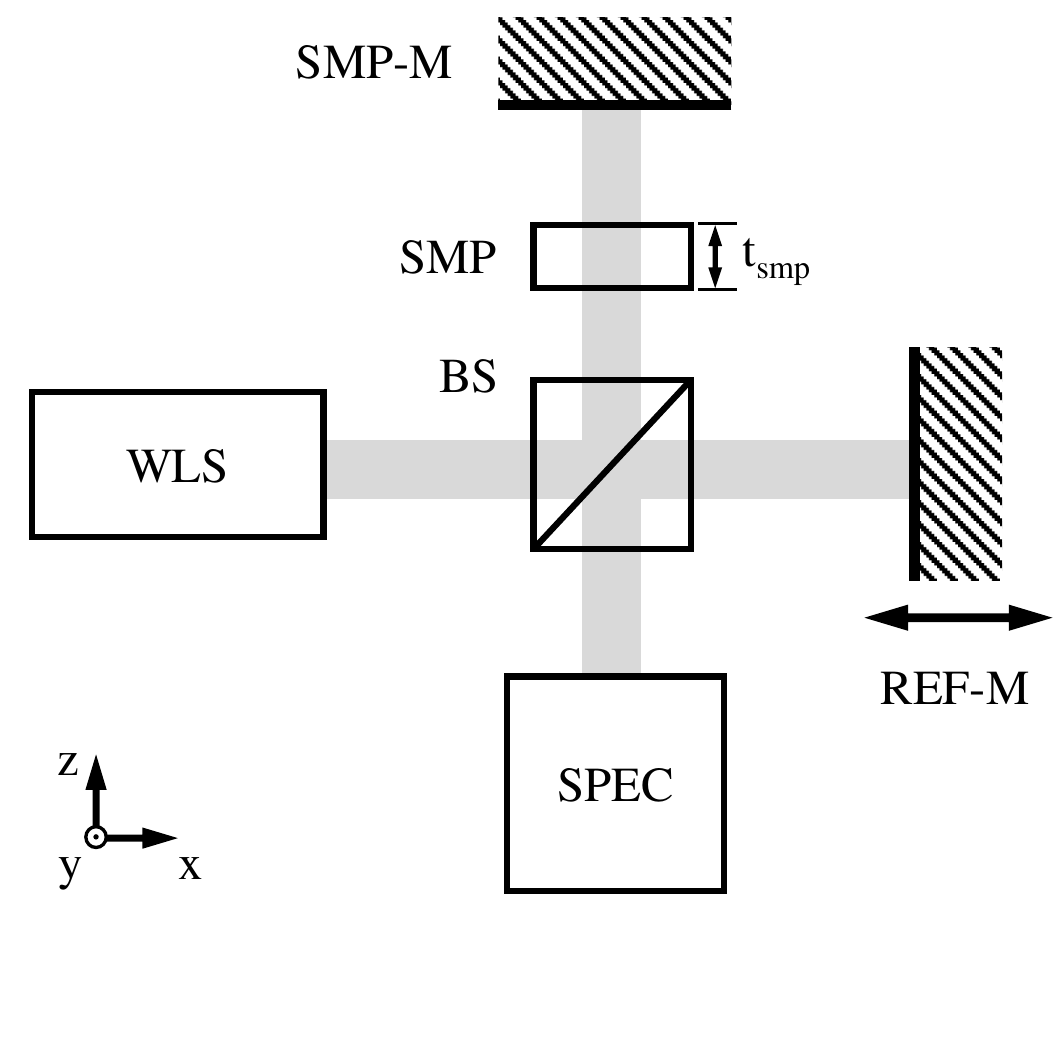}
				\put(1,1){\makebox(0,0){a)}}
			\end{overpic}  
			\begin{overpic}[scale=0.32]{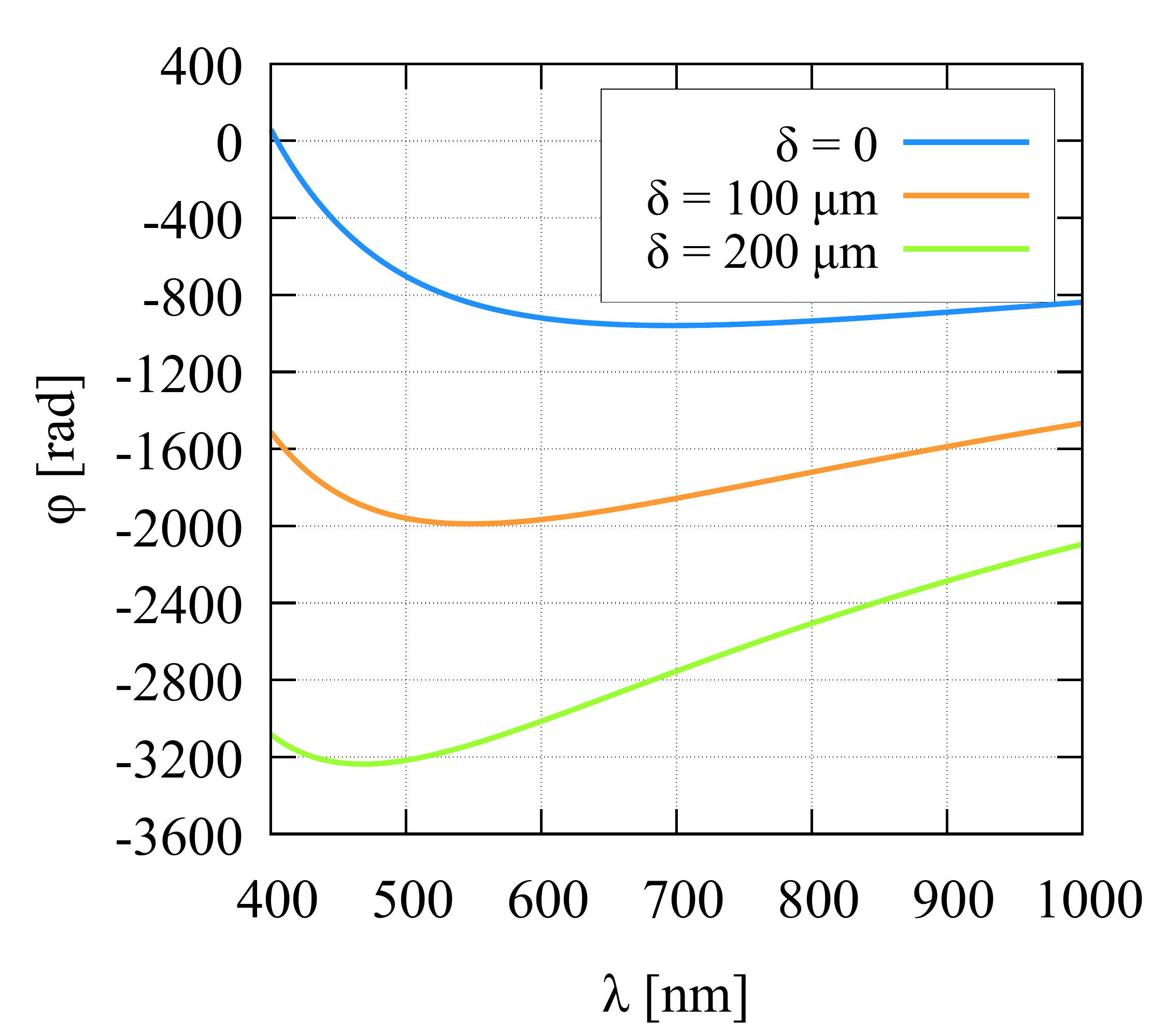}
				\put(1,1){\makebox(0,0){b)}}
			\end{overpic}
		\end{tabular}
	\end{center}
	\caption[Simple setup for the characterization of transmissive samples with regard to the refractive index]{a) Simple setup for the characterization of transmissive samples where a WLS - white light source is splitted by a BS - beam splitter so that in one arm the light transmits through the SMP - sample with the thickness $t_{smp}$ before and after it is reflected from SMP-M - sample mirror while in the second arm the light is reflected from REF-M - reference mirror which can be adjusted in the x-dimension; both signals are analyzed after recombination using a SPEC - spectrometer. b) Simulated, spectrally-resolved phase data from such a setup with a sample of N-BK7 with a thickness of 5 mm for a number of different delays of the reference arm noted in $\delta$.}\label{poylmer_temporal_approach_simple_setup}
\end{figure}

\begin{figure}[h]
	\begin{center}
		\begin{tabular}{c}
			\begin{overpic}[scale=0.31]{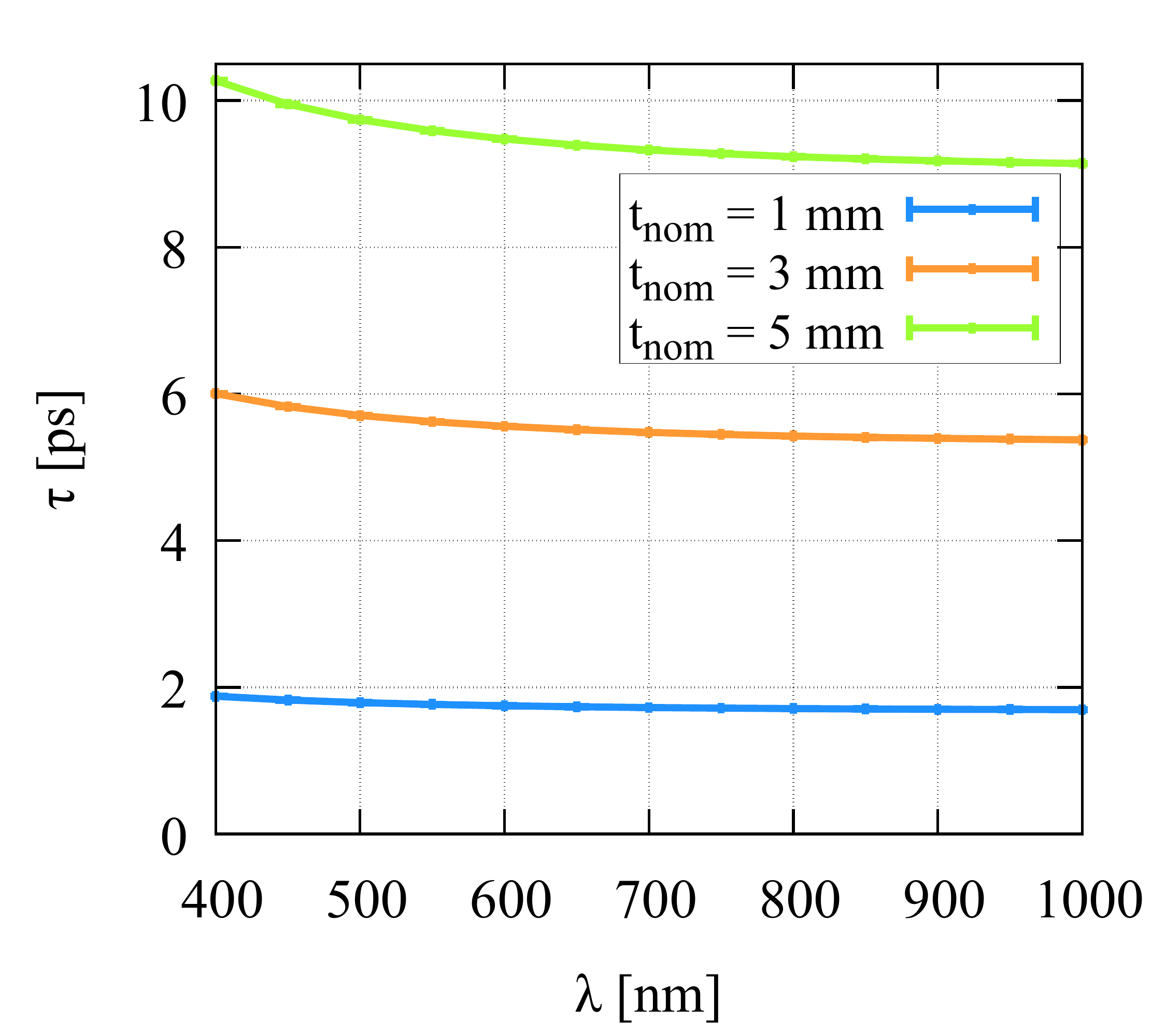}
				\put(1,1){\makebox(0,0){a)}}
			\end{overpic}  
			\begin{overpic}[scale=0.31]{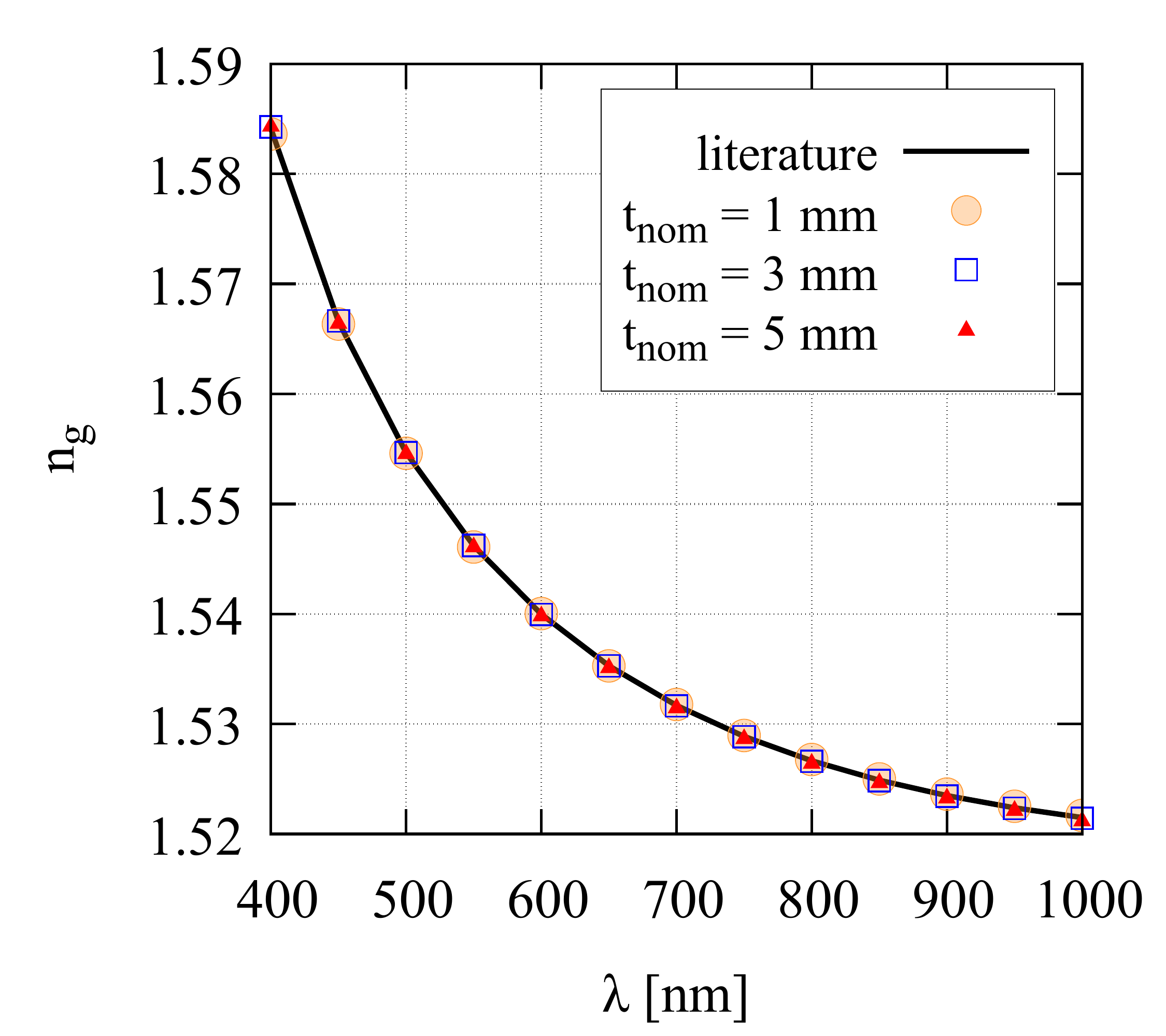}
				\put(1,1){\makebox(0,0){b)}}
			\end{overpic}
		\end{tabular}
	\end{center}
	\caption[Measured temporal delays for N-BK7 samples]{a) Measured temporal delays for N-BK7 samples of 1,3 and 5\,mm nominal thickness respectively in the spectral range from (400\,-\,1000)\,nm and b) calculated group refractive indices for the three nominal thicknesses in relation to the values known from literature according to \cite{SchottKatalog}.}\label{poylmer_temporal_approach_results_delay_nG}
\end{figure}

\begin{figure}[h]
	\begin{center}
		\begin{tabular}{c}
			\begin{overpic}[scale=0.32]{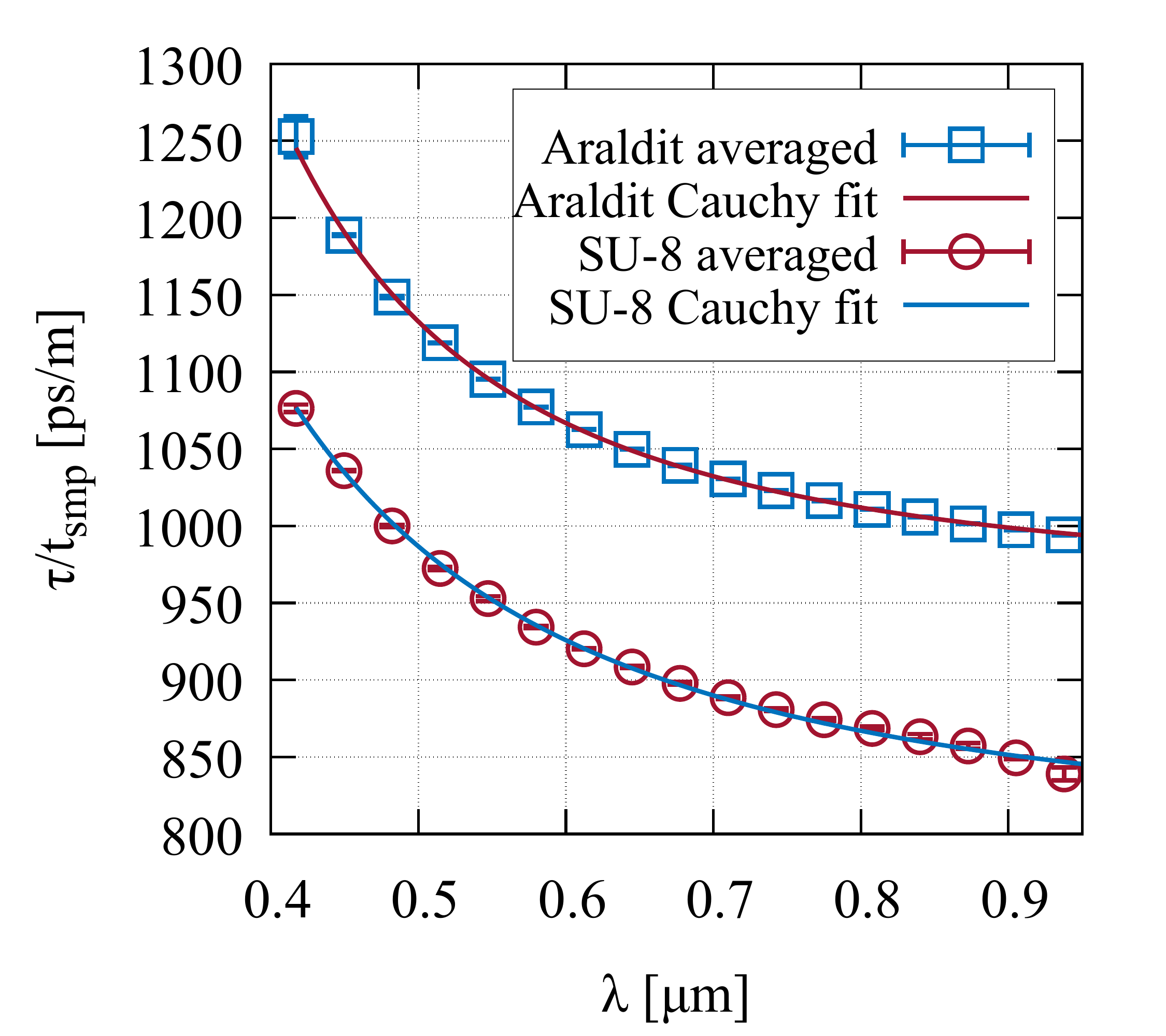}
				\put(1,1){\makebox(0,0){a)}}
			\end{overpic}
			\begin{overpic}[scale=0.32]{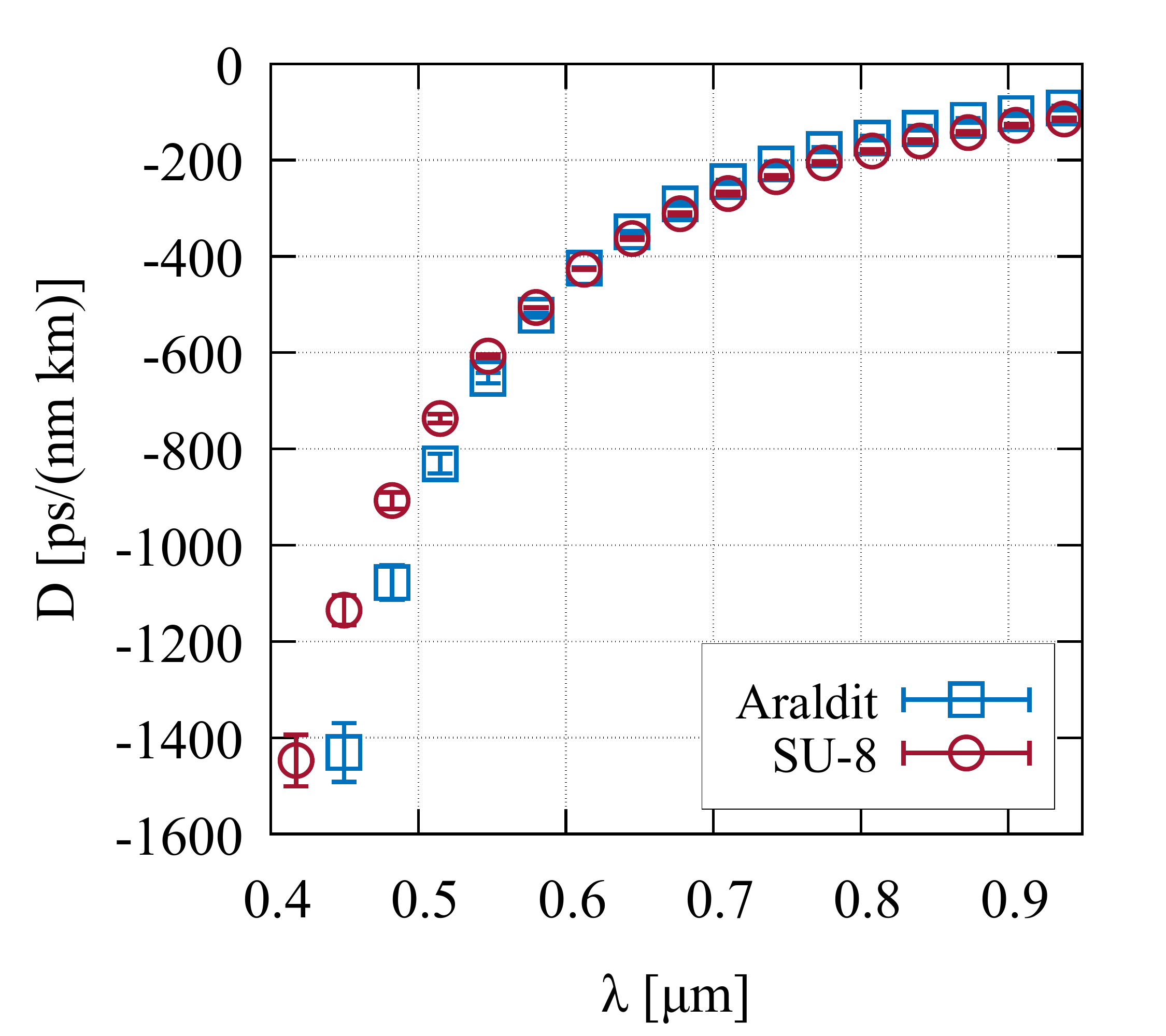}
				\put(1,1){\makebox(0,0){b)}}
			\end{overpic}
		
		\end{tabular}
	\end{center}
	\caption[Results of the temporal approach measurements for Araldite and SU-8 samples]{Results of the temporal approach measurements for Araldite and \gls{SU-8} samples with a) recorded delays due to dispersion and corresponding \textsc{Cauchy} fitted curves as well as b) calculated dispersion $D$ from fitted curves.}\label{poylmer_temporal_approach_results}
\end{figure}

\begin{figure}[h]
	\begin{center}
		\begin{tabular}{c}
			\begin{overpic}[scale=0.32]{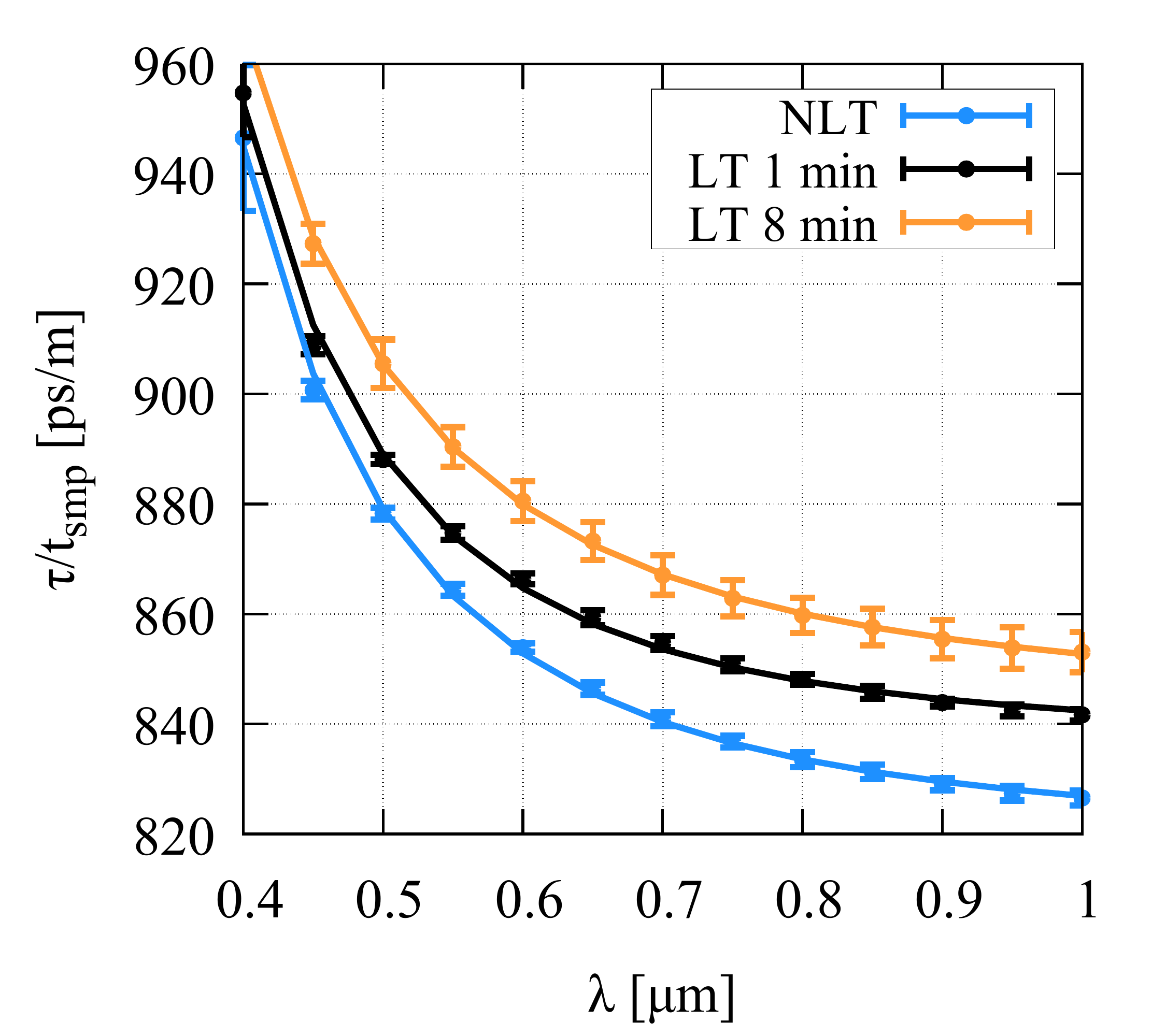}
				\put(1,1){\makebox(0,0){a)}}
			\end{overpic}
			\begin{overpic}[scale=0.32]{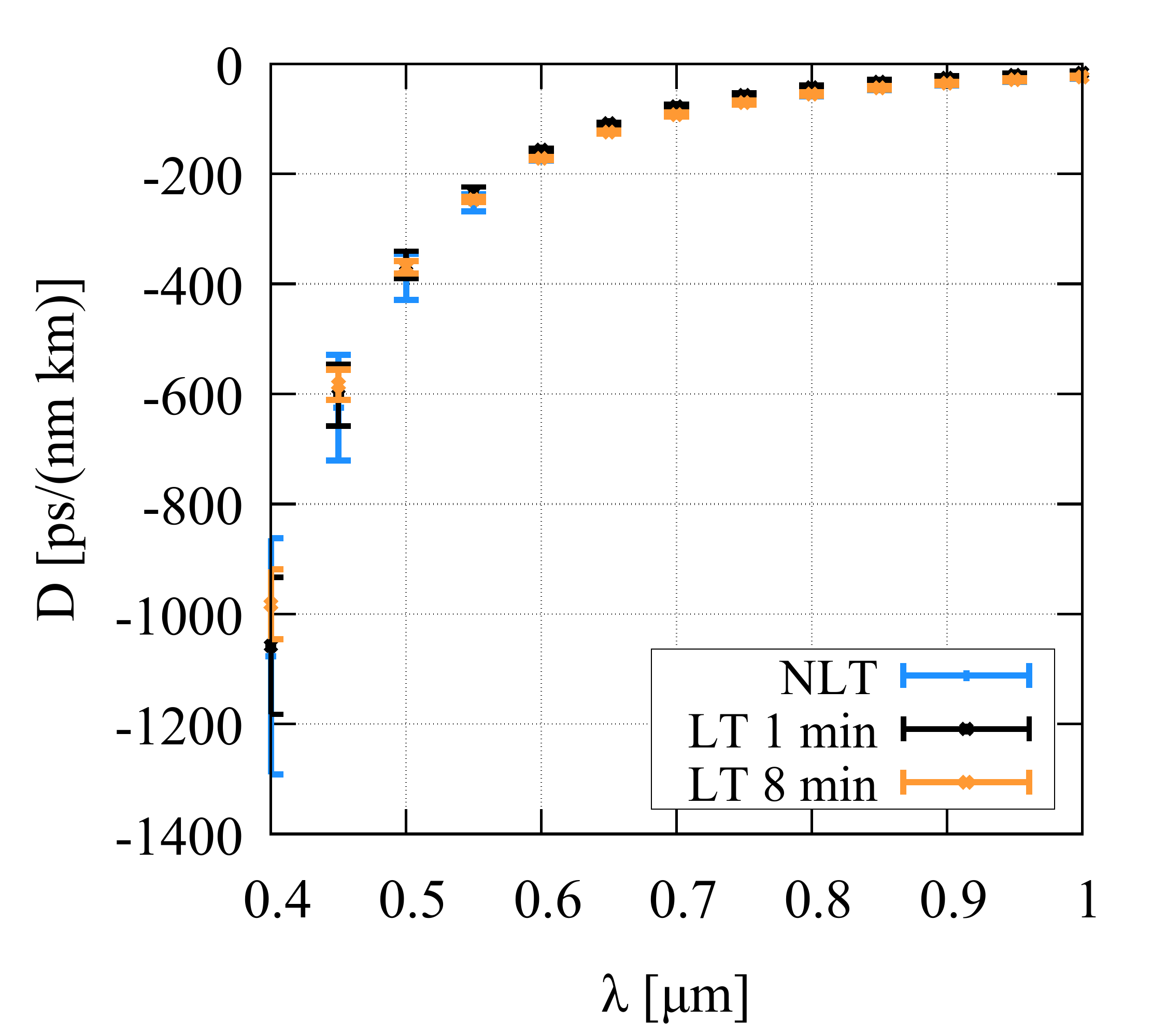}
				\put(1,1){\makebox(0,0){b)}}
			\end{overpic}
		\end{tabular}
	\end{center}
	\caption[Results for the measured temporal delays in relation to the lamination time of EVA samples]{a) Results for the measured temporal delays $\tau$ due to the different lamination times normalized to the material thickness $t_{smp}$ for three representative lamination times with NLT - no lamination, LT 1 min - 1 minute and LT 8 min - 8 minutes lamination and the corresponding fits using a \textsc{Cauchy} model according to Eq.(\ref{Poly:EQ_Cauchy}) and b) derivation of the temporal delay relative to the wavelength according to Eq. (\ref{Poly:EQ_DispersionParam}).}\label{result}
\end{figure}

\begin{figure}[h]
	\begin{center}
		\begin{tabular}{c}
			\includegraphics[origin=c,scale=.32]{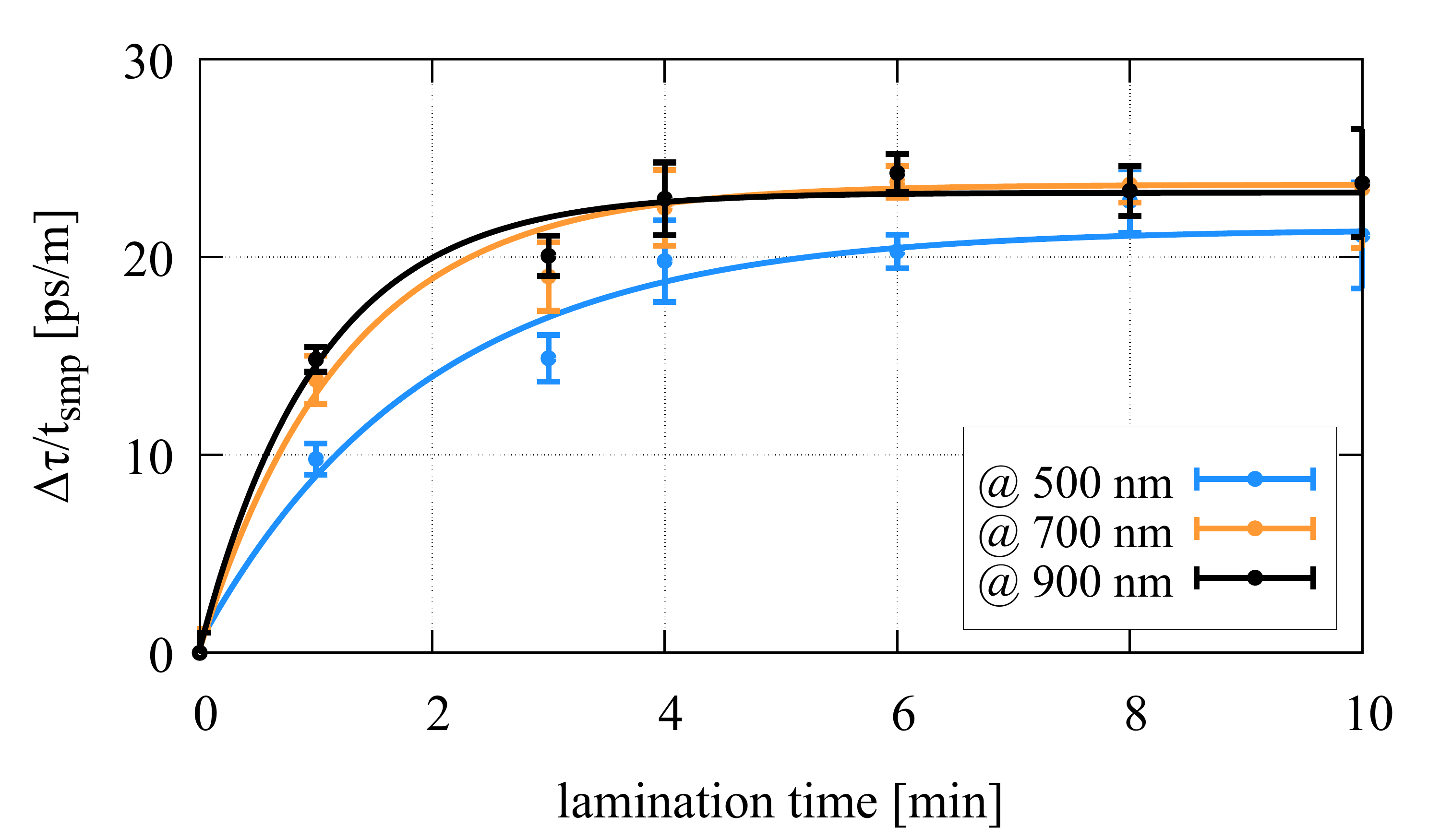}
		\end{tabular}
	\end{center}
	\caption{Plot of the differences in $\Delta \tau/t_{smp}$ over the lamination time selected probing wavelengths.}\label{result_lam_time}
\end{figure}

A temporal approach of the aforementioned two-beam interferometer was realized in a Michelson configuration where the sample is a transmissive part of one interferometric arm while the reference arm mirror is translatable to achieve temporal control, Fig.\,\ref{poylmer_temporal_approach_simple_setup}\,a).
According to Eq.\,(\ref{cross-linking_interferometer_equation}), the signal depends on the wavelength-dependent refractive index $n^{smp}(\lambda)$, the sample thickness $t_{smp}$ and the path difference between the arms $\delta$. As $n^{smp}(\lambda)$ and $t_{smp}$ are material constants for a supposed bulk material, the path difference is the only variable which can be used to evaluate the refractive index. A variation of the path difference leads to a deformation of the phase and most notably, to a shift of the phase minimum according to Eq. (\ref{cross-linking_interferometer_equation}), Fig. \ref{poylmer_temporal_approach_simple_setup} b). The minimum is described by its wavelength, the equalization wavelength  $\lambda_{eq}$ and can be tracked as a function of the path difference $\delta(\lambda)$ or the temporal delay \glssymbol{DelayInterfero}. This information can be used to calculate the group refractive index of the material using 
\begin{equation}\label{EQgroupIndex_temporal}
n_g^{smp}(\lambda) = \frac{\delta(\lambda)}{t_{smp}} = \frac{\tau(\lambda) \cdot c}{t_{smp}}.
\end{equation}
The measurement can be performed either in a relative or absolute way. For relative measurements, the delay introduced relative to a starting position (noted with $\lambda_{eq}^0$) is used to calculate the relative group refractive index $\Delta n_g^{smp}(\lambda)$. Absolute values of $n_g^{smp}(\lambda)$ can be obtained, if the delay is referenced to the stationary phase point of the interferometer in a dispersion-free status.\\
In an initial experiment, a reference mirror was placed onto a precision stage which was then used to introduce defined delays to the signal in form of path differences $\delta_n$. By tuning the delay to reach a certain equalization wavelength $\lambda_{eq}^n$, repeated measurements to calculate $n_g^{smp}(\lambda)$ were possible. In order to quantify the method, measurements on a set of samples of N-BK7 glass with nominal thicknesses \glssymbol{NomThickness} of 1, 3 and 5 mm were performed, Fig. \ref{poylmer_temporal_approach_results_delay_nG} a).
In an experiment with ten measurements per nominal thickness it was found that the mean standard deviation over all equalization wavelengths for $t_{nom}$\,=\,1\,mm was $\overline{\sigma_1(\tau)}$\,=\,\num{1.86e-3}\,ps, for $t_{nom}$\,=\,3\,mm was $\overline{\sigma_3(\tau)}$\,=\,\num{1.99e-3}\,ps and for $t_{nom}$\,=\,5\,mm was $\overline{\sigma_5(\tau)}$\,=\,\num{3.75e-3}\,ps. The mean values of all three delay slopes were used to calculate the group refractive index according to Eq. (\ref{EQgroupIndex_temporal}), Fig. \ref{poylmer_temporal_approach_results_delay_nG} b). The errors of the measured data have been calculated for every equalization wavelength relative to the respective literature values for N-BK7, \cite{SchottKatalog}. The root-mean-square error for the calculated mean group refractive index of the nominal thickness $t_{nom}$\,=\,1\,mm was  $\Delta n_{g1}$\,=\,\num{2.08e-4}, $\Delta n_{g3}$\,=\,\num{6.54e-5}  for $t_{nom}$\,=\,3\,mm and $\Delta n_{g5}$\,=\,\num{1.45e-4} for $t_{nom}$\,=\,5\,mm. The mean standard deviation over all equalization wavelengths was  $\overline{\sigma_1(n_g)}$ = \num{5.72e-4}, $\overline{\sigma_3(n_g)}$ = \num{1.93e-4} and $\overline{\sigma_5(n_g)}$ = \num{2.13e-4} for the three nominal sample thicknesses. As the delay was acquired as primary information, it was used in conjunction with the thickness of the sample in order to calculate the refractive index for each  equalization wavelength. This calculation is not reliant on any knowledge about the underlying material model. In case of a known material composition, a model can be chosen in order to fit the measured data. For the evaluation data of N-BK7, a fit using the basic Sellmeier equation was performed using the parameters \glssymbol{SellmeierCoeffs},
\begin{equation}
n^2 (\lambda) = 1 + \frac{B_1 \lambda^2}{\lambda^2 - C_1} +\frac{B_2 \lambda^2}{\lambda^2 - C_2} + \frac{B_3 \lambda^2}{\lambda^2 - C_3}.
\end{equation}
It was found that the root-mean-square error of the fitted data in relation to the refractive index reported in literature was $\Delta n_{g1}^{fit}$\,=\,\num{2.71e-4}, $\Delta n_{g3}^{fit}$\,=\,\num{2.06e-4} and $\Delta n_{g5}^{fit}$\,=\,\num{2.30e-4} for the three nominal sample thicknesses. These results prove the accuracy of the method to determine the refractive index of transmissive samples which in turn can be utilized to characterize cross-linking of polymers as typical cross-linking differences in the range of $\Delta n$\,=\,0.001\,-\,0.02 are expected, \cite{2Photon_waveguides}. An error propagation for the temporal approach was performed and documented in section \ref{SubSec:ErrorPropTemporal}.\\
For further evaluation, two epoxy-based samples in different states of cross-linking have been investigated with the described method. Both samples were distinctively different in their appearance as one sample was made of bulk material (Araldite epoxy, $t_{smp}^A$ 3.04 mm) while the second one was a thin-layered \gls{SU-8} ($t_{smp}^{SU8}$\,=\,0.23\,mm). The properties of both samples have been studied using the setup described in Fig. \ref{poylmer_temporal_approach_simple_setup} a) over a spectral range from (0.4\,-\,1)\,\textmugreek m. The dispersion related temporal delay was recorded and referenced to the sample thickness for comparison, Fig. \ref{poylmer_temporal_approach_results} a).
The materials show significant differences in their refraction behavior over the spectral range. While Araldite shows a delay ranging from 1000\,ps/m (@900\,nm) to over 1200\,ps/m (@450\,nm), the \gls{SU-8} sample just causes delays of 850\,ps/m to 1050\,ps/m at the same spectral points. Both data sets could be well fitted with a \textsc{Cauchy} model to approximate their behavior using the parameters \glssymbol{CauchyCoeffs}, \cite{Sultanova}
\begin{equation}
\tau(\lambda) = A_1 + \frac{A_2}{\lambda^2}+\frac{A_3}{\lambda^4}
\label{Poly:EQ_Cauchy}
\end{equation}
The difference in the relative temporal delay is an indicator of the different degrees of cross-linking. The first derivative of the temporal delay $\tau$ in relation to the wavelength $\lambda$, was used to calculate the corresponding dispersion \glssymbol{DispParameter}, Fig. \ref{poylmer_temporal_approach_results} b),
\begin{equation}
D(\lambda) = \frac{\partial\tau(\lambda)}{\partial\lambda \cdot t_{smp}} = - \frac{1}{t_{smp}} \left(\frac{6A_2}{\lambda^3}+\frac{20A_3}{\lambda^5}\right).
\label{Poly:EQ_DispersionParam}
\end{equation}
This representation of the measured optical properties focuses on the slope of the dispersion induced delay. The differences between both materials are still evident but much smaller than the relative delay. Especially for shorter wavelengths, the differences are significant which indicates a different absorption behavior, hence a different molecular composition. This results indicates that the two materials are similar in their composition, but show differences in cross-linking. In the spectral region below 600\,nm a higher dispersion of \gls{SU-8} is visible. A reason for that might be the different mechanism of the cross-linking initiation. While the polymerization of \gls{SU-8} is initiated by UV-light radiation, the polymerization of Araldite is chemically initiated. In the wavelength range from 600\,nm to 750\,nm the dispersion of both materials is approximately the same. In the near infrared region, Araldite shows a slightly higher dispersion. Both curves are significantly different in terms of the measured standard deviation.\\ 
In order to support the observations made in temporal low-coherence interferometry, an additional material characterization were undertaken by \textsc{Vickers} micro hardness tests. The results were compared to the obtained optical properties, Tab. \ref{tab:parameters}.
\begin{table}[h]
	\caption[Conclusive representation of key properties measured for polymeric samples]{Conclusive representation of key properties measured for both polymeric samples with the \textsc{Cauchy} coefficients, the micro hardness and the sample thickness.} \label{tab:parameters}
	\begin{center}       
		\begin{tabular}{ccc} 
				\hline
			\rule[-1ex]{0pt}{3.5ex}  & Araldite & SU-8  \\
			\hline
			\rule[-1ex]{0pt}{3.5ex}  \textsc{Cauchy} coefficients & & \\
			
			\rule[-1ex]{0pt}{3.5ex}  $A_1 [\frac{ps}{m}]$ & $956.9 \pm 3.1$ & $793.9 \pm 2.7$   \\
			\rule[-1ex]{0pt}{3.5ex}  $A_2 [\frac{ps \mu m^2}{m}]$ & $9.9 \pm 1.0$ & $15.3 \pm 0.7$  \\
			\rule[-1ex]{0pt}{3.5ex}  $A_3 [\frac{ps\mu m^4}{m}]$ & $0.7\pm 0.1$ & $0.1 \pm 0.1$  \\
			\hline
			\rule[-1ex]{0pt}{3.5ex}  hardness $[HV]$ & $1.40 \pm 0.01$ & $1.57 \pm 0.07$  \\
			\rule[-1ex]{0pt}{3.5ex}  thickness $t_{smp}$  $[mm]$& $3.04 \pm 0.01$ & $0.23 \pm 0.02$  \\
			\hline
		\end{tabular}
	\end{center}
\end{table} 
Hardness measurements have been performed on a micro indenter device operating with a force of 0.245\,N in 5 indentations per sample. The micro hardness tests support the measurements of the optical properties as only minor differences are to be noted. This corresponds with the fact that both materials have a similar chemical composition. The slight differences therefore have to be caused in a difference of the cross-linking. \gls{SU-8} on the one hand is harder than Araldite but on the other hand constantly shows lower refraction over the recorded spectral range. An effect similar to that has been shown in literature, where the refractive index of \gls{SU-8} was determined at different baking steps, \cite{parida2009}. Here, the samples which were subjected to longer baking have shown higher cross-linkage as well as lower overall refraction.\\
However, it should be noted that the experimental approaches for the mechanical and optical properties are fundamentally different. Since micro hardness is measured at distinct points on the samples' surfaces, it can be heavily influenced by residual stresses and local inhomogeneities. In contrast to that, the dispersion measurements integrate the properties spatially over the sample's thickness as well as over the cross-section (dependent on the spot size of the light setup). Therefore, the hardness measurements are only used as a tool for the classification of the determined optical properties. With the aid of these measurements, it was possible to determine the existence of differences in cross-linkage, These should be observable in optical measurements as well.\\
While the presented results were gathered from samples of slightly different materials, further experiments which only examine one material with defined degrees of cross-linking were carried out. A sample set of \gls{EVA} was used for this purpose. The samples where prepared from sheet material. The sheets were exposed to a temperature of 150\,\si{\degree}C in an industrial laminator. The exposure time determined the state of cross-linking. Beforehand, the samples have been analyzed with state-of-the-art methods by Hirschl et al., \cite{Hirschl}.\\
Especially in the field of \gls{PV}, \gls{EVA}, which are used as encapsulants there, have to maintain their properties over an operation time of 20 - 30 years, \cite{Pern}. Mainly, these encapsulants serve as a protection to prevent damage from mechanical, electrical and humid sources. They have to provide high strain and temperature stability to compensate for the different thermal expansion coefficients. Besides, they have to compensate for stresses and prevent cracks of the substrate materials. Another important function is the optical coupling of the light in the desired wavelength region. That demands a transmission of $>$\,90\,\% with tolerated losses of maximal 5\,\% in 20 years, \cite{Czanderna}. \gls{EVA} is a random co-polymer of ethylene and vinyl acetate with a percentage of vinyl acetate typically in the range from 28 to 33 weight-\% for \gls{PV} module applications. The native \gls{EVA} would not fulfill the thermo-mechanical requirements due to its melting range between 60 and 70\,\si{\degree}C. By chemical cross-linking utilizing hydroperoxides during \gls{PV} module lamination, the moldable \gls{EVA} sheet is transformed into a highly transparent elastomer with the required thermo-mechanical stability up to 100\,\si{\degree}C, \cite{Czanderna,Hirschl}. It shows good adhesion, high transmission in the interesting wavelength region and it is sufficiently long-term stable regarding its properties. In order to establish the desired properties of polymers like \gls{EVA}, it is necessary to develop and control appropriate curing processes.\\
The prepared \gls{EVA} samples of different cross-linking states were cut from sheets in proper pieces with a thickness of 400\,\textmugreek m. The temporal delay $\tau$ of each sample in relation to the white-light point was recorded in repeated measurements with 10 repetitions each. The data was then normalized to the corresponding sample thickness $t_{smp}$ and plotted as an averaged curve in relation to the wavelength, Fig. \ref{result} a). 
The data points were also fitted using the \textsc{Cauchy} equation Eq. (\ref{Poly:EQ_Cauchy}). From the plot it becomes obvious that the differences in the delay due to dispersion are $\leq$\,10\,\nicefrac{ps}{m}. It is also obvious that the differences between an un-laminated sample (NLT) and a laminated sample (LT 1 min) are rather high. That is to especially consider in relation to a longer lamination period from LT 1 min to LT 8 min. This fact leads to the assumption that the curing reaction starts fast.\\
In order to gather additional information, the dispersion parameter $D$ was calculated as first derivative of the fitted data according to Eq. (\ref{Poly:EQ_DispersionParam}), Fig. \ref{result} b). The results reveal that there is no particular difference between the varying degrees of cross-linking. Although some differences in the wavelength range of (0.4\,-\,0.6)\,\textmugreek m can be observed, the corresponding errorbar proves that the dispersion slope in relation to the wavelength is constant for different degrees of cross-linking. This leads to the assumption that the magnitude of the temporal delay and therefore also of the group refractive index can be used as a measure for cross-linking differences. The slope of the wavelength-dependent curves is not a suitable measure to make out cross-linking differences.\\
The most important information, the degree of cross-linking, can be extracted by plotting the differences in temporal delay $\Delta \tau / t_{smp}$ versus the lamination time, Fig. \ref{result_lam_time}.
The data acquisition over a broad spectral range enables the analysis at different wavelengths. For comparison, the relative delay at three probing wavelengths (0.5, 0.7 and 0.9)\,\textmugreek m was analyzed. It is visible that the data is equal within the standard deviation of the measurements especially for the spectral probing points of (0.7 and 0.9)\,\textmugreek m. The data at 0.5\,\textmugreek m shows a slight deviation. In order to compare the results to other methods, the measurements have been fitted using a pseudo-first order reaction kinetics model\footnote{In situations where the amount of cross-linker is small compared to the amount of polymer, the reaction can be described as a first order reaction. In this context, the term \textit{pseudo-first order} is used, \cite{Hirschl}.} of the form $y = a + b \cdot e^{-kt}$ where $k$ is the characteristic constant of the reaction process and $a$, $b$ describe the absolute position of the slope. This model is commonly used to describe cross-linking behavior determined by characterization approaches, \cite{Hirschl}. As good fits (R\textsuperscript{2}\,=\,0.976, 0.985 and 0.989) could be obtained at all probed wavelengths, an averaging over a range of wavelengths can be used to determine the degree of cross-linking with higher statistical confidence. In comparison to other technologies, the presented results show reasonable errors in the range of 6.35\,-\,8.39\,\%. Also, the trend of the data is in good analogy to reference technologies like soxhlet-extraction, Raman spectroscopy, DSC or DMA, \cite{Hirschl}. \gls{EVA} samples show  fast cross-linking at the beginning of the lamination process which significantly slows down after 3\,-\,4 minutes. The calibration of the $\Delta \tau / t_{smp}$ data to a degree of cross-linking on a percentage scale can be done by choosing an appropriate reference technology.\\
In consequence, the temporal approach showed the ability to gather model-free refractive index data over a large spectral range, (0.4\,-\,1)\,\textmugreek m, with deviations to literature values in the range of \num{2.06e-4} - \num{2.71e-4}. It could be shown, that cross-linking of polymers for industrial applications can be evaluated on the basis of the refractive index measurements. The results proofed to be comparable and in terms of their standard deviation, more reliable than the results of established technologies such as soxhlet-extraction or Raman spectroscopy. The additional statistical confidence through the measurement over a large spectral range as well as the ability to gather spectrally-resolved refractive index data are advantages of this approach over established methods.

\section{Scan-free approach}

\subsection{Wrapped-phase derivative evaluation (WPDE)}\label{TextWPDE}
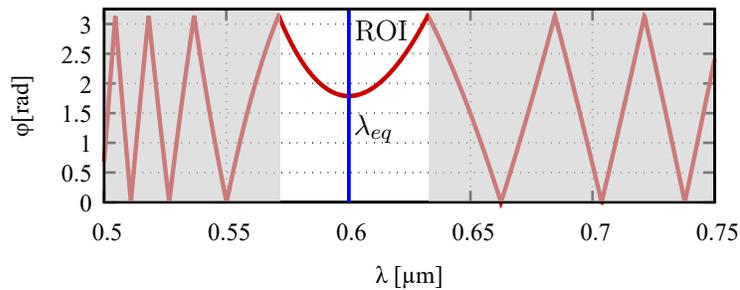
\begin{figure}[h]
\centering
		\begin{tabular}{c}
			\begin{overpic}[scale=0.33, grid=false]{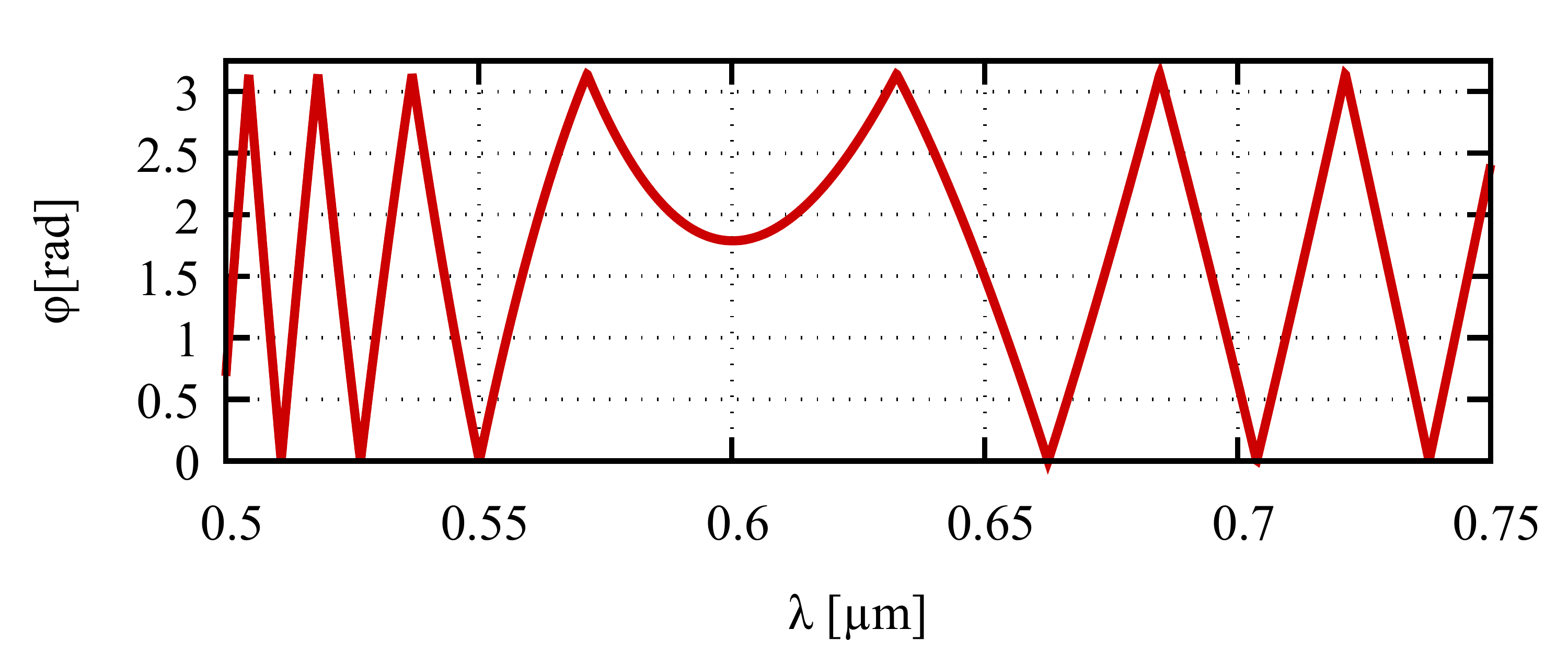}
				\put(50,22){\makebox(0,0){$\lambda_{eq}$}}
				\put(45.22,12){
					\begin{tikzpicture}
					\draw[line width =.5mm,blue] (0,0) -- (0,2.55);
					\end{tikzpicture}
				}
				\put(50,35){\colorbox{white}{\makebox(0,0){ROI}}}

				\put(13.35,12.4){
					\begin{tikzpicture}
					\fill[gray!40!white,fill opacity=0.6] (0,0) rectangle (2.32,2.53);
					\end{tikzpicture}
				}
				
				\put(56,12.4){
					\begin{tikzpicture}
					\fill[gray!40!white,fill opacity=0.6] (0,0) rectangle (3.78,2.53);
					\end{tikzpicture}
				}

			\end{overpic}
		\end{tabular}
	\caption[Principle of the WPDE method]{a) Simulated data of a wrapped phase due to the $cos^{-1}$ operation according to Eq.\,(\ref{cos_operation_cross-linking}) with marked equalization wavelength $\lambda_{eq}$ and ROI for the extraction of $\varphi_{loc}$.}\label{wrapped_phase-scheme_analysis}
\end{figure}

\begin{figure}[h]
	\begin{center}
		\begin{tabular}{c}
				\begin{overpic}[scale=0.68]{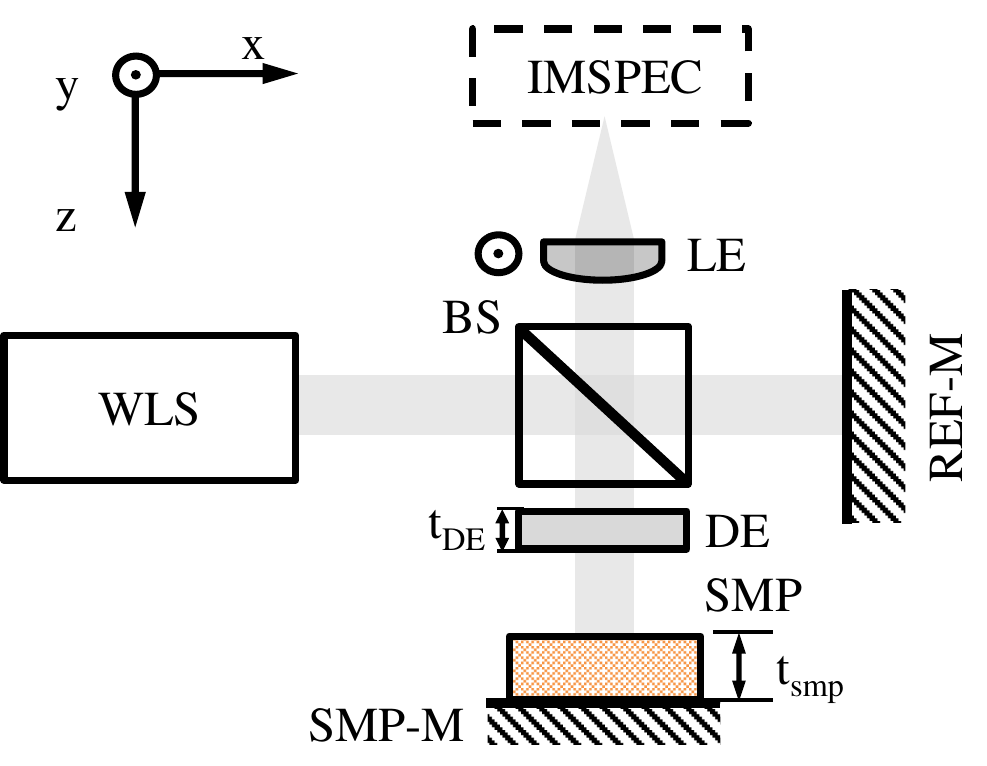}
				\put(1,1){\makebox(0,0){a)}}
			\end{overpic}
			\begin{overpic}[scale=0.3]{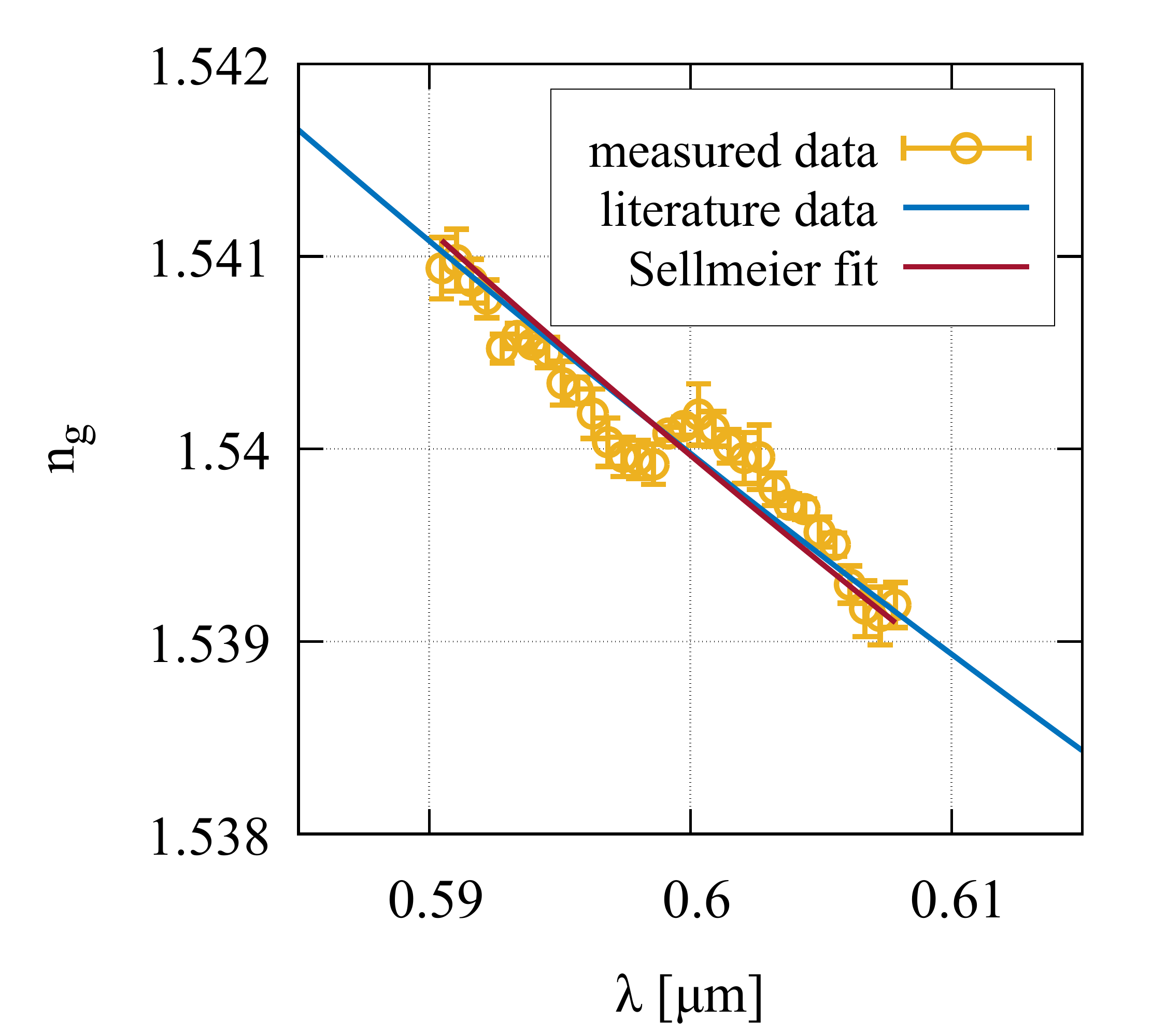}
				\put(1,1){\makebox(0,0){b)}}
			\end{overpic}
		\end{tabular}
	\end{center}
	\caption[Setup for the imaging approach to cross-linking characterization with WPDE]{Setup for the imaging approach to cross-linking characterization with WPDE incorporating a WLS- white light source which is split into a reference arm with a REF-M reference mirror and a sample arm which consists of a SMP - sample of $t_{smp}$ on a SMP-M - sample mirror and an optional DE - dispersive element of $t_{DE}$. The beams of both arms are recombined by the BS - beamsplitter and imaged by a LE - lens onto the IMSPEC - imaging spectrometer. LE can be translatable in the $y$-dimension to gather areal cross-linking information and b) Plot of the averaged measured group refractive index of N-BK7 ($t$\,=\,5\,mm, 10 measurements) which was calculated using the \gls{WPDE} approach and its corresponding \textsc{Sellmeier} fit in comparison to the literature values according to \cite{SchottKatalog}.}\label{PicPoly:SetupAddDispersion}
\end{figure}

One significant drawback of the temporal approach to estimate the degree of cross-linking with the refractive index is the need for mechanical scanning of one interferometric arm. It possibly introduces additional errors and increases the measurement time.\\
From Eq. (\ref{phase_cross-linking_interferometer_equation}) it is known that the wavelength-dependent refractive index $n^{smp}(\lambda)$ is contained in the phase of the interferometer output in combination with the sample thickness $t_{smp}$ and the path difference $\delta$. Under the assumption that $t_{smp}$ as well as $\delta$ are known, the relevant cross-linking information can be found in the refractive index.\\
By rewriting Eq. (\ref{phase_cross-linking_interferometer_equation}), the measured phase-term \glssymbol{PhaseMeas}, containing the refractive index, can be extracted
\begin{equation}
\varphi_{meas} = cos^{-1} \left(\frac{I(\lambda,x)}{I_0(\lambda)} - 1 \right)=  2\pi \frac{(n^{smp}(\lambda)-1)t_{smp}-\delta(x)}{\lambda} + \varphi_{off} \label{wpde_single_eq}.
\end{equation}
Inherent to this approach is the ambiguity of the resulting values as $\varphi$ is not limited to the range of 0\,-\,\textpi. Other works have shown methods to perform the correct quadrant selection in order to resolve this ambiguity, \cite{Calatroni1996}. In contrast, an alternative method to avoid quadrant selection was developed by performing a local signal analysis in the spectral range close to $\lambda_{eq}$, Fig. \ref{wrapped_phase-scheme_analysis} a).
In the first stage, this approach determines the phase minimum and defines a \gls{ROI} around the minimum. For this purpose, the raw measured data is analyzed using a \gls{STFT} where a \gls{FFT} is performed in one small window of the complete data set which is then slid over the signal successively along the wavelength dimension, see subsection \ref{Profilo:SubSec:FreqAna}. This approach accounts for the non-uniform frequency of the signal. As a result, the minimum of the extracted frequency slope can be determined from the power spectrum. It represents the position of the phase minimum which also occurs at $\lambda_{eq}$. The \gls{ROI} is defined as a local wavelength range \glssymbol{LambdaLocal} in the proximity of the detected $\lambda_{eq}$ where only unambiguous phase data is included. This so called local phase, \glssymbol{PhaseLocal}, is subject to a phase offset, \glssymbol{PhaseOffset}, with regard to the absolute phase due to the $\cos^{-1}$-operation, Eq. (\ref{cos_operation_cross-linking}).\\
A second analytical step implements a newly developed approach called \gls{WPDE}, where $\varphi_{loc}$ is differentiated with respect to the wavelength, noted with $\frac{\partial}{\partial \lambda}$,
\begin{equation}
\frac{\partial \varphi_{loc}}{\partial \lambda} = \frac{\partial}{\partial \lambda} \left(2\pi \frac{[n(\lambda_{loc})-1]t_{smp}-\delta(x)}{\lambda_{loc}} + \varphi_{off} \right) \label{derivative_eq_sample-only}
\end{equation}
This eliminates the phase offset $\varphi_{off}$ and enables the evaluation of the cross-linking characteristics in terms of the group refractive index $n_g^{smp}(x,\lambda)$ as well as the \gls{RDOT} \glssymbol{DerivOptThickness}
\begin{eqnarray}
n_g^{smp}(x,\lambda) && = 1 - \frac{\kappa}{t_{smp}} \label{n_g_crosslinking_sample-only} \\
\textrm{with } \kappa && = \frac{\varphi_{loc}^\prime \cdot \lambda^2}{2 \pi} - \delta\\
t^\prime_{OPT} && = n_g^{smp}(x,\lambda) \cdot t_{smp} = t_{smp} -\frac{\varphi_{loc}^\prime \cdot \lambda^2}{2 \pi} - \delta \label{EQ_RDOT_sample-only}
\end{eqnarray}
where \glssymbol{DerivPhaseLocal} is calculated from the measured data using the difference quotient with $\Delta \lambda$ as interval.
This case holds true when experiments, as sketched out in Fig.\,\ref{poylmer_temporal_approach_simple_setup}\,a), are performed  where one simple sample is part of the interferometer as well as the primary source of dispersion. In situations where samples with low dispersion are to be measured or the simultaneous measurement of the samples surface profile should be realized, a modified setup with additional dispersion is favorable, Fig. \ref{PicPoly:SetupAddDispersion}.
In this case, the phase term of Eq. (\ref{wpde_single_eq}) has to be expanded by an appropriate term for the dispersive element,
\begin{equation}
cos^{-1}\left[  \frac{I_{meas}(x,\lambda)}{I_0(\lambda)} -1 \right] = \varphi =  2\pi \frac{ \left[ \left( n^{smp}(x, \lambda)  - 1 \right)t_{smp} \right] + \left[ \left( n^{DE}(\lambda)  - 1 \right)t_{DE} \right] -\delta }{\lambda}\label{cos_operation_cross-linking}.
\end{equation}
According to this equation, the derivative in order to access the group refractive index can be noted as
\begin{equation}\label{derivative_eq}
\varphi_{loc}^\prime =\frac{\partial}{\partial \lambda} \left( 2\pi \frac{ \left[ \left( n^{smp}(x, \lambda)  - 1 \right)t_{smp} \right] + \left[ \left( n^{DE}(\lambda)  - 1 \right)t_{DE} \right] -\delta }{\lambda} + \varphi_{off} \right),
\end{equation}
which leads to a new description of the group refractive index and the \gls{RDOT} $t^\prime_{OPT}$ for the approach with additional dispersion
\begin{eqnarray}
&&n_g^{smp}(x,\lambda) = 1 - \frac{\lambda^2 \cdot \xi}{2\pi \cdot t_{smp}} \label{n_g_crosslinking_add-Dispersion} \\
&& \textrm{with } \xi = \varphi_{loc}^\prime - \frac{2\pi}{\lambda^2}\left[ \left(1 - n_g^{DE} \right)t_{DE} + \delta  \right]\\
&& t^\prime_{OPT} = n_g^{smp}(x,\lambda) \cdot t_{smp} = t_{smp} - \frac{\lambda^2 \xi}{2\pi}.\label{EQ_RDOT_add-Dispersion}
\end{eqnarray}
A detailed derivation of Eq.\,(\ref{n_g_crosslinking_sample-only}) and (\ref{EQ_RDOT_sample-only}) for the sample-only approach as well as for the approach with additional dispersion resulting in Eq.\,(\ref{n_g_crosslinking_add-Dispersion}) and (\ref{EQ_RDOT_add-Dispersion}) can be found in the appendices \ref{APNDX_deriv_sample-only} and \ref{APNDX_deriv_add-Dispersion}.\\
In order to evaluate the algorithm, the group refractive index of a N-BK7 sample with a nominal thicknesses of 5 mm was determined. The averaged standard deviation of 10 consecutive measurements of the sample was found to be \num{9.97e-5}. The averaged group refractive index data of these 10 measurements was fitted using a Sellmeier equation, Fig. \ref{PicPoly:SetupAddDispersion} b). In resemblance to the literature values, \cite{SchottKatalog}, a root-mean-square error of \num{1.65e-4} and of \num{3.36e-5} was achieved for the averaged measured and for the fitted data respectively. Compared to the measurements using the temporal approach, subsection \ref{subsection_temporal_approach}, this demonstrates an improvement as the \gls{RMS} error was $\Delta n_{g5}^{fit}$\,=\,\num{2.30e-4}. An additional advantage over the temporal approach is the ability to gather the wavelength-dependent group refractive index without the need for mechanical scanning. The result shows that the \gls{WPDE} approach achieves a comparable accuracy to state-of-the-art refractive index measurement technologies. Furthermore, the refractive index resolution is sufficient to characterize cross-linking in waveguide polymers, where differences in the range of $\Delta n=$ 0.001 - 0.02 are expected, taking the respective sample thickness into account, \cite{2Photon_waveguides}.\\  
This result is calculated only within the \gls{ROI} and is dependent on the amount of dispersion, represented by $n_g^{smp} (\lambda)$. Therefore, it is valid only within a small spectral range. Different approaches have been considered to gather information over the complete spectral range of the data  set. On the one hand, the \gls{WPDE} analysis algorithm can be applied to other \gls{ROI}s within the data. The advantage is that the group refractive index can be calculated without an a priori knowledge of the underlying material model. On the other hand, one can calculate the group refractive index over the complete spectral range, if the material model of the sample is known.


\subsection{Spatially-resolved approaches}

All approaches described so far have been based on point-wise measurements of spectra and relied on scanning either one arm of the interferometer to gather wavelength-dependent information (temporal approach) or on scanning of the sample in order to gather information from different locations of the sample (\gls{WPDE}).
\subsubsection{Scan-free temporal approach}
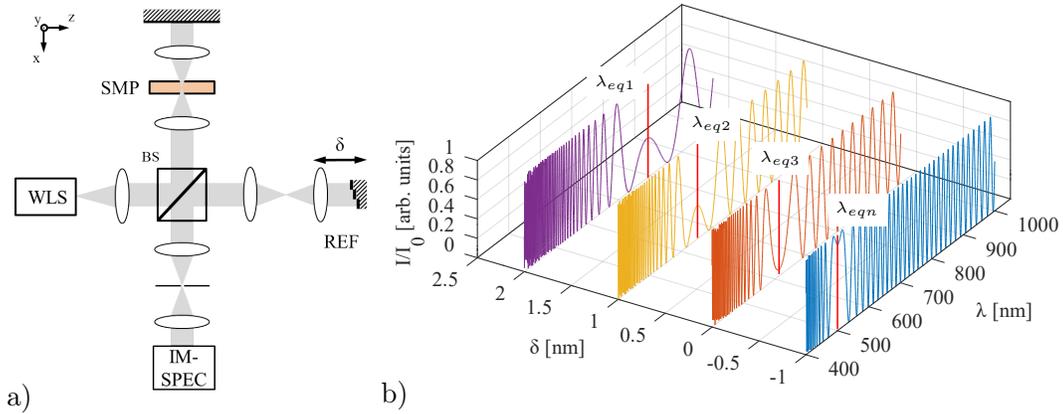
\begin{figure}[h]
	\begin{center}
		\begin{tabular}{c}
			\begin{overpic}[scale=0.4]{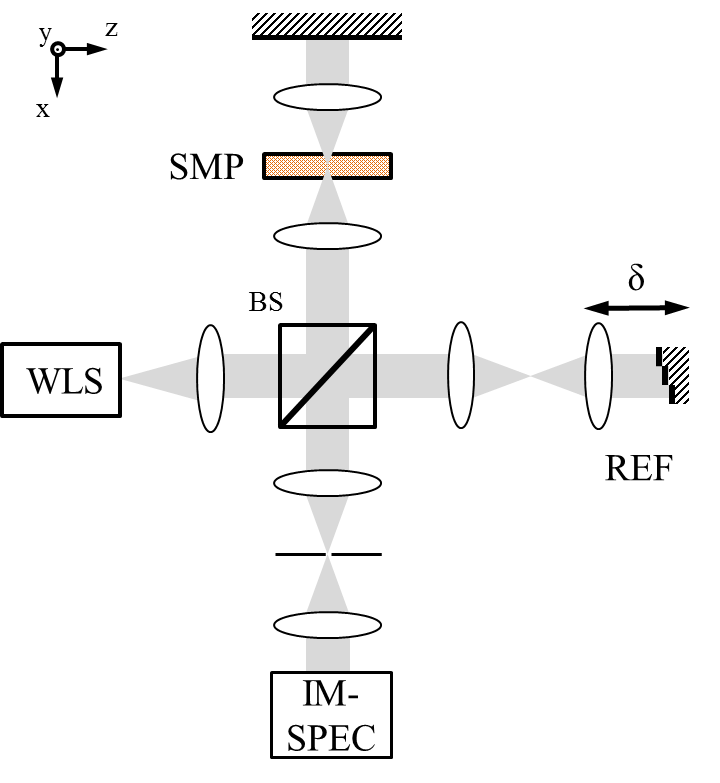}
				\put(1,1){\makebox(0,0){a)}}
			\end{overpic}
			\begin{overpic}[scale=0.3, grid = false]{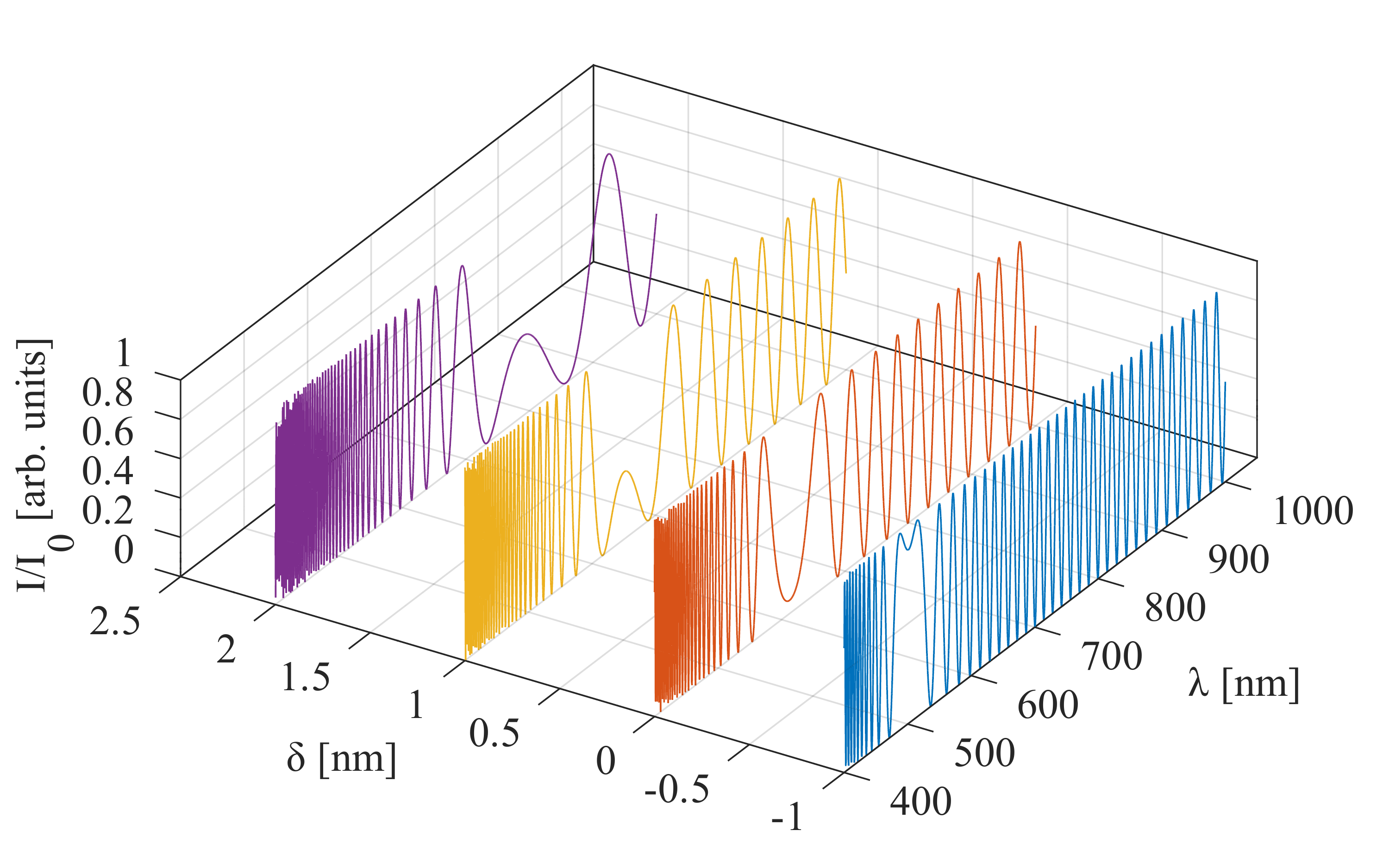}
				\put(1,1){\makebox(0,0){b)}}
				\put(37.7,32.5){\tikz \draw[line width=0.2mm,red] (0,0) -- (0,1.25);}
				\put(44.9,23.75){\tikz \draw[line width=0.2mm,red] (0,0) -- (0,1.25);}
				\put(56.75,18.5){\tikz \draw[line width=0.2mm,red] (0,0) -- (0,1.25);}
				\put(65.25,10.5){\tikz \draw[line width=0.2mm,red] (0,0) -- (0,1.25);}
				\put(29,46){\colorbox{white}{\textcolor{black}{\tiny$\lambda_{eq1}$}}}
				\put(43,40){\colorbox{white}{\textcolor{black}{\tiny$\lambda_{eq2}$}}}
				\put(53,35){\colorbox{white}{\textcolor{black}{\tiny$\lambda_{eq3}$}}}
				\put(64,27.5){\colorbox{white}{\textcolor{black}{\tiny$\lambda_{eqn}$}}}
			\end{overpic}
		\end{tabular}
	\end{center}
	\caption[Proposed setup for the scan-free temporal evaluation of samples]{a) Proposed setup for the scan-free temporal evaluation of samples with a WLS - white light source split by a BS - beam splitter into one arm that focuses light into a SMP - sample which then gets back reflected by a mirror and a second arm, known as REF - reference arm where light is focused in the same way before it gets back reflected on a  mirror which introduces controlled temporal delays in the x-domain $\delta_n$ and a IMSPEC - imaging spectrometer in order to analyze the signal in a spatially-resolved manner and b) a simulated signal of the imaging spectrometer with different temporal delays $\delta_n$ and corresponding equalization wavelengths $\lambda_{eqn}$. }\label{poly_spatial_resolved_temporal_principle}
\end{figure}

The evaluation of the refractive index of a sample and therefore of the degree of cross-linking relied on mechanical scanning of one interferometer arm in order to scan the spectral domain for certain equalization wavelengths $\lambda_{eq}$. The underlying principle is the introduction of different temporal delays to the setup. A possible method to introduce the delays all at once, relies on the spatially encoding of them and the appropriate detection, Fig. \ref{poly_spatial_resolved_temporal_principle}.
For this purpose, the standard, one-dimensional spectrometer is replaced by an imaging spectrometer with appropriate optics. The imaging setup enables the detection of spectral information in one spatial domain. In the proposed setup, the probing beam is focused in the sample volume and re-collimated onto a mirror. Consequently, the beam of the reference arm is also focused and re-collimated without a sample being present. The mirror of the reference arm is designed to introduce temporal delays to the beam with a spatial distribution. After recombination of both beams and their spectral detection, this distribution of delays can be recorded as spectra with different adequate equalization wavelengths. Analogous to section \ref{subsection_temporal_approach}, these wavelengths can be used to calculate the refractive index of the sample at these discrete points. Depending on the number and size of the delays as well as the construction of the spectrometer, a large spectral range can be covered. Model-based fits can be calculated accordingly. The introduction of delays can be done by means of a stepped mirror, a deformable micro-mirror array or a transmissive element with a refractive index gradient.

\subsubsection{Imaging WPDE}
\begin{figure}[h]
	\begin{center}
		\begin{tabular}{c}
			\begin{overpic}[scale=0.58]{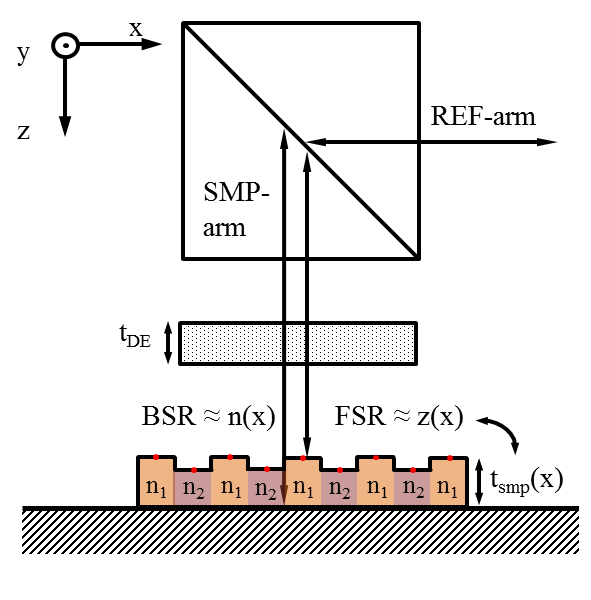}
				\put(1,1){\makebox(0,0){a)}}
			\end{overpic}
			\begin{overpic}[scale=0.30]{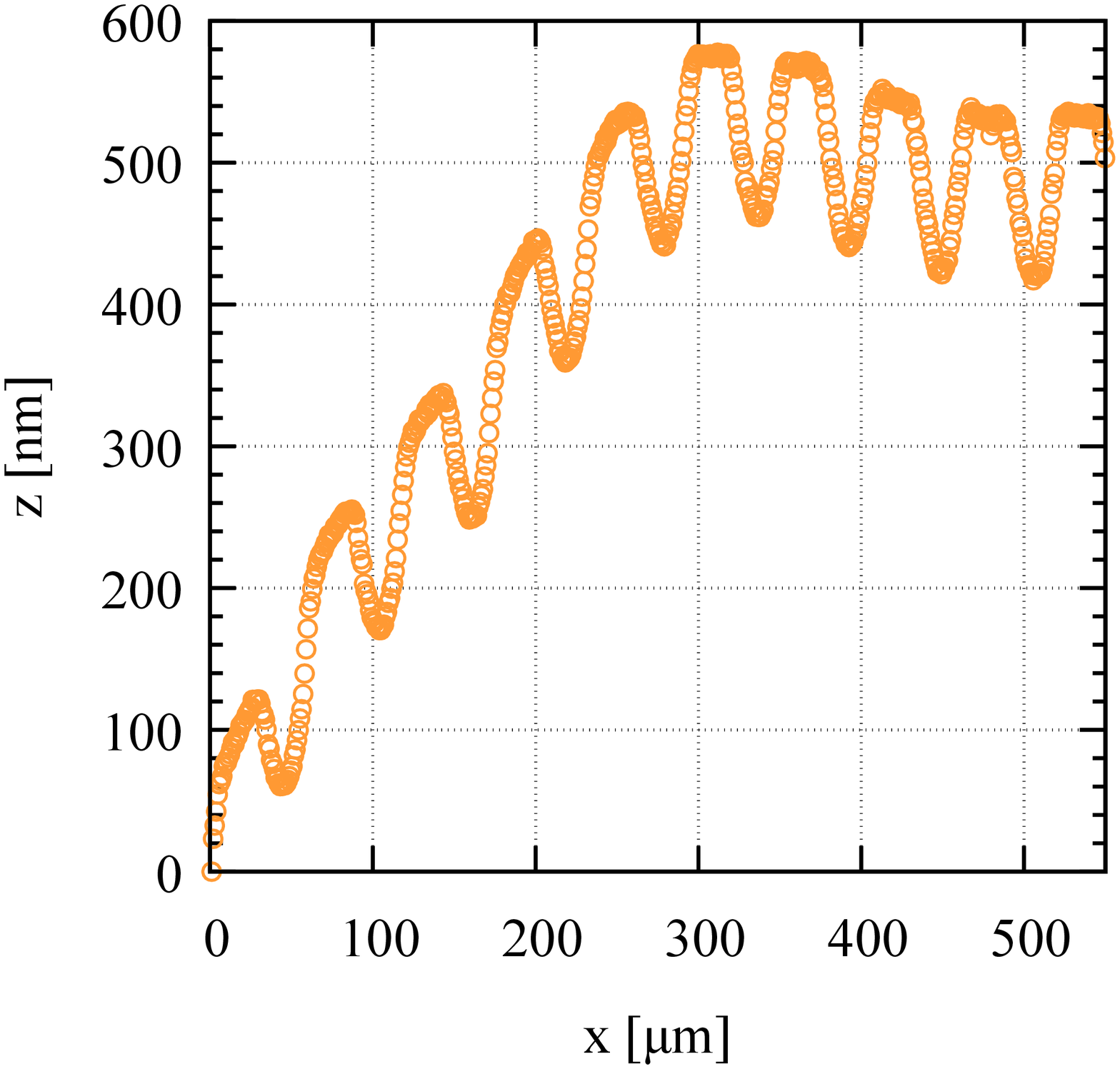}
				\put(1,1){\makebox(0,0){b)}}
			\end{overpic}
		\end{tabular}
	\end{center}
	\caption[Detail of the photo-resist sample under test with a lithographically generated, spatially-dependent refractive index pattern]{a) Detail of the photo-resist sample under test with a lithographically generated, spatially-dependent refractive index pattern and a resulting surface height profile due to shrinkage. After transmission through a DE - dispersive element with $t_{DE}$ a FSR - front-side reflex from the sample can be used to interfere with light from the REF-arm - reference arm in order to calculate the surface height profile z(x) $\sim$ $t_{smp}(x)$  while a BSR - back-side reflex can be used in conjunction with $t_{smp}(x)$ to calculate the refractive index profile n(x) as a measure for the degree of cross-linking across the sample b) plot of the measured surface profile $z(x)$ from a polymer sample under investigation utilizing a wavelength-calibrated imaging spectrometer.}\label{PolyPic:SamplePCCLandSurface}
\end{figure}

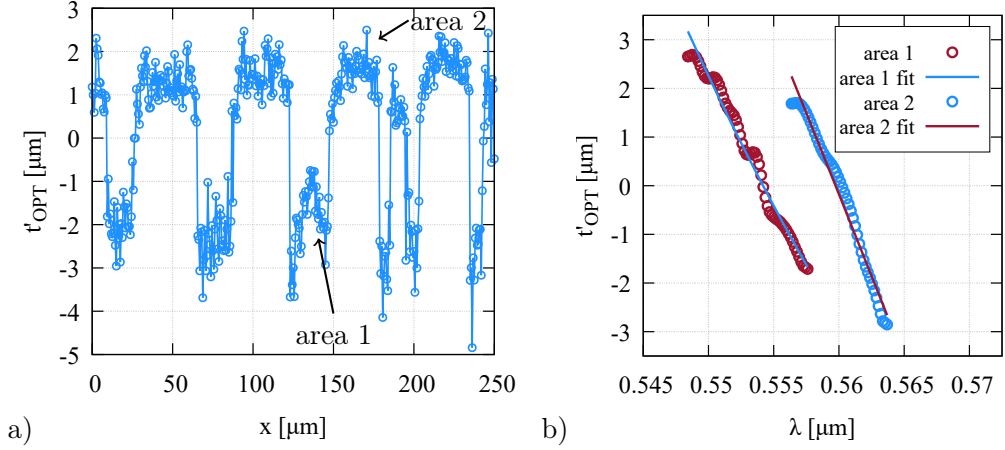
\begin{figure}[h]
	\begin{center}
		\begin{tabular}{c}
			\begin{lpic}{chapters/polymer_characterization/pictures/results_2D_t_rOPT_lateral_3(0.305)}
				\lbl[b,l]{0,0;a)}
				\lbl[b,l]{125.5,44;area 1}
				\lbl[b,l]{175.5,181;area 2}
				\begin{tikzpicture}
				\draw[step=1cm,gray,very thin,opacity=0] (-2,-2) grid (6,6);
				\draw[step=1cm,thick,black,->] (2.2,-.25) -- (2,.75);
				\draw[step=1cm,thick,black,->] (3.2,3.65) -- (2.8,3.4);
				\end{tikzpicture}
			\end{lpic}
			\begin{lpic}{chapters/polymer_characterization/pictures/results_2D_t_rOPT_lambda_3(0.3)}
				\lbl[b,l]{0,0; b)}
			\end{lpic}
		\end{tabular}
	\end{center}
	\caption[Results of the measured relative derived optical thickness (RDOT)]{Results of the measured RDOT a) spatially resolved along one sample dimension of a lithographically structured photo-resist layer with structures having a nominal pitch of 50 \textmugreek m on a Si-substrate at a wavelength of 557 nm and b) mean values and fitted data for two marked areas with different degrees of cross-linking over a spectral range.}\label{PolyPic:ImagingWPDEresult}
\end{figure}

In a second approach, the aforementioned implementation of an imaging spectrometer was also used to perform \gls{WPDE} with spatial resolution using the setup described in Fig.\,\ref{PicPoly:SetupAddDispersion}\,a).
Within the setup, a sample with the thickness $t_{smp}$ is placed on a reflecting substrate and is used as a mirror for one interferometer arm. Correspondingly, the reference arm only compromises a mirror and no additional dispersive element. Depending on the thickness and amount of dispersion of the sample, an element with additional dispersion with the thickness $t_{DE}$ might be necessary in the sample arm in order to enhance the measurability. A detailed explanation on the usage of an additional dispersive element is given in subsection \ref{subsection_influences_limitations}. The setup is also equipped with a translation stage for the imaging lens LE. This lens enables the recording of areal cross-linking information. As the imaging spectrometer allows capturing refractive index data along a line, the translation of the imaging lens in the $y$-dimension enables the stacking of these line profiles in order to receive information on the whole two-dimensional plane, refer also to Fig.\,\ref{Profilo:Pic:Result_3D_standard_mum}\,a). Using this method neither the sample nor the reference arm have to be moved during measurements which prevents obstructions of the interferometric measurement that due to movement. The data analysis was performed analogous to the \gls{WPDE} approach described in section \ref{TextWPDE}. As already pointed out, in case of the usage of an additional dispersive element, a modified set of equations, Eq. (\ref{n_g_crosslinking_add-Dispersion}), has to be applied.\\
This approach was used to characterize lithographically generated structures in a photo-resist, Fig. \ref{PolyPic:SamplePCCLandSurface}.
The resist was spin-coated on a Si-wafer with a thickness of $t_{smp}$\,=750\,\textmugreek m. Afterwards, it was exposed to visible light, (400\,-\,420)\,nm, for primary cross-linking and to UV-radiation, (300\,-\,360)\,nm, in a secondary cross-linking process. The secondary cross-linking was performed through a mask to generate structures of rectangular refractive index patterns which also lead to the shrinkage of the cross-linked areas.  The goal of the investigations was the determination the surface height profile that is altered due to shrinkage as well as the refractive index profile which is due to different degrees of cross-linking.\\
The surface height profile of the sample was characterized with the described setup using the front-surface reflex and the profilometry approach described in chapter \ref{ChapterProfilometry}, Fig. \ref{PolyPic:SamplePCCLandSurface} b). It is obvious, that apart from a slight overall waviness, the sample shows a regular height pattern with the expected pitch of 50\,\textmugreek m. The depth of the shrunken areas is about 120 nm, which lies in the expected range. In consequence, these calculated height profiles enable the separation of shrinkage from the refractive index information for every sample individually and simultaneously.\\
With the knowledge of the surface height profile of the sample due to shrinkage, the correct thickness along the spatial domain, $t_{smp} = t_{smp}(x) \sim z(x)$, can be calculated. Therefore, the surface height profile was measured in relation to the substrate. By the application of either Eq. (\ref{n_g_crosslinking_add-Dispersion}) or (\ref{EQ_RDOT_add-Dispersion}), the group refractive index or the relative derived optical thickness can be calculated corresponding to its position on the sample, Fig. \ref{PolyPic:ImagingWPDEresult} a).
For the results pictured above, the \gls{RDOT} profile of the sample was calculated for a single wavelength of 557\,nm. The spatial profile allows a resolution of cross-linking differences of 4\,\textmugreek m in the lateral domain. Although the results are affected by noise and batwing-effects, \cite{Xie_PE16}, a dynamic range of $\pm$\,1.5\,\textmugreek m in the \gls{RDOT} for the given sample was revealed over a lateral range of nearly 550\,\textmugreek m, while a section of 250\,\textmugreek m is displayed here. Furthermore, it also has to be noted that the plateaus do not show completely flat \gls{RDOT} profiles. This behavior was attributed to a mixture of effects ranging from diffraction during exposure of the structures to deformation during shrinkage and diffraction during measurements. As the profile was taken at a specific wavelength, it represents only a fraction of the captured information, which was originally analyzed over a spectral range of 20\,nm. \\
In order to estimate the effect of cross-linking, the \gls{RDOT} differences  have been measured over the complete spectral range as a mean value of two different exposed areas, Fig. \ref{PolyPic:ImagingWPDEresult} b). An \gls{RDOT} difference of about 3\,\textmugreek m between the differently cross-linked areas could be resolved while the \gls{RDOT} slope for every area was determined over 10\,nm. The results are affected by noise in the original data which is amplified by the process of taking the derivative. Some smoothing with a Gaussian filter was applied to the data.\\
One of the main advantages of the described approach is the lack of necessity for a model in order to calculate the spectrally-resolved refractive index. As some compromise towards the size of the spectral measurement range was made by the choice of the dispersive element, subsection \ref{TextWPDE}, the application of a refractive index model might become interesting in post-processing. In the context of (photo-)polymers, a variant of \textsc{Cauchy}'s equation was selected, \cite{Soave,Sultanova}. Using this model, the group refractive index $n_g^{smp} (\lambda)$ can be calculated according to Delbarre et al. \cite{Delbarre} with
\begin{equation}
n_g^{smp} (\lambda) = n(\lambda) - \frac{dn(\lambda)}{d\lambda} \cdot \lambda= A_1 + \frac{3 A_2}{\lambda^2} + \frac{5A_3}{\lambda^4}.
\end{equation}
By appropriate fitting, the \textsc{Cauchy} coefficients $A_1$, $A_2$ and $A_3$ were determined which enable the calculation of the refractive index and the group refractive index over any given spectral range where the \textsc{Cauchy} model is valid.

\section{Influences and limitations}\label{subsection_influences_limitations}

\begin{figure}[h]
	\begin{center}
		\begin{tabular}{c}
			\begin{overpic}[scale=0.35]{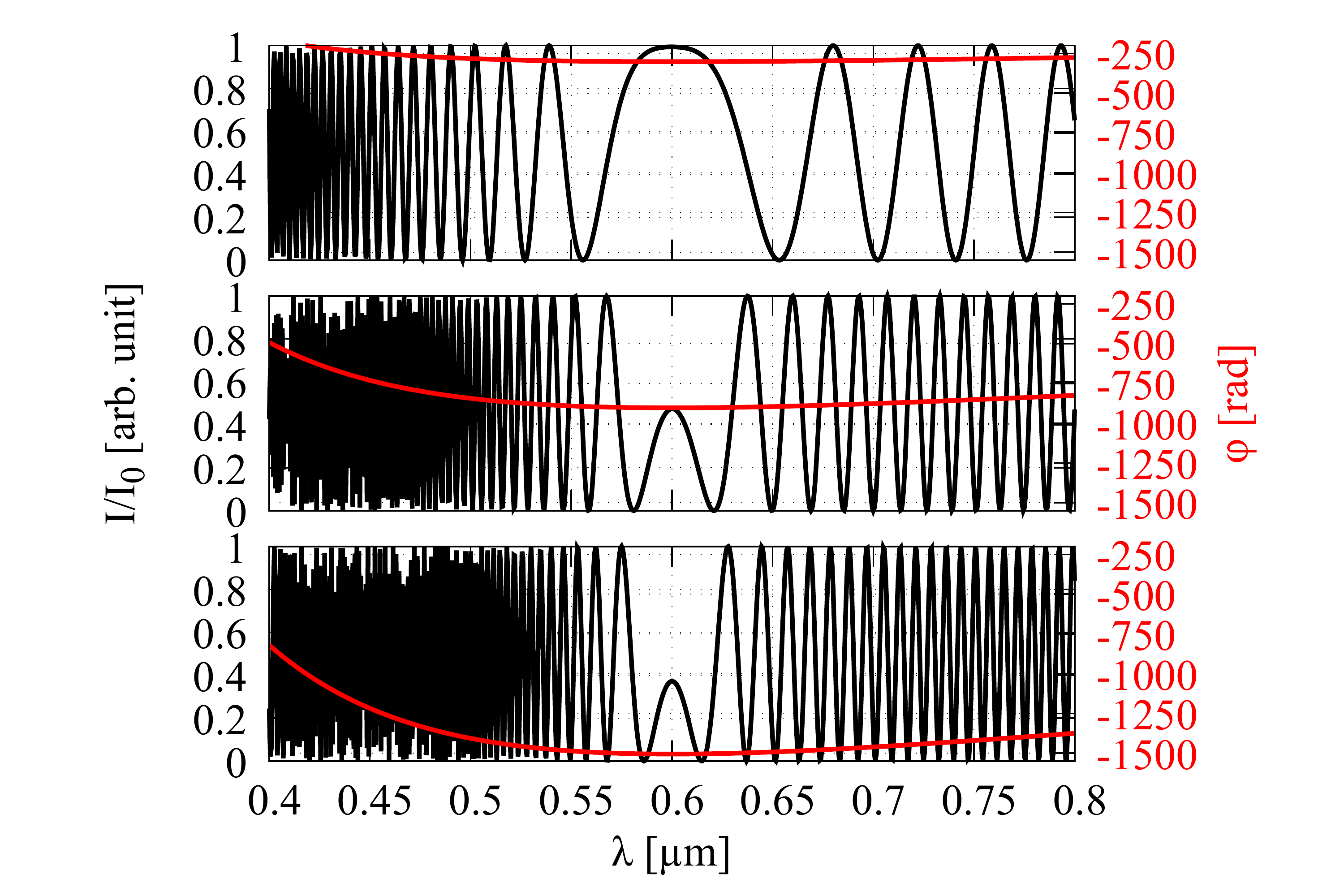}
				\put(1,55){\makebox(0,0){a)}}
				\put(1,36){\makebox(0,0){b)}}
				\put(1,17){\makebox(0,0){c)}}
			\end{overpic}
		\end{tabular}
	\end{center}
	\caption[Simulated signal in order to demonstrate the influence of dispersion on the relative signal intensity]{Simulated signal in order to demonstrate the influence of dispersion on the relative signal intensity and phase slope at an equalization wavelength of $\lambda_{eq}$\,=\,0.6\,\textmugreek m and sample of N-BK7 with a)\,$t_{smp}$\,=\,1 mm, b)\,$t_{smp}$\,=\,3\,mm and c)\,$t_{smp}$\,=\,5 mm}\label{PolyPic:Limit:phases_intensities}
\end{figure}

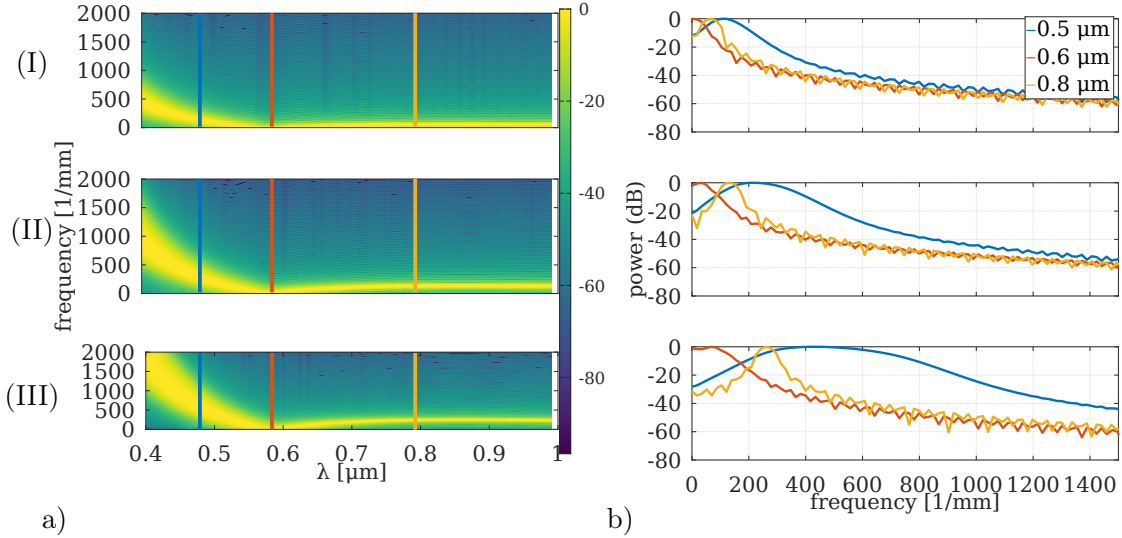
\begin{figure}[h]
	\begin{center}
		\begin{tabular}{c}
			\begin{overpic}[scale=0.091, grid = false]{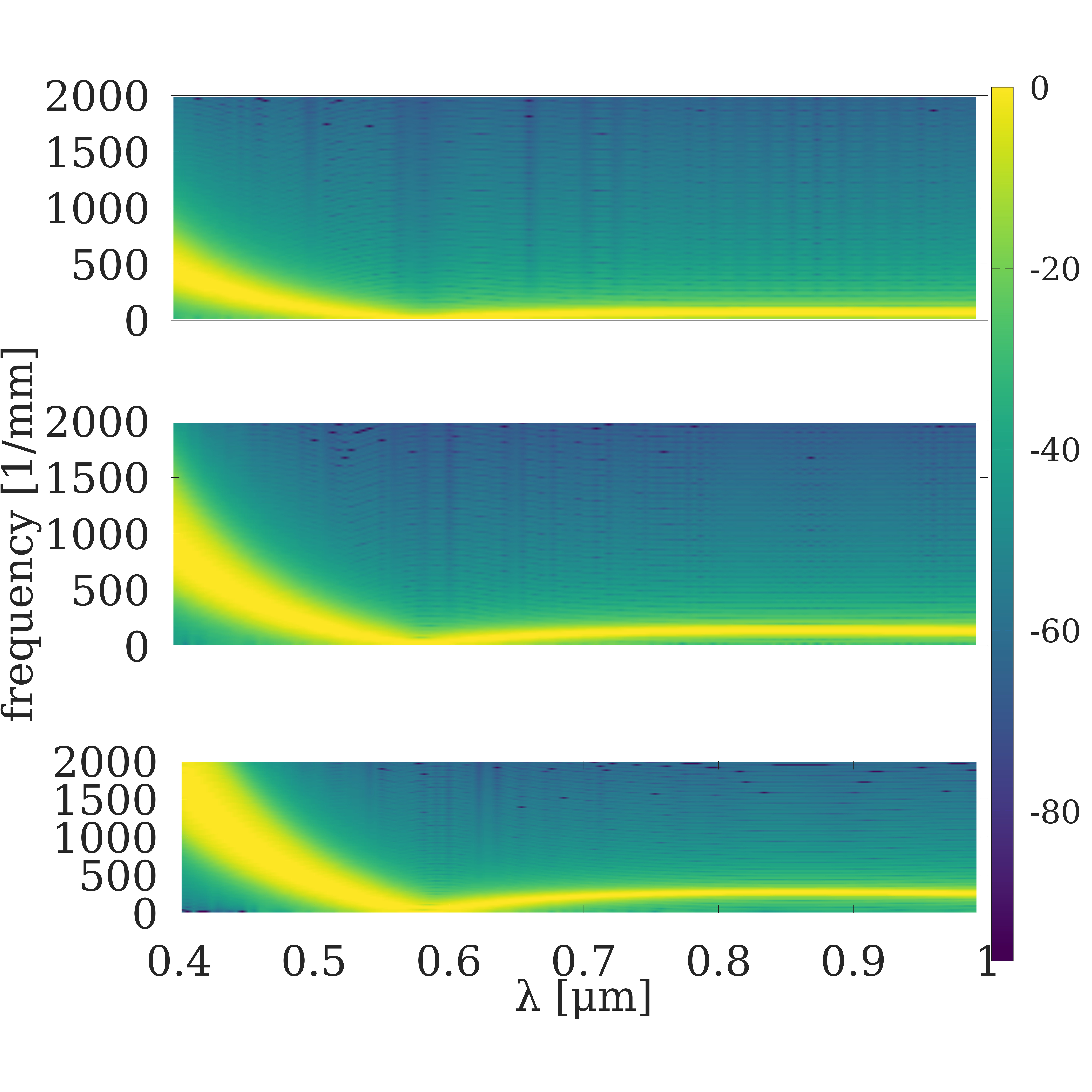}
				\definecolor{M_blue}{rgb}{0,0.4470,0.7410}
				\definecolor{M_orange}{rgb}{0.8500,0.3250,0.0980}
				\definecolor{M_yellow}{rgb}{0.9290,0.6940,0.1250}

				\put(0,0){\makebox(0,0){a)}}
				\put(-4,82){\makebox(0,0){(I)}}
				\put(-4,52){\makebox(0,0){(II)}}
				\put(-4,22){\makebox(0,0){(III)}}
				
				\put(25.8,16){\tikz \draw[line width=0.5mm,M_blue] (0,0) -- (0,1.02);}
				\put(38.75,16){\tikz \draw[line width=0.5mm,M_orange] (0,0) -- (0,1.02);}
				\put(64.5,16){\tikz \draw[line width=0.5mm,M_yellow] (0,0) -- (0,1.02);}
				
				\put(25.8,40.7){\tikz \draw[line width=0.5mm,M_blue] (0,0) -- (0,1.51);}
				\put(38.75,40.7){\tikz \draw[line width=0.5mm,M_orange] (0,0) -- (0,1.51);}
				\put(64.5,40.7){\tikz \draw[line width=0.5mm,M_yellow] (0,0) -- (0,1.51);}
			
				\put(25.8,70.5){\tikz \draw[line width=0.5mm,M_blue] (0,0) -- (0,1.51);}
				\put(38.75,70.5){\tikz \draw[line width=0.5mm,M_orange] (0,0) -- (0,1.51);}
				\put(64.5,70.5){\tikz \draw[line width=0.5mm,M_yellow] (0,0) -- (0,1.51);}
			\end{overpic}
		
		\begin{overpic}[scale=0.09]{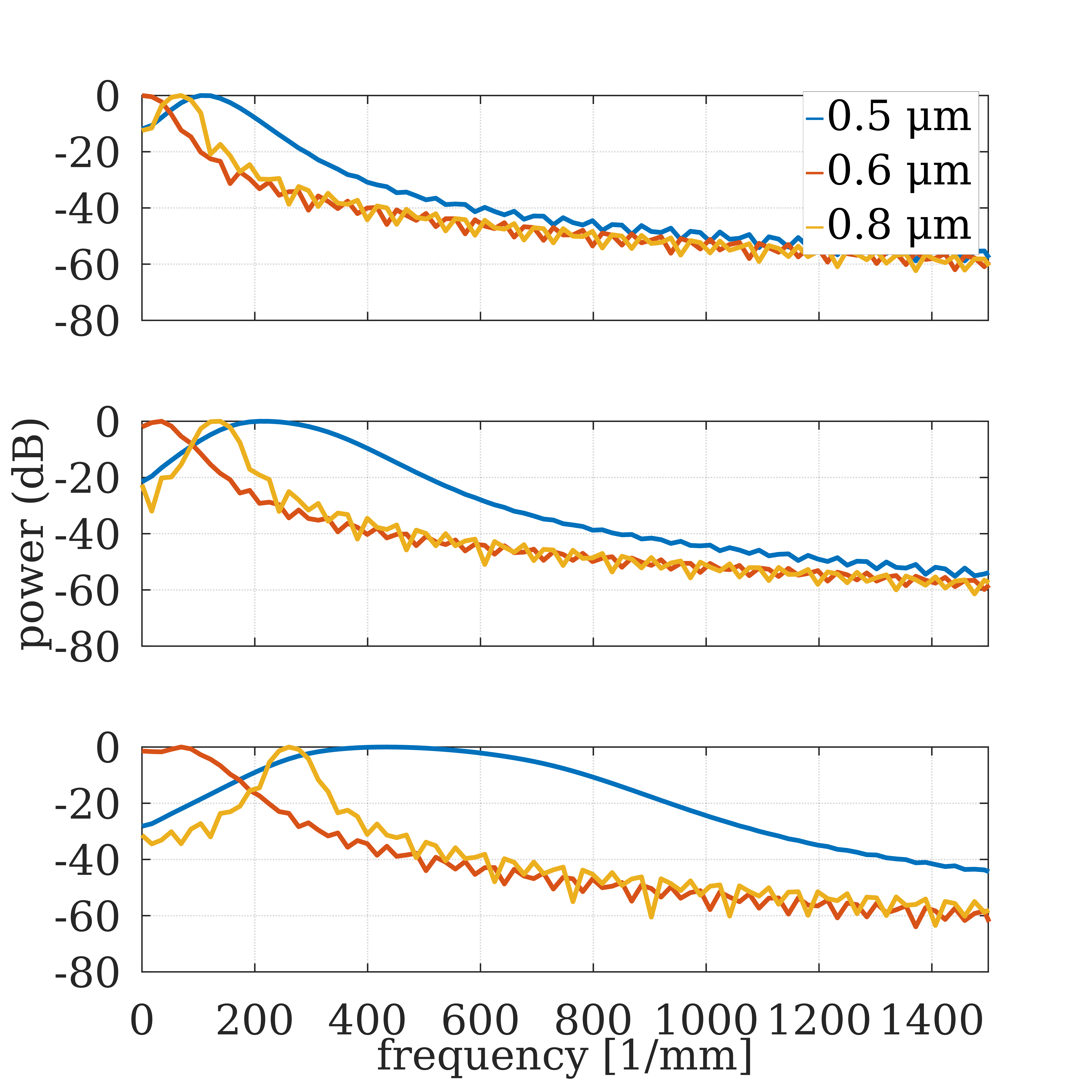}
			\put(0,0){\makebox(0,0){b)}}
			
		\end{overpic}
\end{tabular}
	\end{center}
	\caption[STFT analysis of the simulated signals in order to demonstrate the influence of dispersion]{a) STFT analysis of the simulated signals in order to demonstrate the influence of dispersion at an equalization wavelength of $\lambda_{eq}$\,=\,0.6\,\textmugreek m and a sample of N-BK7 with (I)\,$t_{smp}$\,=\,1\,mm, (II)\,$t_{smp}$\,=\,3\,mm and (III)\,$t_{smp}$\,=\,5\,mm with colored markings to indicate the data used for b) visualization of the limitation to determine $\lambda_{eq}$ for setups with dispersive elements $<$ 1\,mm where the frequency resolution of the STFT is not sufficient to resolve the peak for the phase minimum (here at \,0.6\,\textmugreek m) from the rest of the phase signal.}\label{PolyPic:Limit:phases_intensities_STFT}
\end{figure}

The measurement range as well as the accuracy of the cross-linking determination is dependent on the accuracy of the determination of the equalization wavelength $\lambda_{eq}$, the sample thickness $t_{smp}$ and the path difference of the interferometer $\delta$ amongst other parameters, Eq.\,(\ref{cross-linking_interferometer_equation}) and Eq.\,(\ref{wpde_single_eq}). While the position of the equalization wavelength depends on the path difference between both interferometer arms, the width of the fringe around $\lambda_{eq}$ is determined by the amount of optical dispersion. Hence, thicker materials show tighter fringe spacing than thinner samples of the same refractive index, Fig. \ref{PolyPic:Limit:phases_intensities}.\\
The different measurement approaches discussed in this chapter demand different signal types for analysis. While the detection of just the equalization wavelength with the temporal approach works with very little dispersion, hence a very wide fringe spacing, the \gls{WPDE} method requires a tighter fringe spacing in order to resolve the phase minimum in the power spectrum, Fig.\,\ref{PolyPic:Limit:phases_intensities_STFT}.
The estimation of the phase minimum is performed as tracking for the minimal frequency of the component carrying the most power in the spectrogram. An analysis of the frequency content for three distinct wavelengths, $\lambda_{eq}$\,=\,0.6\,\textmugreek m as well as 0.5 and 0.8\,\textmugreek m, visualizes the problem of low dispersion, Fig.\,\ref{PolyPic:Limit:phases_intensities_STFT}\,b). The lower the dispersion in the setups, the closer the not infinitely sharp peaks of the power spectrum move towards the lower frequencies. For samples of N-BK7 with thicknesses below 1\,mm, a distinct separation is not possible anymore.  Potential solutions to this problem can be the application of an adaptive windows size during the \gls{STFT} which would decrease the peak width for frequency peaks as well as the introduction of additional dispersive elements. The exact thickness which necessitates additional dispersion is dependent on the sample's refractive index.  
Also, it has to be noted, that the increase in dispersion reduces the spectral range as only the data within one phase jump around $\lambda_{eq}$ is used in the algorithm. In order to compensate for the loss in resolution due to the additional dispersion, the before mentioned precision fit of the spectral data around the equalization wavelength can be utilized.\\
A detailed description of significant influences and error sources is given in the following section.

\subsubsection{Influence of shrinkage}
\begin{figure}[h]
	\begin{center}
		\begin{tabular}{c}
			\begin{overpic}[scale=0.27]{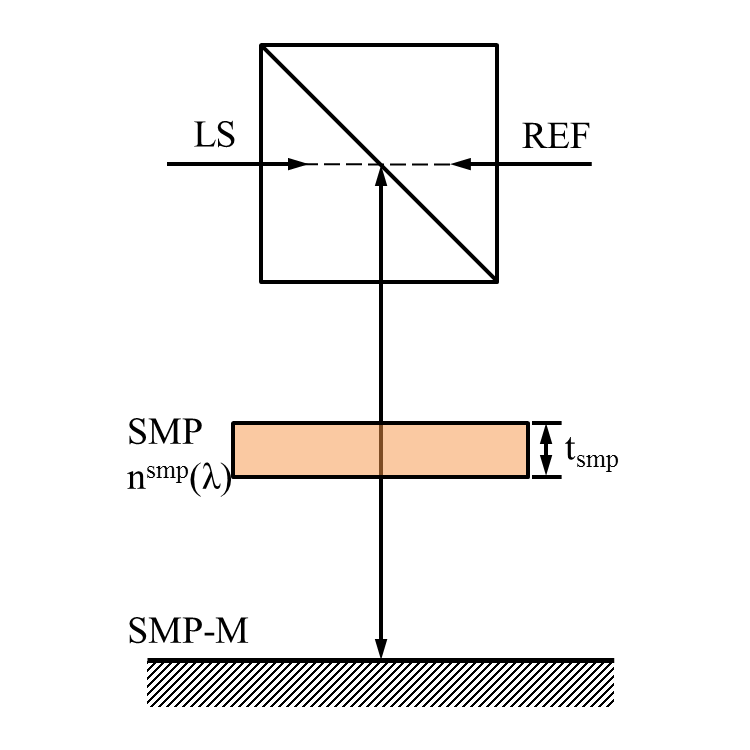}
				\put(1,1){\makebox(0,0){a)}}
			\end{overpic}
			\begin{overpic}[scale=0.27]{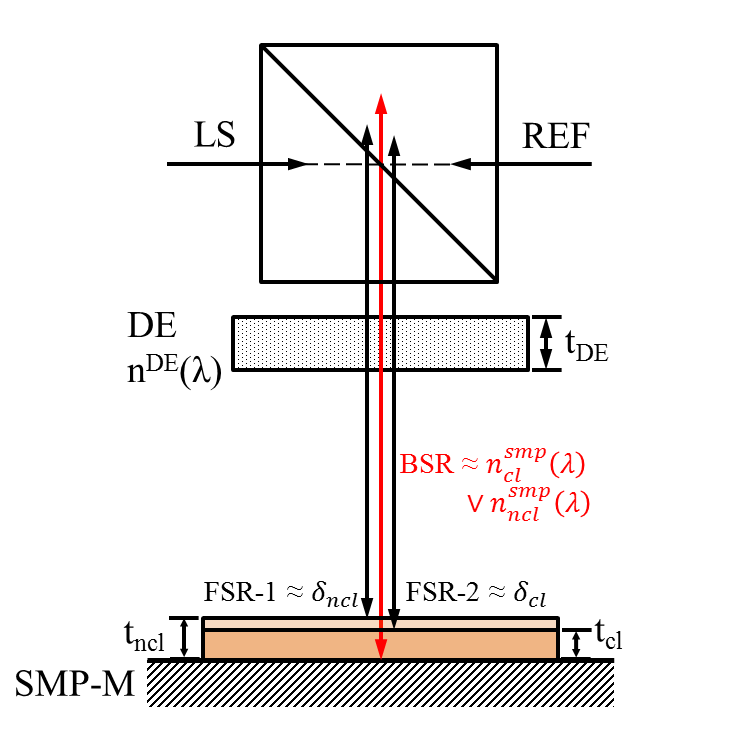}
				\put(1,1){\makebox(0,0){b)}}
			\end{overpic}
		\end{tabular}
	\end{center}
	\caption[Setup for the simulation of the influence of a thickness or refractive index change on the signal in transmission mode]{a) Setup for the simulation of the influence of a thickness or refractive index change on the signal in transmission mode of case (A) with LS - light source path, SMP - sample with refractive index $n^{smp}(\lambda)$ and thickness $t_{smp}$, SMP-M - sample mirror and REF - reference beam path; b) setup for the simulation of shrinkage and cross-linking of a polymer on a reflective sample of case (B) with DE - dispersive element having a refractive index $n^{DE}(\lambda)$ and a thickness $t_{DE}$, non-cross-linked sample with thickness $t_{ncl}$ and cross-linked sample with thickness $t_{ncl}$ as well as the respective FSR - front-side reflexes which influence the path difference $\delta_{ncl}$ and $\delta_{cl}$ and the BSR - back-side reflex which is influenced by $n_{ncl}^{smp}(\lambda)$ or $n_{cl}^{smp}(\lambda)$.}\label{polymer_signal_setups}
\end{figure}

\begin{figure}[h]
	\begin{center}
		\begin{tabular}{c}
			\begin{overpic}[scale=0.3]{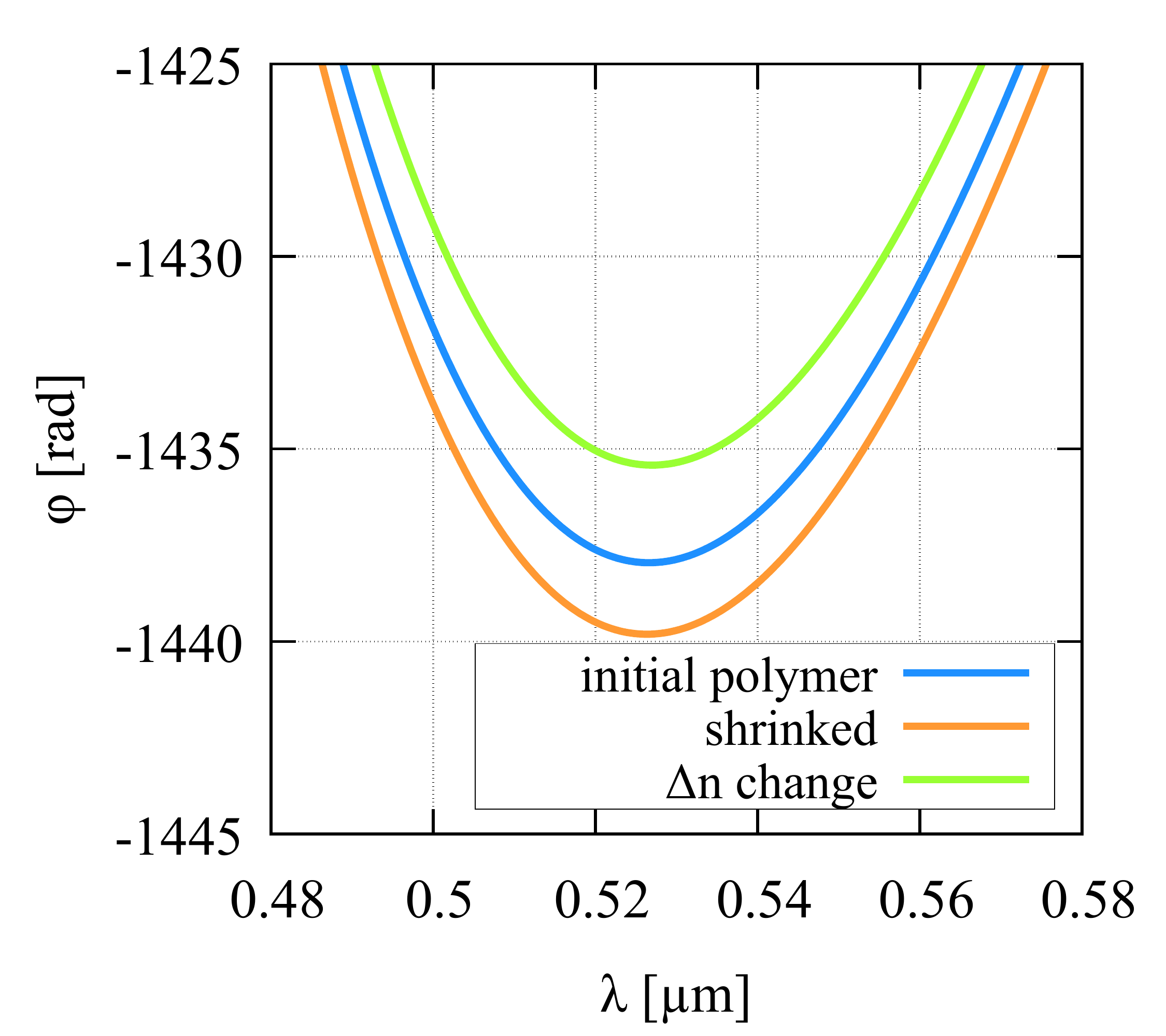}
				\put(1,1){\makebox(0,0){a)}}
			\end{overpic}
			\begin{overpic}[scale=0.3]{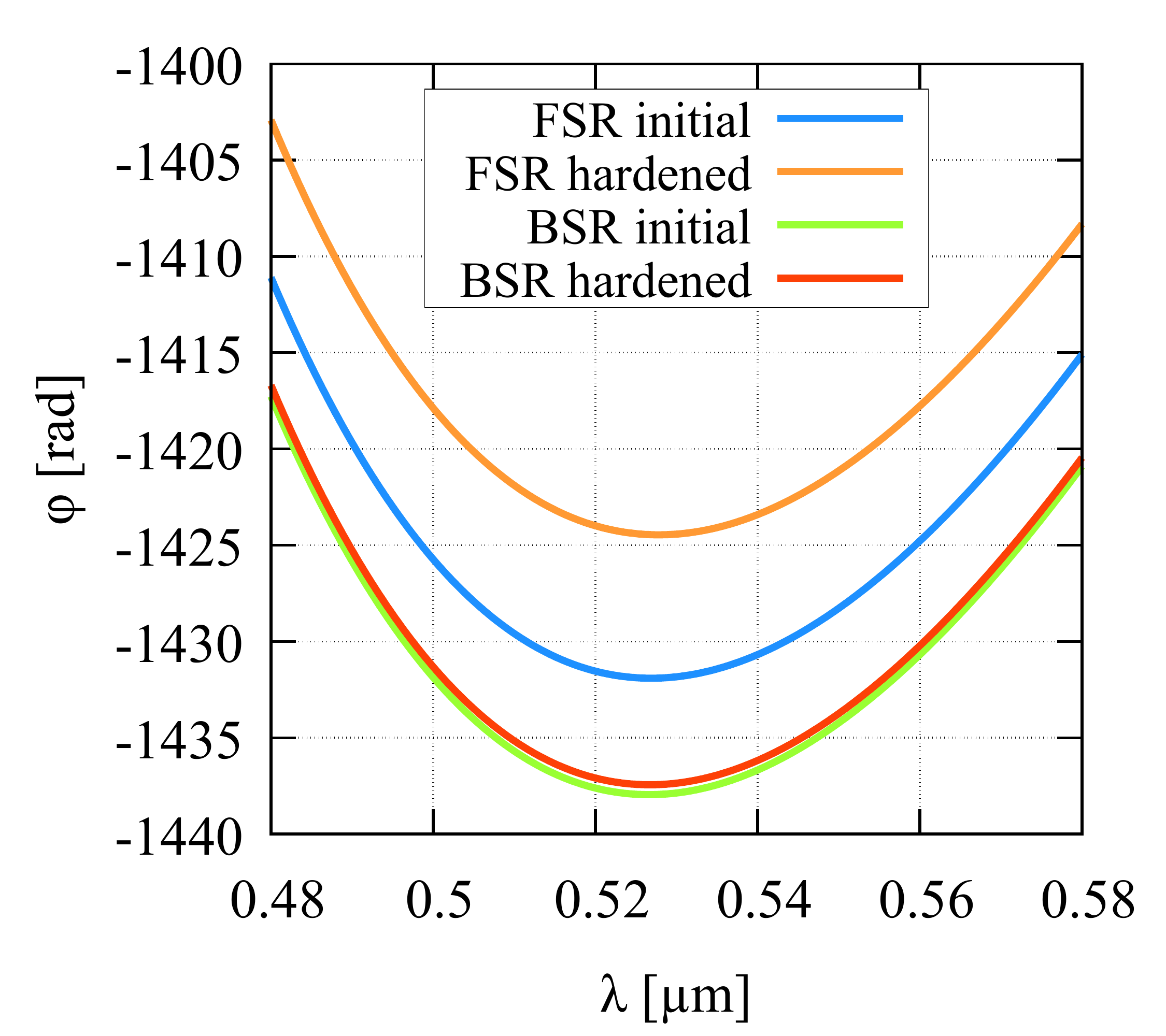}
				\put(1,1){\makebox(0,0){b)}}
			\end{overpic}
		\end{tabular}
	\end{center}
	\caption[Simulated phase signals of the influence of changes in sample thickness and refractive index]{Simulated phase signals of the influence of changes in sample thickness and refractive index for a) study of case (A) where a sample is transmitted and both changes are analyzed separately and b) study of case (B) where front side (FSR) and back side reflections (BSR) from a non-hardened and hardened sample have been investigated.}\label{PolyPic_simulation_case_A_and_B}
\end{figure}

The curing and cross-linking of polymeric materials results in a change of the refractive index as well as in dimensional changes like shrinkage. As pointed out in Eq.\,(\ref{cross-linking_interferometer_equation}), both properties influence the measurable signal in low-coherence interferometry. In order to provide a meaningful metrology tool, the influence of both properties on the measured data was to be studied. For this purpose, simulations using an industrial photo-resist material system were performed. The material properties after different processing, e.g. softbake/hardbake and UV-hardening, were determined with the reference techniques tactile profilometry and spectral ellipsometry, [H. Aßmann\textbackslash Th. Albrecht, personal communication, 12.07. \& 30.08.2018], Tab.\,\ref{tab:resist_specification_IFD}.\\  
\begin{table}[h]
	\caption[List of key properties of an exemplary positive photo-resist before and after processing]{List of key properties of an exemplary positive photo-resist before and after processing in a UV-harden cross-linking step which were used during the simulation, [H. Aßmann\textbackslash Th. Albrecht, personal communication, 12.07. \& 30.08.2018].} \label{tab:resist_specification_IFD}
\centering    
		\begin{tabular}{ccc} 
			\hline
			\rule[-1ex]{0pt}{3.5ex}  property & value  \\
			\hline
			\rule[-1ex]{0pt}{3.5ex}  resist type & positive photo-resist \\
			
			\rule[-1ex]{0pt}{3.5ex}  $t_{smp}$ before processing& 1878.32\,nm\\
			\rule[-1ex]{0pt}{3.5ex}  $t_{smp}$ after processing& 1755.49\,nm&  \\
			\rule[-1ex]{0pt}{3.5ex}  $\Delta t_{smp}/ t_{smp}$& 6.5\,\% \\
			\hline
			\rule[-1ex]{0pt}{3.5ex}  $n^{smp}$ before processing (@675\,nm)& 1.6168\\
			\rule[-1ex]{0pt}{3.5ex}  $n^{smp}$ after processing (@675\,nm)& 1.6702\\
			\rule[-1ex]{0pt}{3.5ex}  $\Delta n^{smp}/n^{smp}$& 3.2\,\% \\
			\hline
		\end{tabular}
\end{table} 
It can be seen, that significant changes in thickness as well as in refractive index occur in a counteracting fashion. In order to calculate the influence of these two properties on the measurement data, two simple simulation cases have been studied.
\begin{description}
	\item[(A) transmissive sample:] A transmissive sample with a given thickness $t_{smp}$ and a reflective index $n^{smp}$ is placed in one arm of the interferometer. By changing one of the parameters during simulation while leaving the second fixed, the individual influences can be estimated, Fig.\,\ref{polymer_signal_setups}\,a).
	\item[(B) reflective sample:] A defined layer of photo resist is spin-coated on a reflective surface (e.g. Si or glass). In this configuration, the expected measurement signal can be composed of the different components such as the front side reflex of a shrinked and non-shrinked sample as well as of the back side reflex from a sample with different states of cross-linking, Fig.\,\ref{polymer_signal_setups}\,b).
\end{description}
When analyzing case (A), a thickness change due to shrinkage of 6.5\,\% was used to calculate the influence on the phase signal and equalization wavelength by also using a fixed refractive index of $n^{smp}(\lambda)$ for the sample., Fig.\,\ref{PolyPic_simulation_case_A_and_B}\,a).
The resulting phase signal showed a relative change of about 0.13\,\% in its amplitude for the equalization wavelength, whereas the equalization wavelength showed a change of about -0.038\,\%. It can be noted, that the changes are small relative to the thickness change. Furthermore, the equalization wavelength change is an order of magnitude smaller than the change of the phase signal. In order to evaluate the influence of a change in refractive index $\Delta n$, the thickness of the simulated material was kept constant, while the refractive index was changed by 3.2\,\%, Fig.\,\ref{PolyPic_simulation_case_A_and_B} a). In this case, the phase change was found to be -0.18\,\%, while the change in equalization wavelength was 0.074\,\%. These results make clear, that the influence of the refractive index change alone is stronger than a thickness-induced change. Furthermore, as both effects are counteracting, a separation of the changes during a measurement might be obscured. Although shrinkage-induced changes of thickness and refractive index are expected to be in the single percentage regime, \cite{2Photon_waveguides}, this effect has to be taken into account. This result also reveals the necessity of a more profound data analysis as shown in chapter \ref{ChapterProfilometry}. Especially the fit of the measured data in the region of the equalization wavelength, enables high precision in the measurement of both thickness and refractive index of a (polymeric) sample.\\
A more realistic assessment of the signals has been performed by the study of case (B), Fig.\,\ref{polymer_signal_setups}\,b). This case describes four relevant signal components in the analysis of a single layer of polymer, where in one part the sample is assumed to be non-cross-linked (\textit{initial}) and in the other fully cross-linked (\textit{hardened}). As described before,  all signals passed through a known dispersive element. In case (B).1 the front side reflex (FSR) of the non-cross-linked material interface interferes with the reference signal while in case (B).2 the front side reflex of the cross-linked material is analyzed. In this case, as pure shrinkage was studied, a change in the equalization wavelength of 0.19\,\% and a change in the phase signal of -0.52\,\% could be observed. In the cases (B).3 and (B).4 the effect of the refractive index is studied for both samples while the shrinkage effect is included in the signal of the cross-linked material. By analyzing the back reflected signal from both, the cross-linked and the non-cross-linked sample, it is possible to investigate not only the effect of shrinkage but also the effect of refractive index variation due to cross-linking. The results show clearly the counteracting nature of shrinkage and refractive index alternations during cross-linking with the change in the equalization wavelength being 0.037\,\% and a change in the phase signal being -0.036\,\%. This highlights, that a refractive index induced change can be nearly obscured by the influence of the shrinkage of a material. For this reason, all experiments within this work performed an evaluation of the FSR as well as of the BSR signal. This approach enables separating signal changes due thickness variations from those that are caused by refractive index variations and can deliver information on the surface profile of the sample. During the evaluation of thin materials ($t\,<\,$100\,\textmugreek m), the signals of front- and back side reflection mix. In order to perform the correct interpretation of the measured data from cross-linked and non-cross-linked samples, both signal parts have been separated. This can be done by filtration or temporal separation. 

\subsection{Error parameters of the temporal approach}\label{SubSec:ErrorPropTemporal}
The determination of a sample's group refractive index, based on the temporal approach, relies on the translation of the reference mirror in order to capture the path length difference for a number of equalization wavelengths, see subsection \ref{subsection_temporal_approach}. The final calculation relies on the path length difference $\delta_{\lambda}$ and the thickness of the sample $t_{smp}$  
\begin{equation}\label{EQgroupIndex_temporal_err}
n_g(\lambda) = \frac{\delta(\lambda)}{t_{smp}}.
\end{equation}
In order to estimate the error for this measurement, an error propagation was performed with respect to both relevant features
\begin{equation}
\Delta n_g = \sqrt{ \left( \frac{\partial n_g}{\partial \delta(\lambda)} \cdot \Delta \delta(\lambda) \right)^2 + \left( \frac{\partial n_g}{\partial t_{smp}}  \cdot \Delta t_{smp} \right)^2   }.
\end{equation}
The calculation of the derivative of the respective terms leads therefore to
\begin{equation}
\Delta n_g = \sqrt{ \left( \frac{1}{t_{smp}}  \cdot \Delta \delta(\lambda) \right)^2  + \left( -\frac{ \delta(\lambda) }{ t_{smp}^2} \cdot \Delta t_{smp} \right)^2 }.
\end{equation}
Based on the parameters used for the experiments, the error limits were calculated for the dispersive elements of $t_{smp}$\,=\,(1,\,3\,and\,5)\,mm while the measurement accuracy on these thicknesses was $\Delta t_{smp}$ = 20 nm when measured with a tactile profilometer, \cite{TH_iSeries}. The covered path length differences were $\delta(\lambda)$\,=\,(49.8,\,185.5 and 345)\,\textmugreek m respectively while the resolution of the translation stage was $\Delta \delta(\lambda)$\,=\,1\,nm, \cite{Smaract_stage}. This resulted in values of $\Delta n_g^1$\,=\,\num{1.41e-6}, $\Delta n_g^3$\,=\,\num{5.30e-7} , $\Delta n_g^5$\,=\,\num{3.41e-7} for the different measured samples in section \ref{subsection_temporal_approach}.

\subsection{Error propagation in WPDE}
Independently of the kind of \gls{WPDE}, sample-only or with an additional dispersive element, the accuracy of the refractive index calculation is dependent on the deviation of the measured input parameters such as the sample thickness $t_{smp}$, the wavelength $\lambda$ or the thickness of the dispersive element $t_{DE}$. Therefore, a propagation of deviations of these parameters was performed to estimate their relative influences following the scheme\\
\begin{equation}
\Delta n_g = \sqrt{ \left( \frac{\partial n_g}{\partial t_{smp}} \cdot \Delta t_{smp} \right)^2 + \left( \frac{\partial n_g}{\partial t_{DE}}  \cdot \Delta t_{DE} \right)^2 + \left( \frac{\partial n_g}{\partial \lambda}  \cdot \Delta \lambda \right)^2  }.
\end{equation}
For all investigations on this topic, samples are supposed to have thicknesses varying from (0.1\,-\,5)\,mm which were measured with a deviation ranging from (0.16\,-\,20)\,nm, \cite{TH_iSeries}. The dispersive element was made of N-BK7 glass with the given thicknesses. Wavelength dependencies were investigated in a spectral range of (400\,-\,1000)\,nm which was determined with a deviation of 0.1\,nm, \cite{Avantes_spec}.

\subsubsection{Sample-only WPDE}\label{SubSub:smp_only}
\begin{figure}[h]
	\begin{center}
		\begin{tabular}{c}
			\begin{overpic}[scale=0.31]{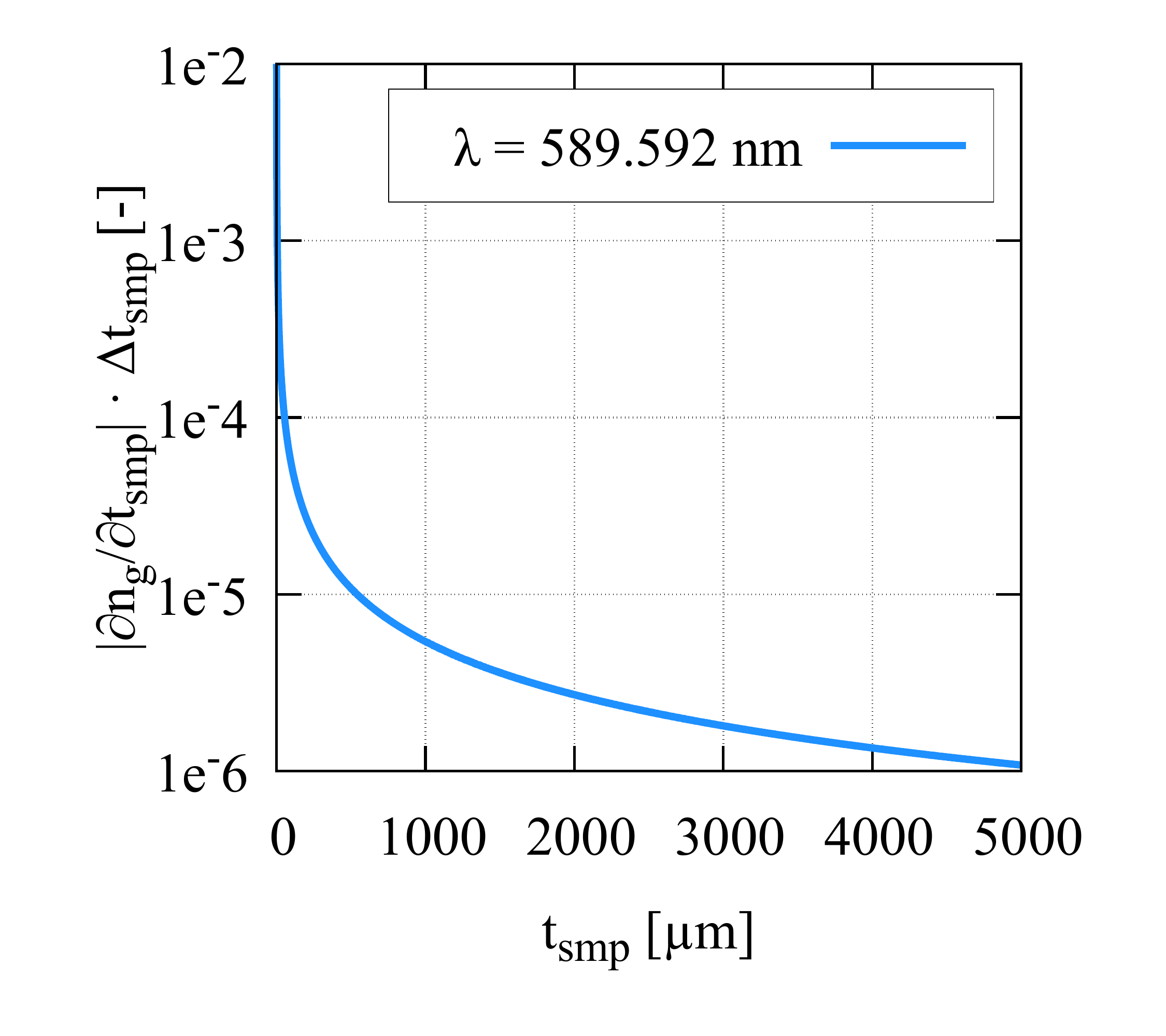}
				\put(1,1){\makebox(0,0){a)}}
			\end{overpic}  
			\begin{overpic}[scale=0.31]{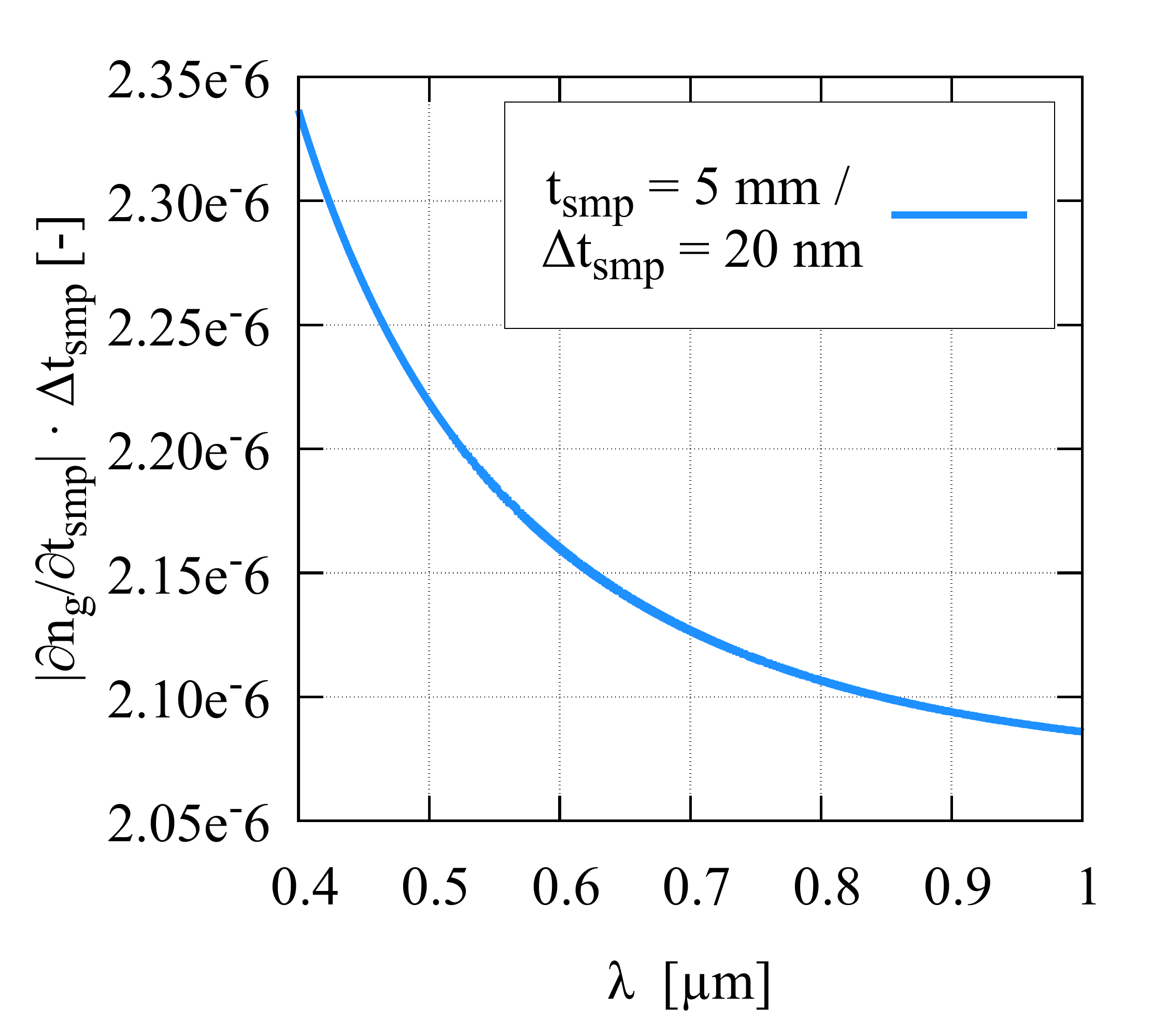}
				\put(1,1){\makebox(0,0){b)}}
			\end{overpic}
		\end{tabular}
	\end{center}
	\caption[Error contribution of the sample thickness to the group refractive index $\Delta n_g(t_{smp})$ for the sample-only WPDE approach]{Error contribution of the sample thickness to the group refractive index $\Delta n_g(t_{smp})$ for the sample-only WPDE approach with a) over a given thickness range at the sodium D1 line ($\lambda$\,=\,589.592\,nm) and b) over a spectral range with $t_{smp}$\,=\,5\,mm and the measurement error for $\Delta t_{smp}$\,=\,20\,nm .}\label{PolyPic:ErrTsmpSingle}
\end{figure}

\begin{figure}[h]
	\begin{center}
		\begin{tabular}{c}
			\begin{overpic}[scale=0.31]{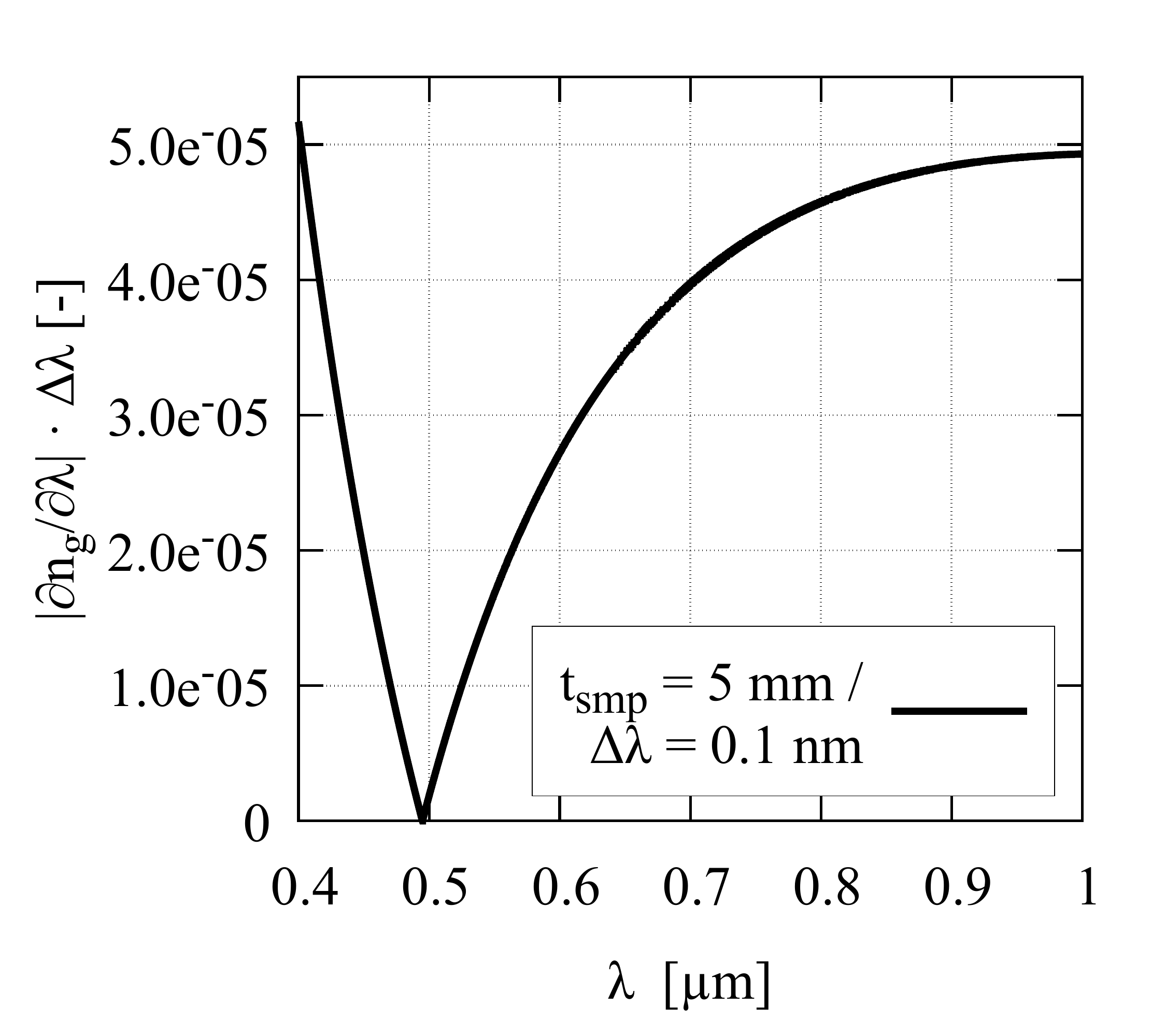}
				\put(1,1){\makebox(0,0){a)}}
			\end{overpic}  
			\begin{overpic}[scale=0.31]{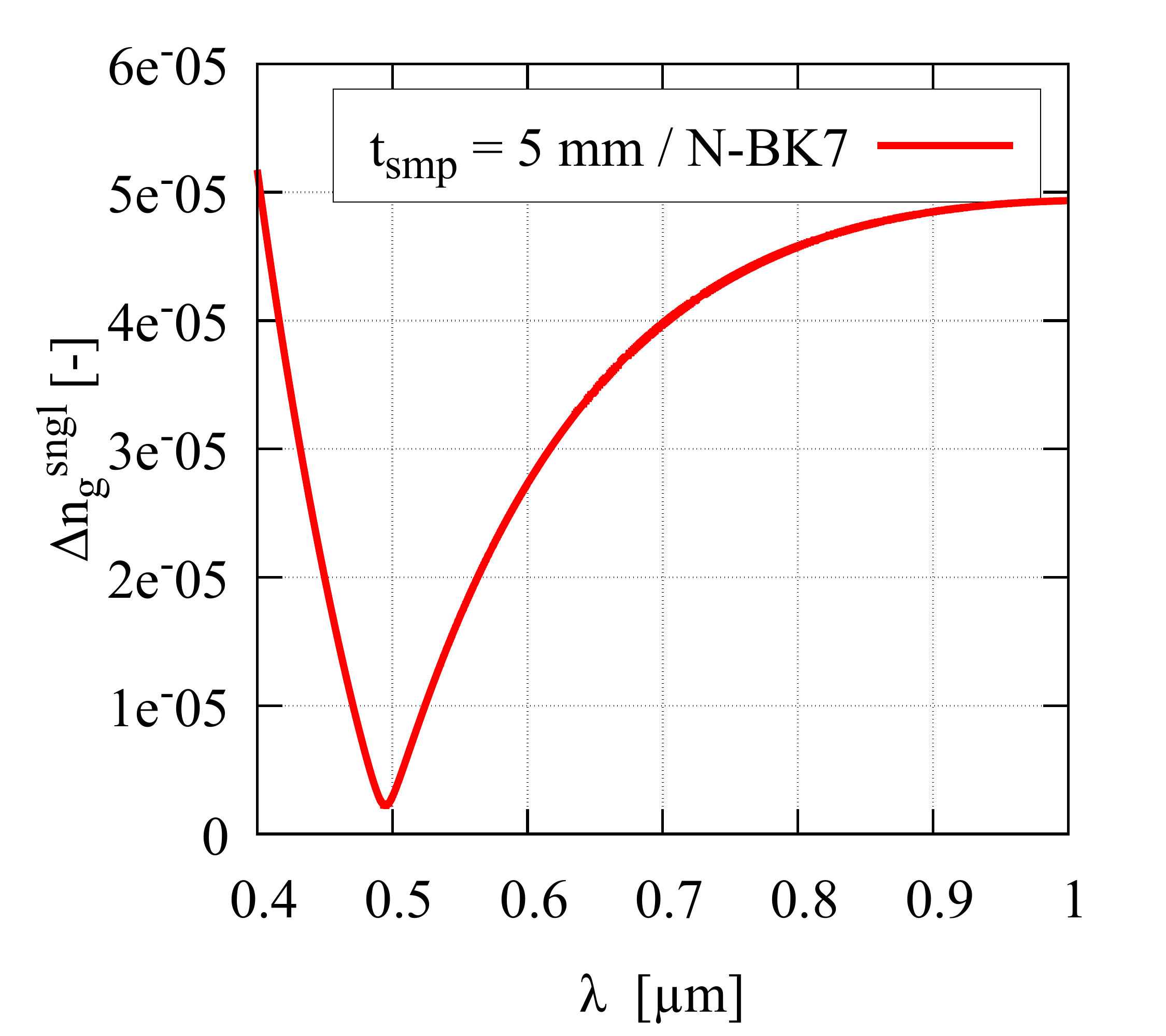}
				\put(1,1){\makebox(0,0){b)}}
			\end{overpic}
		\end{tabular}
	\end{center}
	\caption[Spectral dependency in a range from 0.4 - 1\,\textmugreek m to the group refractive index $\Delta n_g(t_{smp})$ for the sample-only \gls{WPDE} approach]{Spectral dependency in a range from 0.4 - 1\,\textmugreek m with a) error contribution of the wavelength to the group refractive index $\Delta n_g^{sngl}(\lambda)$ for the sample-only \gls{WPDE} approach with $t_{smp}$\,=\,5\,mm and the measurement error for $\Delta \lambda$\,=\,0.1\,nm and b) total error on the calculated group refractive index of the sample with all discussed error contributions according to Eq. (\ref{TotalErrSingleMat}) .}\label{PolyPic:ErrLambdaSingle}
\end{figure}

When using the \gls{WPDE} method for simple transmissive measurements of bulk materials, see section \ref{TextWPDE}, the refractive index depends only on the deviation of the sample thickness $\Delta t_{smp}$ and the wavelength $\Delta \lambda$
\begin{equation}\label{TotalErrSingleMat}
\Delta n_g^{sngl} = \sqrt{ \left( \frac{\partial n_g^{sngl}}{\partial t_{smp}} \cdot \Delta t_{smp} \right)^2 + \left( \frac{\partial n_g^{sngl}}{\partial \lambda} \cdot \Delta \lambda \right)^2  }.
\end{equation}
According to this notation the partial derivative of the group refractive index relative to the sample thickness
\begin{equation}
\frac{\partial n_g^{sngl}}{\partial t_{smp}} \cdot \Delta t_{smp} = \frac{\partial}{\partial t_{smp}} \left(  1 -\frac{\frac{\varphi_{loc}^\prime \lambda^2}{2\pi} - \delta}{t_{smp}}  \right) \cdot \Delta t_{smp} = \frac{(1 - n_g^{smp})t_{smp}^{eff}}{t_{smp}^2} \cdot \Delta t_{smp} 
\end{equation}
reveals a simple quadratic dependency which is countered by the effective thickness $t_{smp}^{eff}$ and the relative path difference $\delta$ that contribute to the measured phase signal $\varphi_{loc}^\prime$.
In the given range of sample thicknesses the deviation of $\Delta n_g(t_{smp})$ could be estimated at the sodium D1 line ($\lambda$\,=\,589.592\,nm), Fig. \ref{PolyPic:ErrTsmpSingle} a).
The quadratic dependence becomes visible. The data here was plotted on a semi-logarithmic scale, it has to be noted that the error is smaller than \num{1e-5} for samples thicker than 500\,\textmugreek m. In a measurement situation this might be considered as a limiting factor for the characterization of thin samples or layers.\\
Furthermore, the contribution to the error for a sample of N-BK7 having the thickness $t_{smp}$\,=\,5\,mm was analyzed over a spectral range from (0.4\,-\,1)\,\textmugreek m, Fig. \ref{PolyPic:ErrTsmpSingle} b). The calculated error for this sample at the sodium D1 line is \num{2.16e-6}. It changes less than \num{0.3e-6} over the given wavelength range. In consequence, for high precision measurements the analysis wavelength should be as low as possible in order to achieve results with minor error although the gain in accuracy is small.\\
Accordingly, the error contribution resulting from the wavelength measurement uncertainty
\begin{equation}
\frac{\partial n_g^{sngl}}{\partial \lambda} \cdot \Delta \lambda = \frac{\partial}{\partial \lambda} \left( 1 - \frac{\frac{\varphi_{loc}^\prime \lambda^2}{2\pi} - \delta}{t_{smp}} \right) \cdot \Delta \lambda=- \frac{2 \lambda \left[ (1-n_g^{smp})t_{smp}^{eff} + \delta \right]}{\lambda_{eff}^2 \cdot t_{smp }} \cdot \Delta \lambda
\end{equation}
was estimated for a N-BK7 sample having the thickness $t_{smp}$\,=\,5\,mm over the spectral range of 0.4 to 1\,\textmugreek m, Fig.\,\ref{PolyPic:ErrLambdaSingle}\,a).   
Interestingly, the error contribution is quiet high with $\pm$\num{5e-5} but becomes zero at a wavelength of 494.9\,nm. This results from the ratio of the group refractive index to the wavelength squared which in case of N-BK7 results in a zero error within the wavelength range of interest. In this case, experiments should be designed to measure the group refractive index with the \gls{WPDE} method in a wavelength range close to this minimum. The error varies from \num{5.17e-5} at 0.4\,\textmugreek m to \num{4.93e-5} at 1\,\textmugreek m. In case of different materials, simulations have to be performed to find an optimized probing wavelength.\\
Consequently, the overall error according to Eq.\,(\ref{TotalErrSingleMat}) shows a minimum at this wavelength, Fig.\,\ref{PolyPic:ErrLambdaSingle} b). It can be seen, that for a sample thickness of $t_{smp}$\,=\,5\,mm, the contribution of the wavelength is quite significant and introduces the main error. Furthermore, it is known from the simulations, that the sample material as well as its thickness can have an even more significant influence on the error of the calculated group refractive index. This behavior should be considered when designing experiments for arbitrary materials.


\subsubsection{WPDE with additional DE}\label{SubSubSec:WPDEaddDE} 
\begin{figure}[h]
	\begin{center}
		\begin{tabular}{c}
			\begin{overpic}[scale=0.31]{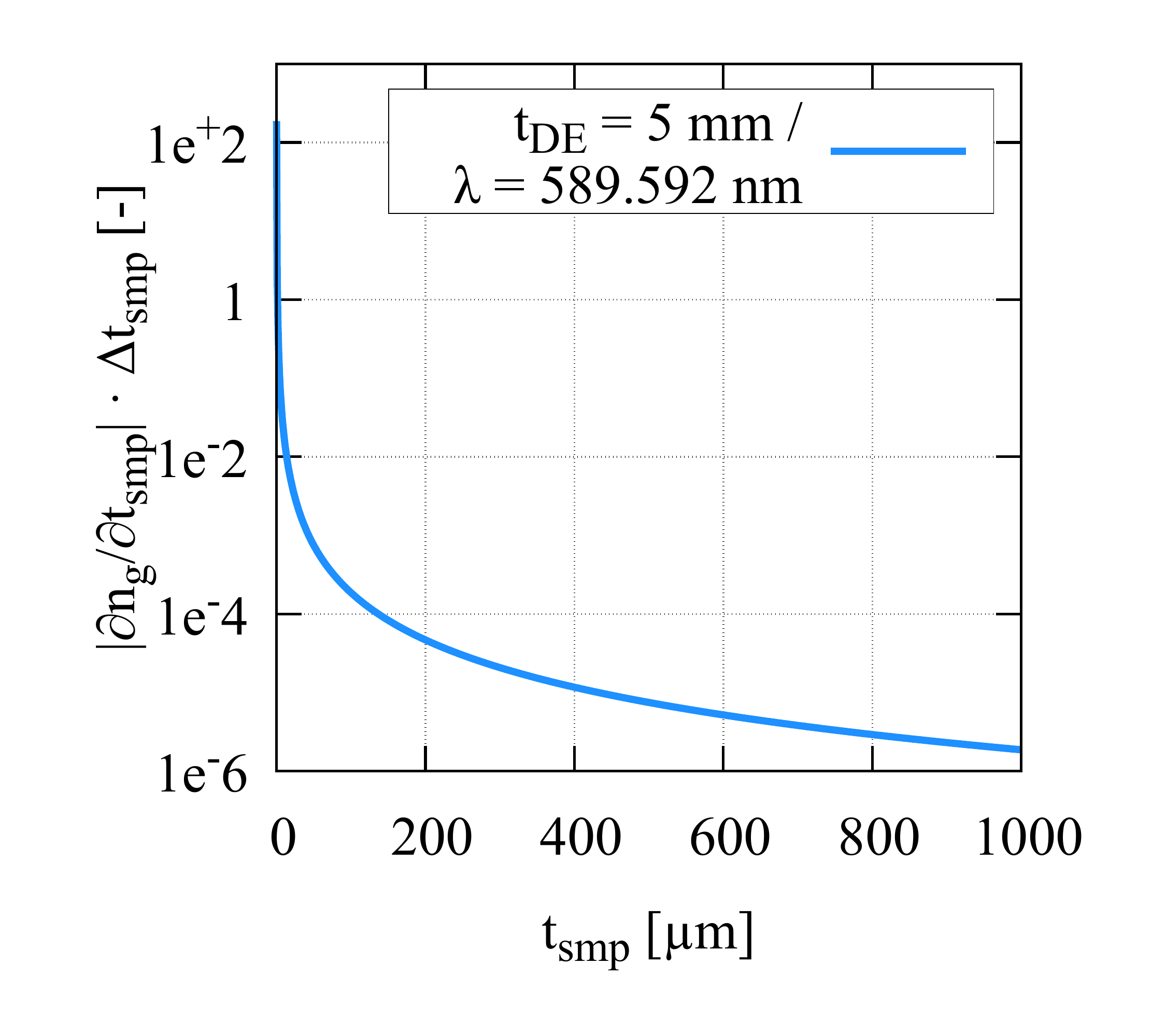}
				\put(1,1){\makebox(0,0){a)}}
			\end{overpic}  
			\begin{overpic}[scale=0.31]{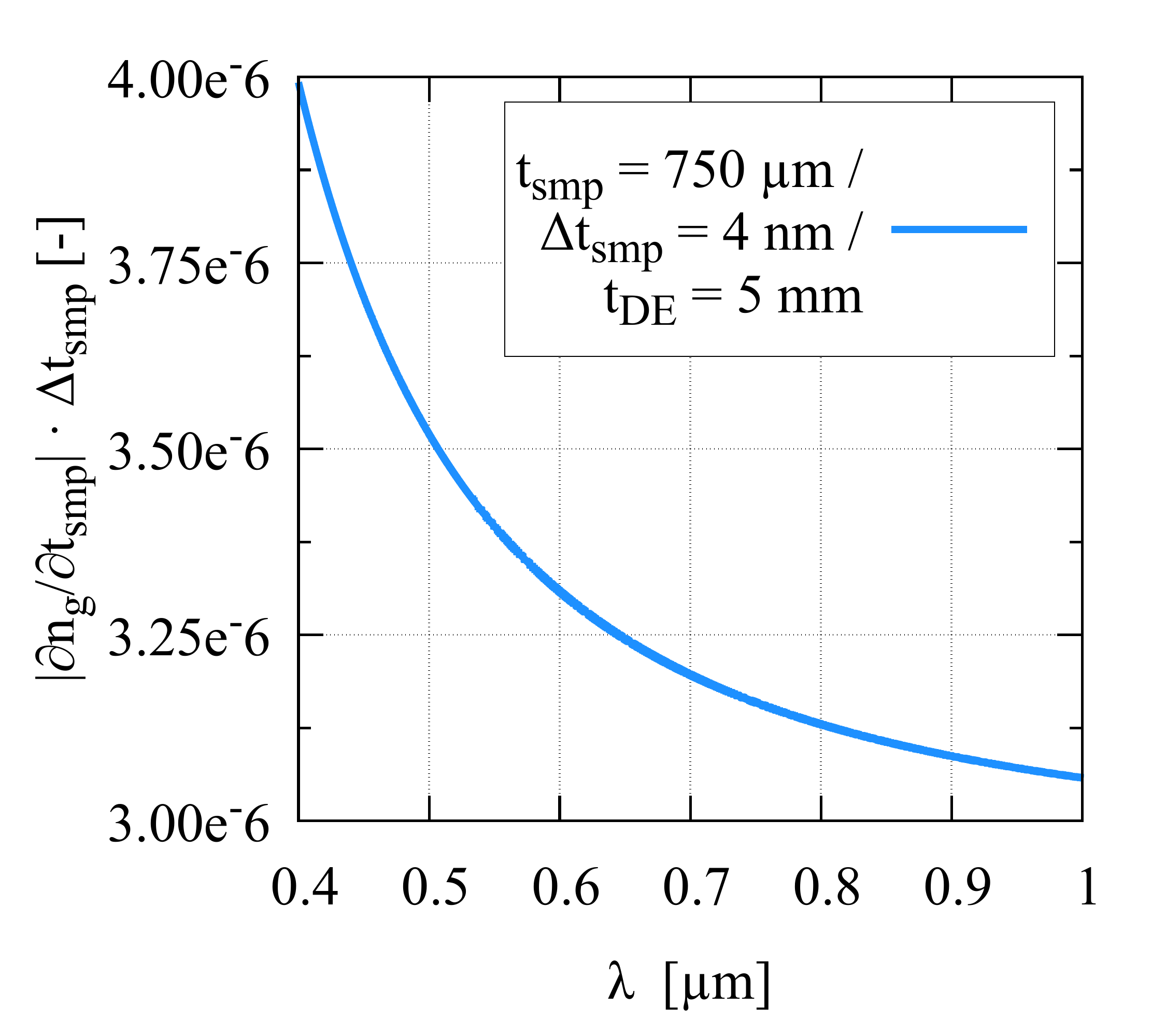}
				\put(1,1){\makebox(0,0){b)}}
			\end{overpic}
		\end{tabular}
	\end{center}
	\caption[Error contribution of the sample to the group refractive index $\Delta n_g(t_{smp})$ for the WPDE approach with additional dispersion]{Error contribution of the sample to the group refractive index $\Delta n_g(t_{smp})$ for the WPDE approach with additional dispersion a) over a given sample thickness range at the sodium D1 line ($\lambda$\,=\,589.592\,nm) and b) over a spectral range with $t_{DE}$\,=\,5\,mm, $t_{smp}$\,=\,750\,\textmugreek m and the measurement error for $\Delta t_{smp}$\,=\,4\,nm.}\label{PolyPic:ErrtsmpAddDisp}
\end{figure}

\begin{figure}[h]
	\begin{center}
		\begin{tabular}{c}
			\begin{overpic}[scale=0.31]{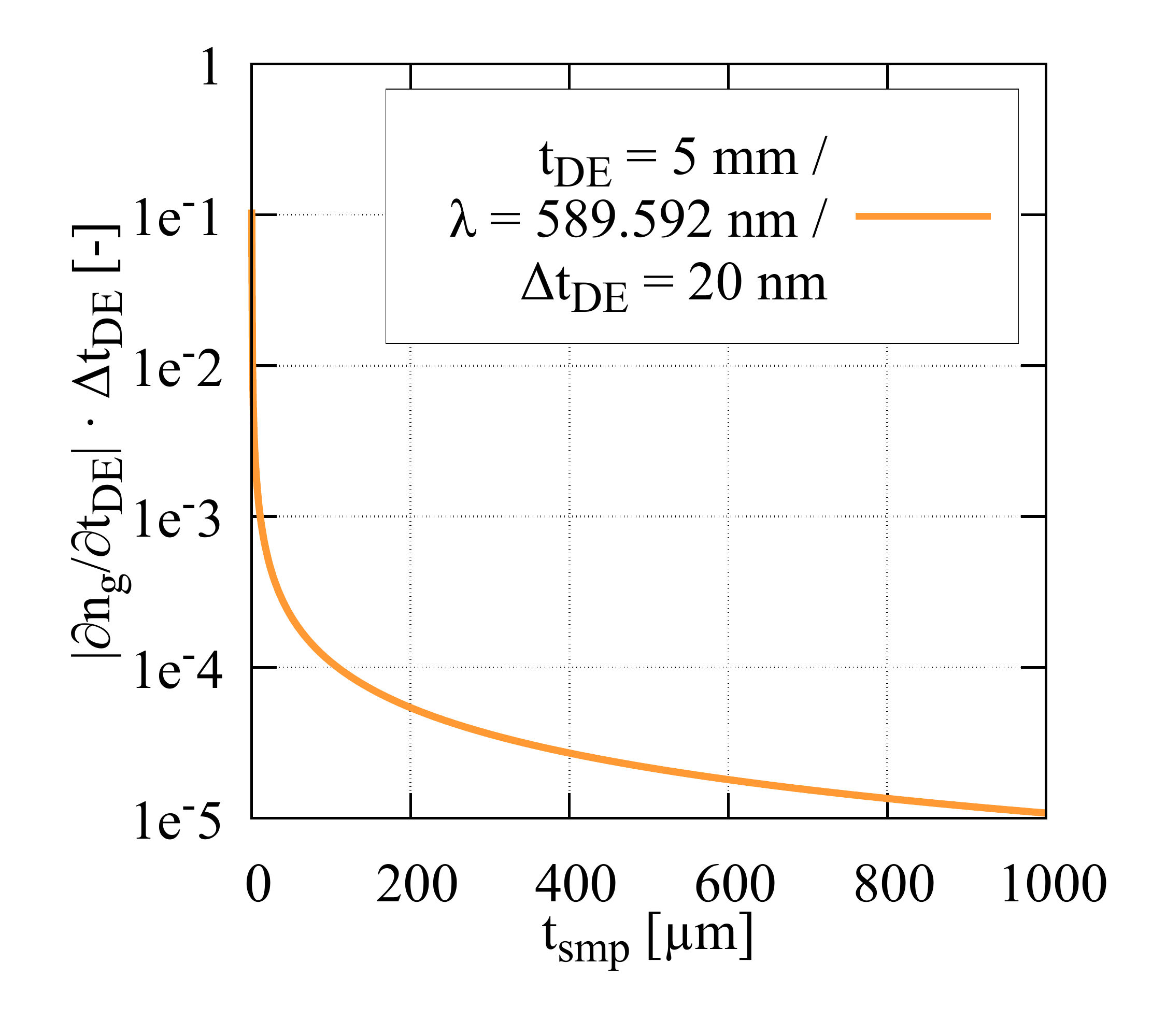}
				\put(1,1){\makebox(0,0){a)}}
			\end{overpic}  
			\begin{overpic}[scale=0.31]{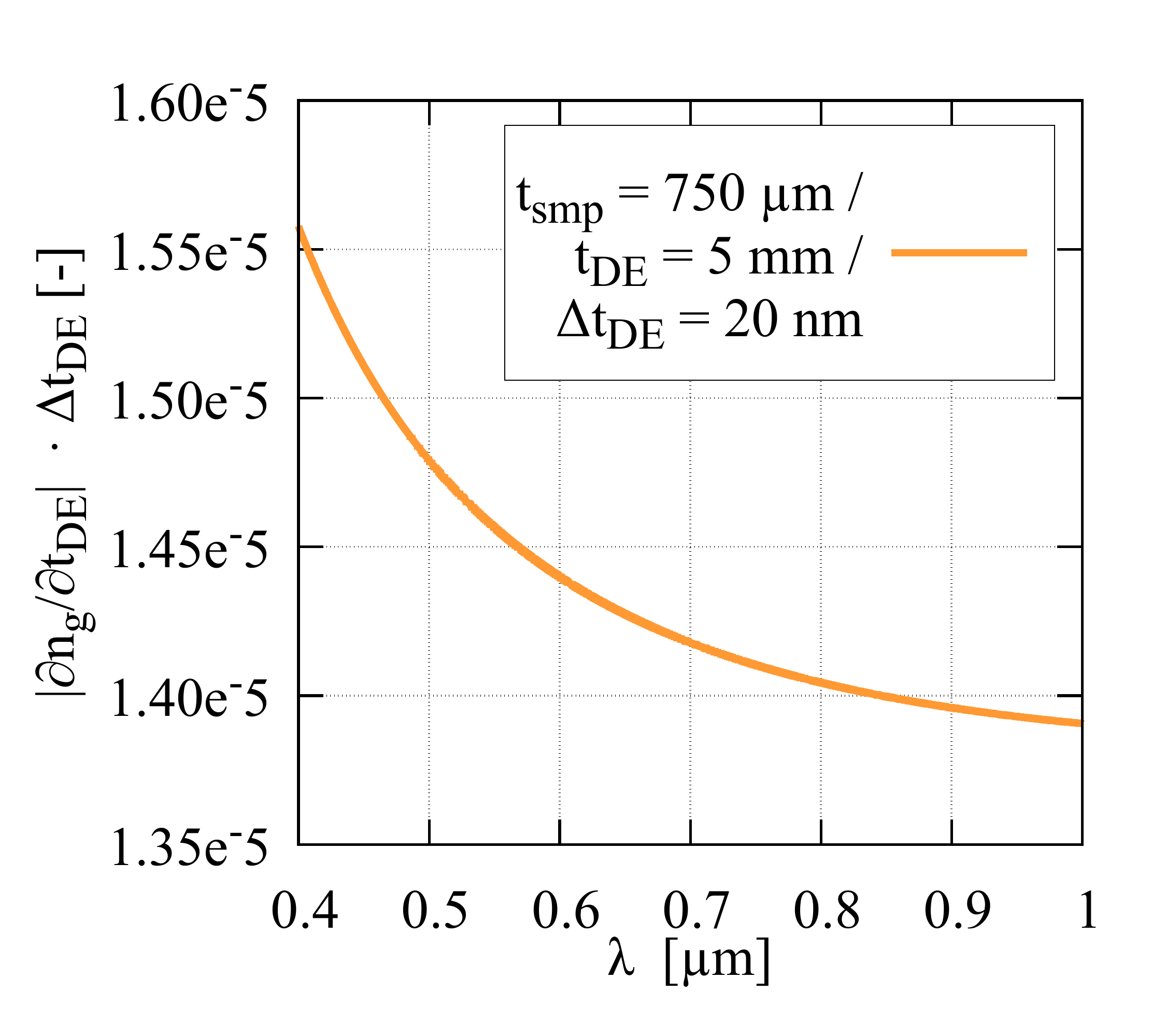}
				\put(1,1){\makebox(0,0){b)}}
			\end{overpic}
		\end{tabular}
	\end{center}
	\caption[Error contribution of the dispersive element to the group refractive index $\Delta n_g(t_{DE})$ for the WPDE approach with additional dispersion]{Error contribution of the dispersive element to the group refractive index $\Delta n_g(t_{DE})$ for the WPDE approach with additional dispersion a) over a given sample thickness range at the sodium D1 line ($\lambda$\,=\,589.592\,nm) and b) over a spectral range with $t_{DE}$\,=\,5\,mm, $t_{smp}$\,=\,750\,\textmugreek m and the measurement error for $\Delta t_{DE}$\,=\,20\,nm.}\label{PolyPic:ErrtDEAddDisp}
\end{figure}

\begin{figure}[h]
	\begin{center}
		\begin{tabular}{c}
			\begin{overpic}[scale=0.31]{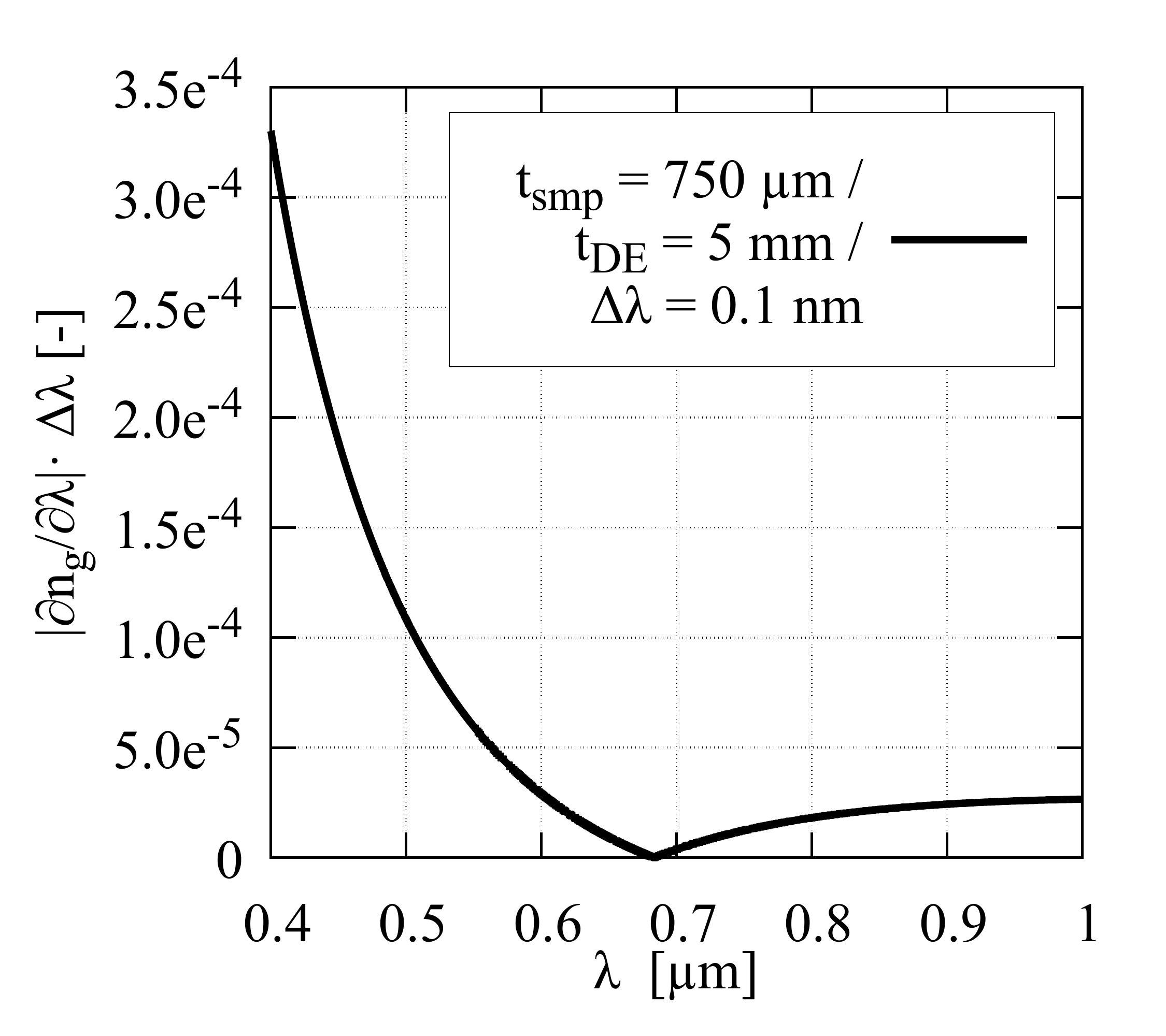}
				\put(1,1){\makebox(0,0){a)}}
			\end{overpic}  
			\begin{overpic}[scale=0.31]{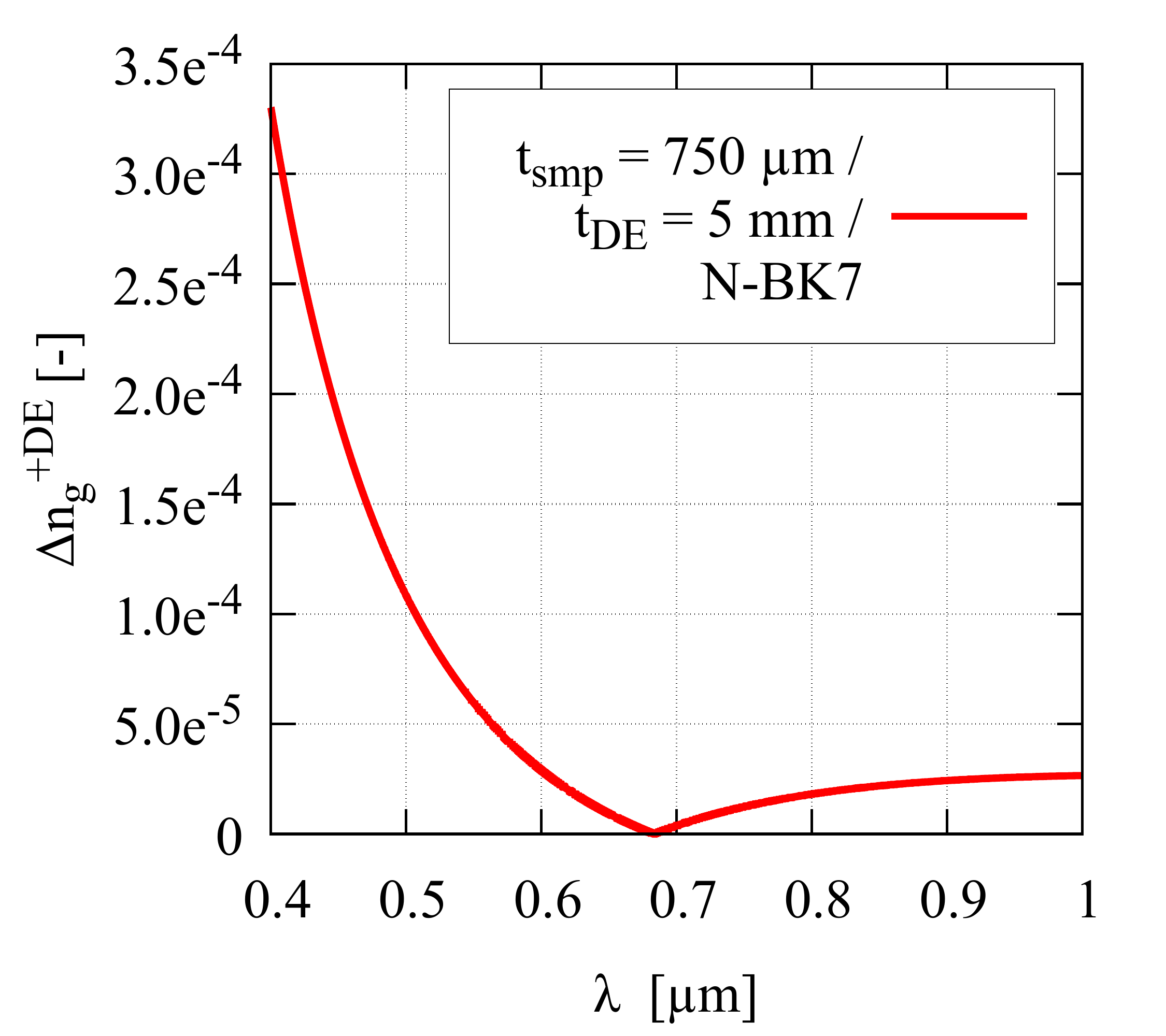}
				\put(1,1){\makebox(0,0){b)}}
			\end{overpic}
		\end{tabular}
	\end{center}
	\caption[Spectral dependency in a range from (0.4\,-\,1)\,\textmugreek m to the group refractive index $\Delta n_g(t_{DE})$ for the WPDE approach with additional dispersion]{Spectral dependency in a range from 0.4\,-\,1\,\textmugreek m with a) error contribution of the wavelength to the group refractive index $\Delta n_g^{+DE}(\lambda)$ for the WPDE approach with additional dispersion with $t_{DE}$\,=\,5\,mm, $t_{smp}$\,=\,750\,\textmugreek m and the measurement error for $\Delta \lambda$\,=\,0.1\,nm and b) total error on the calculated group refractive index of the sample with all discussed error contributions according to Eq.\,(\ref{EqErrNGaddDis}).}\label{PolyPic:ErrLambaNGAddDisp}
\end{figure}

As discussed within this chapter, \gls{WPDE} can be applied in some cases with an additional dispersive element in the setup, \ref{PicPoly:SetupAddDispersion}. In this case, the propagation of deviations is based on a different equation for the determination of the group refractive index $n_g^{+DE}$, Eq.\,(\ref{cos_operation_cross-linking})
\begin{eqnarray}\label{nG_addDis}
&&n_g^{+DE}(x,\lambda) = 1 - \frac{\lambda^2 \cdot \xi}{2\pi \cdot t_{smp}} \label{n_g_crosslinking} \\
&& \textrm{with } \xi = \varphi_{loc}^\prime - \frac{2\pi}{\lambda^2}\left[ \left(1 - n_g^{DE} \right)t_{DE} + \delta  \right].
\end{eqnarray}
Furthermore, it additionally depends on the measurement uncertainty of the thickness of the dispersive element $\Delta t_{DE}$
\begin{equation}\label{EqErrNGaddDis}
\Delta n_g^{+DE} = \sqrt{ \left( \frac{\partial n_g^{+DE}}{\partial t_{smp}} \cdot \Delta t_{smp} \right)^2 + \left( \frac{\partial n_g^{+DE}}{\partial t_{DE}} \cdot \Delta t_{DE} \right)^2 + \left( \frac{\partial n_g^{+DE}}{\partial \lambda} \cdot \Delta \lambda \right)^2  }.
\end{equation}
Similar to the case in subsection \ref{SubSub:smp_only}, the deviation relative to the error of the sample thickness is quadratically dependent on the sample thickness itself\\
\begin{eqnarray}
\begin{split}
\frac{\partial n_g^{+DE}}{\partial t_{smp}} \cdot \Delta t_{smp} &= \frac{\partial}{\partial t_{smp}} \left( \frac{\lambda^2 \cdot \xi}{2\pi \cdot t_{smp}^2} \right) \cdot \Delta t_{smp}\\
& =\frac{1}{t_{smp}^2}\left(\frac{\lambda^2 \cdot \varphi_{loc}^\prime}{2\pi} - \left[ \left(1 - n_g^{DE} \right)t_{DE} + \delta  \right] \right) \cdot \Delta t_{smp}.
\end{split}\\
\textrm{with } \varphi_{loc}^\prime = \frac{2\pi}{\lambda^2}  \left[ (1 - n_g^{smp}) \cdot t_{smp}^{eff} + (1 - n_g^{DE} ) \cdot t_{DE}^{eff} +  \delta \right]\\
\frac{\partial n_g^{+DE}}{\partial t_{smp}} \cdot \Delta t_{smp} = \frac{1}{t_{smp}^2} \left[ (1 - n_g^{smp}) \cdot t_{smp}^{eff} + (1 - n_g^{DE} ) \left( t_{DE}^{eff} - t_{DE}\right) \right] \cdot \Delta t_{smp}.
\end{eqnarray}
It has to be noted that the calculation of $\varphi_{loc}^\prime$ using the effective thicknesses $ t_{smp}^{eff}$ and $ t_{DE}^{eff}$ is performed only in context of the error propagation in order to simulate the signal which is usually measured. Using the same range of sample thicknesses, with a dispersive element of $t_{DE}$\,=\,5\,mm N-BK7, a thickness-dependent error can be calculated for the sodium D1 line ($\lambda$\,=\,589.592\,nm), Fig.\,\ref{PolyPic:ErrtsmpAddDisp} a).
It is clear from the equation and the plot that the contribution of the dispersive element to this particular error is minimal. The measured thickness $t_{DE}$ as well as the effective thickness $t_{DE}^{eff}$, which contributes to the composition of the phase signal $\varphi_{loc}^\prime$, only affect the error with their difference to each other. Therefore, analogous to the approach with a single sample material, the error contribution with respect to the sample thickness dominates. It is smaller than \num{1e-5} for sample thicknesses above 420\,\textmugreek m for measurements at the sodium D1 line. With regard to the spectral range of the measurement, the 1/$t_{smp}^2$ behavior is smaller than \num{3e-5} and thus neglectable in comparison to the influence of the sample thickness, Fig. \ref{PolyPic:ErrtsmpAddDisp} b). An error variation of \num{1e-6} can be observed over a spectral range of 0.6\,\textmugreek m. In consequence, the measurement wavelength range should be as low as possible in order to optimize the error contribution of the sample thickness.\\
When partially deriving Eq. (\ref{nG_addDis}) with respect to the thickness of the dispersive element,
\begin{equation}
\frac{\partial n_g^{+DE}}{\partial t_{DE}} \cdot \Delta t_{DE} = - \frac{(1-n_g^{DE})}{t_{smp}} \cdot \Delta t_{DE}
\end{equation}
it becomes obvious that the error contribution of the dispersive element is determined by its refractive index and not by the thickness itself. But however, the thickness of the sample shows a 1/$t_{smp}$ influence on this error contribution, Fig. \ref{PolyPic:ErrtDEAddDisp} a).
The error contribution becomes significant for samples thinner than 200\,\textmugreek m for the sodium D1 line where it is larger than  \num{5.41e-5}. The differences in the error contribution for different wavelengths are in the range of $\pm$\num{.09e-5} over the spectral range of 0.4\,-\,1\,\textmugreek m while its maximum value is \num{1.56e-5} for a wavelength of 0.4\,\textmugreek m, Fig. \ref{PolyPic:ErrtDEAddDisp} b). This leads to the same consequence as for the error contribution of the sample thickness alone. In order to minimize the error in the calculation of the group refractive index, the spectral range for measurements should be as low as possible. This can be achieved by the choice of the light source, the spectrometer and by tuning the path length difference of both arms as to shift the ROI appropriately, see Fig. \ref{PicPoly:SetupAddDispersion}.\\
Furthermore, the group refractive index was also partially derived with respect to the wavelength $\lambda$ in order to estimate its influence on the error. For this operation, the equation was rewritten and broken down into its main wavelength dependent components X and Y
\begin{eqnarray}\label{EqErrNGLambdaStart}
\frac{\partial n_g^{+DE}}{\partial \lambda} \cdot \Delta \lambda =\frac{\partial}{\partial \lambda} \left( 1 - \frac{ \varphi_{loc}^\prime \cdot \lambda^2}{2\pi \cdot t_{smp}} - \frac{ [1-n_g^{DE}(\lambda)]t_{DE}}{t_{smp}}  - \frac{\delta}{t_{smp}} \right) \cdot \Delta \lambda\\
\textrm{where } \frac{\varphi_{loc}^\prime \cdot \lambda^2}{2\pi \cdot t_{smp}} = X\\
\textrm{and } \frac{ [1-n_g^{DE}(\lambda)]t_{DE}}{t_{smp}} = Y.
\end{eqnarray}
In this notation the derivative of X with respect to $\lambda$ can be written as $\frac{\partial X}{\partial \lambda}$
\begin{equation}\label{EqErrSolutiondX}
\frac{\partial X}{\partial \lambda}  = \frac{\varphi_{loc}^\prime \cdot \lambda}{\pi \cdot t_{smp}},
\end{equation}
while the derivative of Y with respect to $\lambda$ can be written as $\frac{\partial Y}{\partial \lambda}$
\begin{equation}\label{EqErrSolutiondY}
\frac{\partial Y}{\partial \lambda}  =\frac{\partial }{\partial \lambda} \left( \frac{ [1-n_g^{DE}(\lambda)]t_{DE}}{t_{smp}}\right) = - \frac{t_{DE}}{t_{smp}} \frac{\partial n_g^{DE}}{\partial \lambda}. 
\end{equation}
In the case examined here, the dispersive element is made of N-BK7 glass which leads to the use of the Sellmeier equation for the refractive index, \cite{SchottKatalog}, 
\begin{equation}
n^{DE} = \sqrt{\frac{A_1 \lambda^2}{\lambda^2 - B_1} + \frac{A_2 \lambda^2}{\lambda^2 - B_2} + \frac{A_3 \lambda^2}{\lambda^2 - B_3}  + 1}.
\end{equation}
This equation is the basis for the calculation of $\frac{\partial n_g^{DE}}{\partial \lambda}$ under the assumption that the group refractive index is computed using, \cite{Delbarre}, 
\begin{equation}
n_g = n - \frac{dn}{d\lambda} \cdot \lambda.
\end{equation}
By calculating the derivative of the refractive index for glass with respect to the wavelength $\frac{dn}{d\lambda}$, the group refractive index for N-BK7 can be formulated as
\begin{eqnarray}
n_g^{DE} = n^{DE} - \frac{G}{2 \cdot n^{DE}} \cdot \lambda\\
\textrm{with } G = \sum\limits_{i=1}^3 \frac{-2A_i B_i \lambda}{(\lambda^2 - B_i)^2}. 
\end{eqnarray}
In consequence the partial derivative of the group refractive index can be formulated as
\begin{eqnarray}\label{EqErrdNGdLambda}
\frac{\partial n_g^{DE}}{\partial \lambda} = \left( \frac{G}{2n^{DE}} - \frac{\left( \frac{\partial G}{\partial \lambda}  \lambda + G\right)\cdot  n^{DE} - \frac{G^2}{2  n^{DE}} \cdot \lambda}{2  (n^{DE})^2} \right) \\
\begin{split}
\textrm{with }\frac{\partial G}{\partial \lambda} =&  \sum\limits_{i=1}^3 \frac{2A_i B_i \left(3\lambda^2 + B_i \right)}{(\lambda^2 - B_i)^3}.
\end{split}
\end{eqnarray}
Eq. (\ref{EqErrdNGdLambda}) is used to eventually determine the error contribution of the wavelength dependency to the calculation of the sample's group refractive index with Eq. (\ref{EqErrNGLambdaStart}), the solution of $\frac{\partial X}{\partial \lambda}$ from Eq. (\ref{EqErrSolutiondX}) and $\frac{\partial Y}{\partial \lambda}$ from Eq. (\ref{EqErrSolutiondY}) in conjunction with $\frac{\partial n_g^{DE}}{\partial \lambda}$ from Eq.\,(\ref{EqErrdNGdLambda})
\begin{eqnarray}
\frac{\partial n_g^{+DE}}{\partial \lambda} \cdot \Delta \lambda = \left( - \frac{\varphi_{loc}^\prime \cdot \lambda}{\pi \cdot t_{smp}} - \frac{t_{DE}}{t_{smp}}\frac{\partial n_g^{DE}}{\partial \lambda} \right) \cdot \Delta \lambda \label{EQPolyWPDE-error_lambda}\\
\textrm{with } \varphi_{loc}^\prime = \frac{2\pi}{\lambda^2}  \left[ (1 - n_g^{smp}) \cdot t_{smp}^{eff} + (1 - n_g^{DE} ) \cdot t_{DE}^{eff} +  \delta \right]\label{PolyEqPhi_simulation}\\
\begin{split}
\frac{\partial n_g^{+DE}}{\partial \lambda} \cdot \Delta \lambda = &- \left( \frac{ \pi}{\lambda \cdot t_{smp}} \left[ (1 - n_g^{smp}) \cdot t_{smp}^{eff} \right. \right.\\
& \left. \left. + (1 - n_g^{DE} ) \cdot t_{DE}^{eff} +  \delta \right] - \frac{t_{DE}}{t_{smp}}\frac{\partial n_g^{DE}}{\partial \lambda}  \right) \cdot \Delta \lambda.
\end{split}
\end{eqnarray}
It has to be noted that the calculation of $\varphi_{loc}^\prime$ is only performed in context of the error propagation with the effective thicknesses $ t_{smp}^{eff}$ and $ t_{DE}^{eff}$. In an experiment this would be the signal to measure. The wavelength contribution to the error was studied in a spectral range of (400\,-\,1000)\,nm for a dispersive element of $t_{DE}$\,=\,5\,mm and a sample of N-BK7 with $t_{smp}$\,=\,750\,\textmugreek m, Fig. \ref{PolyPic:ErrLambaNGAddDisp} a).
It is obvious, that the error contribution of the wavelength to the group refractive index $\Delta n_g^{+DE}(\lambda)$ is one order of magnitude larger than the influences of the sample and the DE thickness. Furthermore, it shows a significant minimum at a wavelength of $\lambda$\,=\,683.75\,nm. This behavior is similar to the sample-only approach of \gls{WPDE} and is due to the spectral dependence of the refractive index of N-BK7 as sample and DE material.\\
Finally, the total error of the group refractive index calculation for a 750\,\textmugreek m thick sample of N-BK7 and a 5\,mm thick dispersive element of the same material has been determined as an absolute combination of all error contributions according to Eq.(\ref{EqErrNGaddDis}), Fig. \ref{PolyPic:ErrLambaNGAddDisp}. Due to the large influence of the error contribution of the refractive index measurement, the total error is dominated by it. Therefore, it shows the same characteristic minimum at  $\lambda$\,=\,683.75\,nm with a value of \num{1.46e-5}. It is important to note, that the error increases exponentially for probing wavelengths smaller than the minimum with values larger than \num{3e-4}, while the error slowly tends towards a boundary value for probing wavelengths above the minimum. The boundary value is significantly smaller than \num{5e-5}. As this behavior is dominated by the refractive index of the materials present in the setup, an error estimation should be performed for every experiment. \\
The newly developed approach for the characterization of cross-linking in polymers differs from state-of-the-art technologies in the accuracy of measurement as well as in the variety and quality of possible measurements. In comparison to classical approaches such as Soxhlet extraction, DSC and DMA, the interferometric approach determines cross-linking in an indirect manner, but works non-destructive. This results in the ability to measure samples in in-line situations as well as during the lifetime of a product. It was shown that the typical measurement error was about 0.1\,\% in the scan-free approach. These errors are significantly lower than an error of 2-4\,\% that classical approaches posses. Compared to other, recently researched approaches, such as Raman or luminescence spectroscopy, the developed approach shows significant advantages relating its measurement speed where single profiles can be gathered in about 50 ms compared to multiple minutes. In summary, the developed approach combines high accuracy in the determination of cross-linking with high acquisition speeds and the ability to work in a scan-free fashion over lateral measurement ranges of multiple millimeters. Additionally, it allows non-destructive evaluations and combines material characterization with surface profilometry. In comparison with the identified state-of-the-art technologies for the determination of the degree of cross-linking, the DE-LCI approach shows significant advantages regarding accuracy, measurement time, inline capabilities and others, Tab. \ref{PolyTab:comparison}.
\begin{table}[h]
	\caption[Comparison of technologies for the characterization of the degree of cross-linking characterization in polymers]{Comparison of state-of-the-art technologies with the developed DE-LCI approach for the degree of cross-linking characterization in polymers} \label{PolyTab:comparison}
	\centering    
	\begin{tabular}{p{2cm}p{1.5cm}p{1.2cm}p{1.cm}p{1cm}p{1.5cm}p{3.75cm}} 
		\hline
		\rule[-1ex]{0pt}{3.5ex}  approach & meas. time & error & spatial resolution & inline & non-destruc-tive & comments \\
		\hline
		\rule[-1ex]{0pt}{3.5ex}  Soxhlet extraction \cite{Oreski,Hirschl} & 42 h & 2-4\,\% & no & no & no & gold standard\\
		\rule[-1ex]{0pt}{3.5ex}  DSC \cite{Hirschl2013} & 90 min. & 10\,\% & no & no & no & usable with many polymer types\\
		\rule[-1ex]{0pt}{3.5ex}  DMA \cite{Hirschl2013,Stark,Bradler2018} & 2-6 h & 10,\% & no & no & no & direct determination of degree of cross-linking\\
		\rule[-1ex]{0pt}{3.5ex}  \raggedright spectroscopic \cite{Peike,Schlothauer} & 50-100\,s & 4-6\,\% & yes & yes & yes & single point measurement, time consuming\\
		\rule[-1ex]{0pt}{3.5ex}  DE-LCI & 50\,ms & 0.01\,\% & yes & yes & yes & indirect but very fast and precise\\

		\hline
	\end{tabular}
\end{table}

\chapter{Thin-film characterization}\label{Chapter:ThinFilm}
The third intended application for the proposed dispersion-encoded low-coherence interferometry is the evaluation of thin-film characteristics on substrate materials. Due to the usage of thin-film technologies in high-volume production in e.g. the photovoltaics and semiconductor industry, process monitoring becomes relevant in order to ensure functional parameters such as solar cell efficiency, \cite{Wang2020}. In this context, film thickness as well as film homogeneity over large areas are important criteria for quality assurance. This section describes the modifications and developments of the \gls{DE-LCI} approach in order to measure film thickness of nm-sized films on comparatively thick substrates. Furthermore, results of different sample measurements are presented.

\section{Setup considerations}
\begin{figure}[h]
\centering
		\begin{tabular}{c}
			\begin{overpic}[scale=0.6]{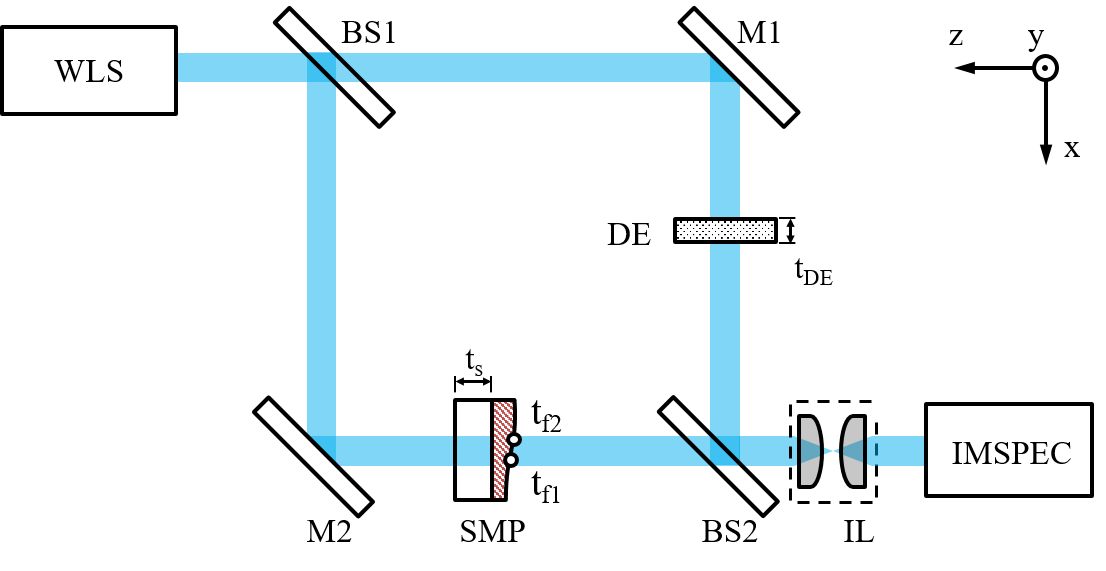}
			\end{overpic}
		
		\end{tabular}
	\caption[Experimental setup based on a Mach-Zehnder configuration for thin-film characterization]{Experimental setup based on a Mach-Zehnder configuration with a WLS - white light source which is splitted 50:50 by BS1 - first beamsplitter into a reference beam which is directed by M1 - reference arm mirror and manipulated by a DE - dispersive element (N-BK7, t\textsubscript{DE} = 6.23 mm) and the sample beam which is directed by M2 - sample arm mirror through the SMP - sample with thin-film (thickness slope t\textsubscript{f}(x)) on a substrate. Both are recombined by BS2 - second beamsplitter and imaged using the IL - imaging lens on a IMSPEC - imaging spectrometer where the analysis is performed.}\label{ThinFilm:Pic:ExpSetup_single}
\end{figure}

\begin{figure}[h]
\centering
		\begin{tabular}{c}
			\begin{overpic}[scale=0.8]{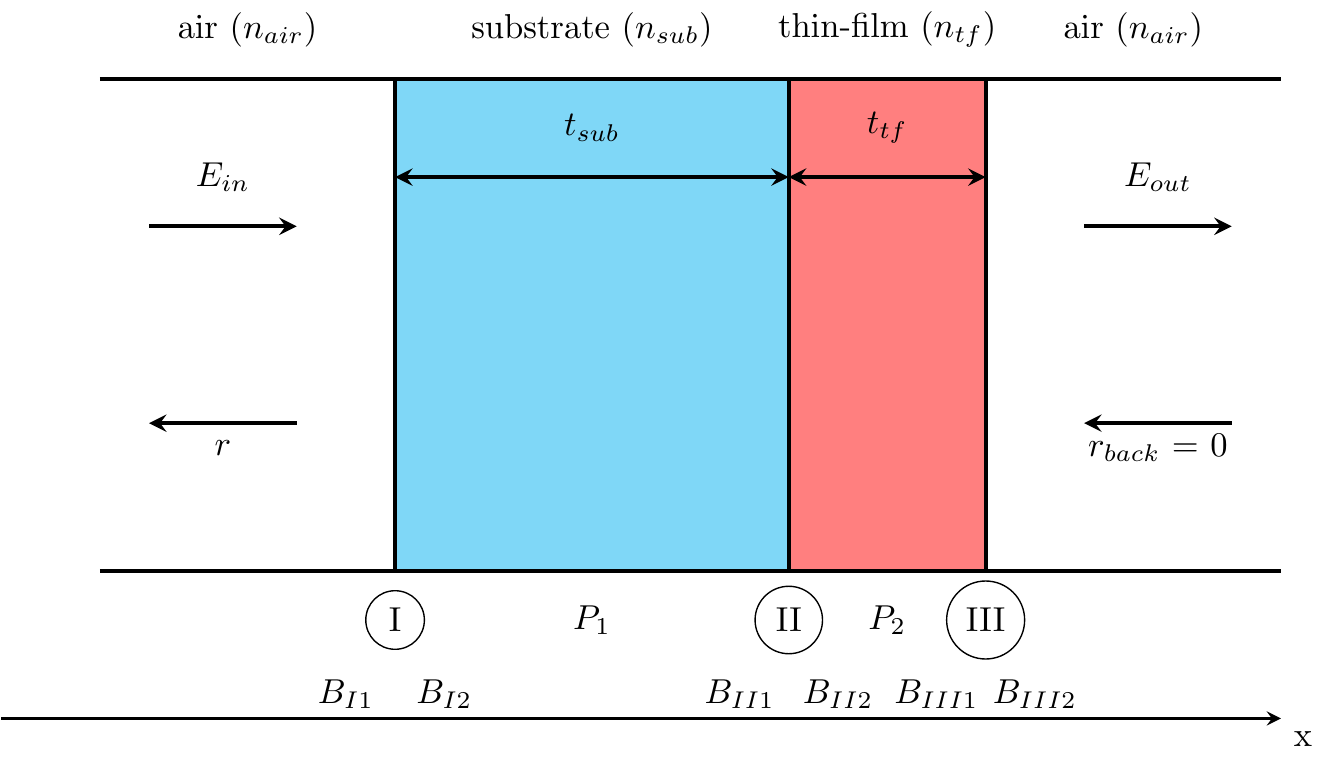}
			\end{overpic}
		\end{tabular}
	\caption{Simplified schema of the transfer-matrix formalism used to calculate the film thickness of a layer $t_{tf}$ on a substrate using the information of the present boundaries $B_{ij}$ and materials the electric field propagates through $P_{j}$, \cite{TransferMatrix}.}\label{PicSigTransfer}
\end{figure}

In order to achieve the goal of scan-free, spatially resolved film-thickness measurements of transmissive samples, a Mach-Zehnder interferometer was combined with an imaging spectrometer analogous to the approaches used in surface profilometry (see chapter \ref{ChapterProfilometry}) and polymer characterization (see chapter \ref{ChapterPolymer}), Fig.\ref{ThinFilm:Pic:ExpSetup_single}.
The collimated beam of a white-light source ($\Delta \lambda$\,=\,(400\,-\,1000)\,nm) was divided by a broadband plate beamsplitter into sample and reference arm. The reference arm contained a dispersive element (N-BK7, $t_{DE}$\,=\,6.23\,mm) while the sample arm was equipped with the particular sample which was a thin-film on a transparent substrate. If the sample contained a film thickness gradient, it was mounted in such a way, that the thickness gradient was aligned with the x-axis. The thickness and refractive index properties of the sample were dependent on the composition of the substrate material (\glssymbol{ThickSubstrate}, \glssymbol{RefIndSubstrate}) and on the number of film layers (\glssymbol{ThickFilm}, \glssymbol{RefIndFilm}) so that $t_{smp} = f(t_{sub},t_{fn})$ and $n^{smp} = f(n_{sub},n_{fn})$. The transmitted light was superimposed with light from the reference arm after passing the secondary beamsplitter BS2 and imaged onto the slit of an imaging spectrometer. The used imaging magnification was typically M\,=\,0.6.  The spectrally decomposed signal was detected with a two-dimensional \gls{CMOS}-array of a camera. \\
The signal formation and analysis were largely based on the theory of dispersion-controlled low-coherence interferometry established for precision profilometry and cross-linking characterization in polymers, \ref{ChapterProfilometry}. However, due to the composition of the sample as a multilayered system, a mathematically correct attribution if all signal components to the sample composition has to be performed. Typically, a transfer-matrix formalism is used to describe the transmission and reflection of light at every material interface, \cite{TransferMatrix}, Fig. \ref{PicSigTransfer}.
The basic idea of this formalism is that one singular matrix is used to describe the propagation of the electric field through a whole system of multiple layers. As of the design of the experiment within this work, only light with normal incidence regarding the front surface of the material system was taken into account. The formalism calculates the electric field components for a forward traveling wave (positive $x$-direction) \glssymbol{Ein} and for a backwards traveling wave $E_{out}$ in order to account for transmission and reflection at every material interface due to the material properties by using the wavenumber \glssymbol{Wavenumber} and the propagation vector $x$
\begin{equation}\label{Ei_E_o}
E_{in} = \Xi e^{-ik x}+ \Pi e^{+ik x}, \qquad	E_{out}=\Upsilon e^{-ik x}+ \Omega e^{+ik x},
\end{equation}
where the propagation coefficients \glssymbol{FieldCoeff} hold information on the materials and interfaces. Considering the fact that the tangential components of the electric field must be continuous at a material interface as a boundary condition and that the matrices of every single layer have to be multiplied, a notation of the electric field component of every material and subsequent material boundary can be found as a matrix describing the electric field when entering, \glssymbol{Min}, and exiting the material system, \glssymbol{Mout}, 
\begin{equation}\label{EQ:ThinFilm_TFmatrix_general}
\underbrace{\left(\begin{array}{cc} e^{-ik x} & e^{+ik x}\\ -ik e^{-ik x} & +ik e^{+ik x} \end{array}\right)}_{M_{i}}
\left({\begin{array}{cc} \Xi \\ \Pi \end{array}}\right)
=
\underbrace{\left(\begin{array}{cc} e^{-ik x} & e^{+ik x}\\ -ike^{-ik x} & +ik e^{+ik x} \end{array}\right)}_{M_{o}}
\left({\begin{array}{cc} \Upsilon \\ \Omega \end{array}}\right)
\end{equation}
 which leads to the final transfer matrix notation \glssymbol{MTrans},
\begin{equation}\label{EQ:ThinFilm_FWP_general}
M_T=M_{0i}^{-1} \cdot M_{0o}^{} \cdot M_{0o}^{-1} \cdot M_{1i}^{} \cdot ....M_{(j-1)o}^{-1} \cdot M_{ji}^{}.
\end{equation}
The description of the electric field in this notation represents the resulting wave in the sample arm. In combination with a notation for the reference arm, the complete propagation of light through the Mach-Zehnder interferometer can be described. In combination with the estimated equalization wavelength, measured data sets could be fitted according to the methods mentioned before, see section \ref{SecProfiloSigAn}. A detailed derivation of a simple, one-layer sample material which was used for analysis of the majority of measurements presented in this chapter, can be found in the appendix \ref{APNDX:ThinFilm_TransferMat}.\\
One important requirement in the practical implementation of this approach is the consideration of the substrate material as it shows incoherent behavior. Otherwise, the calculated signal inhibits higher order interferences which could reduce the signal-to-noise ratio significantly. Methods to minimize the influence of this problem would either be averaging the propagation in the substrate material by applying random phases or utilize a net-radiation method to get rid of the disturbing interference overlays, \cite{Troparevsky2010,Santbergen2013}.

\section{Characterization of thin-films on bulk substrates}
\begin{figure}[h]
	\begin{center}
		\begin{tabular}{c}
			\begin{overpic}[scale=0.3]{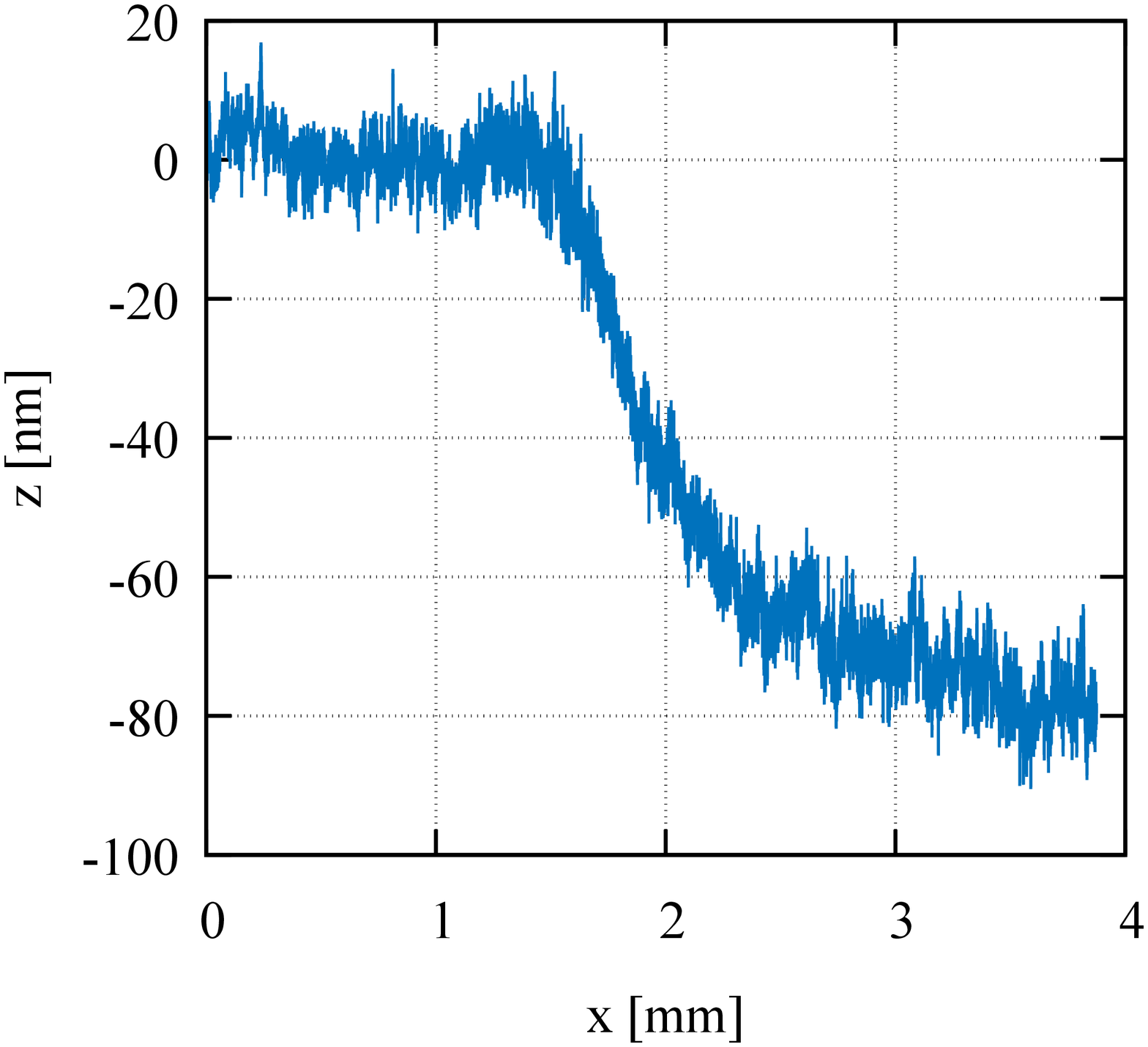}
				\put(1,1){\makebox(0,0){a)}}
				
			\end{overpic}
			\begin{overpic}[scale=0.3]{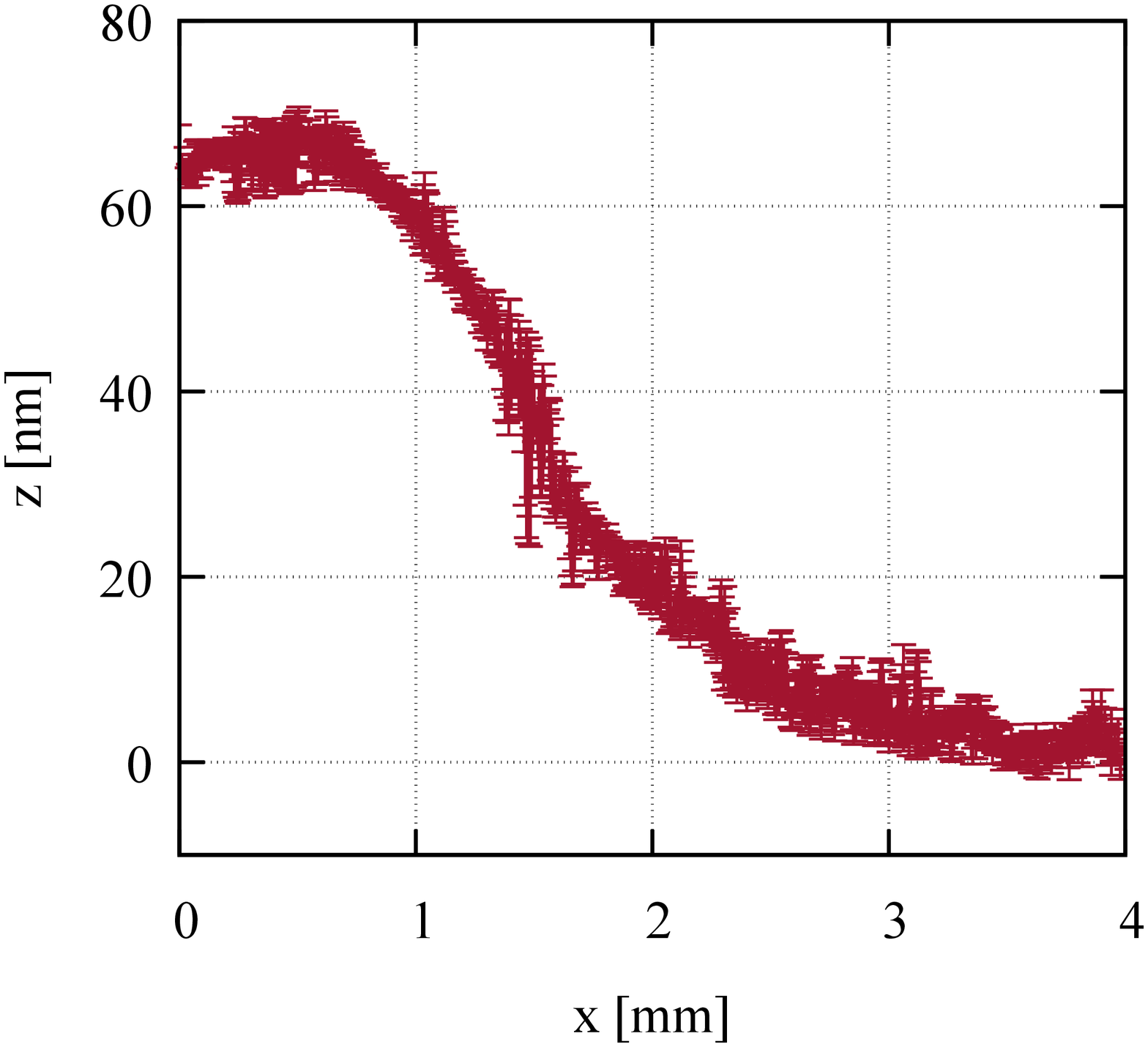}
				\put(1,1){\makebox(0,0){b)}}
			\end{overpic}
		\end{tabular}
	\end{center}
	\caption[Plot of the thickness gradient -  comparison of DE-LCI and tactile profilometer]{a) Plot of the thickness gradient, measured with a stylus profilometer as well as b) plot of the thickness gradient by DE-LCI.}\label{ThinFilm:Pic:Result_bulk_material}
\end{figure}

In order to evaluate the characteristics of the setup regarding its capabilities in thin-film characterization, samples of single-layer \gls{ITO} coatings on polished float glass substrates (CEC020S, PGO GmbH, Germany) were prepared. The samples were half-sided chemical etched in a ferric chloride bath at 230\,K for 3 hrs.\\
The determination of height profiles on substrates with coated thin-films can be performed by classical methods like spectral photometry or tactile profilometry either in a point wise or scanning fashion, Fig.\,\ref{ThinFilm:Pic:Result_bulk_material}\,a).  
The acquisition of surface profiles using a tactile profilometer over a range of 4\,mm can be performed within about 30\,s, depending on the desired lateral resolution. The results show that the thickness gradient was captured with a mean height of 65.9\,$\pm$\,12.3\,nm. Additionally, some significant noise with a value of $\pm$\,3.6\,nm was observed in the data. In contrast to the scanning acquisition, the thickness gradient could be determined in a single acquisition along the lateral domain using the \gls{DE-LCI} approach, Fig. \ref{ThinFilm:Pic:Result_bulk_material} b). In direct comparison, the captured profile, representing the average of ten successive data acquisitions (10\,x\,30\,ms), shows significantly lower noise of about 1.8\,nm. Accordingly, the thickness gradient over 4\,mm lateral measurement range was determined with (63.0\,$\pm$\,1.6)\,nm. The lateral resolution was about 4.2\,\textmugreek m, determined by the designed magnification of M\,=\,0.6. The significantly lower noise as well as the higher resolution and measurement speeds are clear advantages of the \gls{DE-LCI} approach in thin-film characterization.

\section{Characterization of flexible substrate materials}
\begin{figure}[h]
	\begin{center}
		\begin{tabular}{c}
			\begin{overpic}[scale=0.5]{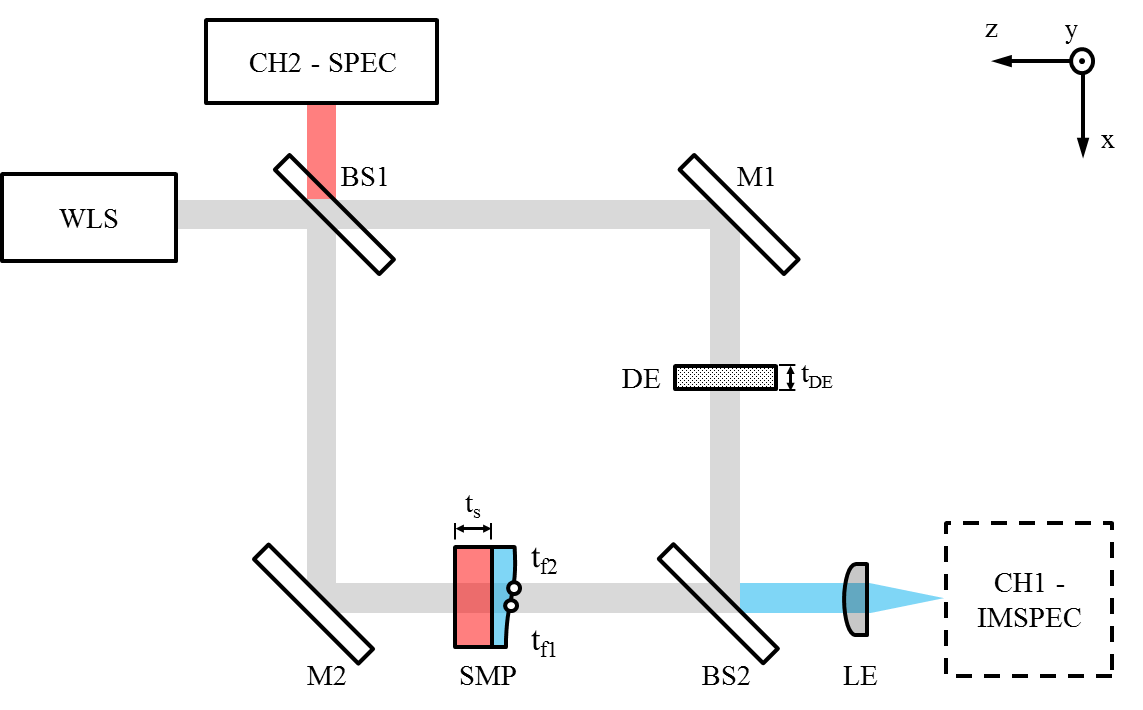}
			\end{overpic}
			
		\end{tabular}
	\end{center}
	\caption[Dual-channel setup based on a Mach-Zehnder configuration for thin-film characterization]{Dual-channel setup based on a Mach-Zehnder configuration with a WLS - white light source which is splitted 50:50 by BS1 - first beamsplitter into a reference beam which is directed by M1 - reference arm mirror and manipulated by a DE - dispersive element (t\textsubscript{DE} = 6.23 mm) and the sample beam which is directed by M2 - sample arm mirror through the SMP - sample with thin-film (thickness slope t\textsubscript{f}(x)) on a substrate (t\textsubscript{s} = 135 \textmugreek m) and both are recombined by BS2 - second beam splitter. The analysis is performed using a CH1 - IMSPEC - primary, imaging spectrometer on which the recombined beam is imaged to with a LE - lens, here marked in blue as well as with the CH2 - SPEC - secondary spectrometer which records the interference of the back-reflected signal from the substrate, here marked in red.}\label{ThinFilm:Pic:ExpSetup_dual}
\end{figure}

\begin{figure}[h]
	\begin{center}
		\begin{tabular}{c}
			\begin{overpic}[scale=0.32]{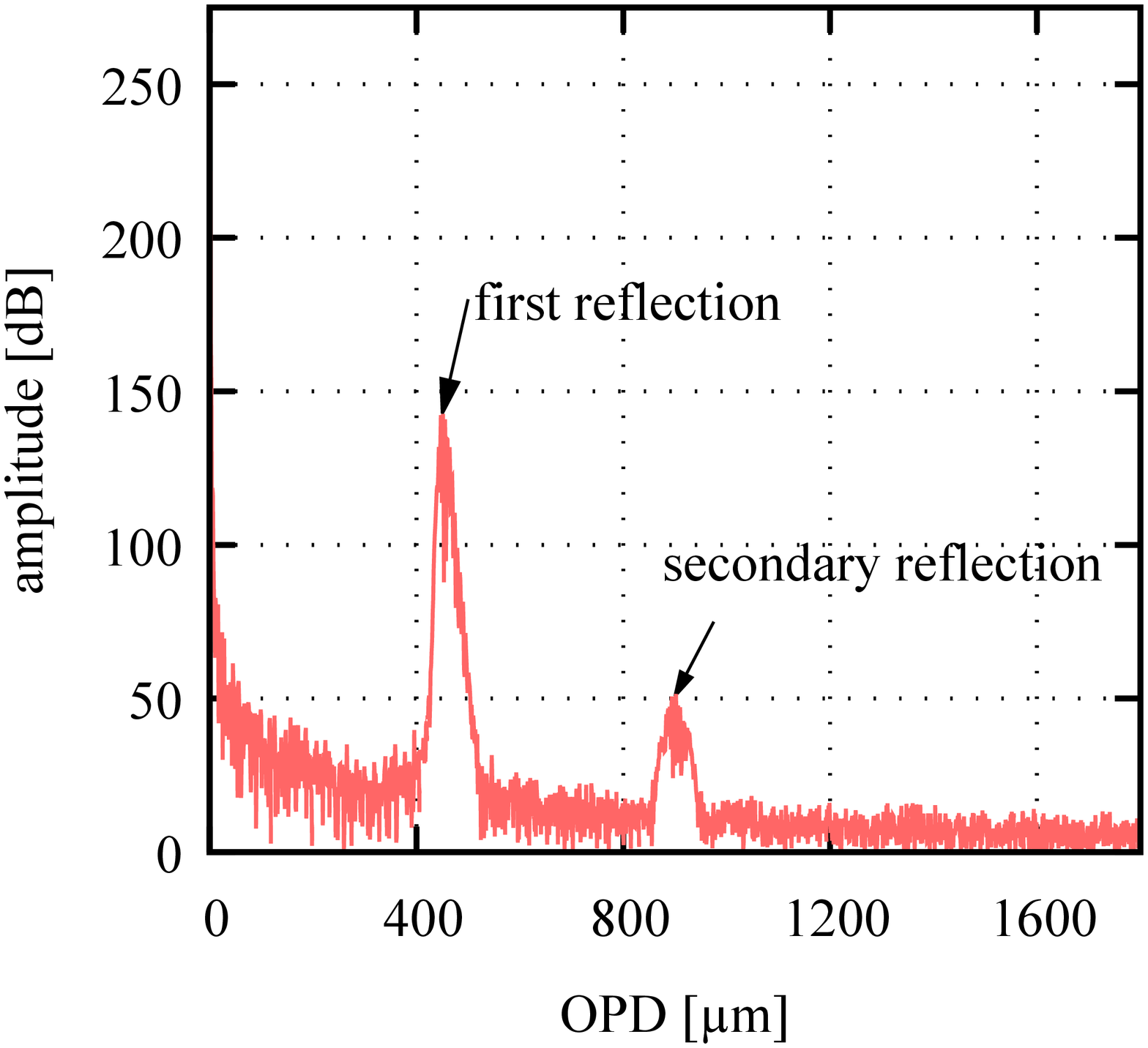}
				\put(1,1){\makebox(0,0){a)}}
				
			\end{overpic}
			\begin{overpic}[scale=0.32]{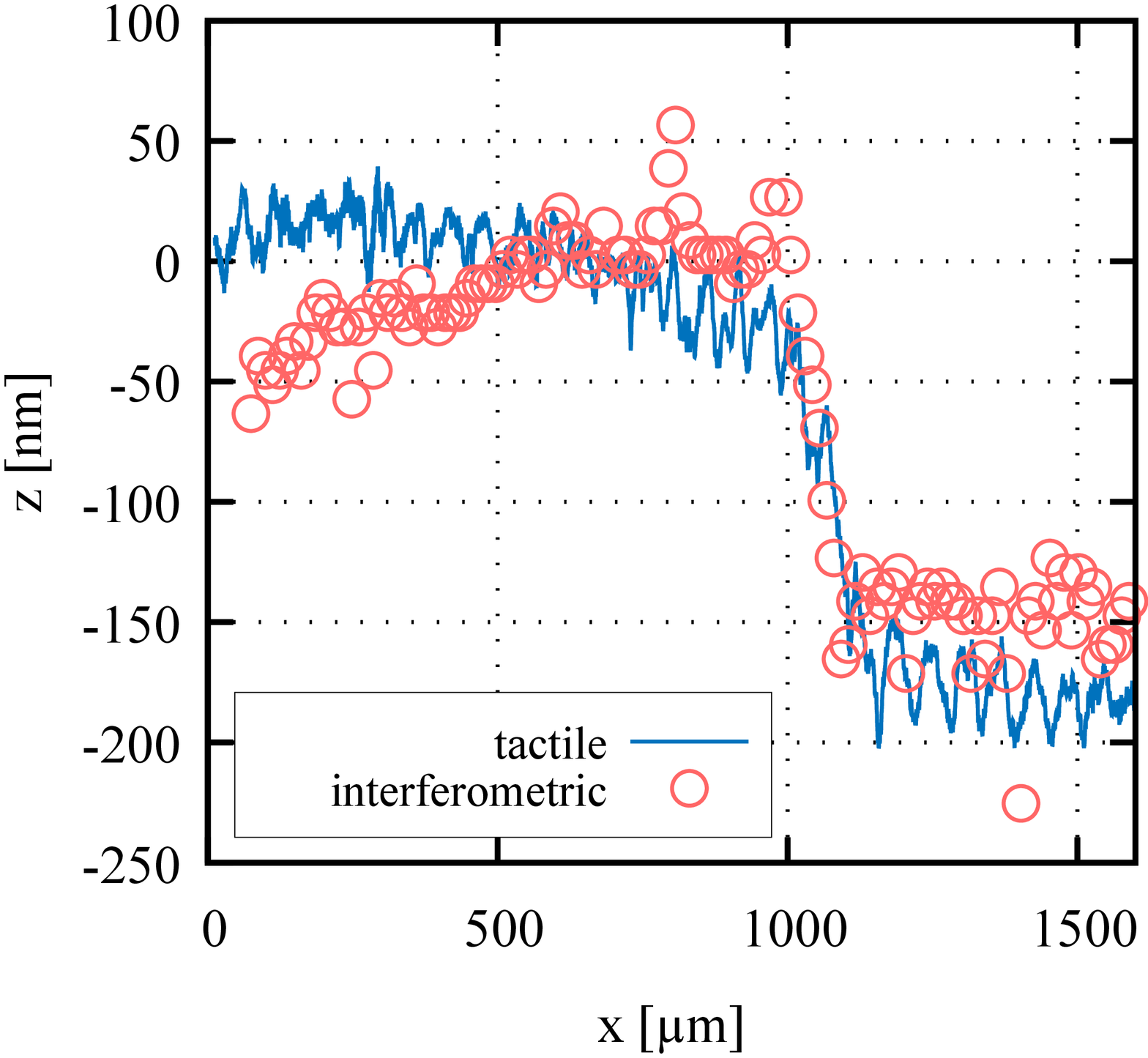}
				\put(1,1){\makebox(0,0){b)}}
			\end{overpic}
		\end{tabular}
	\end{center}
	\caption[Results of film thickness measurements using the the dual-channel interferometer]{Results of film thickness measurements using a dual-channel interferometer with a) Fourier-analyzed data of the secondary channel with the x-axis scaled to the optical path difference (OPD) showing peaks from interference of two reflections which were used to calculate the substrate thickness as 137.3 and 135.1 \textmugreek m respectively and b) slope of the ITO coating on this substrate measured in the primary channel having a ITO thickness of 151.6 nm in comparison to the slope measured on a tactile profilometer where the height was measured with 152.4 nm.}\label{ThinFilm:Pic:Result_dual_channel}
\end{figure}

As described above, the transfer-matrix formalism is used as a mathematical basis for the fitting of experimentally acquired spectra. The influence of the substrate material parameters is usually low when analyzing thin-films on bulk substrates. Some applications of thin-films are fabricated on thin, flexible substrates such as polymer sheets, \cite{Wang2020}. Thus, a more accurate knowledge of the material properties, especially of the thickness of these substrate materials is necessary. In order to account for this, a dual-channel variation of the originally developed Mach-Zehnder interferometer was developed, Fig. \ref{ThinFilm:Pic:ExpSetup_dual}.
In this setup, light is guided as before, see Fig. \ref{ThinFilm:Pic:ExpSetup_single}. Additionally, light from all interfaces but specifically that from the substrate was collected in a secondary detection path at the first beamsplitter. This path was equipped with a high-resolution grating spectrometer ($\sigma_{spec}$ = 0.3 nm, Avaspec ULS3648 VB, Avantes BV, Apeldoorn, The Netherlands). In this context, the surfaces of the substrate material can be considered as a resonator where interference occurs dependent on the distance of the surfaces, hence the thickness. By utilizing the spectrometer to record the spectral modulations of this interference signal, the actual thickness was calculated according to the principles of Fourier-domain optical coherence tomography, \cite{octBook}.\\
Sheets of \gls{PET} foil substrate ($t_{s}$ = 135 \textmugreek m) with a coated \gls{ITO} layer (nominal thickness $t_{fn}^{ITO}$ = 150 nm), typically applied in the photovoltaics industry, were used as samples. The \gls{ITO} coating was partly removed from the sample by means of chemical etching to generate a film-thickness gradient for spatial investigations.
From the measured back-reflected interference signal, a Fourier analysis with appropriate x-axis scaling could be performed, Fig. \ref{ThinFilm:Pic:Result_dual_channel}\,a).
The data shows a significant amount of DC-signal components and noise, but also very clear peaks. These peaks could be attributed to the first and second order reflections of the substrate surfaces. The peak information was extracted by applying a Blackman-Harris window function and fitted with a Gaussian function. From the fitted data, the optical path distances of the reflections could be estimated with (450.4 and 443.2)\,\textmugreek m respectively. With the knowledge of the refractive index of the substrate material, the thickness was calculated for both reflections with (137.3 and 135.1)\,\textmugreek m. The thickness of the substrate material could be confirmed by measurements on a tactile profilometer. With this result, an appropriate start value for the calculation of the film thickness with the imaging spectrometer is given. The method only allows the thickness calculation of the substrate material as the thin-film generates high-frequent interference which is not resolvable with a standard spectrometer.\\
By making use of the measured substrate thickness, data from the transmission measurement could be analyzed with the transfer-matrix formalism and appropriate fitting according to the model described with Eq. (\ref{Profilo:EQ:basic_interferometer}) and (\ref{disp_phase}), Fig. \ref{ThinFilm:Pic:Result_dual_channel} b). The results show that the slope of an \gls{ITO} film on a \gls{PET} substrate can be sampled with the appropriate resolution. From this slope the film thickness was measured with 151.6 nm. This corresponds well with a measurement of the sample with a tactile profilometer (Talysurf i-Series, Taylor Hobson Ltd, UK) with which the thickness was measured as 152.4 nm. In the data , it is noticeable that the tactile profilometer shows significant, periodic noise which can be attributed to the surface roughness of the substrate material. In direct comparison the surface roughness is less pronounced in the interferometric data. Furthermore, the interferometric system was capable of resolving the slope as well as the film thickness with high precision. This simultaneous measurement helped to improve the accuracy of the fit model for the thin-film thickness determination without compromising the actual measurement.\\
While ellipsometric approaches can be error-prone on e.g. flexible substrates, the demonstrated approach is capable to measure film thickness in this setting with high accuracy. In comparison to spectral reflectometry, another state-of-the-art technology, the developed approach enables comparable resolutions on film thickness measurements while also maintaining a far larger measurement range of about 80\textmugreek m. This enables the simultaneous capture of the film thickness as well as the thickness of substrate materials. In summary, the DE-LCI approach shows significant advantages regarding accuracy, measurement time and the variety of possible samples, Tab. \ref{ThinTab:comparison}.
\begin{table}[h]
	\caption[Comparison of technologies for the characterization thin-films]{Comparison of state-of-the-art technologies with the developed DE-LCI approach for thin-film characterization}\label{ThinTab:comparison}
	\centering    
	\begin{tabular}{p{2cm}p{1.1cm}p{1.4cm}p{1.cm}p{1.2cm}p{1.5cm}p{3.75cm}} 
		\hline
		\rule[-1ex]{0pt}{3.5ex}  approach & meas. time & resolution [nm] & spatial resolution & inline & flexible substrates & comments \\
		\hline
		\rule[-1ex]{0pt}{3.5ex}  reflectometry \cite{Osten} & 60\,s & 10\,nm & no & yes & yes & single point measurements, large integration times necessary\\
		\rule[-1ex]{0pt}{3.5ex}  ellipsometry \cite{Ogieglo2020} & 8\,s/$\lambda$ & 0.01\,nm & yes & yes & no & gold standard, high accuracy\\
		
		\rule[-1ex]{0pt}{3.5ex}  DE-LCI & 50\,ms & 0.1\,nm & yes & yes & yes & combines advantages\\
		
		\hline
	\end{tabular}
\end{table}

\chapter{Conclusion}
The characterization of features such as geometrical dimensions or functional parameters like roughness or degree of cross-linking plays an important role during production in industries such as semiconductors, organic-electronics and the photovoltaics industry. The ability to gather measurements in a process-integrated fashion is of high interest.\\
The scope of this work, in this context, was the development of an optical metrology tool which is capable to provide information on surface features and material properties in a fast and versatile way. In contrast to existing technologies, a novel approach was developed which delivered information on surface line profiles without the need for mechanical scanning and with higher resolution. Furthermore, in some aspects the approach was also able to gather information which was not accessible in a spatially-resolved fashion before. The developed setup was based on a modified, dispersion-enhanced low-coherence interferometer (DE-LCI). The properties of interest were characterized in the three major applications surface profilometry, polymer and thin-film characterization.\\
In the main part of this work, the DE-LCI approach was designed and implemented for surface profilometry. Here, it was shown how controlled dispersion can be used to encode path length differences in an interferometer and therefore surface height information in the spectral domain. The dispersion was controlled by a dispersive element, i.e. a glass window, which was used as one possibility to adjust the axial measurement range of the setup. Additionally, it was demonstrated how an appropriately designed imaging spectrometer can be utilized to gather the surface height information along a line profile in a single data acquisition without the need for mechanical scanning. The introduction of an additional, movable lens after the interferometric recombination extended the setups capabilities and allowed the acquisition of  three-dimensional surface height information. The development of a custom data analysis and fitting routine led to an estimation of the axial measurement range of $\Delta z$\,=\,79.91 \textmugreek m while a theoretical axial resolution of 0.088\,nm was calculated for a selected experimental configuration. With these values, a dynamic range in the axial dimension of DR\,=\,\num{9.08e5} was estimated.\\
The characterization of a setup with the calculated properties was performed by analyzing the surface profiles of different measured step height standards. Most notably, the determination of the profile and the height of (101.8\,$\pm$\,0.1)\,nm on a silicon height standard demonstrated the capabilities of the setup. It could be shown that the setup has a repeatability of $\overline{\sigma_z}$\,=\,0.13\,nm while the axial resolution was found to be $\Delta z_{min}$\,=\,0.1\,nm. In relation to the available measurement range of $\Delta z$\,=\,79.91 \textmugreek m, the experimentally determined dynamic range was DR\,=\,\num{7.99e5}. This value is about 6 times higher than comparable, current approaches known from literature, \cite{Reichold2019}.\\
Furthermore, it could be demonstrated that functional parameters such as surface roughness are measurable with the same axial resolution like surface profiles although this data was gathered over a lateral measurement range of up to 1.5\,mm. This separates the novel approach distinctively from established technologies such as tactile profilometry or confocal laser scanning microscopy which rely on time consuming and error-prone methods of scanning or stitching to enable lateral measurement ranges of the same order.\\
The capabilities of the setup regarding the acquisition of three-dimensional surface height information were evaluated with a measurement of a \textmugreek m-sized, PTB-calibrated height standard. By imaging an area of 1.5\,x\,0.25\,mm\textsuperscript{2} with sub-nm resolution, the measurement of steps with heights of (971.26\,$\pm$\,0.31), (4951.40\,$\pm$\,0.28) and (19924.00\,$\pm$\,0.36)\,nm was performed while additional features such as the roughness of each step could be acquired simultaneously. \\
An extension of the the setup with an imaging spectrometer for the NIR spectral range ($\Delta \lambda$\,=\,(1133\,-\,1251)\,nm) enabled tomographic measurements of a silicon sample. Specifically, the front and back surface of a thinned wafer could be investigated.\\
In summary, the developed dispersion-encoded approach to interferometry for surface profilometry proved to be of high resolution in the axial dimension while capturing  large measurement ranges in the lateral dimension. Due to the high-dynamic range and fast data processing, an application can be envisioned in process-integrated metrology for industrial production.\\
The determination of the degree of cross-linking is an important criterion in quality assurance of polymer processing. The degree of cross-linking determines important mechanical properties of the fabricated products as well as their long-term durability. Additionally, the cross-linking process bears potential for optimization during production in regard to speed and properties. Therefore, production accompanying monitoring is desirable.\\
Based on these conditions, the DE-LCI approach was adapted accordingly and tested for its ability to characterize polymers. The characterization was based on the measurement of the wavelength-dependent refractive index $n(\lambda)$ as a measure for the degree of cross-linking. The DE-LCI approach was utilized in a temporal scanning as well as in a scan-free configuration. Using the temporal scanning configuration, $n(\lambda)$ could be analyzed over large spectral ranges of $\Delta \lambda$\,=\,(400\,-\,1000)\,nm while a resolution in terms of the group refractive index of $\overline{\sigma_5(n_g)}$\,=\,\num{2.13e-4} was achieved. The measurement took several seconds and no spatial resolution could be accomplished. In contrast, the scan-free configuration was capable to measure the refractive index with a resolution of \num{3.36e-5} on a profile of 250\,\textmugreek m length in 50\,ms. The spectral range was about 20\,nm. In context of the scan-free configuration, a novel mathematical method for the analysis of phase data based on wrapped-phase derivative evaluation (WPDE) was developed and qualified. Both configurations were tested using typical samples from the photovoltaics and semiconductor industry respectively.\\
In contrast to existing technologies to determine the degree of cross-linking, the novel approach is fast, non-destructive and capable to perform spatially-resolved measurements with a lateral resolution of about 5\,\textmugreek m. The combination of characteristics is unique to the novel approach.\\
The precise control of single-layer thickness in the production of thin-film systems is crucial for the performance of these systems e.g. in the organic electronics or photovoltaics industry.\\
The DE-LCI setup was adapted as a Mach-Zehnder type interferometer in order to characterize the thickness of single-layer thin-films. By measuring a layer of \gls{ITO} on a bulk glass substrate, it was demonstrated that the layer thickness can be measured with a resolution of 1.6\,nm. The modification of the setup by using a secondary spectrometric detection channel allowed to capture back-reflected light of the sample. This way, it was possible to in-situ evaluate the substrate thickness of a flexible substrate ($t_{sub}$ = 135 \textmugreek m) while measuring the film thickness of a layer of ITO ($t_{ITO}$\,=\,151.6\,nm) simultaneously. This improved the robustness of the underlying transfer-matrix model which was utilized to calculate the film thickness from recorded spectral data.\\
The ability to capture film thickness data with spatial resolution within a single frame acquisition separates the DE-LCI approach from existing technologies and determines its usability as a process-integratable tool in future applications.\\
As demonstrated, the scope of this work was to show a range of possible applications of the developed DE-LCI approach while examining each application in depth. While answering the major questions by experiments and analysis, some additional questions arose which could not be addressed within the scope of this work. Most notably, alternative approaches to analyze the captured data might be addressed in future works. Preliminary experiments have shown that methods of image processing and statistical analysis are promising with regard to resolution, speed and robustness. Furthermore, it can be noted that the analysis of the captured data by means of FD-OCT is possible, \cite{Moon2018}. Here, the different dispersion in the interferometer arms has to be compensated by multiplying the spectral data with an appropriate phase function, \cite{Wojtkowski2004}. After re-sampling and Fourier transformation, the data contains three components, the desired peak in depth space, second the broadened mirror term and third the broadened DC term. Due to the signal construction in an DE-LCI approach, these terms overlap and need more sophisticated compensation in order to perform analysis known from full-range OCT approaches, \cite{KoettigDEFR}. Nevertheless, this approach bears potential for future work as it will increase the possible axial measurement range of the approach. Additionally, the development of fitting routines which can deal with data where the equalization wavelength lies not within the spectral range of the detector are interesting to enhance with respect to the axial measurement range. Beside these approaches to data analysis, advanced design and construction efforts with regard to mechanical stability and thermal management are interesting to further improvements of e.g. the repeatability of the setup. In terms of hardware aspects, the utilization of the triggerable high-power NIR supercontinuum light source makes the characterization of dynamic surface profilometry of oscillating samples in a stroboscopic acquisition mode possible. The experimental validation of theoretical designed methods to gather three-dimensional surface information in a scan-free fashion is of high interest as well.\\
Furthermore, the characterization of cross-linking mechanisms during actual processes with high temporal resolution as well as the monitoring of internal stresses during cross-linking by polarization enhanced DE-LCI is an interesting topic for future research.\\
The extension of the data analysis in thin-film characterizations towards multi-layered systems can be envisioned.\\
The combination of advantageous features such as the versatility of applications, the high dynamic range in the axial as well as the lateral domain, the capabilities to tune the measurement ranges easily and the high resolution characterize the developed approach. In conclusion, it can be said that the DE-LCI approach was developed and qualified in its main features and bears the potential for interesting applications in research as well as in industry.

\appendix
\chapter{Appendix surface profilometry}
\section{Design, Calculation and Construction of an imaging spectrometer}\label{Profilo:APNDX:Sec:imaging_spectrometer}
This sections aims to deliver additional information on the developed imaging spectrometer used for core experiments within this work.

\subsection{Design and Calculation}\label{Profilo:APNDX_Calc_ImagSpec}
\begin{figure}[h]
	\begin{center}
		\begin{tabular}{c}
			\begin{overpic}[scale=0.75]{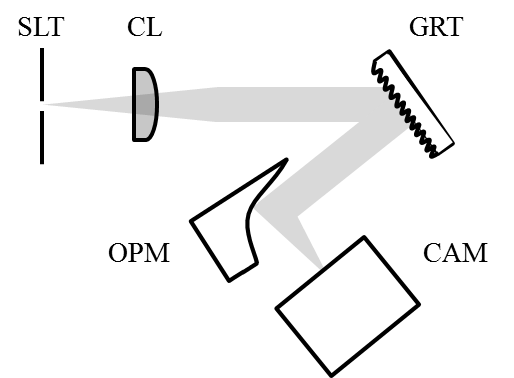}
				\put(1,1){\makebox(0,0){a)}}
			\end{overpic} 
		 	\begin{overpic}[scale=0.65]{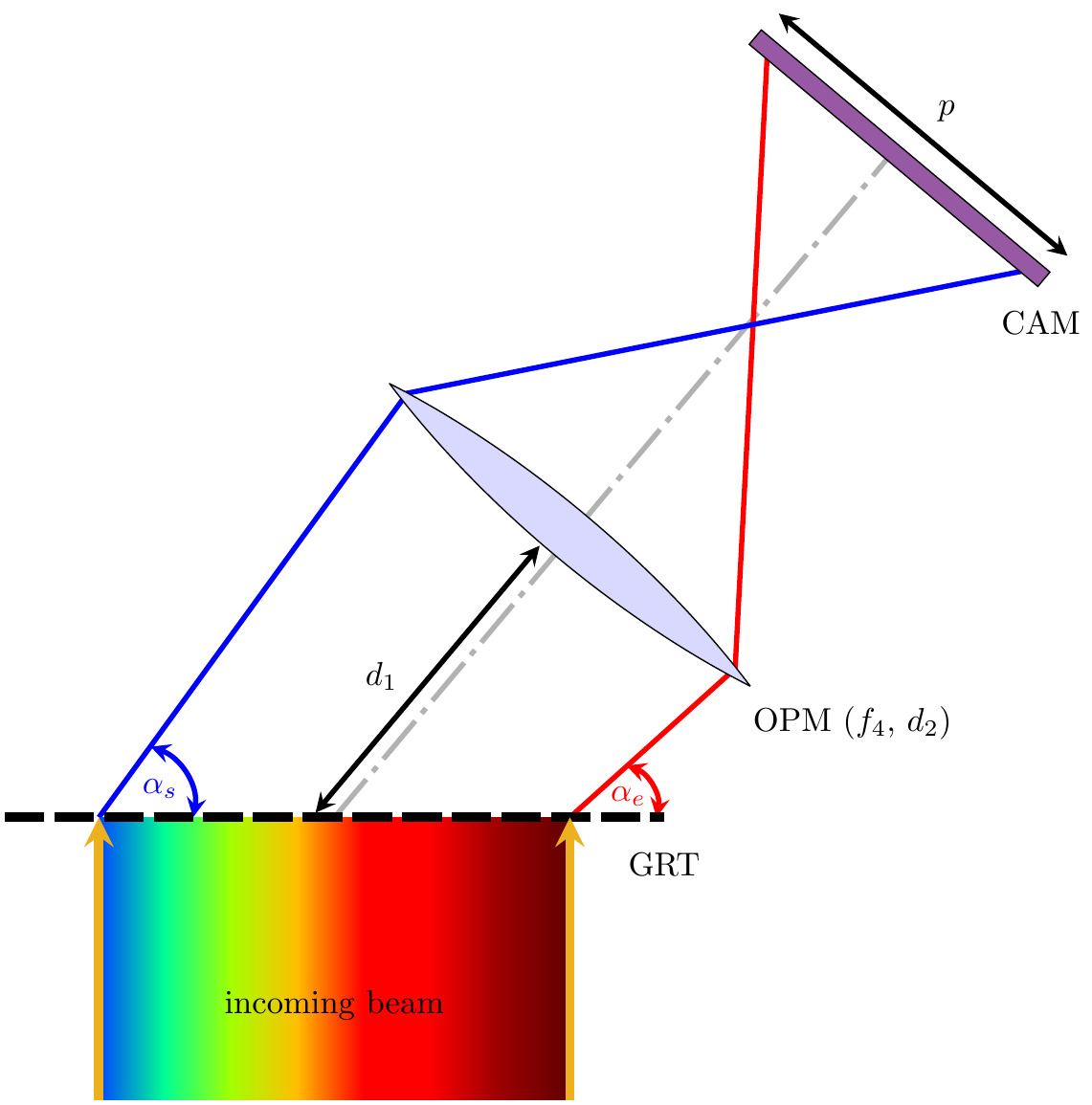}
		 		\put(1,1){\makebox(0,0){b)}}
		 	\end{overpic} 
		\end{tabular}
	\end{center}
	\caption[Simplified Czerny-Turner design for the planned imaging spectrometer and ray tracing approach]{a) Simplified Czerny-Turner design for the planned imaging spectrometer where light form the SLT - slit is collimated using a CL - collimation lens onto the GRT - grating and imaged by an OPM - off-axis parabolic mirror onto a CAM - camera and b) Detailed schematic of the ray tracing approach where the incoming beam is diffracted by the grating dependent on the wavelength ($\lambda_{s}$ and $\lambda_{e}$ under $\alpha_s$ and $\alpha_e$) and the grating constant; later it is focused by the OPM having a focal length of $f_4$ and a clear aperture of $d_2$ in a distance $d_1$ away from the grating to the plane of the CAM having the size  $p$.}\label{ProfiloPicAPNDX:czerny_turner_setup}
\end{figure}

\begin{figure}[h]
	\begin{center}
		\begin{tabular}{c}
			\begin{overpic}[scale=0.65]{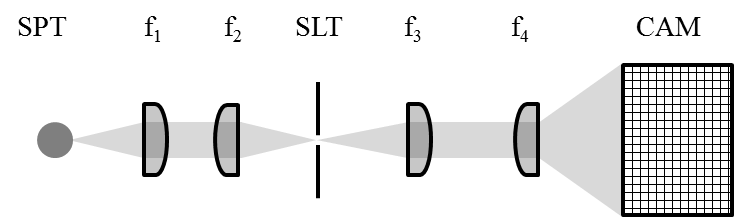}
			\end{overpic} 
		\end{tabular}
	\end{center}
	\caption[Simplified schematic for the calculation of the imaging elements of the imaging spectrometer]{Simplified schematic for the calculation of the imaging elements where the SPT - measurement spot is imaged by two optical elements having focal lengths of f\textsubscript{1} and f\textsubscript{2} onto the SLT - slit. The slit is further imaged onto the CAM - camera using the optical elements with focal lengths f\textsubscript{3} and f\textsubscript{4}.}\label{ProfiloPicAPNDX:imaging_components}
\end{figure}

As initial experiments showed the necessity of a scan-free approach in order to reach high axial measurement accuracy with out artifacts from translational movements, the design of an imaging spectrometer began with appropriate calculations and simulations. As a starting point, it was assumed to build a system based on a Czerny-Turner design, Fig. \ref{ProfiloPicAPNDX:czerny_turner_setup} a).
In order to choose necessary optical components for the construction of the setup, a simple ray tracing model based on the Czerny-Turner design was derived. For the purpose of this calculation some assumptions were done
\begin{itemize}
	\item a spectral range is considered as given (depending on the desired measurement range in height, determined by the dispersive element used)
	\item a collimated beam with a given width illuminates the grating
	\item the grating constant, which is to be determined, leads to a different diffraction angle for every wavelength
	\item the subsequent element images the different spectral parts on the imaging plane of the camera which has a finite size
	\item the camera chip was considered as a fixed parameter having the dimensions of 2048\,x\,2048 pixels and physical dimensions of 11.8\,x\,11.8\,mm\textsuperscript{2} (Basler acA2040-90um-NIR, Basler GmbH, Germany)
\end{itemize}
Based on this, the core section of the imaging spectrometer, consisting of the grating and imaging part, was modeled, Fig. \ref{ProfiloPicAPNDX:czerny_turner_setup} b).
In this regard, the chosen spectral range has to be imaged in the combination of the grating constant, the focusing optics distant to the grating and its focal length, \glssymbol{FocalLen}, to the finite size of the camera. For the purpose of calculating variants and finding the most suitable solution, a simulation tool for the ray tracing process was written using the assumptions made here. It has to be noted that for simplicity of the calculations it was considered to have a setup with a transmissive grating and a lens as focusing element. In reality this will rather be a reflective grating and an off-axis parabolic mirror. This assumption has no effect on the calculations whatsoever. The goal was to find a solution where the spectral range (i.e. the measurement range) was maximized while the size of the sensor area was utilized best. The resolution was not set as priority as previous experiments have shown that the resolution is more dependent on the quality of the fitting algorithm than on the physical limitations. Furthermore, the resolution increases with an increase in the grating constant. At the same time, this leads to a larger divergence of the light after the grating. The setup has to be more compact in size in order to detect a large spectral range. At some point, the geometrical conditions limit the spectral range which can be imaged. Therefore a trade-off between spectral range and resolution has to be made, which was done in favor for the spectral range in this application. After the simulation it was chosen to use a grating with a resolution of 300\,\nicefrac{lines}{mm} and a focusing element with a focal length of approximately 100 mm. The actual chosen parts are summarized in Tab. \ref{ProfiloAPNDX:Tab:ImSpecVISparameters}.\\
As the ray tracing simulation only considers the design regarding the axial measurement range, a separate calculation for the lateral  measurement range and resolution had to be done. For this purpose, the optical model can be simplified to only the imaging components, Fig. \ref{ProfiloPicAPNDX:imaging_components}.
From this schematic, one can derive that the imaging magnification outside of the spectrometer M\textsubscript{1} as well as the spectrometers magnification M\textsubscript{2} with
\begin{eqnarray}
M_1 &= \frac{f_2}{f_1}\\
M_2 &= \frac{f_4}{f_3}.
\end{eqnarray}
In this case, the magnification of the system is given by
\begin{equation}
M_{sys} = M_1 \cdot M_2.
\end{equation}
As the focal length of the last element inside the spectrometer (f\textsubscript{4}\,=\,101.6\,mm) is already given by the spectral calculation it can be used to determine the focal length f\textsubscript{3} with the size of the sensor to be \glssymbol{SensorSize}\,=\,11.8\,mm and magnification which can be chosen. In order to compare results regarding measurement time and dynamic range to other approaches, \cite{RuizHSI,CLSMReview}, a lateral measurement range \glssymbol{LatMeasRange} of at least 2\,mm was favored. In consequence the necessary system magnification should not exceed $M_{sys} = \frac{p}{\Delta l} = 5.9$. Due to geometrical constraints during the construction of the spectrometer, f\textsubscript{3} was chosen to 25\,mm which sets the spectrometers internal magnification to M\textsubscript{2}\,=\,4.06. In order to reach the goal for the lateral measurement range the external magnification was set to M\textsubscript{1}\,=\,1.25 by choosing f\textsubscript{1}\,=\,200\,mm and f\textsubscript{2}\,=\,250 mm which results in $\Delta l$\,=\,2.32\,mm.\\ 
Using these parameters, real components with values as close as possible to the calculation have been chosen, Tab. \ref{ProfiloAPNDX:Tab:ImSpecVISparameters}.
\begin{table}[h]
	\captionabove{Calculated parameter values and and chosen parts for the imaging spectrometer in the VIS spectral range.} \label{ProfiloAPNDX:Tab:ImSpecVISparameters}
	\begin{center}       
		\begin{tabular}{p{5cm}p{3cm}p{5cm}} 
			\hline
			\rule[-1ex]{0pt}{3.5ex} name & parameter & part trade name  \\
			\hline
			\rule[-1ex]{0pt}{3.5ex}  slit &  10 \textmugreek m & Thorlabs S10RD \\
			\rule[-1ex]{0pt}{3.5ex}  collimation lens f\textsubscript{3} & 25 mm & Thorlabs AC127-025-A-ML   \\
			\rule[-1ex]{0pt}{3.5ex}  grating & 300 lines/mm, 500 nm blaze & Thorlabs GR25-0305 \\
			\rule[-1ex]{0pt}{3.5ex}  off-axis parabolic mirror f\textsubscript{4} & 101.6 mm &  Thorlabs MPD149-G01\\
			\rule[-1ex]{0pt}{3.5ex}  CMOS camera p& 2048 x 2048 px\textsuperscript{2} / 11.8 x 11.8 mm\textsuperscript{2} & Basler acA2040-90um-NIR\\
			\hline
		\end{tabular}
	\end{center}
\end{table}
For some experiments in the subject of surface profilometry experiments were performed in the near-infrared spectral regime. Simulations and a construction of an appropriate imaging spectrometer were done using the same approach. The simulation lead to a different set of components for this device, Tab. \ref{ProfiloAPNDX:Tab:ImSpecNIRparameters}.
\begin{table}[h]
	\caption{Calculated parameter values and and chosen parts for the imaging spectrometer in the NIR spectral range.} \label{ProfiloAPNDX:Tab:ImSpecNIRparameters}
	\begin{center}       
		\begin{tabular}{p{5cm}p{3.25cm}p{4cm}} 
			\hline
			\rule[-1ex]{0pt}{3.5ex} name & parameter & part trade name  \\
			\hline
			\rule[-1ex]{0pt}{3.5ex}  slit &  20 \textmugreek m & Thorlabs S20RD \\
			\rule[-1ex]{0pt}{3.5ex}  off-axis parabolic mirror f\textsubscript{3} & 25.4 mm & Thorlabs MPD019-P01   \\
			\rule[-1ex]{0pt}{3.5ex}  grating & 600 lines/mm, 1.25\,\textmugreek m blaze & Thorlabs GR25-0613 \\
			\rule[-1ex]{0pt}{3.5ex}  off-axis parabolic mirror f\textsubscript{4} & 101.6 mm &  Thorlabs MPD149-G01\\
			\rule[-1ex]{0pt}{3.5ex}  InGAs camera & 640 x 512 px\textsuperscript{2} / 12.84 x 10.24 mm\textsuperscript{2} & Xenics Bobcat 640\\
			\hline
		\end{tabular}
	\end{center}
\end{table}

\subsection{Construction}
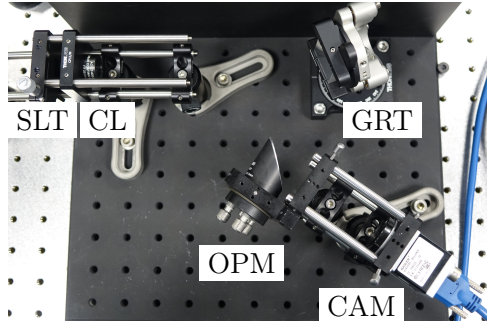
\begin{figure}[h]
	\begin{center}
		\begin{tabular}{c}
			\begin{overpic}[scale=0.32]{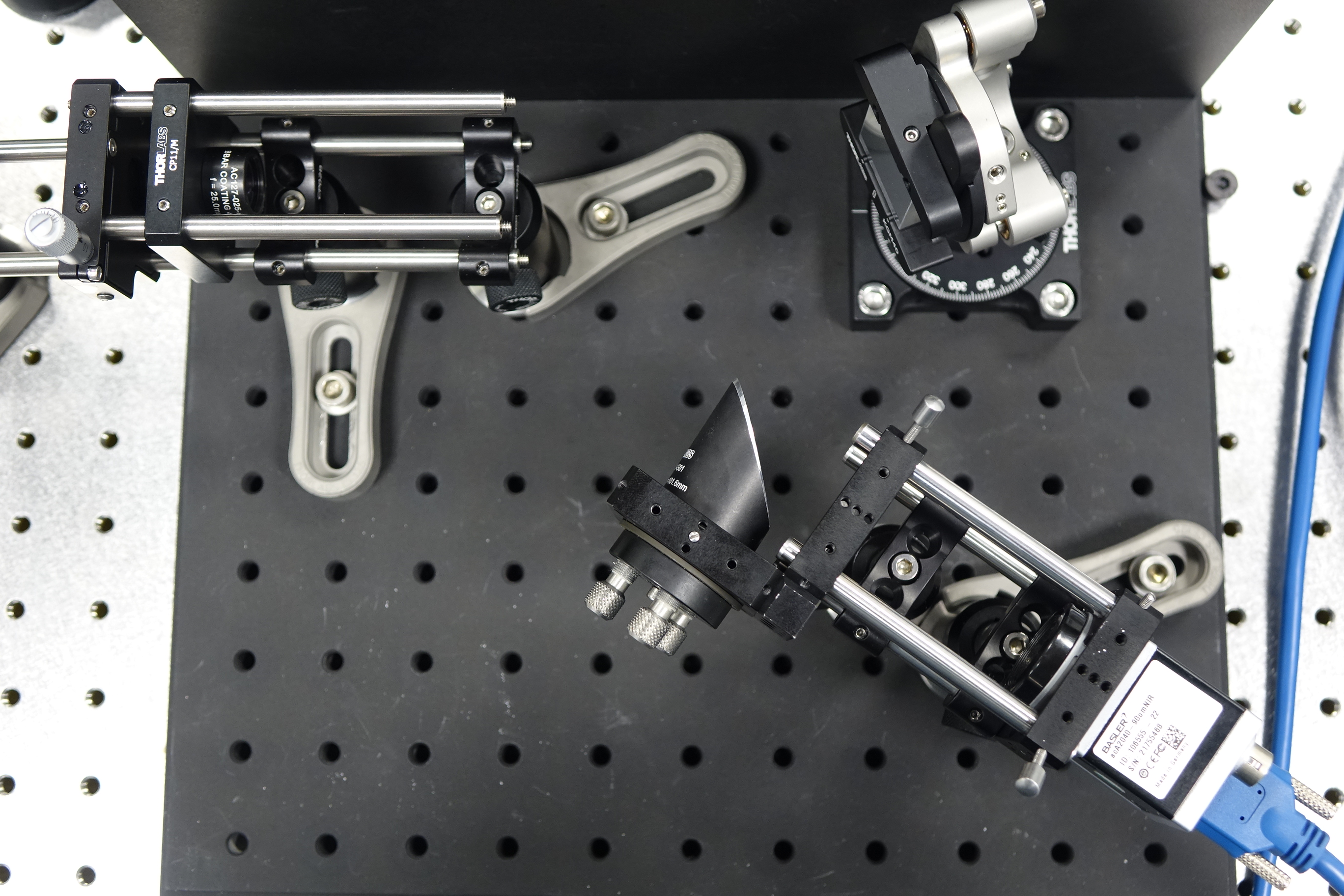}
				\put(0,40){\colorbox{white}{\textcolor{black}{SLT}}}
				\put(15,40){\colorbox{white}{\textcolor{black}{CL}}}
				\put(70,40){\colorbox{white}{\textcolor{black}{GRT}}}
				\put(40,10){\colorbox{white}{\textcolor{black}{OPM}}}
				\put(65,1){\colorbox{white}{\textcolor{black}{CAM}}}
			\end{overpic} 
		\end{tabular}
	\end{center}
	\caption[Constructed imaging spectrometer]{Constructed imaging spectrometer with SLT- slit, CL – collimation lens, GRT – grating, OPM – off-axis parabolic mirror, CAM - camera.}\label{ProfiloPicAPNDX:Real_ImSpec}
\end{figure}

The calculation of the spectrometer setup directly led to the order of appropriate parts and an initial setup. For the purpose, a light tight robust housing was necessary in order to prevent the camera to collect stray light which lowers the signal-to-noise ratio. As a hosing, an in-house construction with modular elements and compatibility to typical optical components could be used, Fig. \ref{ProfiloPicAPNDX:Real_ImSpec}.
The construction of a spectrometer requires very robust and long-term stable components in order to ensure performance of calibrated values such as the wavelength calibration during operation. Therefore, all components where picked with regard to that. The initial setup was adjusted and calibrated using a well-known gas discharge lamp using the methods described below.

\subsection{Calibration methods}
\begin{figure}[h]
	\begin{center}
		\begin{tabular}{c}
			\begin{overpic}[scale=0.185]{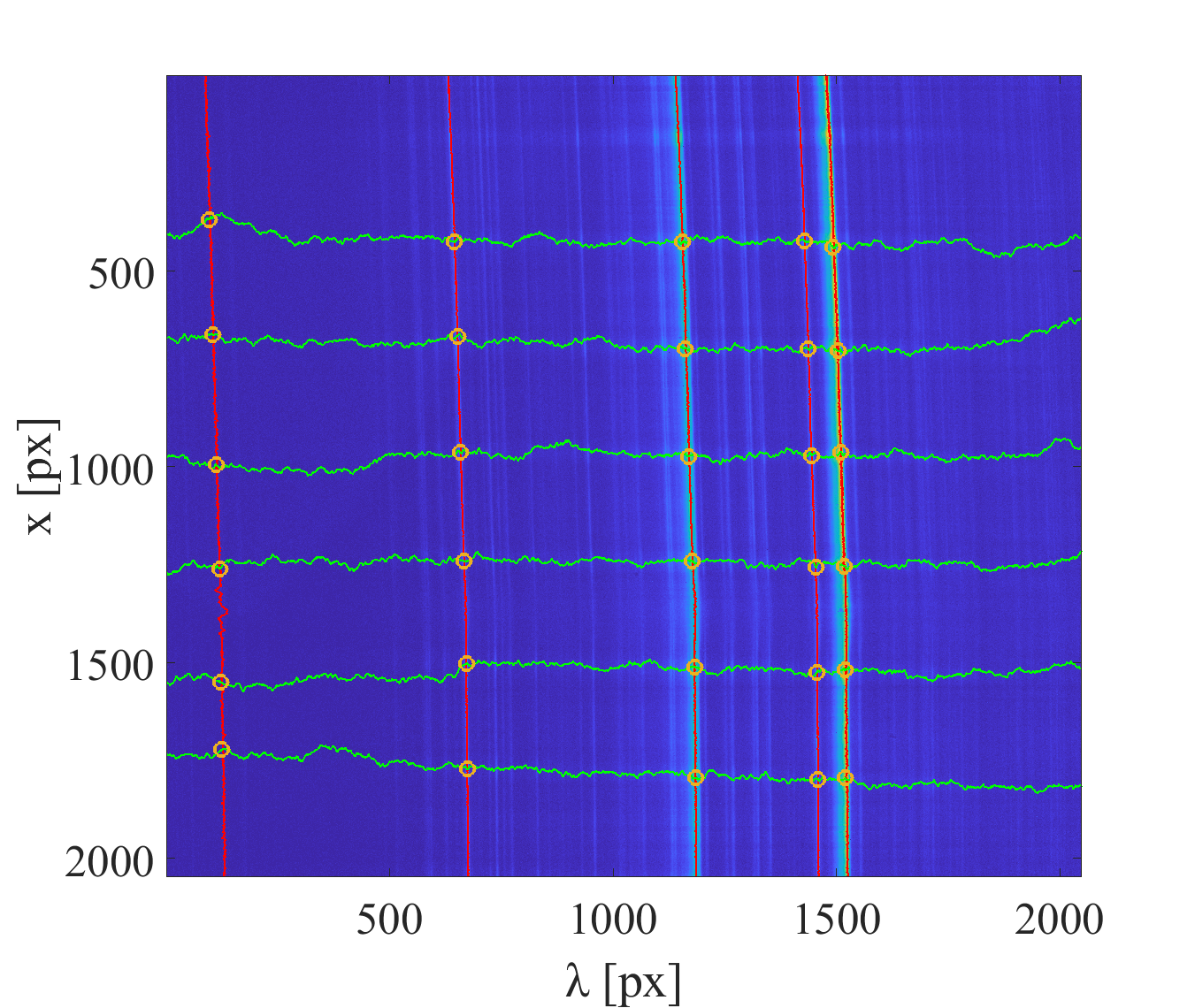}
				\put(1,1){\makebox(0,0){a)}}
			\end{overpic} 
			\begin{overpic}[scale=0.185]{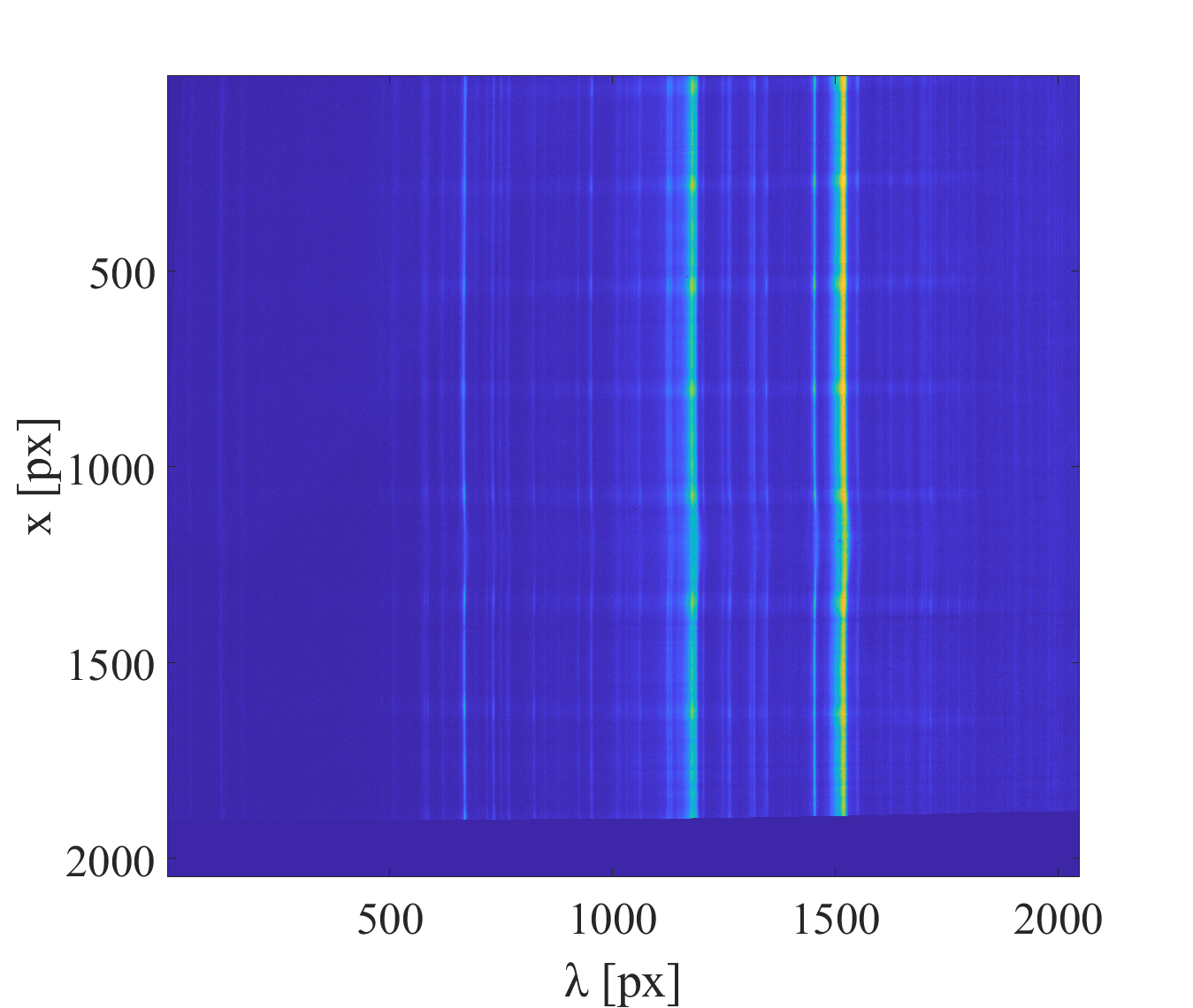}
				\put(1,1){\makebox(0,0){b)}}
			\end{overpic} 
		\end{tabular}
	\end{center}
	\caption[Calibration images and routines applied in imaging spectrometer measurements]{a) Recorded calibration data where the spectral lines of a gas-discharge lamp are running vertically (marked in red) and an imaged pattern of lines in the spatial dimension runs horizontally (marked in green). Together, the intersection between these lines (orange circles) could be used to calculate a correction matrix of spherical and chromatic aberrations which was used to compute the b) corrected calibration image which was used for spectral calibration.}\label{ProfiloPicAPNDX:calibration_images}
\end{figure}

The calibration of the spectrometer is most important as the calculation of height information relies on spectral data. The calibration routine was designed to address spectral calibration as well as correction for spherical and chromatic aberrations. In order to reach optimal results, the spectrum of a well known gas discharge lamp was recorded with the imaging spectrometer while a regularly spaced pattern of lines (G391122000, Qioptiq Photonics GmbH \& Co. KG, Germany) was imaged in the spatial domain, Fig. \ref{ProfiloPicAPNDX:calibration_images} a).
The subsequent calibration consisted of two steps. Using the spectral lines in the vertical dimension (marked in red) and the spatial pattern in the horizontal dimension (marked in green) the intersections between these lines could be acquired. Based on this, the aberrations could be determined and corrected for, Fig. \ref{ProfiloPicAPNDX:calibration_images} a). The aberration information was saved as a correction matrix which was used to correct every initially captured measurement raw data. Furthermore, the position of the five most intense spectral peaks ($\lambda_I$\,-\,$\lambda_{V}$) was used for spectral calibration. For this purpose, the spectrum was recorded beforehand using a wavelength calibrated standard spectrometer (Avantes AvaSpec ULS2048XL) to determine the wavelengths with $\lambda_I$\,=\,536.33\,nm, $\lambda_{II}$\,=\,546.65\,nm, $\lambda_{III}$\,=\,590.56\,nm, $\lambda_{IV}$\,=\,670.64\,nm and $\lambda_{V}$\,=\,755.49\,nm. With the help of \cite{Coseti_calibration}
\begin{equation}\label{ProfiloAPNDX:EQmapping}
\lambda_{calc} = a + bp_{px} + cp_{px}^2
\end{equation}
the recorded pixel positions of the peaks \glssymbol{PixPos} could be mapped to positions on the spectral range \glssymbol{LambCalc} using the fit parameters \glssymbol{CalCoeff}. It is visible from the recorded calibration image that a chromatic aberration occurs as the spectral lines show slight deviations along the $x$-dimension where they should run straight. In order to correct for this, the spectral mapping according to Eq.\,(\ref{ProfiloAPNDX:EQmapping}) was performed for every line in the $x$-domain individually. The calculated spectral mapping was saved and applied to every data set measured subsequently.

\subsection{Noise evaluation}\label{ANPDX:Profilo_noise}
\begin{figure}[h]
	\begin{center}
		\begin{tabular}{c}
			\begin{overpic}[scale=.32]{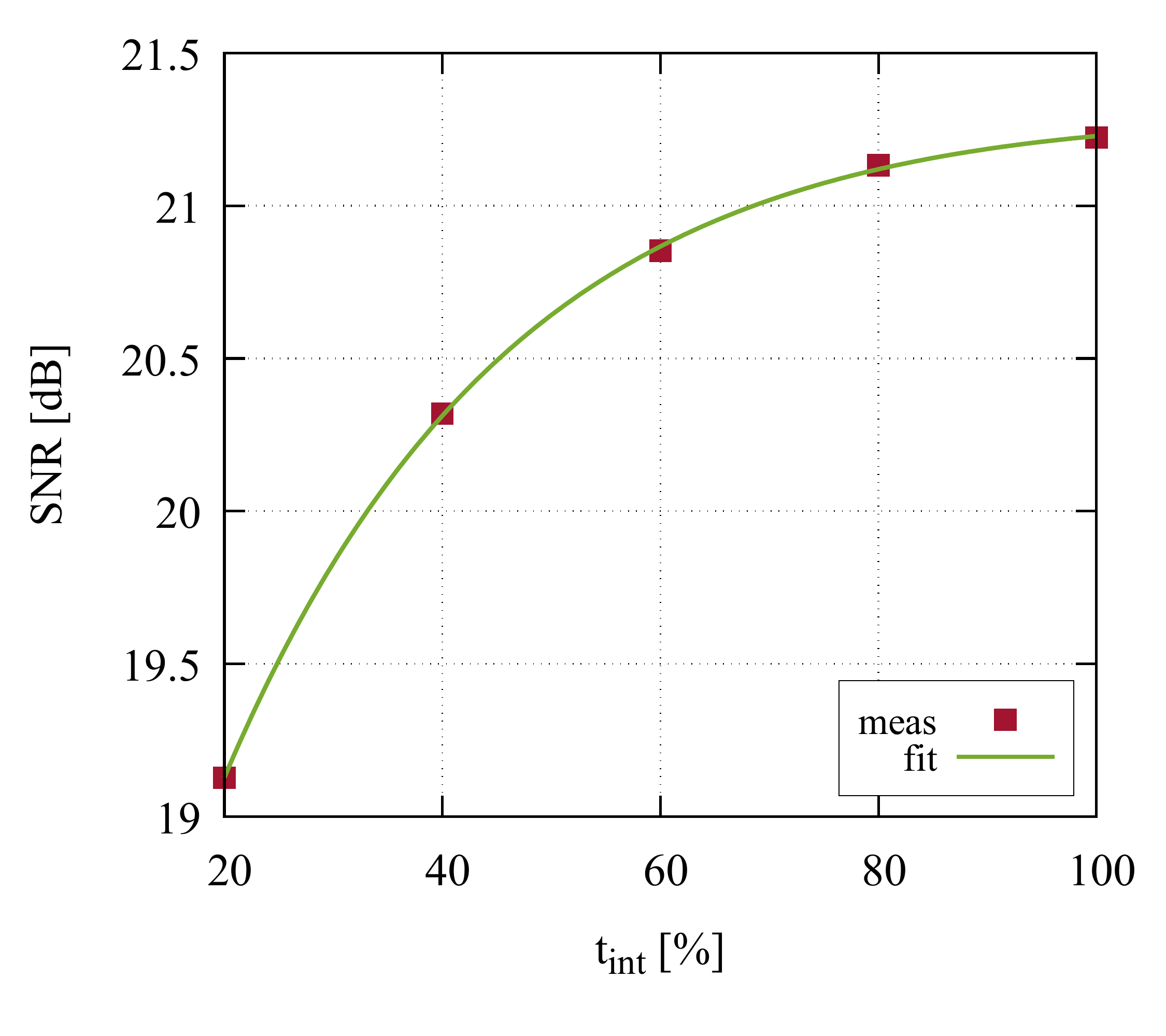}
				\put(1,1){\makebox(0,0){a)}}
			\end{overpic}
			\begin{overpic}[scale=.32]{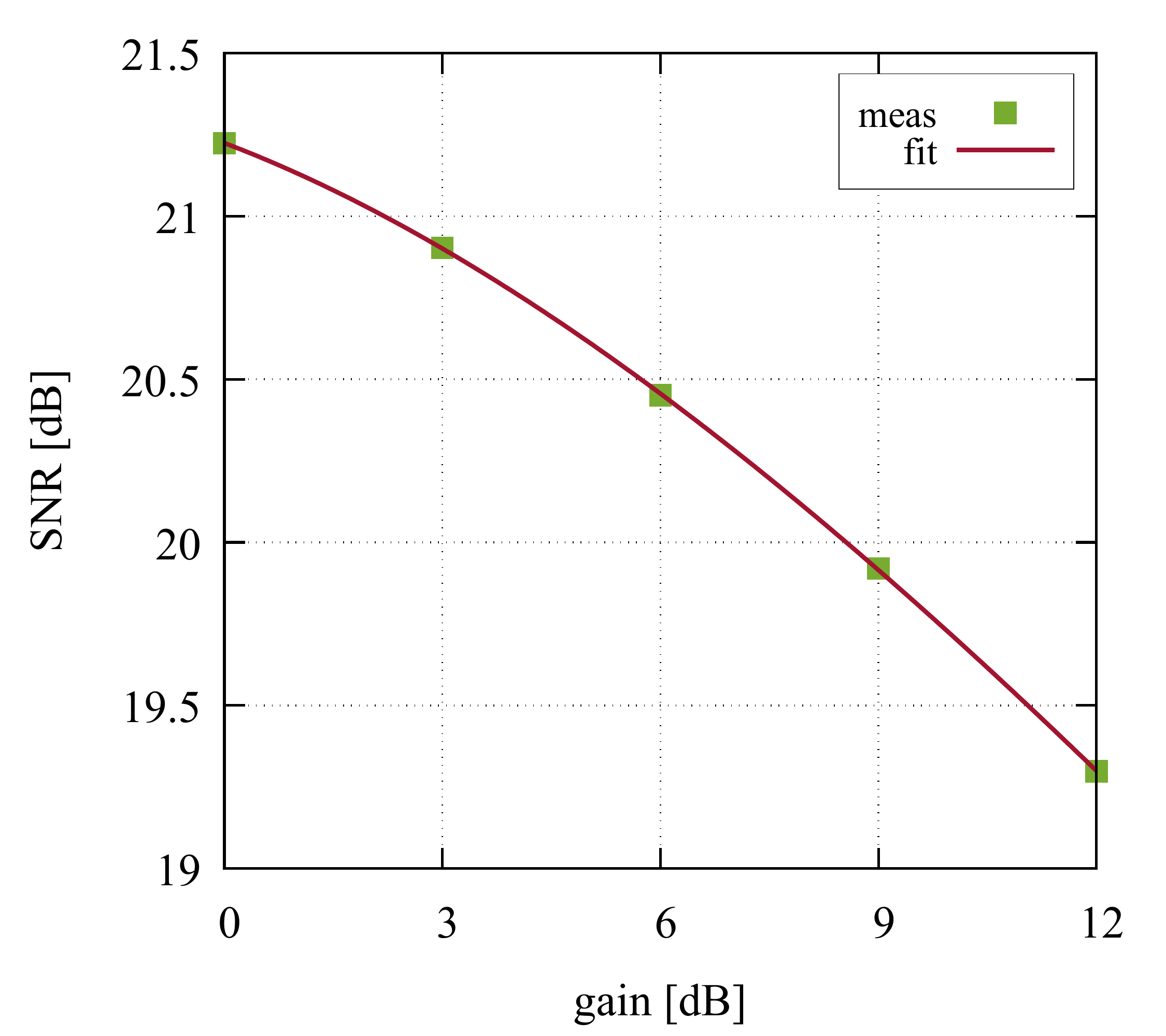}
				\put(1,1){\makebox(0,0){b)}}
			\end{overpic}
		\end{tabular}
	\end{center}
	\caption[Result of roughness measurements used for the determination of de-focus influence]{Results of the noise analysis of the interferometer with imaging spectrometer with a) Dependency of the noise level from the used integration time $t_{int}$ as well as b) Dependency of the noise from the used camera gain which both follow an inverse logarithmic trend (see fit curves).}\label{APNDXProfilo:Pic:Result_noise_cam}
\end{figure}

As already outlined in Section \ref{Profilo:Sec:Meas_range}, the spectrometer was attached to the complete interferometric system in order to perform an analysis of the noise behavior. The gathered results were used as a basis in order to calculate the noise-induced resolution limit using Eq.\,(\ref{noise_resolution}). For the purpose of this investigation the noise level was determined for a range of integration times. As a common reference, an integration time was chosen where the camera was just not overexposed. The subsequent integration times were reduced by a factor of \nicefrac{1}{2}, Fig.\,\ref{APNDXProfilo:Pic:Result_noise_cam}\,a).
It can be seen that the decrease in SNR can be modeled using an inverse logarithmic function. Similarly, the influence of the electronic gain of the camera was evaluated. For this examination, the integration time was kept at 100\,\% while the gain was increased in steps of 3\,dB, Fig. \ref{APNDXProfilo:Pic:Result_noise_cam}\,b). The SNR equally shows logarithmic decrease with an increase in gain. In all experiments within this work, the camera was used with gain at 0\,dB and an exposure time set to the maximum possible value without overexposing.  
\section{Error estimation of focus variations}
\begin{figure}[h]
	\begin{center}
		\begin{tabular}{c}
			\begin{overpic}[scale=.37]{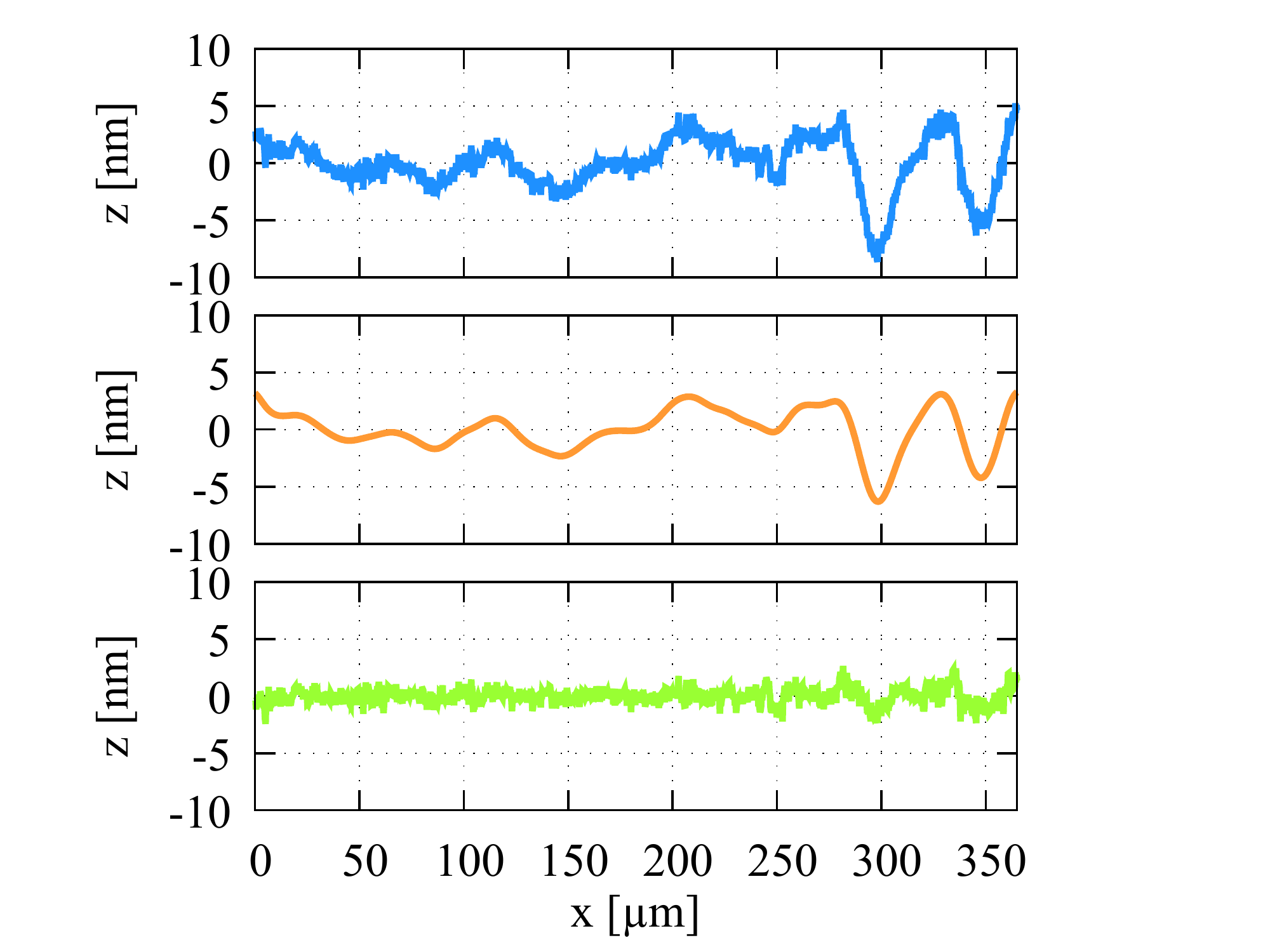}
				\put(1,1){\makebox(0,0){(I)}}
				\put(1,63){\makebox(0,0){a)}}
				\put(1,42){\makebox(0,0){b)}}
				\put(1,21){\makebox(0,0){c)}}
			\end{overpic}
			\begin{overpic}[scale=.37]{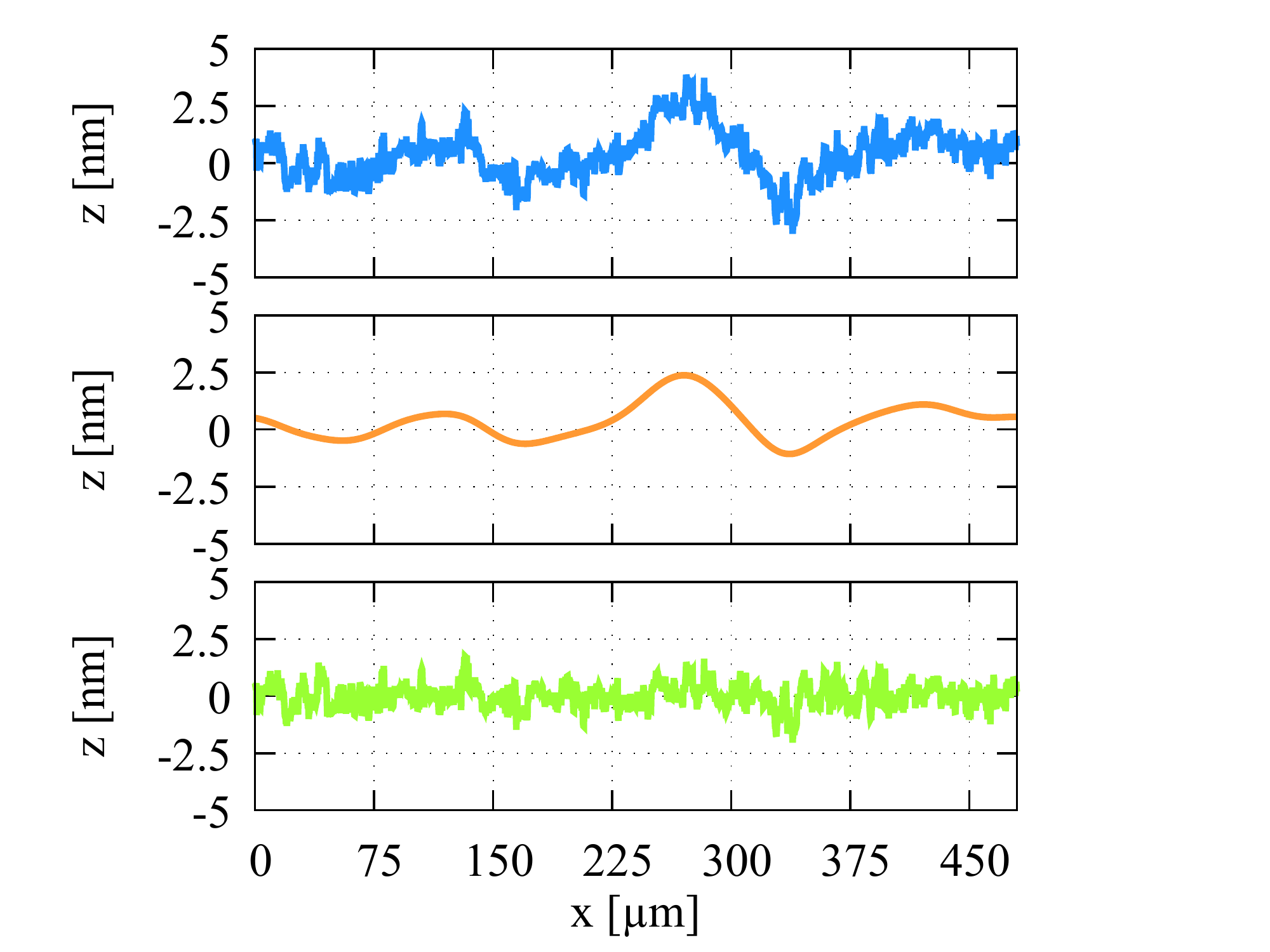}
				\put(1,1){\makebox(0,0){(II)}}
			\end{overpic}
		\end{tabular}
	\end{center}
	\caption[Result of roughness measurements used for the determination of de-focus influence]{Result of roughness measurements on (I) a polished glass substrate and (II) a Si-sample both with a) primary, form corrected profile, b) waviness profile (Gaussian low-pass filtration of a)) and c) roughness profile (Gaussian high-pass filtration of a) ) using a cut-off wavelength of $\Lambda_c$\,=\,25\,\textmugreek m and $\Lambda_c$\,=\,80\,\textmugreek m respectively.}\label{Profilo:Pic:Result_roughness_glass_si}
\end{figure}

\begin{figure}[h]
	\begin{center}
		\begin{tabular}{c}
			\begin{overpic}[scale=.32]{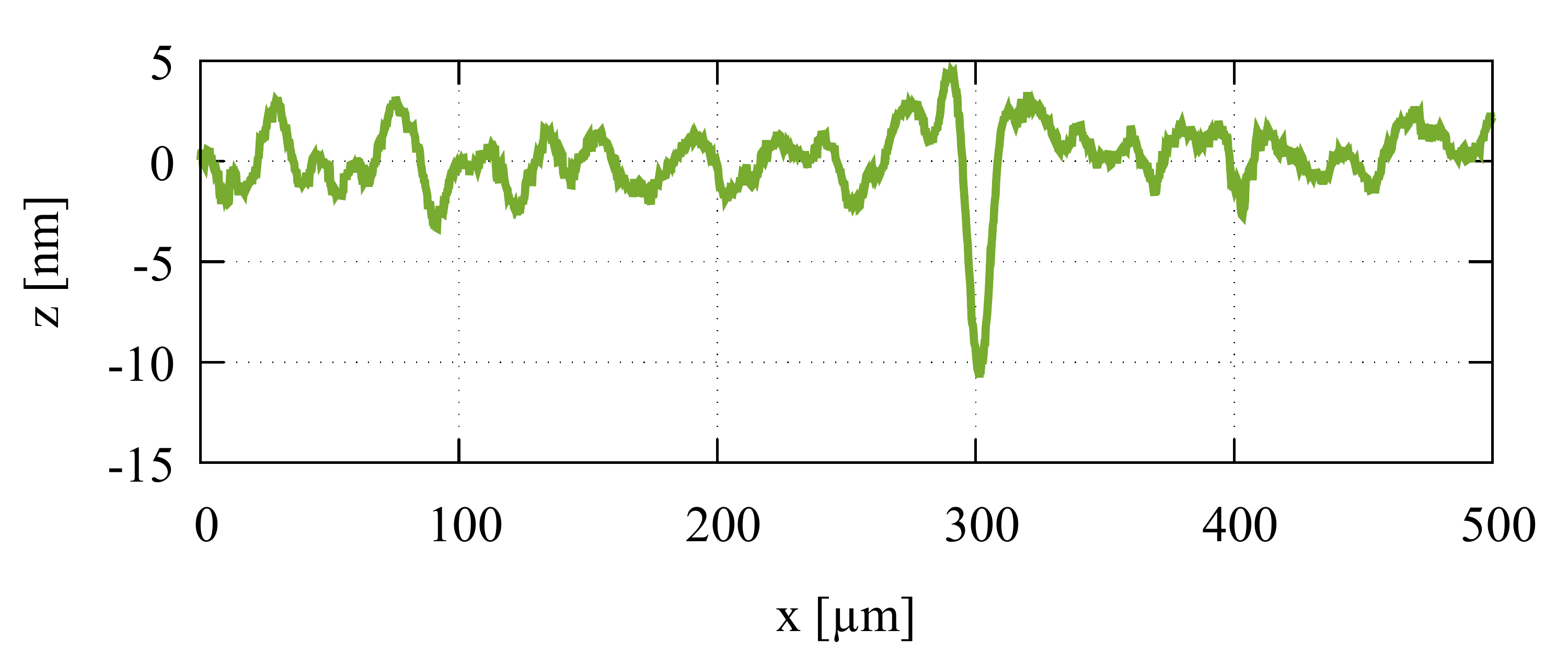}
				\put(1,1){\makebox(0,0){a)}}
			\end{overpic}
		\end{tabular}
	\end{center}
	\caption[Example data set of a Si-surface for the investigation of the de-focus influence]{Example data set of a Si surface for the investigation of the focus influence on features like depth and FWHM of the scratch as well as on roughness and waviness.}\label{Profilo:Pic:Result_example_influence_focus}
\end{figure}

\begin{figure}[h]
	\begin{center}
		\begin{tabular}{c}
			\begin{overpic}[scale=.32]{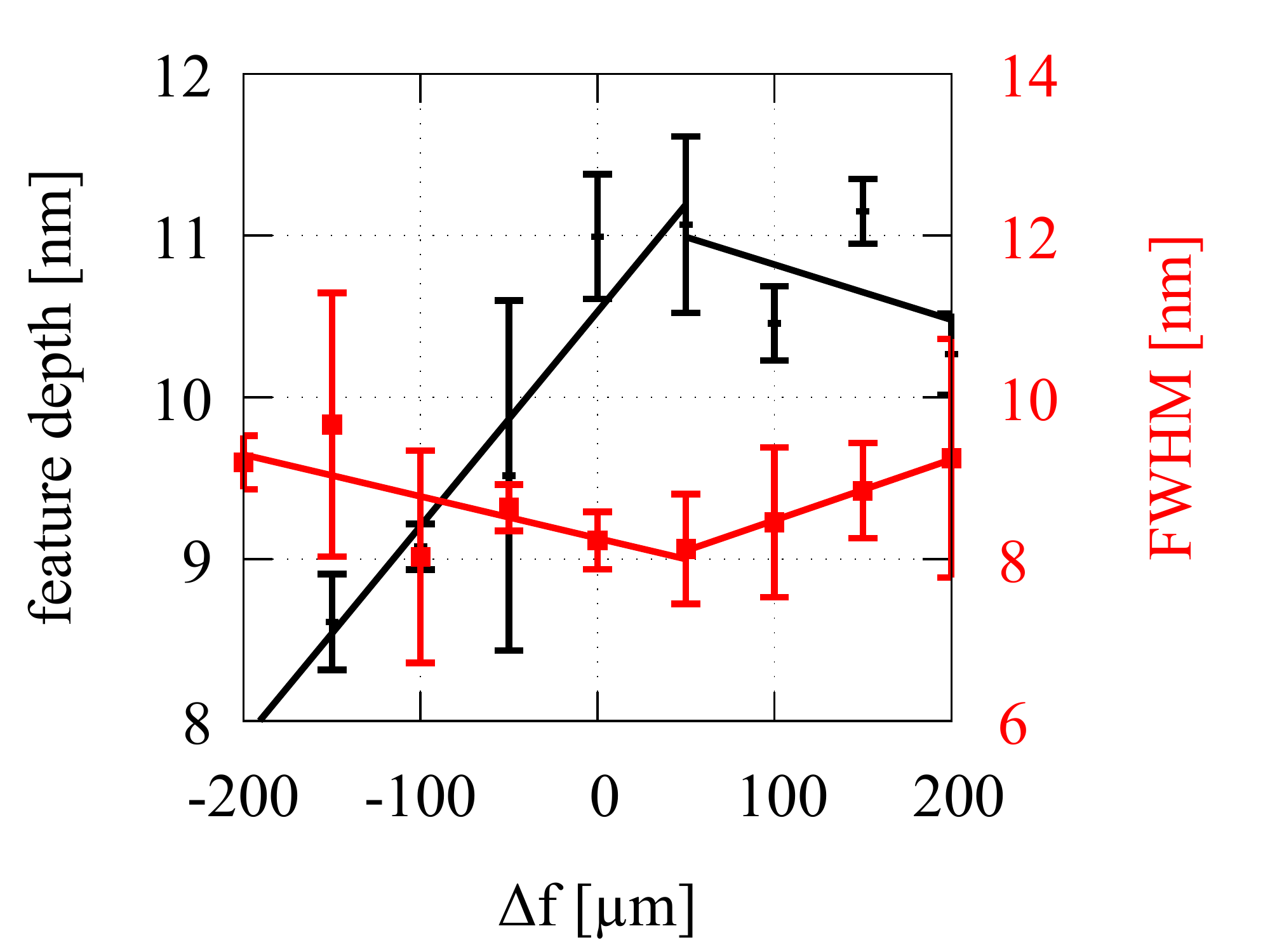}
				\put(1,1){\makebox(0,0){a)}}
			\end{overpic}
			\begin{overpic}[scale=.32]{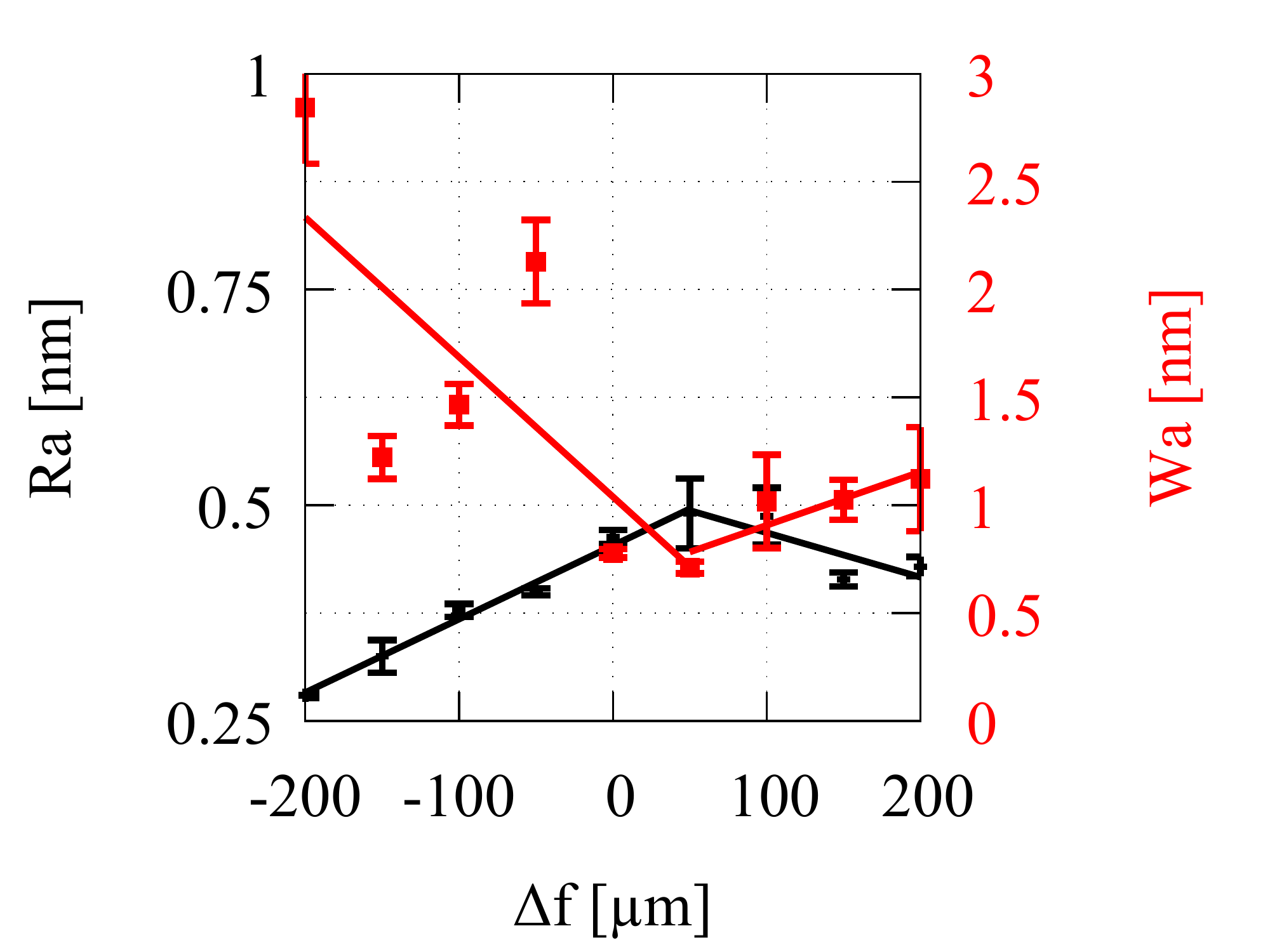}
				\put(1,1){\makebox(0,0){b)}}
			\end{overpic}
		\end{tabular}
	\end{center}
	\caption[Investigation on the influence of the axial focus position in DE-LCI]{Investigation on the influence of the axial focus position in DE-LCI with a) Plot of the feature height and the features FWHM in relation to a change in focus in steps of $\Delta f$\,=\,50\,\textmugreek m as well as b) Plot of the roughness and waviness measured at the same focus positions where dots with errorbars are measured values and solid lines are linear fits.}\label{Profilo:Pic:Result_roughness_influence_focus}
\end{figure}

Based on the initial validation, further measurements of surfaces with supposedly very low roughness like polished glass and silicon were performed with $\Lambda_c$\,=\,25 and 80\,\textmugreek m respectively, Fig. \ref{Profilo:Pic:Result_roughness_glass_si}.
For the glass sample a mean Ra\,=\,(0.35\,$\pm$\,0.03)\,nm and a RMS roughness of Rq\,=\,(0.44\,$\pm$\,0.05)\,nm were calculated from a set of 10 measurements. The measurements showed a mix of very high frequent features and some low-frequent, waviness features. The measured roughness is significantly above of the estimated RMS noise of the system which was RMS\,=\,0.088\,nm and close to the value of 3\,\si{\angstrom} quoted by the manufacturer from an AFM measurement. The analysis of an equal amount of measurements from the silicon sample lead to values of Ra\,=\,(0.40\,$\pm$\,0.01)\,nm and Rq=\,(0.51\,$\pm$\,0.01)\,nm. This is one order of magnitude larger than the noise. From measurements on different parts of the sample it became obvious that the roughness is equally distributed. Some irregularities such as polishing marks and scratches in the size of up to 10 nm can be found randomly distributed, Fig. \ref{Profilo:Pic:Result_example_influence_focus}.
The presence of prominent features like this raised the question about the influence focus on the roughness evaluation and the determination of small-sized features. For the purpose of closer examination, a series of profiles at a position with such a feature was captured. The focus was changed in steps of \glssymbol{DeltaFocus}\,=\,50\,\textmugreek m over a range of $\pm$\,200\,\textmugreek m. Over this data set the analysis of the depth of an irregularity and the full-width half maximum (FWHM) was used to gather information on the focus influence, Fig. \ref{Profilo:Pic:Result_roughness_influence_focus} a).
The results make clear that the feature depth follows a small but linear slope in relation to focus positions. It became clear by performing linear fits to the data for positive and negative focus changes that the point of optimal focus was at about  $f$\,=\,+50\,\textmugreek m. Furthermore, it became visible that the slopes were not perfectly symmetric for both focus ranges. Although significant error bars show that some of this change can be attributed noise, an overall change of 0.013\,nm/\textmugreek m and 0.003\,nm/\textmugreek m in relation to the focus shift was measurable using a linear fit. In contrast, the width of this feature showed a linear slope with a opposing behavior. The minimal feature width was found at the point of best focus. Both of these findings correspond well with the basic imaging theory known from microscopy.\\
Additional to this basic features, the influence of the focus on functional parameters such as roughness and waviness was examined, Fig. \ref{Profilo:Pic:Result_roughness_influence_focus} b). It can be seen that analogous to the increase in feature depth the Ra roughness increases linearly with the focus. This is a logical consequence as larger single features will contribute to an overall increase in roughness. However, due to the calculation of Ra as the mean value of all points of the profile, the change depends on the surface structure of the specific sample. Depending on the distribution of peaks and valleys and increase in feature size can keep the Ra roughness constant. In the case of this sample, the slope of the Ra value follows the behavior of the feature depth, Fig. \ref{Profilo:Pic:Result_roughness_influence_focus} a) and a slight increase in Ra due to a change of focus was visible. The change in waviness hast to be discussed in correspondence with the FWHM of features examined before. When imaging of the location of best focus, higher frequent surface features will be smoothed which was visible in the increase in FWHM. This leads, dependent on the filter settings applied, to a decrease in roughness and an increase in waviness. For the analyzed sample, this can be observed as the measurements $\pm \Delta f$ away from the zero-focus position show a minor increase in waviness, similar to the behavior of the FWHM.

\chapter{Appendix polymer characterization}
\section{Derivation of the group refractive index for the sample-only WPDE approach }\label{APNDX_deriv_sample-only}
This sections aims to deliver additional information on the derivation of Eq. (\ref{n_g_crosslinking_sample-only}) in the subsection \ref{TextWPDE} based on Eq. (\ref{derivative_eq_sample-only}), 

\begin{equation}
\frac{\partial \varphi}{\partial \lambda} = \frac{\partial}{\partial \lambda} \left(2\pi \frac{(n-1)t_{smp}-\delta}{\lambda} + \Delta \varphi \right).
\end{equation}	
For the purpose of readability, the spatially and spectrally dependent refractive indices of the sample $n^{smp}(x, \lambda)$ and  $n_g^{smp}(x, \lambda)$ and  will be noted as $n^{smp}$ and $n_g^{smp}$ respectively in the following equations.
In the case of the first derivative, the quotient rule applies with
\begin{eqnarray}
\frac{\partial \varphi}{\partial \lambda} && = \frac{2 \pi \cdot \frac{\partial n}{\partial \lambda} \cdot t_{smp} \cdot \lambda -2\pi \left[ \left( n^{smp} - 1\right) \cdot t_{smp} - \delta \right] \cdot 1}{\lambda^2}\\
&& = \frac{2\pi}{\lambda^2} \left( \frac{\partial n}{\partial \lambda} \cdot t_{smp} \cdot \lambda - \left[ \left( n^{smp} - 1\right) \cdot t_{smp} - \delta \right]   \right)\\
\textrm{with } \frac{\partial n}{\partial \lambda} && = \frac{n^{smp} - n_g^{smp}}{\lambda} \textrm{ according to \cite{Delbarre},}\\
\frac{\partial \varphi}{\partial \lambda} && = \frac{2\pi}{\lambda^2} \left[ \left( 1 - n_g^{smp}\right) \cdot t_{smp} + \delta \right].
\end{eqnarray}
This equation can be rewritten and partially substituted with $\kappa$ for readability in order to solve for the group refractive index of the sample $n_g^{smp}$ which equals Eq.\,(\ref{n_g_crosslinking_sample-only}) and the relative derived optical thickness (RDOT) $t^\prime_{OPT}$ which equals Eq.\,(\ref{EQ_RDOT_sample-only}) in the main text, whereas it has to be noted that  $n^{smp} =  n^{smp}(x, \lambda)$ and  $n_g^{smp} =  n_g^{smp}(x, \lambda)$
\begin{eqnarray}
&&n_g^{smp} = 1 - \frac{\kappa}{t_{smp}} \\
\textrm{with } && \kappa = \frac{\varphi_{loc}^\prime \cdot \lambda^2}{2 \pi} - \delta\\
&& t^\prime_{OPT} = n_g^{smp} \cdot t_{smp} = t_{smp} -\frac{\varphi_{loc}^\prime \cdot \lambda^2}{2 \pi} - \delta .
 \end{eqnarray}
\section{Derivation of the group refractive index for the WPDE approach with additional dispersion}\label{APNDX_deriv_add-Dispersion}
This section aims to deliver additional information on the derivation of Eq.\,(\ref{n_g_crosslinking}) in  section \ref{TextWPDE} based on Eq.\,(\ref{derivative_eq}), 
\begin{equation}
\varphi_{loc}^\prime = \frac{\partial}{\partial \lambda} \left( 2\pi \frac{ \left[ \left( n^{smp}(x, \lambda)  - 1 \right)t_{smp} \right] + \left[ \left( n^{DE}(\lambda)  - 1 \right)t_{DE} \right] -\delta }{\lambda} + \Delta \varphi \right).
\end{equation}
For the purpose of readability, the spatially and spectrally dependent refractive index of the sample $n^{smp}(x, \lambda)$ will be noted as $n^{smp}$ and the spectrally dependent refractive index of the dispersive element $n^{DE}(\lambda)$ will be noted as $n^{DE}$ in the following equations.
In the case of the first derivative, the quotient rule applies with
\begin{eqnarray}
&& f^\prime(x) = \frac{g^\prime(x) \cdot k(x) - g(x) \cdot k^\prime(x)}{k^2(x)} \\
&& g(x) = 2\pi \cdot \left[ \left( n^{smp}  - 1 \right)t_{smp} + \left( n^{DE}  - 1 \right)t_{DE} -\delta \right] \\
&& g^\prime(x) = 2\pi \cdot \left[ \frac{dn^{smp}}{d\lambda} \cdot t_{smp} + \frac{dn^{DE}}{d\lambda} \cdot t_{DE} \right] \\
&& k(x) = \lambda\\
&& k^\prime(x) = 1\\
&& k^2(x) = \lambda^2.
\end{eqnarray}
This leads to 
\begin{equation}
\begin{split}
\varphi_{loc}^\prime & = \frac{ 2\pi \cdot \left[ \frac{dn^{smp}}{d\lambda} \cdot t_{smp} + \frac{dn^{DE}}{d\lambda} \cdot t_{DE} \right] \cdot \lambda }{\lambda^2} \\ 
& - \frac{2\pi \cdot \left[ \left( n^{smp}  - 1 \right)t_{smp} + \left( n^{DE}  - 1 \right)t_{DE} -\delta \right]}{\lambda^2}
\end{split}
\end{equation}
which can be simplified and separated into two terms, $A$ and $B$, for the sake of individual assessment
\begin{eqnarray}
&&
\begin{split}
\varphi_{loc}^\prime & = \frac{ 2\pi \cdot \left[ \frac{dn^{smp}}{d\lambda} \cdot t_{smp} + \frac{dn^{DE}}{d\lambda} \cdot t_{DE} \right]}{\lambda} \\
& - \frac{2\pi \cdot \left[ \left( n^{smp}  - 1 \right)t_{smp} + \left( n^{DE}  - 1 \right)t_{DE} -\delta \right]}{\lambda^2}\\
& = A - B \label{AminusB}
\end{split}
\\
&& A \equiv \frac{ 2\pi \cdot \left[ \frac{dn^{smp}}{d\lambda} \cdot t_{smp} + \frac{dn^{DE}}{d\lambda} \cdot t_{DE} \right]}{\lambda}\\
&& B \equiv \frac{2\pi \cdot \left[ \left( n^{smp}  - 1 \right)t_{smp} + \left( n^{DE}  - 1 \right)t_{DE} -\delta \right]}{\lambda^2}.
\end{eqnarray}
Under the assumption that the derivative of the refractive index $n$, in relation to the wavelength $\lambda$, can be described as a difference of the index and the group index $n_g$ following \cite{Delbarre}
\begin{equation}
\frac{dn}{d\lambda} = \frac{n - n_g}{\lambda}, 
\end{equation}
where the term $A$ can be rewritten as
\begin{eqnarray}
\begin{split}
A & = \frac{2\pi}{\lambda} \left[ \frac{n^{smp} - n_g^{smp}}{\lambda} \cdot t_{smp} + \frac{n^{DE}-n_g^{DE}}{\lambda} \cdot t_{DE} \right]\\
& = \frac{2\pi}{\lambda^2} \left[ (n^{smp} - n_g^{smp}) \cdot t_{smp} + ( n^{DE}-n_g^{DE} ) \cdot t_{DE} \right],\\
\end{split}
\end{eqnarray}
and the term $B$ can be rewritten as
\begin{eqnarray}
B = \frac{2\pi}{\lambda^2} \left[ (n^{smp} - 1) \cdot t_{smp} + ( n^{DE} - 1 ) \cdot t_{DE} - \delta \right].\\
\end{eqnarray}
In consequence Eq. (\ref{AminusB}) can be expressed with
\begin{eqnarray}
\begin{split}
\varphi_{loc}^\prime & = A - B \\
& = \frac{2\pi}{\lambda^2} \left[ (n^{smp} - n_g^{smp}) \cdot t_{smp} - (n^{smp} - 1) \cdot t_{smp} \right.\\
& \left. + (n^{DE}-n_g^{DE} ) \cdot t_{DE} - (n^{DE} - 1 ) \cdot t_{DE} +  \delta \right] \\
& = \frac{2\pi}{\lambda^2} \left[ (n^{smp} - n_g^{smp} - n^{smp} + 1) \cdot t_{smp} \right.\\
& \left. + (n^{DE}-n_g^{DE} -n^{DE} +1 ) \cdot t_{DE} +  \delta \right]\\
& = \frac{2\pi}{\lambda^2}  \left[ (1 - n_g^{smp}) \cdot t_{smp} + (1 - n_g^{DE} ) \cdot t_{DE} +  \delta \right]\\
& = \frac{2\pi}{\lambda^2}  (1 - n_g^{smp}) \cdot t_{smp} + \frac{2\pi}{\lambda^2}  \left[ (1 - n_g^{DE} ) \cdot t_{DE} +  \delta \right]\\
\varphi_{loc}^\prime - \frac{2\pi}{\lambda^2}  \left[ (1 - n_g^{DE} ) \cdot t_{DE} +  \delta \right] &= \frac{2\pi}{\lambda^2}  (1 - n_g^{smp}) \cdot t_{smp}.
\end{split}
\end{eqnarray}
For readability, the left part of the equation is substituted with $\xi$ and solved for the group refractive index of the sample, $n_g^{smp}$, which equals Eq. (\ref{n_g_crosslinking}) and (\ref{EQ_RDOT_add-Dispersion}) in the main text of the paper, whereas it has to be noted that  $n^{smp} =  n^{smp}(x, \lambda)$ and $n^{DE} =  n^{DE}(\lambda)$: 
\begin{eqnarray}
&& \xi = \varphi_{loc}^\prime - \frac{2\pi}{\lambda^2}  \left[ (1 - n_g^{DE} ) t_{DE} +  \delta \right]\\
&& \text{and } n_g^{smp} = 1 - \frac{\lambda^2 \cdot \xi}{2\pi \cdot t_{smp}}.
\end{eqnarray}

\section{Derivation of the group refractive index in error propagation}
This section provides the detailed derivation of $\frac{\delta n}{\delta \lambda}$ and $\frac{\delta G}{\delta \lambda}$ both used in the error propagation of the group refractive index relative to the wavelength $\frac{\partial n_g^{+DE}}{\partial \lambda}$ in section \ref{SubSubSec:WPDEaddDE} especially in Eq. (\ref{EQPolyWPDE-error_lambda}). This error propagation has two important parts which are labeled $X$ and $Y$ in equations Eq. (\ref{EqErrSolutiondX}) and Eq. (\ref{EqErrSolutiondY}) respectively. As the derivation of $\frac{\delta X}{\delta \lambda}$ is straight forward, $\frac{\delta Y}{\delta \lambda}$ is slightly more complex, 
\begin{equation}
\frac{\partial Y}{\partial \lambda}  =\frac{\partial }{\partial \lambda} \left( \frac{ [1-n_g^{DE}(\lambda)]t_{DE}}{t_{smp}}\right) = - \frac{t_{DE}}{t_{smp}} \frac{\partial n_g^{DE}}{\partial \lambda}.\label{EQ:APNDX:ErrSolutiondY} 
\end{equation}
It is obvious that the derivative of the DEs group refractive index has to be performed. From \cite{Delbarre} follows, 
\begin{equation}
n_g = n - \frac{\delta n}{\delta \lambda} \cdot \lambda.
\end{equation}
In a general case, it is assumed, that the DE is made of glass, so that the Sellmeier equation according to  \cite{SchottKatalog} can be used to model the refractive index as follows
\begin{equation}
n^{DE} = \sqrt{  \sum\limits_{i=1}^3 \frac{A_i \lambda^2}{\lambda^2 - B_i}  + 1}
\end{equation}
In this case, $\frac{\delta n}{\delta \lambda}$ can be calculated using the chain rule
\begin{eqnarray}
\frac{\partial n}{\delta \lambda} = \frac{1}{2}f^{-\frac{1}{2}} \cdot \frac{\partial f}{\partial \lambda} \label{EQ:APNDX:nGDeriv:dN_dLambda_in-out}\\
\textrm{with } f = \sum\limits_{i=1}^3 \frac{A_i \lambda^2}{\lambda^2 - B_i}  + 1.
\end{eqnarray}
For the derivative of the inner function the quotient rule can be used, so that
\begin{eqnarray}
&& \frac{\partial f}{\partial \lambda} = f^\prime(x) = \frac{g^\prime(x) \cdot k(x) - g(x) \cdot k^\prime(x)}{k^2(x)} \\
&& g(x) = \sum\limits_{i=1}^3 A_i \lambda^2 \\
&& g^\prime(x) = \sum\limits_{i=1}^3 2A_i \lambda\\
&& k(x) =  \sum\limits_{i=1}^3 \lambda^2 - B_i\\
&& k^\prime(x) =  \sum\limits_{i=1}^3 2\lambda\\
&& k^2(x) = \sum\limits_{i=1}^3 (\lambda^2 - B_i)^2.
\end{eqnarray}
Which leads to 
\begin{eqnarray}
\frac{\partial f}{\partial \lambda} &= \sum\limits_{i=1}^3 \left( \frac{ 2A_i \lambda \cdot (\lambda^2 - B_i) - A_i \lambda^2 \cdot  2\lambda}{(\lambda^2 - B_i)^2} \right)\\
&= \sum\limits_{i=1}^3 \left( \frac{ 2A_i \lambda (\lambda^2 - B_i - \lambda^2) }{(\lambda^2 - B_i)^2} \right)\\
&= \sum\limits_{i=1}^3 \left( \frac{ -2A_i \lambda  B_i}{(\lambda^2 - B_i)^2} \right) = G
\end{eqnarray}
Inserting this result into Eq. (\ref{EQ:APNDX:nGDeriv:dN_dLambda_in-out}), it can be written
\begin{eqnarray}
\frac{\partial n}{\partial \lambda} = \frac{G}{2 n}  \label{EQ:APNDX:nGDeriv:dN_dLambda}\\
\textrm{and } n_g = n - \frac{G }{2 n} \cdot \lambda.
\end{eqnarray}
With this result, the derivative of $n_g$ with respect to the wavelength $\lambda$, $\frac{\partial n_g^{+DE}}{\partial \lambda}$ which is needed to solve the initial equation for the error propagation Eq. (\ref{EQ:APNDX:ErrSolutiondY}), can be calculated accordingly with
\begin{eqnarray}
\frac{\partial n_g}{\partial \lambda} = \frac{\partial n}{\partial \lambda} - \frac{\partial }{\partial \lambda} \left(\frac{G }{2 n} \cdot \lambda \right)\\
\textrm{with } \frac{\partial }{\partial \lambda} \left(\frac{G }{2 n} \cdot \lambda \right) = Q.
\end{eqnarray}
In order to solve for $Q$, again the quotient rule was applied with
\begin{eqnarray}
&& g(x) =  G \cdot \lambda\\
&& g^\prime(x) = G^\prime \cdot \lambda + G \\
&& k(x) = 2n \\
&& k^\prime(x) = 2 \frac{\partial n}{\partial \lambda}  \\
&& k^2(x) = 4n^2.
\end{eqnarray}
Which leads to 
\begin{eqnarray}
Q &= \frac{(G^\prime \cdot \lambda + G) \cdot 2n - G \lambda \cdot 2 \frac{\partial n}{\partial \lambda }}{ 4n^2.}  \\
&= \frac{(G^\prime \cdot \lambda + G) \cdot n - G \lambda \cdot \frac{\partial n}{\partial \lambda }}{ 2n^2.}\\
\textrm{where } \frac{\partial n}{\partial \lambda } &= \frac{G}{2 n}\\
Q &= \frac{(G^\prime \cdot \lambda + G) \cdot n - \frac{G^2}{2n} \cdot \lambda}{ 2n^2}. \label{EQ:APNDX:nGDeriv:Q}\\
\end{eqnarray}
Therefore, under the usage of Eq. (\ref{EQ:APNDX:nGDeriv:dN_dLambda}) and Eq. (\ref{EQ:APNDX:nGDeriv:Q}), the derivative of $n_g$ with respect to the wavelength $\lambda$ can be written as
\begin{eqnarray}
\frac{\partial n_g}{\partial \lambda} = \frac{G}{2 n} - \frac{(G^\prime \cdot \lambda + G) \cdot n - \frac{G^2}{2n} \cdot \lambda}{ 2n^2}.\label{EQ:APNDX:nGDeriv:dN_dLambda_solution}
\end{eqnarray}
This result shows the necessity of a final derivation in order to fully solve Eq. (\ref{EQ:APNDX:ErrSolutiondY}), where $G^\prime$ of Eq. (\ref{EQ:APNDX:nGDeriv:dN_dLambda_solution}) has to be calculated with
\begin{eqnarray}
G^\prime = \frac{\partial G}{\partial \lambda} = \frac{\partial}{\partial \lambda} \left( \sum\limits_{i=1}^3 \left( \frac{ -2A_i B_i \lambda}{(\lambda^2 - B_i)^2} \right) \right)
\end{eqnarray}
Once again, the quotient rule applies with
\begin{eqnarray}
&& g(x) = \sum\limits_{i=1}^3 -2A_i B_i\lambda \\
&& g^\prime(x) = \sum\limits_{i=1}^3 -2A_i B_i\\
&& k(x) = \sum\limits_{i=1}^3  (\lambda^2 - B_i)^2 \\
&& k^\prime(x) = \sum\limits_{i=1}^3 4\lambda(\lambda^2 - B_i)\\
&& k^2(x) = \sum\limits_{i=1}^3 (\lambda^2 - B_i)^4.
\end{eqnarray}
Which can be combined to
\begin{eqnarray}
\frac{\partial G}{\partial \lambda} &= \sum\limits_{i=1}^3 \left( \frac{-2A_i B_i \cdot (\lambda^2 - B_i)^2 - -2A_i B_i\lambda \cdot 4\lambda(\lambda^2 - B_i)}{(\lambda^2 - B_i)^4} \right)\\
&= \sum\limits_{i=1}^3 \left( \frac{-2A_i B_i \cdot (\lambda^2 - B_i) \left[(\lambda^2 - B_i)- 4 \lambda^2  \right] }{(\lambda^2 - B_i)^4} \right)\\
&= \sum\limits_{i=1}^3 \left( \frac{2A_i B_i \left[3\lambda^2 + B_i \right] }{(\lambda^2 - B_i)^3} \right)
\end{eqnarray}

\chapter{Appendix thin-film characterization}
\section{Transfer matrix formulation for a single layer model}\label{APNDX:ThinFilm_TransferMat}
\begin{figure}[h]
	\begin{center}
	\begin{tabular}{c}
		\begin{overpic}[scale=0.75]{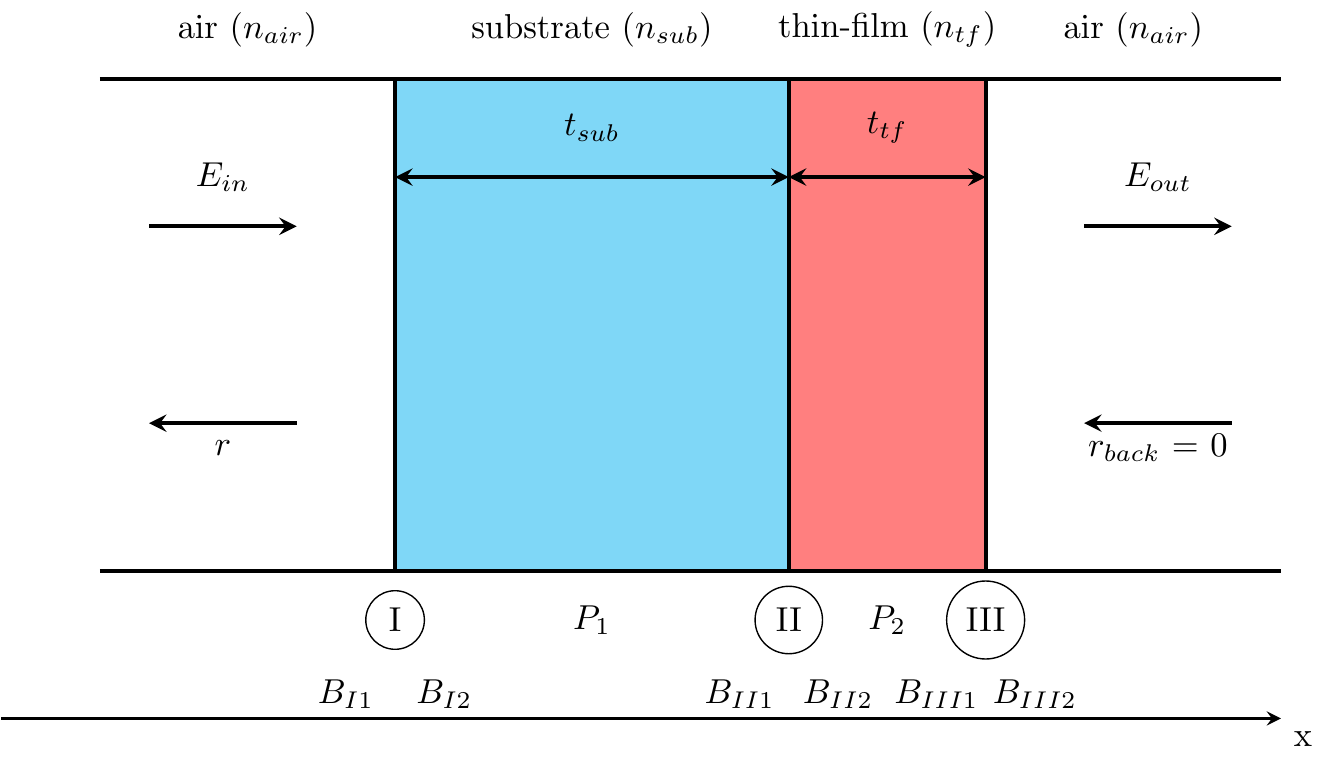}
			\put(0,0){\makebox(0,0){a)}}
		\end{overpic}
		\end{tabular}
	\caption[Simplified depiction of a single thin-film on a substrate which was used to formulate a transfer matrix]{Simplified depiction of a single thin-film on a substrate which was used to formulate a transfer matrix where an incoming electric field $E_{in}$ is propagated along the axis $x$ using the boundary conditions $B_{ij}$ and the propagators for every material $P_j$.}\label{APNDX_fit_model_sketch}
\end{center}
\end{figure}

This section aims to deliver additional information on the transfer matrix used to describe a simple one-layer model for a thin film on a substrate. The formulated transfer matrix was the basis for the film thickness analysis in chapter \ref{Chapter:ThinFilm}.\\
All experiments performed to measure the film thickness of thin films within this work were carried out on a sample which consisted of a simple single-layer system, Fig. \ref{APNDX_fit_model_sketch}.
Where the incoming electric field $E_{in}$ is propagated along the axis $x$ using the boundary conditions \glssymbol{BoundaryTFM} on the boundaries between the materials air, substrate and thin-film. Additionally, the propagators for every material \glssymbol{PropagatorTFM} with individual refractive indices ($n_{air}$, $n_{sub}$, $n_{tf}$) were used to describe the propagation. As the idea of the transfer matrix formalism was expressed in a rather generalized form with Eq. (\ref{EQ:ThinFilm_TFmatrix_general}), it can be re-formulated according to the real conditions described with Fig.\,\ref{APNDX_fit_model_sketch}. According to the involved materials, the wavenumber components can be described for every material
\begin{eqnarray}
k_0 = \frac{2\pi}{\lambda}\\
k_1 = k_0 \cdot n_{air}\\
k_2 = k_0 \cdot n_{sub}\\
k_3 = k_0 \cdot n_{tf}\\
k_4 = k_0 \cdot n_{air} = k_1.
\end{eqnarray}
Here, $k_0$ represents the basic wavenumber component and it is multiplied by the appropriate refractive index per material where the indices \textsubscript{air}, \textsubscript{tf} and \textsubscript{sub} represent the materials air, thin-film and substrate respectively. Analogously, these materials have the thickness t\textsubscript{air}, t\textsubscript{tf} and t\textsubscript{sub}. The propagation of an incoming electric field E\textsubscript{in} is performed as an alternation between a modification of the field at the material boundary and its propagation through a material. For the modification caused by the boundary between air and the thin-film material, the components
\begin{eqnarray}
B_{I1} = \left(\begin{array}{cc} 1 & 1\\ -ik_1 & +ik_1 \end{array}\right)\\
B_{I2} = \left(\begin{array}{cc} 1 & 1\\ -ik_2 & +ik_2 \end{array}\right)
\end{eqnarray}
can be noted. The propagator which describes the influence of the thin-film material on the incoming electric field can be described with
\begin{equation}
P_1 = \left(\begin{array}{cc} e^{-ik_2t_{sub}} & 0\\ 0 & e^{+ik_2t_{sub}} \end{array}\right).
\end{equation}
The second material boundary, between thin-film and substrate material can be described with
\begin{eqnarray}
B_{II1} = \left(\begin{array}{cc} 1 & 1\\ -ik_2 & +ik_2 \end{array}\right)\\
B_{II2} = \left(\begin{array}{cc} 1 & 1\\ -ik_3 & +ik_3 \end{array}\right).
\end{eqnarray}
Consequently, the propagator of the substrate material is noted as
\begin{equation}
P_2 = \left(\begin{array}{cc} e^{-ik_3t_{tf}} & 0\\ 0 & e^{+ik_3t_{tf}} \end{array}\right).
\end{equation}
Finally, the third boundary describes the electric field leaving the substrate material towards air with
\begin{eqnarray}
B_{III1} = \left(\begin{array}{cc} 1 & 1\\ -ik_3 & +ik_3 \end{array}\right)\\
B_{III2} = \left(\begin{array}{cc} 1 & 1\\ -ik_4 & +ik_4 \end{array}\right).
\end{eqnarray}
Using the noted components, a so called \gls{FWP} can be modeled describing the electric field traveling through the complete material setup following the idea expressed with Eq. (\ref{EQ:ThinFilm_FWP_general}) 
\begin{equation}
FWP = B_{III2}^{-1} \cdot B_{III1} \cdot P_2 \cdot B_{II2}^{-1} \cdot B_{II1} \cdot P_1 \cdot B_{I2}^{-1} \cdot B_{I1}.
\end{equation}
As a result, the \gls{FWP} is a square matrix of the type $\left(\begin{array}{cc} a & b\\ c & d \end{array}\right)$ which can be used in order to calculate the resulting electric field containing the transmission and the reflection components, \glssymbol{ReflectionTFM} and \glssymbol{BackReflectionTFM} in the thin-film system alike
\begin{equation}
\left(\begin{array}{c} E_{out} \\ r_{back} \end{array}\right) = \left(\begin{array}{cc} a & b\\ c & d \end{array}\right) \left(\begin{array}{c} E_{in} \\ r \end{array}\right),
\end{equation}
where $E_{out}$ describes the transmission through the system which is relevant in the Mach-Zehnder setup used within this work. The incoming electric field $E_{in}$ can be normalized to $1$ while the back-reflection at the end of the system, which could be used to model multiple-internal reflections within the material system, is assumed to be $0$. This equation was solved for $E_{out}$ as
\begin{equation}
E_{out} = a - b \cdot \frac{c}{d}
\end{equation}
and was used as a model to fit experimentally captured spectra with respect to the thickness of the thin-film under investigation $t_{tf}$.

\printbibliography

\addchap[Publications]{Publications}
\markboth{Publications}{Publications}
	\subsection*{Peer-reviewed journals}
	
	\begin{itemize}
		\item \textbf{Taudt, Ch.}; Nelsen, B.; Baselt, T.; Koch, E.; Hartmann, P., "High-dynamic-range areal profilometry using an imaging, dispersion-encoded low-coherence interferometer", \textit{Opt. Express} 28, 17320-17333 (2020); \href{https://doi.org/10.1364/OE.389839}{doi: 10.1364/OE.389839}. 
		\item \textbf{Taudt, Ch.}; Nelsen, B.; Rossegger, E.; Schlögl, S.; Koch, E.; Hartmann, P., "Spatially Resolved Cross-Linking Characterization by Imaging Low-Coherence Interferometry", \textit{Sensors 2019}, 19, 1152; \href{https://dx.doi.org/10.3390/s19051152}{doi: 10.3390/s19051152}.
		\item Baselt, T.; \textbf{Taudt, Ch.}; Nelsen, B.; Lasagni, A.F.;  Hartmann, P., "Development of a method to overcome the power threshold during supercontinuum generation based on an Yb-doped photonic crystal fiber", \textit{ Opt. Eng.} 57(2), 021207 (2018);\linebreak \href{https://dx.doi.org/10.1117/1.OE.57.2.021207}{doi: 10.1117/1.OE.57.2.021207}.
	\end{itemize}

	\subsection*{Conference proceedings}
	\begin{itemize}
		\item \textbf{Taudt, Ch.}; Baselt, T.; Nelsen, B.; Koch, E.; Hartmann, P., "One-shot roughness measurements based on dispersion-encoded low coherence interferometry", \textit{Proc. SPIE 11352}, Optics and Photonics for Advanced Dimensional Metrology, 1135212 (1 April 2020); \href{https://doi.org/10.1117/12.2555807}{doi: 10.1117/12.2555807}.
		\item  Baselt, T.; \textbf{Taudt, Ch.}; Nelsen, B.; Lasagni, A.F.;  Hartmann, P., "Supermode supercontinuum generation in the cladding of a photonic crystal fiber", \textit{Proc. SPIE 11264}, Nonlinear Frequency Generation and Conversion: Materials and Devices XIX, 112641Z (2 March 2020); \href{https://doi.org/10.1117/12.2545633}{doi: 10.1117/12.2545633}.
		\item \textbf{Taudt, Ch.}; Baselt, T.; Nelsen, B.; Koch, E.; Hartmann, P., "Nano-3D-\linebreak Profilometrie auf Basis dispersionskodierter Kurzkohärenzinterferometrie", \textit{Wissenschaft-\linebreak liche Berichte / Scientific Reports HS Mittweida}, 01/2019, pp. 66-69, ISSN 1437-7624.
		\item \textbf{Taudt, Ch.}; Preuss, M., Nelsen, B.; Baselt, T.; Koch, E.; Hartmann, P., "Thin-film characterization with a dual-channel dispersion-encoded imaging low-coherence interferometry approach", \textit{Proc. SPIE 10925}, Photonic Instrumentation Engineering VI, 109250L (7 March 2019); \href{https://dx.doi.org/10.1117/12.2509676}{doi: 10.1117/12.2509676}.
		\item Nelsen, B.; Kabardiadi-Virkovski, A.; Baselt, T.; \textbf{Taudt, Ch.}; Hartmann, P., "Kilohertz dynamic Fourier filter for synthetic-aperture binary hologram reconstruction",\textit{Proc. SPIE 10944}, Practical Holography XXXIII: Displays, Materials, and Applications, 109440U (1 March 2019); \href{https://dx.doi.org/10.1117/12.2509955}{doi: 10.1117/12.2509955}.
		\item  Jacobs, P.-P.; Nelsen, B.; \textbf{Taudt, Ch.}; Rudek, F.; Hartmann, P.,	"In-process monitoring of laser ablation on thin steel membranes by multispectral shape from-shading", \textit{Proc. SPIE 10925}, Photonic Instrumentation Engineering VI, 109250E (4 March	2019); \href{https://dx.doi.org/10.1117/12.2511075}{doi: 10.1117/12.2511075}.
		\item Nelsen, B.; Jacobs, P.; \textbf{Taudt, Ch.}; Rudek, F.; Hartmann, P., "Holographic characterization and laser structuring of a microphone membrane", \textit{Proc. SPIE 10678}, Optical Micro- and Nanometrology VII, 106780F (24 May 2018); \href{https://dx.doi.org/10.1117/12.2307487}{doi: 10.1117/12.2307487}.
		\item Preuß, M.; \textbf{Taudt, Ch.}; Nelsen, B.; Hartmann, P.,"Dispersion-controlled low-coherent interferometry for thin-film characterization", \textit{Proc. SPIE 10678}, Optical Micro- and Nanometrology VII, 106780R (24 May 2018); \href{https://dx.doi.org/10.1117/12.2307166}{doi: 10.1117/12.2307166}.
		\item \textbf{Taudt, Ch.}; Nelsen, B.; Schlögl, S.; Koch, E.; Hartmann, P., "Fast Cross-Linking-Characterization of Waveguide-Polymers on Wafers by Imaging Low-Coherence Interferometry", \textit{Proceedings 2018}, 2, 1046., \href{https://dx.doi.org/10.3390/proceedings2131046}{doi: 10.3390/proceedings2131046}.
		\item \textbf{Taudt, Ch.}; Baselt, T.; Nelsen, B.; Aßmann, H.; Greiner, A.; Koch, E.; Hartmann, P., "Characterization of edge effects in precision low-coherence interferometry using broadband light sources", \textit{Proc. SPIE 10329}, Optical Measurement Systems for Industrial Inspection X, 1032932 (26 June	2017); \href{https://dx.doi.org/10.1117/12.2270318}{doi: 10.1117/12.2270318}.
		\item \textbf{Taudt, Ch.}; Baselt, T.; Nelsen, B.; Aßmann, H.; Greiner, A.; Koch, E.; Hartmann, P., "Evaluation of the thermal stability of a low-coherence interferometer for precision surface profilometry", \textit{Proc. SPIE 10110}, Photonic Instrumentation Engineering IV, 1011015 (20 February 2017); \href{https://dx.doi.org/10.1117/12.2252375}{doi: 10.1117/12.2252375}.
		\item Taccheo, S.; Schuster, K.; Ferrari, M.; Seddon, A.; Marciniak, M.; \textbf{Taudt, Ch.}; Troles, J.; Valentini, G.; Dorosz, D.; Prudenzano, F.; Jaeger, M.; Dandrea, C.; Ivanda, M.; Chiasera, A.; Sujecki, S.; Nazabal, V.; Comelli, D.; Baghdasaryan, H.; Baselt, T.; Hartmann, P.; Lucianetti, A.;  Peterka, P.; Klotzbach, A.; Adam, J.-L.; Gebavi, H., "Challenges and Future Trends in Fiber Lasers",\textit{IEEE Proc. of the 2016 18th International Conference on Transparent Optical Networks (ICTON)}, \href{https://dx.doi.org/10.1109/ICTON.2016.7550715}{doi: 10.1109/ICTON.2016.7550715}.
		\item \textbf{Taudt, Ch.}; Baselt, T.; Nelsen, B.; Aßmann, H.; Greiner, A.; Koch, E.; Hartmann, P., "Two-dimensional low-coherence interferometry for the characterization of nanometer wafer topographies", \textit{Proc. SPIE 9890}, Optical Micro- and Nanometrology VI, 98900R (3 May 2016); \href{https://dx.doi.org/10.1117/12.2227887}{doi: 10.1117/12.2227887}.
		\item \textbf{Taudt, Ch.}; Baselt, T.; Nelsen, B.; Aßmann, H.; Greiner, A.; Koch, E.; Hartmann, P., "Measurement of surface topographies in the nm-range for power chip technologies by a modified low-coherence interferometer", \textit{Proc. SPIE 9754}, Photonic Instrumentation Engineering III, 97540H (16 March 2016);\linebreak \href{https://dx.doi.org/10.1117/12.2212913}{doi: 10.1117/12.2212913}.
		\item \textbf{Taudt, Ch.}; Baselt, T.; Oreski, G.; Hirschl, Ch.; Koch, E.; Hartmann, P., "Cross-linking characterization of polymers based on their optical dispersion utilizing a white-light interferometer", \textit{Proc. SPIE 9525}, Optical Measurement Systems for Industrial Inspection IX, 95250P (22 June 2015); \href{https://dx.doi.org/10.1117/12.2184814}{doi: 10.1117/12.2184814}.
		\item Baselt, T.; \textbf{Taudt, Ch.}; Lasagni, A.F.;  Hartmann, P., "Experimental measurement of group velocity dispersion during operation in cladding-pumped large-mode-area Yb-doped fibers", \textit{Proc. SPIE 9525}, Optical Measurement Systems for Industrial Inspection IX, 95253Z (22 June 2015); \href{https://dx.doi.org/10.1117/12.2184803}{doi: 10.1117/12.2184803}.		
		\item \textbf{Taudt, Ch.}; Augenstein, A.; Baselt, T.; Aßmann, H.; Greiner, A.; Koch, E.; Hartmann, P., "Characterization of a dispersion-controlled approach to surface profilometry on wafers using a white-light interferometer", \textit{Proc. SPIE 9517}, Smart Sensors, Actuators, and MEMS VII; and Cyber Physical Systems, 95170W (21 May 2015); \href{https://dx.doi.org/10.1117/12.2179022}{doi: 10.1117/12.2179022}.
		\item \textbf{Taudt, Ch.}; Baselt, T.; Koch, E.; Hartmann, P., "Investigation of a novel approach for the cross-linking characterization of SU-8 photoresist materials by means of optical dispersion measurements", \textit{Proc. SPIE 8975}, Reliability, Packaging, Testing, and Characterization of	MOEMS/MEMS, Nanodevices, and Nanomaterials XIII, 89750A (7 March 2014); \href{https://dx.doi.org/10.1117/12.2039052}{doi: 10.1117/12.2039052}.
		\item Meissner, S.; Cimalla, P.; Fischer, B.; \textbf{Taudt, Ch.}; Baselt, T.; Hartmann, P.; Koch, E., "A new small-package super-continuum light source for optical	coherence tomography", \textit{Proc. SPIE 8611}, Frontiers in Ultrafast	Optics: Biomedical, Scientific, and Industrial Applications XIII, 86110K (15 March 2013); \href{https://dx.doi.org/10.1117/12.2004080}{doi: 10.1117/12.2004080}.
	
	\end{itemize}

	\subsection*{Patents}
	\begin{itemize}
		\item Hartmann, P.; Taudt, Ch.; Baselt, T., Optische Anordnung zur Bestimmung von Lageänderungen, Positionen, Verformung, Bewegungen, Beschleunigungen und Geschwindigkeiten, \textit{DE102013019774A1}, 2019.
		\item Hartmann, P.; Taudt, Ch.; Baselt, T., Scanfreie 3D Oberflächentopograhieerfassung mit einem modifizierten Kurzkohärenzinterferometer, \textit{in submission process}, 2020.
	\end{itemize}


\addchap[Acknowledgments]{Acknowledgments}
\markboth{Acknowledgments}{Acknowledgments}

This work as well as the accompanying process of conceptualization, experimentation and revision was only possible with the support of many important people whom I want to thank.\\ 
First and foremost I want to express gratitude to my two advisors Prof.\,Dr.\,rer.\,nat.\,habil. Peter Hartmann and Prof.\,Dr.\,rer.\,nat.\,Edmund Koch for giving me the opportunity to work on a scientifically challenging and significant project as well as for their critical support. Especially, I want to thank Prof.\,Hartmann for his long-term support in all scientific questions as well as for the opportunity to participate in various scientific-technical projects within the Arbeitsgruppe Optische Technologien of the Westsächsische Hochschule Zwickau (WHZ) and in the Fraunhofer Anwendungszentrum für Optische Messtechnik und Ober-\linebreak flächentechnologien (AZOM). The groups have always provided me with an interesting and fruitful environment for all my work. I also want to thank Prof.\,Koch for his always helpful critique in scientific questions and writing as well as his introduction into all topics which relate to OCT. He gave me the chance to conduct this work as a collaboration between TU Dresden and WHZ.\\
Within the mentioned groups of WHZ and AZOM everyone was always welcoming and helpful; I am grateful to all of my coworkers. In particular, I want to take the opportunity to thank Mr. Tobias Baselt for all the scientific-technical advice he shared as well as for the many topics we were able to work on collaboratively. Furthermore, I want to thank Dr. Bryan Nelsen for the numerous advice he gave, the ideas he triggered, for enduring all my questions, the proof reading and the friendship which evolved. I also thank Mr. Alexander Kabardiadi-Virkovski for the endless discussions we had, which in turn led to new discussions that broadened my mind. Many of my coworkers became friends over the years and supported me in one or the other way; I want to thank especially Mr. Alexander Augenstein, Mr. Paul P. Jacobs, Mr. Benjamin Lempe, Mr. Ronny Maschke, Mr. Marco Preuß, Mr. Florian Rudek and Mr. Marcus Wittig. I also want to thank Ms. Antje Schuschies and Ms. Sarah Schenke-Simon for proof reading the original manuscript.\\
Some of the experimental work presented here could not have come into existence without some samples, discussions and technical support. I want to thank Dr.\,Sandra Schlögl and Dr.\,Gernot Oreski of Polymer Compentence Center Leoben for the rich collaboration on various topics regarding polymer cross-linking. I want to thank Mr. Heiko Aßmann, Mr. Thomas Albrecht and Mr. Andreas Greiner of Infineon Technologies Dresden GmbH who provided various support in form of on-site measurement opportunities and discussions to help me understand a lot of aspects regarding semiconductor manufacturing processes. I also want to thank Mr. Malte Hapich (Ametek GmbH, business unit Taylor Hobson) and Mr. Philipp Hanke (Layertec GmbH) who provided samples and expertise which helped me to work on the topic of surface roughness evaluation. Furthermore, I want to thank Prof.\,Dr.\,rer.\,nat.\,Stefan Braun for the preparation of the 3\,nm step sample which helped me to characterize the resolution of the surface profilometry setup.\\
This work could not have been written with just scientific and technical support. I want to thank my family and friends who supported me unconditionally throughout the years. I am especially grateful to Manja who is and was the most stable and helpful person on my side! Although I could not spark the same euphoria in you for the results that thrilled me, you helped me over the many failed experiments and moments of disappointment. I will always be thankful for your support! Of all the numerous friends who shared their time with me over the years, I want to thank Sarah, Konrad and Klaus in particular for their warmth and encouragement.
Some long lab hours and extensive work periods were only possible through a constant supply of sugar; thanks to Bäckerei Kunze, there was never a lack of the much needed cake and pastry.

\addchap[Declaration of Authorship]{Declaration of Authorship}
\markboth{Declaration of Authorship}{Declaration of Authorship}

Hiermit versichere ich, dass ich die vorliegende Arbeit ohne unzulässige Hilfe Dritter und ohne Benutzung anderer als der angegebenen Hilfsmittel angefertigt habe; die aus fremden Quellen direkt oder indirekt übernommenen Gedanken sind als solche kenntlich gemacht. Bei der Auswahl und Auswertung des Materials sowie bei der Herstellung des Manuskripts habe ich Unterstützungsleistungen von folgenden Personen erhalten: \textit{Ich habe im Rahmen dieser Arbeit Publikationen mit Co-Autoren entsprechend der Publikationsliste erarbeitet. Mir wurden außerdem Proben zur Verfügung gestellt von Herrn M. Hapich (Fa. Ametek GmbH/Taylor Hobson; Rauheitsnormale), Herr Philipp Hanke (Fa. Layertec GmbH; Spiegelsubstrat mit Beschichtung) sowie durch Herrn Prof.\,Dr.\,rer.\,nat.\,Stefan Braun (Westsächsische Hochschule Zwickau; 3\,nm Höhenstufe auf Si). Daneben habe ich keine Unterstützungsleistungen erhalten.}\\
\\
Weitere Personen waren an der geistigen Herstellung der vorliegenden Arbeit nicht beteiligt. Insbesondere habe ich nicht die Hilfe eines kommerziellen Promotionsberaters in Anspruch genommen. Dritte haben von mir weder unmittelbar noch mittelbar geldwerte Leistungen für Arbeiten erhalten, die im Zusammenhang mit dem Inhalt der vorgelegten Dissertation stehen.\\
\\
Die Arbeit wurde bisher weder im Inland noch im Ausland in gleicher oder ähnlicher Form einer anderen Prüfungsbehörde vorgelegt und ist auch noch nicht veröffentlicht worden.\\
\\
Mir ist bekannt, dass die Nichteinhaltung dieser Erklärung zum nachträglichen Entzug des Doktortitels führen kann.\\
\\
Ich bestätige, dass ich die Promotionsordnung der Fakultät Elektrotechnik und Informationstechnik vom 08. August 1994 in der geänderten Fassung 12.05.2012 und die Änderungssatzung vom 18.06.2018 der Technischen Universität Dresden anerkenne.\\
\\
\\
\\
Zwickau, den 28.05.2020\\
\\
\\
\\
Christopher Taudt

\end{document}